\documentclass[12pt]{article}
\usepackage{graphicx}
\usepackage{epsfig}
\usepackage{epstopdf}
\DeclareGraphicsExtensions{.pdf,.eps,.png,.jpg,.mps,.tif}
\usepackage{amssymb,graphicx}

\usepackage{graphicx}
\usepackage{wrapfig}
\usepackage{lscape}
\usepackage{rotating}

\usepackage[
singlelinecheck=false 
]{caption}

\usepackage{cite}

\def\lsim{\mathrel{\rlap{\lower3pt\hbox{\hskip0pt$\sim$}}
     \raise1pt\hbox{$<$}}}         
\def\gsim{\mathrel{\rlap{\lower4pt\hbox{\hskip1pt$\sim$}}
     \raise1pt\hbox{$>$}}}         

\usepackage{amsmath}
\usepackage{amsfonts}

\begin{document}
\begin{titlepage}

\centerline{\Large \bf *K-means and Cluster Models for Cancer Signatures}
\medskip

\centerline{Zura Kakushadze$^\S$$^\dag$\footnote{\, Zura Kakushadze, Ph.D., is the President of Quantigic$^\circledR$ Solutions LLC,
and a Full Professor at Free University of Tbilisi. Email: zura@quantigic.com} and Willie Yu$^\sharp$\footnote{\, Willie Yu, Ph.D., is a Research Fellow at Duke-NUS Medical School. Email: willie.yu@duke-nus.edu.sg}}
\bigskip

\centerline{\em $^\S$ Quantigic$^\circledR$ Solutions LLC}
\centerline{\em 1127 High Ridge Road \#135, Stamford, CT 06905\,\,\footnote{\, DISCLAIMER: This address is used by the corresponding author for no
purpose other than to indicate his professional affiliation as is customary in
publications. In particular, the contents of this paper
are not intended as an investment, legal, tax or any other such advice,
and in no way represent views of Quantigic$^\circledR$ Solutions LLC,
the website \underline{www.quantigic.com} or any of their other affiliates.
}}
\centerline{\em $^\dag$ Free University of Tbilisi, Business School \& School of Physics}
\centerline{\em 240, David Agmashenebeli Alley, Tbilisi, 0159, Georgia}
\centerline{\em $^\sharp$ Centre for Computational Biology, Duke-NUS Medical School}
\centerline{\em 8 College Road, Singapore 169857}
\medskip
\centerline{(January 30, 2017)}

\bigskip
\medskip

\begin{abstract}
{}We present *K-means clustering algorithm and source code by expanding statistical clustering methods applied in https://ssrn.com/abstract=2802753 to quantitative finance. *K-means is statistically deterministic without specifying initial centers, etc. We apply *K-means to extracting cancer signatures from genome data without using nonnegative matrix factorization (NMF). *K-means' computational cost is a fraction of NMF's. Using 1,389 published samples for 14 cancer types, we find that 3 cancers (liver cancer, lung cancer and renal cell carcinoma) stand out and do not have cluster-like structures. Two clusters have especially high within-cluster correlations with 11 other cancers indicating common underlying structures. Our approach opens a novel avenue for studying such structures. *K-means is universal and can be applied in other fields. We discuss some potential applications in quantitative finance.
\end{abstract}
\medskip
\end{titlepage}

\newpage
\section{Introduction and Summary}

{}Every time we can learn something new about cancer, the motivation goes without saying. Cancer is different. Unlike other diseases, it is not caused by ``mechanical" breakdowns, biochemical imbalances, etc. Instead, cancer occurs at the DNA level via somatic alterations in the genome structure. A common type of somatic mutations found in cancer is due to single nucleotide variations (SNVs) or alterations to single bases in the genome, which accumulate through the lifespan of the cancer via imperfect DNA replication during cell division or spontaneous cytosine deamination \cite{Goodman}, \cite{Lindahl}, or due to exposures to chemical insults or ultraviolet radiation \cite{Loeb}, \cite{Ananthaswamy}, etc. These mutational processes leave a footprint in the cancer genome characterized by distinctive alteration patterns or mutational signatures.

{}If we can identify all underlying signatures, this could greatly facilitate progress in understanding the origins of cancer and its development. Therapeutically, if there are common underlying structures across different cancer types, then a therapeutic for one cancer type might be applicable to other cancers, which would be a great news.\footnote{\, Another practical application is prevention by pairing the signatures extracted from cancer samples with those caused by known carcinogens (e.g., tobacco, aflatoxin, UV radiation, etc).} However, it all boils down to the question of usefulness, i.e., is there a small enough number of cancer signatures underlying all (100+) known cancer types, or is this number too large to be meaningful or useful? Indeed, there are only 96 SNVs,\footnote{\, In brief, DNA is a double helix of two strands, and each strand is a string of letters A, C, G, T corresponding to adenine, cytosine, guanine and thymine, respectively. In the double helix, A in one strand always binds with T in the other, and G always binds with C. This is known as base complementarity. Thus, there are six possible base mutations C $>$ A, C $>$ G, C $>$ T, T $>$ A, T $>$ C, T $>$ G, whereas the other six base mutations are equivalent to these by base complementarity. Each of these 6 possible base mutations is flanked by 4 possible bases on each side thereby producing $4 \times 6 \times 4 = 96$ distinct mutation categories.} so we cannot have more than 96 signatures.\footnote{\, Nonlinearities could undermine this argument. However, again, it all boils down to usefulness.} Even if the number of true underlying signatures is, say, of order 50, it is unclear whether they would be useful, especially within practical applications. On the other hand, if there are only a dozen or so underlying signatures, then we could hope for an order of magnitude simplification.

{}To identify mutational signatures, one analyzes SNV patterns in a cohort of DNA sequenced whole cancer genomes. The data is organized into a matrix $G_{is}$, where the rows correspond to the $N=96$ mutation categories, the columns correspond to $d$ samples, and each element is a nonnegative occurrence count of a given mutation category in a given sample. Currently, the commonly accepted method for extracting cancer signatures from $G_{is}$ \cite{Alexandrov.NMF} is via nonnegative matrix factorization (NMF) \cite{Paatero}, \cite{LeeSeung}. Under NMF the matrix $G$ is approximated via $G \approx W~H$, where $W_{iA}$ is an $N\times K$ matrix, $H_{As}$ is a $K\times d$ matrix, and both $W$ and $H$ are nonnegative. The appeal of NMF is its biologic interpretation whereby the $K$ columns of the matrix $W$ are interpreted as the weights with which the $K$ cancer signatures contribute into the $N=96$ mutation categories, and the columns of the matrix $H$ are interpreted as the exposures to the $K$ signatures in each sample. The price to pay for this is that NMF, which is an iterative procedure, is computationally costly and depending on the number of samples $d$ it can take days or even weeks to run it. Furthermore, it does not automatically fix the number of signatures $K$, which must be either guessed or obtained via trial and error, thereby further adding to the computational cost.\footnote{\, Other issues include: i) out-of-sample instability, i.e., the signatures obtained from non-overlapping sets of samples can be dramatically different; ii) in-sample instability, i.e., the signatures can have a strong dependence on the initial iteration choice; and iii) samples with low counts or sparsely populated samples (i.e., those with many zeros -- such samples are ubiquitous, e.g., in exome data) are usually deemed not too useful as they contribute to the in-sample instability.}

{}Some of the aforesaid issues were recently addressed in \cite{BioFM}, to wit: i) by aggregating samples by cancer types, we can greatly improve stability and reduce the number of signatures;\footnote{\, As a result, now we have the so-aggregated matrix $G_{is}$, where $s=1,\dots,d$, and $d=n$ is the number of cancer types, not of samples. This matrix is much less noisy than the sample data.} ii) by identifying and factoring out the somatic mutational noise, or the ``overall" mode (this is the ``de-noising" procedure of \cite{BioFM}), we can further greatly improve stability and, as a bonus, reduce computational cost; and iii) the number of signatures can be fixed borrowing the methods from statistical risk models \cite{StatRM} in quantitative finance, by computing the effective rank (or eRank) \cite{RV} for the correlation matrix $\Psi_{ij}$ calculated across cancer types or samples (see below). All this yields substantial improvements \cite{BioFM}.

{}In this paper we push this program to yet another level. The basic idea here is quite simple (but, as it turns out, nontrivial to implement -- see below). We wish to apply clustering techniques to the problem of extracting cancer signatures. In fact, we argue in Section \ref{sec.2} that NMF is, to a degree, ``clustering in disguise". This is for two main reasons. The prosaic reason is that NMF, being a nondeterministic algorithm, requires averaging over many local optima it produces. However, each run generally produces a weights matrix $W_{iA}$ with columns (i.e., signatures) not aligned with those in other runs. Aligning or matching the signatures across different runs (before averaging over them) is typically achieved via nondeterministic clustering such as k-means. So, not only is clustering utilized at some layer, the result, even after averaging, generally is both noisy\footnote{\, By ``noise" we mean the statistical errors in the weighs obtained by averaging. Typically, such error bars are not reported in the literature on cancer signatures. Usually they are large.} and nondeterministic! I.e., if this computationally costly procedure (which includes averaging) is run again and again on the same data, generally it will yield different looking cancer signatures every time!

{}The second, not-so-prosaic reason is that, while NMF generically does not produce exactly null weights, it does produce low weights, such that they are within error bars. For all practical purposes we might as well set such weights to zero. NMF requires nonnegative weights. However, we could as reasonably require that the weights should be, say, outside error bars (e.g., above one standard deviation -- this would render the algorithm highly recursive and potentially unstable or computationally too costly) or above some minimum threshold (which would still further complicated as-is complicated NMF), or else the non-compliant weights are set to zero. As we increase this minimum threshold, the matrix $W_{iA}$ will start to have more and more zeros. It may not exactly have a binary cluster-like structure, but it may at least have some substructures that are cluster-like. It then begs the question: are there cluster-like (sub)structures present in $W_{iA}$ or, generally, in cancer signatures?

{}To answer this question, we can apply clustering methods directly to the matrix $G_{is}$, or, more, precisely, to its de-noised version $G^\prime_{is}$ (see below) \cite{BioFM}. The na\"ive, brute-force approach where one would simply cluster $G_{is}$ or $G^\prime_{is}$ does not work for a variety of reasons, some being more nontrivial or subtle than others. Thus, e.g., as discussed in \cite{BioFM}, the counts $G_{is}$ have skewed, long-tailed distributions and one should work with log-counts, or, more precisely, their de-noised versions. This applies to clustering as well. Further, following a discussion in \cite{StatIndClass} in the context of quantitative trading, it would be suboptimal to cluster de-noised log-counts. Instead, it pays to cluster their normalized variants (see Section \ref{sec.2} hereof). However, taking care of such subtleties does not alleviate one big problem: nondeterminism!\footnote{\, Deterministic (e.g., agglomerative hierarchical) algorithms have their own issues (see below).} If we run a vanilla nondeterministic algorithm such as k-means on the data however massaged with whatever bells and whistles, we will get random-looking disparate results every time we run k-means with no stability in sight. We need to address nondeterminism!

{}Our solution to the problem is what we term {\em *K-means}. The idea behind *K-means, which essentially achieves determinism {\em statistically}, is simple. Suppose we have an $N\times d$ matrix $X_{is}$, i.e., we have $N$ $d$-vectors ${\bf X}_i$. If we run k-means with the input number of clusters $K$ but initially unspecified centers, every run will generally produce a new local optimum. *K-means reduces and in fact essentially eliminates this indeterminism via two levels. At level 1 it takes clusterings obtained via $M$ independent runs or samplings. Each sampling produces a binary $N\times K$ matrix $\Omega_{iA}$, whose element equals 1 if ${\bf X}_i$ belongs to the cluster labeled by $A$, and 0 otherwise. The aggregation algorithm and the source code therefor are given in \cite{StatIndClass}. This aggregation -- for the same reasons as in NMF (see above) -- involves aligning clusters across the $M$ runs, which is achieved via k-means, and so the result is nondeterministic. However, by aggregating a large number $M$ of samplings, the degree of nondeterminism is greatly reduced. The ``catch" is that sometimes this aggregation yields a clustering with $K^\prime < K$ clusters, but this does not pose an issue. Thus, at level 2, we take a large number $P$ of such aggregations (each based on $M$ samplings). The occurrence counts of aggregated clusterings are not uniform but typically have a (sharply) peaked distribution around a few (or manageable) number of aggregated clusterings. So this way we can pinpoint the ``ultimate" clustering, which is simply the aggregated clustering with the highest occurrence count. This is the gist of *K-means and it works well for genome data.

{}So, we apply *K-mean to the same genome data as in \cite{BioFM} consisting of 1,389 (published) samples across 14 cancer types (see below). Our target number of clusters is 7, which was obtained in \cite{BioFM} using the eRank based algorithm (see above). We aggregated 1,000 samplings into clusterings, and we constructed 150,000 such aggregated clusterings (i.e., we ran 150 million k-means instances). We indeed found the ``ultimate" clustering with 7 clusters. Once the clustering is fixed, it turns out that within-cluster weights can be computed via linear regressions (with some bells and whistles) and the weights are automatically positive. That is, we do not need NMF at all! Once we have clusters and weights, we can study reconstruction accuracy and within-cluster correlations between the underlying data and the fitted data that the cluster model produces.

{}We find that clustering works well for 10 out the 14 cancer types we study. The cancer types for which clustering does not appear to work all that well are Liver Cancer, Lung Cancer, and Renal Cell Carcinoma. Also, above 80\% within-cluster correlations arise for 5 out of 7 clusters. Furthermore, remarkably, one cluster has high within-cluster correlations for 9 cancer types, and another cluster for 6 cancer types. These appear to be the leading clusters. Together they have high within-cluster correlations in 11 out of 14 cancer types. So what does all this mean?

{}Additional insight is provided by looking at the within-cluster correlations between signatures Sig1 through Sig7 extracted in \cite{BioFM} and our clusters. High within-cluster correlations arise for Sig1, Sig2, Sig4 and Sig7, which are precisely the signatures with ``peaks" (or ``spikes" -- ``tall mountain landscapes"), whereas Sig3, Sig5 and Sig6 do not have such ``peaks" (``flat" or ``rolling hills landscapes"); see Figures 14 through 20 of \cite{BioFM}.
The latter 3 signatures simply do not have cluster-like structures. Looking at Figure 21 in \cite{BioFM}, it becomes evident why clustering does not work well for Liver Cancer -- it has a whopping 96\% contribution from Sig5! Similarly, Renal Cell Carcinoma has a 70\% contribution from Sig6. Lung Cancer is dominated by Sig3, hence no cluster-like structure. So, Liver Cancer, Lung Cancer and Renal Cell Carcinoma have little in common with other cancers (and each other)! However, 11 other cancers, to wit, B Cell Lymphoma, Bone Cancer, Brain Lower Grade Glioma, Breast Cancer, Chronic Lymphocytic Leukemia, Esophageal Cancer, Gastric Cancer, Medulloblastoma, Ovarian Cancer, Pancreatic Cancer and Prostate Cancer, have 5 (with 2 leading) cluster structures substantially embedded in them.

{}In Section \ref{sec.2} we i) discuss why applying clustering algorithms to extracting cancer signatures makes sense, ii) argue that NMF, to a degree, is ``clustering in disguise", and iii) give the machinery for building cluster models via *K-means, including various details such as what to cluster, how to fix the number of clusters, etc. In Section \ref{sec.3} we discuss i) cancer genome data we use, ii) our application of *K-means to it, and iii) the interpretation of our empirical results. Section \ref{sec.4} contains some concluding remarks, including a discussion of potential applications of *K-means in quantitative finance, where we outline some concrete problems where *K-means can be useful. Appendix \ref{app.code} contains R source code for *K-means and cluster models.

\newpage
\section{Cluster Models}\label{sec.2}

{}The chief objective of this paper is to introduce a novel approach to identifying cancer signatures using clustering methods. In fact, as we discuss below in detail, our approach is more than just clustering. Indeed, it is evident from the get-go that blindly using nondeterministic clustering algorithms,\footnote{\, Such as k-means \cite{Steinhaus}, \cite{Lloyd1957}, \cite{Forgy}, \cite{MacQueen}, \cite{Hartigan}, \cite{HartWong}, \cite{Lloyd1982}.}  which typically produce (unmanageably) large numbers of local optima, would introduce great variability into the resultant cancer signatures.\footnote{\, As we discuss below, in this regard NMF is not dissimilar.} On the other hand, deterministic algorithms such as agglomerative hierarchical clustering\footnote{\, E.g., SLINK \cite{SLINK}, etc. (see, e.g., \cite{HAC}, \cite{StatIndClass}, and references therein).} typically are (substantially) slower and require essentially ``guessing" the initial clustering,\footnote{\, E.g., splitting the data into 2 initial clusters.} which in practical applications\footnote{\, Such as quantitative trading, where out-of-sample performance can be objectively measured. There empirical evidence suggests that such deterministic algorithms underperform so long as nondeterministic ones are used thoughtfully \cite{StatIndClass}.} can often turn out to be suboptimal. So, both to motivate and explain our new approach employing clustering methods, we first -- so to speak -- ``break down" the NMF approach and argue that it is in fact a clustering method in disguise!

\subsection{``Breaking Down" NMF}

{}The current ``lore" -- the commonly accepted method for extracting $K$ cancer signatures from the occurrence counts matrix $G_{is}$ (see above) \cite{Alexandrov.NMF} -- is via nonnegative matrix factorization (NMF) \cite{Paatero}, \cite{LeeSeung}. Under NMF the matrix $G$ is approximated via $G \approx W~H$, where $W_{iA}$ is an $N\times K$ matrix of weights, $H_{As}$ is a $K\times d$ matrix of exposures, and both $W$ and $H$ are nonnegative. However, not only is the number of signatures $K$ not fixed via NMF (and must be either guessed or obtained via trial and error), NMF too is a nondeterministic algorithm and typically produces a large number of local optima. So, in practice one has no choice but to execute a large number $N_S$ of NMF runs -- which we refer to as samplings -- and then somehow extract cancer signatures from these samplings. Absent a guess for what $K$ should be, one executes $N_S$ samplings for a range of values of $K$ (say, $K_{\min} \leq K \leq K_{\max}$, where $K_{min}$ and $K_{max}$ are basically guessed based on some reasonable intuitive considerations), for each $K$ extracts cancer signatures (see below), and then picks $K$ and the corresponding signatures with the best overall fit into the underlying matrix $G$. For a given $K$, different samplings generally produce different weights matrices $W$. So, to extract a single matrix $W$ for each value of $K$ one {\em averages} over the samplings. However, before averaging, one must match the $K$ cancer signatures across different samplings -- indeed, in a given sampling X the columns in the matrix $W_{iA}$ are not necessarily aligned with the columns in the matrix $W_{iA}$ in a different sampling Y. To align the columns in the matrices $W$ across the $N_S$ samplings, once often uses a clustering algorithm such as k-means. However, since k-means is nondeterministic, such alignment of the $W$ columns is not guaranteed to -- and in fact does not -- produce a unique answer. Here one can try to run multiple samplings of k-means for this alignment and aggregate them, albeit such aggregation itself would require another level of alignment (with its own nondeterministic clustering such as k-means).\footnote{\, We should point out that at some level of alignment one may employ a deterministic (e.g., agglomerative hierarchical -- see above) clustering algorithm to terminate the malicious circle, which can be a reasonable approach assuming there is enough stability in the data. However, this too adds a(n often hard to quantify and therefore hidden) systematic error to the resultant signatures.\label{fn.alignment}} And one can do this {\em ad infinitum}. In practice, one must break the chain at some level of alignment, either {\em ad hoc} (essentially by heuristically observing sufficient stability and ``convergence") or via using a deterministic algorithm (see fn. \ref{fn.alignment}). Either way, invariably all this introduces (overtly or covertly) systematic and statistical errors into the resultant cancer signatures and often it is unclear if they are meaningful without invoking some kind empirical biologic ``experience" or ``intuition" (often based on already well-known effects of, e.g., exposure to various well-understood carcinogens such as tobacco, ultraviolet radiation, aflatoxin, etc.). At the end of the day it all boils down to how useful -- or {\em predictive} -- the resultant method of extracting cancer signatures is, including signature stability. With NMF, the answer is not at all evident...

\subsection{Clustering in Disguise?}

{}So, in practice, under the hood, NMF already uses clustering methods. However, it goes deeper than that. While NMF generically does not produce vanishing weights for a given signature, some weights are (much) smaller than others. E.g., often one has several ``peaks" with high concentration of weights, with the rest of the mutation categories having relatively low weights. In fact, many weights can even be within the (statistical plus systematic) error bars.\footnote{\, And such error bars are rarely displayed in the prevalent literature...} Such weights can for all practical purposes be set to zero. In fact, we can take this further and ask whether proliferation of low weights adds any explanatory power. One way to address this is to run NMF with an additional constraint that the weights (obtained via averaging -- see above) should be higher than either i) some multiple of the corresponding error bars\footnote{\, This would require a highly recursive algorithm.} or ii) some preset fixed minimum weight. This certainly sounds reasonable, so why is this not done in practice? A prosaic answer appears to be that this would complicate the already nontrivial NMF algorithm even further, require additional coding and computation resources, etc. However, {\em arguendo}, let us assume that we require, say, that the weights be higher than a preset fixed minimum weight $w_{min}$ or else the weights are set to zero. As we increase $w_{min}$, the so-modified NMF would produce more and more zeros. This does not mean that the resulting matrix $W_{iA}$ would have a {\em binary} cluster structure, i.e., that $W_{iA} = w_i~\delta_{G(i), A}$, where $\delta_{AB}$ is a Kronecker delta and $G:\{1,\dots,N\}\mapsto\{1,\dots,K\}$ is a map from $N=96$ mutation categories to $K$ clusters. Put another way, this does not mean that in the resulting matrix $W_{iA}$ for a given $i$ (i.e., mutation category) we would have a nonzero element for one and only one value of $A$ (i.e., signature). However, as we gradually increase $w_{min}$, generally the matrix $W_{iA}$ is expected to look more and more like having a binary cluster structure, albeit with some ``overlapping" signatures (i.e., such that in a given pair of signatures there are nonzero weights for one or more mutations). We can achieve a binary structure via a number of ways. Thus, a rudimentary algorithm would be to take the matrix $W_{iA}$ (equally successfully before or after achieving some zeros in it via nonzero $w_{min}$) and for a given value of $i$ set all weights $W_{iA}$ to zero except in the signature $A$ for which $W_{iA} = \mbox{max}(W_{iA} | A = 1,\dots,K)$. Note that this might result in some empty signatures (clusters), i.e., signatures with $W_{iA} = 0$ for all values of $i$. This can be dealt with by i) ether simply dropping such signatures altogether and having fewer $K^\prime < K$ signatures (binary clusters) at the end, or ii) augmenting the algorithm to avoid empty clusters, which can be done in a number of ways we will not delve into here. The bottom line is that NMF essentially can be made into a clustering algorithm by reasonably modifying it, including via getting rid of ubiquitous and not-too-informative low weights. However, the downside would be an even more contrived algorithm, so this is not what we are suggesting here. Instead, we are observing that clustering is already intertwined in NMF and the question is whether we can simplify things by employing clustering methods directly.

\subsection{Making Clustering Work}

{}Happily, the answer is yes. Not only can we have much simpler and apparently more stable clustering algorithms, but they are also computationally much less costly than NMF. As mentioned above, the biggest issue with using popular nondeterministic clustering algorithms such as k-means\footnote{\, Which are preferred over deterministic ones for the reasons discussed above.} is that they produce a large number of local optima. For definiteness in the remainder of this paper we will focus on k-means, albeit the methods described herein are general and can be applied to other such algorithms. Fortunately, this very issue has already been addressed in \cite{StatIndClass} in the context of constructing statistical industry classifications (i.e., clustering models for stocks) for quantitative trading, so here we simply borrow therefrom and further expand and adapt that approach to cancer signatures.

\subsubsection{K-means}

{}A popular clustering algorithm is k-means \cite{Steinhaus}, \cite{Lloyd1957}, \cite{Forgy}, \cite{MacQueen}, \cite{Hartigan}, \cite{HartWong}, \cite{Lloyd1982}. The basic idea behind k-means is to partition $N$ observations into $K$ clusters such that each observation belongs to the cluster with the nearest mean. Each of the $N$ observations is actually a $d$-vector, so we have an $N \times d$ matrix $X_{is}$, $i=1,\dots,N$, $s=1,\dots,d$. Let $C_a$ be the $K$ clusters, $C_a = \{i| i\in C_a\}$, $a=1,\dots,K$. Then k-means attempts to minimize\footnote{\, Below we will discuss what $X_{is}$ should be for cancer signatures.}
\begin{equation}\label{k-means}
 g = \sum_{a=1}^K \sum_{i \in C_a} \sum_{s=1}^d  \left(X_{is} - Y_{as}\right)^2
\end{equation}
where
\begin{equation}\label{centers}
 Y_{as} = {1\over n_a} \sum_{i\in C_a} X_{is}
\end{equation}
are the cluster centers (i.e., cross-sectional means),\footnote{\, Throughout this paper ``cross-sectional" refers to ``over the index $i$".} and $n_a = |C_a|$ is the number of elements in the cluster $C_a$. In (\ref{k-means}) the measure of ``closeness" is chosen to be the Euclidean distance between points in ${\bf R}^d$, albeit other measures are possible.

{}One ``drawback" of k-means is that it is not a deterministic algorithm. Generically, there are copious local minima of $g$ in (\ref{k-means}) and the algorithm only guarantees that it will converge to a local minimum, not the global one. Being an iterative algorithm, unless the initial centers are preset, k-means starts with a random set of the centers $Y_{as}$ at the initial iteration and converges to a different local minimum in each run. There is no magic bullet here: in practical applications, typically, trying to ``guess" the initial centers is not any easier than ``guessing" where, e.g., the global minimum is. So, what is one to do? One possibility is to simply live with the fact that every run produces a different answer. In fact, this is acceptable in many applications. However, in the context of extracting cancer signatures this would result in an exercise in futility. We need a way to eliminate or greatly reduce indeterminism.

\subsubsection{Aggregating Clusterings}\label{sub.aggr}

{}The idea is simple. What if we {\em aggregate} different clusterings from multiple runs -- which we refer to as samplings -- into one? The question is how. Suppose we have $M$ runs ($M \gg 1$). Each run produces a clustering with $K$ clusters. Let $\Omega^r_{ia} = \delta_{G^r(i),a}$, $i=1,\dots,N$, $a=1,\dots,K$ (here $G^r:\{1,\dots,N\} \mapsto \{1,\dots,K\}$ is the map between -- in our case -- the mutation categories and the clusters),\footnote{\, Note that here the superscript $r$ in $\Omega^r_{ia}$, $G^r(i)$ and $N^r_a$ (see below) is an index, not a power.} be the binary matrix from each run labeled by $r=1,\dots,M$, which is a convenient way (for our purposes here) of encoding the information about the corresponding clustering; thus, each row of $\Omega^r_{ia}$ contains only one element equal 1 (others are zero), and $N^r_a = \sum_{i=1}^N \Omega^r_{ia}$ (i.e., column sums) is nothing but the number of mutations belonging to the cluster labeled by $a$ (note that $\sum_{a=1}^K N^r_a = N$). Here we are assuming that somehow we know how to properly order (i.e., align) the $K$ clusters from each run. This is a nontrivial assumption, which we will come back to momentarily. However, assuming, for a second, that we know how to do this, we can aggregate the binary matrices $\Omega^r_{ia}$ into a single matrix ${\widetilde \Omega}_{ia} = \sum_{r=1}^M \Omega^r_{ia}$. Now, this matrix does not look like a binary clustering matrix. Instead, it is a matrix of occurrence counts, i.e., it counts how many times a given mutation was assigned to a given cluster in the process of $M$ samplings. What we need to construct is a map $G$ such that one and only one mutation belongs to each of the $K$ clusters. The simplest criterion is to map a given mutation to the cluster in which ${\widetilde\Omega}_{ia}$ is maximal, i.e., where said mutation occurs most frequently. A caveat is that there may be more than one such clusters. A simple criterion to resolve such an ambiguity is to assign said mutation to the cluster with most cumulative occurrences (i.e., we assign said mutation to the cluster with the largest ${\widetilde N}_a = \sum_{i=1}^N {\widetilde\Omega}_{ia}$). Further, in the unlikely event that there is still an ambiguity, we can try to do more complicated things, or we can simply assign such a mutation to the cluster with the lowest value of the index $a$ -- typically, there is so much noise in the system that dwelling on such minutiae simply does not pay off.

{}However, we still need to tie up a loose end, to wit, our assumption that the clusters from different runs were somehow all aligned. In practice each run produces $K$ clusters, but i) they are not the same clusters and there is no foolproof way of mapping them, especially when we have a large number of runs; and ii) even if the clusters were the same or similar, they would not be ordered, i.e., the clusters from one run generally would be in a different order than the clusters from another run.

{}So, we need a way to ``match" clusters from different samplings. Again, there is no magic bullet here either. We can do a lot of complicated and contrived things with not much to show for it at the end. A simple pragmatic solution is to use k-means to align the clusters from different runs. Each run labeled by $r=1,\dots,M$, among other things, produces a set of cluster centers $Y^r_{as}$. We can ``bootstrap" them by row into a $(KM) \times d$ matrix ${\widetilde Y}_{{\widetilde a}s} = Y^r_{as}$, where ${\widetilde a} = a + (r - 1)K$ takes values ${\widetilde a}=1,\dots,(KM)$. We can now cluster ${\widetilde Y}_{{\widetilde a}s}$ into $K$ clusters via k-means. This will map each value of ${\widetilde a}$ to $\{1,\dots,K\}$ thereby mapping the $K$ clusters from each of the $M$ runs to $\{1,\dots,K\}$. So, this way we can align all clusters. The ``catch" is that there is no guarantee that each of the $K$ clusters from each of the $M$ runs will be uniquely mapped to one value in $\{1,\dots,K\}$, i.e., we may have some empty clusters at the end of the day. However, this is fine, we can simply drop such empty clusters and aggregate (via the above procedure) the smaller number of $K^\prime < K$ clusters. I.e., at the end we will end up with a clustering with $K^\prime$ clusters, which might be fewer than the target number of clusters $K$. This is not necessarily a bad thing. The dropped clusters might have been redundant in the first place. Another evident ``catch" is that even the number of resulting clusters $K^\prime$ is not deterministic. If we run this algorithm multiple times, we will get varying values of $K^\prime$. Malicious circle?

\subsubsection{Fixing the ``Ultimate" Clustering}\label{sub.ultimate}

{}Not really! There is one other trick up our sleeves we can use to fix the ``ultimate" clustering thereby rendering our approach essentially deterministic. The idea above is to aggregate a large enough number $M$ of samplings. Each aggregation produces a clustering with some $K^\prime\leq K$ clusters, and this $K^\prime$ varies from aggregation to aggregation. However, what if we take a large number $P$ of aggregations (each based on $M$ samplings)? Typically there will be a relatively large number of different clusterings we get this way. However, assuming some degree of stability in the data, this number is much smaller than the number of {\em a priori} different local minima we would obtain by running the vanilla k-means algorithm. What is even better, the occurrence counts of aggregated clusterings are not uniform but typically have a (sharply) peaked distribution around a few (or manageable) number of aggregated clusterings. In fact, as we will see below, in our empirical genome data we are able to pinpoint the ``ultimate" clustering! So, to recap, what we have done here is this. There are myriad clusterings we can get via vanilla k-means with little to no guidance as to which one to pick.\footnote{\, This is because things are pretty much random and the only ``distribution" at hand is flat.} We have reduced this proliferation by aggregating a large number of such clusterings into our aggregated clusterings. We then further zoom onto a few or even a unique clustering we consider to be the likely ``ultimate" clustering by examining the occurrence counts of such aggregated clusterings, which turns out to have a (sharply) peaked distribution. Since vanilla k-means is a relatively fast-converging algorithm, each aggregation is not computationally taxing and running a large number of aggregations is nowhere as time consuming as running a similar number (or even a fraction thereof) of NMF computations (see below).

\subsection{What to Cluster?}

{}So, now that we know how to make clustering work, we need to decide what to cluster, i.e., what to take as our matrix $X_{is}$ in (\ref{k-means}). The na\"{\i}ve choice $X_{is} = G_{is}$ is suboptimal for multiple reasons (as discussed in \cite{BioFM}).

{}First, the elements of the matrix $G_{is}$ are populated by nonnegative occurrence counts. Nonnegative quantities with large numbers of samples tend to have skewed distributions with long tails at higher values. I.e., such distributions are not normal but (in many cases) roughly log-normal. One simple way to deal with this is to identify $X_{is}$ with a (natural) logarithm of $G_{is}$ (instead of $G_{is}$ itself). A minor hiccup here is that some elements of $G_{is}$ can be 0. We can do a lot of complicated and even convoluted things to deal with this issue. Here, as in \cite{BioFM}, we will follow a pragmatic approach and do something simple instead -- there is so much noise in the data that doing convoluted things simply does not pay off. So, as the first cut, we can take
\begin{equation}\label{log.def}
 X_{is} = \ln\left(1 + G_{is}\right)
\end{equation}
This takes care of the $G_{is} = 0$ cases; for $G_{is}\gg 1$ we have $R_{is}\approx\ln(G_{is})$, as desired.

{}Second, the detailed empirical analysis of \cite{BioFM} uncovered what is termed therein the ``overall" mode\footnote{\, In finance the analog of this is the so-called ``market" mode (see, e.g., \cite{CFM} and references therein) corresponding to the overall movement of the broad market, which affects all stocks (to varying degrees) -- cash inflow (outflow) into (from) the market tends to push stock prices higher (lower). This is the market risk factor, and to mitigate it one can, e.g., hold a dollar-neutral portfolio of stocks (i.e., the same dollar holdings for long and short positions).} unequivocally present in the occurrence count data. This ``overall" mode is interpreted as somatic mutational {\em noise} unrelated to (and in fact obscuring) the true underlying cancer signatures and must therefore be factored out somehow. Here is a simple way to understand the ``overall" mode. Let the correlation matrix $\Psi_{ij} = \mbox{Cor}(X_{is}, X_{js})$, where $\mbox{Cor}(\cdot,\cdot)$ is serial correlation.\footnote{\, Throughout this paper ``serial" refers to ``over the index s".} I.e., $\Psi_{ij} = C_{ij}/\sigma_i\sigma_j$, where $\sigma_i^2 = C_{ii}$ are variances, and the serial covariance matrix\footnote{\, The overall normalization of $C_{ij}$, i.e., $d-1$ (unbiased estimate) vs. $d$ (maximum likelihood estimate) in the denominator in the definition of $C_{ij}$ in (\ref{cov}), is immaterial for our purposes here.}
\begin{equation} \label{cov}
 C_{ij} = \mbox{Cov}(X_{is}, X_{js}) = {1\over {d-1}}\sum_{s=1}^d Z_{is}~Z_{js}
\end{equation}
where $Z_{is} = X_{is} - {\overline X}_i$ are serially demeaned, while the means ${\overline X}_i = {1\over d}\sum_{s=1}^d X_{is}$. The average pair-wise correlation $\rho = {1\over N(N-1)}\sum_{i,j=1;~i\neq j}^N\Psi_{ij}$ between different mutation categories is nonzero and is in fact high for most cancer types we study. This is the aforementioned somatic mutational noise that must be factored out. If we aggregate samples by cancer types (see below) and compute the correlation matrix $\Psi_{ij}$ for the so-aggregated data (across the $n = 14$ cancer types we study -- see below),\footnote{\, So, in this case $d = n = 14$ in (\ref{cov}).} the average correlation $\rho$ is over whopping 96\%. Another way of thinking about this is that the occurrence counts in different samples (or cancer types, if we aggregate samples by cancer types) are not normalized uniformly across all samples (cancer types). Therefore, running NMF, a clustering or any other signature-extraction algorithm on the vanilla matrix $G_{is}$ (or its ``log" $X_{is}$ defined in (\ref{log.def})) would amount to mixing apples with oranges thereby obscuring the true underlying cancer signatures.

{}Following \cite{BioFM}, factoring out the ``overall" mode (or ``de-noising" the matrix $G_{is}$) therefore most simply amount to cross-sectional (i.e., across the 96 mutation categories) demeaning of the matrix $X_{is}$. I.e., instead of $X_{is}$ we use $X^\prime_{is}$, which is obtained from $X_{is}$ by demeaning its columns:\footnote{\, For the reasons discussed above, we should demean $X_{is}$, not $G_{is}$.}
\begin{equation}\label{Xprime}
 X^\prime_{is} = X_{is} - {\overline X}_s = X_{is} - {1\over N} \sum_{j=1}^N X_{js}
\end{equation}
We should note that using $X^\prime_{is}$ instead of $X_{is}$ in (\ref{k-means}) does not affect clustering. Indeed, $g$ in (\ref{k-means}) is invariant under the transformations of the form $X_{is} \rightarrow X_{is} + \Delta_s$, where $\Delta_s$ is an arbitrary $d$-vector, as thereunder we also have $Y_{as}\rightarrow Y_{as} + \Delta_s$, so $X_{is} - Y_{as}$ is unchanged. In fact, this is good: this means that de-noising does not introduce any additional errors into clustering itself. However, the actual {\em weights} in the matrix $W_{iA}$ are affected by de-noising. We discuss the algorithm for fixing $W_{iA}$ below. However, we need one more ingredient before we get to determining the weights, and with this additional ingredient de-noising does affect clustering.

\subsubsection{Normalizing Log-counts}\label{sub.norm.log.counts}

\begin{figure}[ht]
\centering
\includegraphics[scale=0.7]{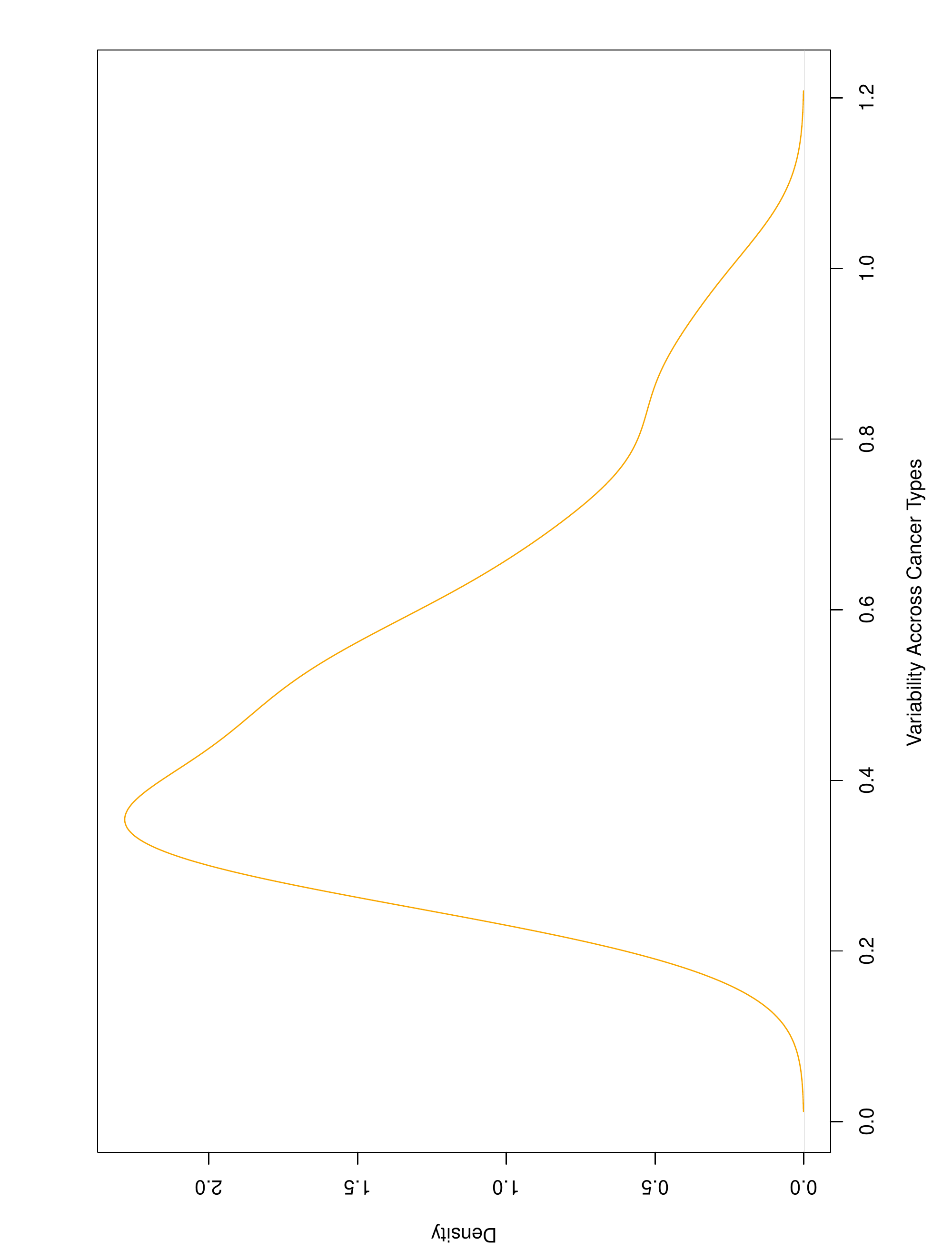}
\caption{Horizontal axis: serial standard deviation $\sigma^\prime_i$ for $N=96$ mutation categories ($i=1,\dots,N$) of cross-sectionally demeaned log-counts $X^\prime_{is}$ across $n=14$ cancer types (for samples aggregated by cancer types, so $s=1,\dots,d$, $d=n$). Vertical axis: density using R function {\tt\small density()}. See Subsection \ref{sub.norm.log.counts} for details.}
\label{FigureVolDensity}
\end{figure}

{}As was discussed in \cite{StatIndClass}, clustering $X_{is}$ (or equivalently $X^\prime_{is}$) would be suboptimal.\footnote{\, More precisely, the discussion of \cite{StatIndClass} is in the financial context, to wit, quantitative trading, which has its own nuances (see below). However, some of that discussion is quite general and can be adapted to a wide variety of applications.} The issue is this. Let $\sigma^\prime_i$ be serial standard deviations, i.e., $(\sigma_i^\prime)^2 = \mbox{Cov}(X^\prime_{is}, X^\prime_{is})$, where, as above, $\mbox{Cov}(\cdot,\cdot)$ is serial covariance. Here we assume that samples are aggregated by cancer types, so $s=1,\dots,d$ with $d = n = 14$. Now, $\sigma^\prime_i$ are not cross-sectionally uniform and vary substantially across mutation categories. The density of $\sigma_i^\prime$ is depicted in Figure \ref{FigureVolDensity} and is skewed (tailed). The summary of $\sigma_i^\prime$ reads:\footnote{\, Qu. = Quartile, SD = Standard Deviation, MAD = Mean Absolute Deviation.} Min = 0.2196, 1st Qu. = 0.3409, Median = 0.4596, Mean = 0.4984, 3rd Qu. = 0.6060, Max = 1.0010, SD = 0.1917, MAD = 0.1859, Skewness = 0.8498. If we simply cluster $X^\prime_{is}$, this variability in $\sigma_i^\prime$ will not be accounted for.

{}A simple solution is to cluster normalized demeaned log-counts ${\widetilde X}^\prime_{is} = X^\prime_{is}/\sigma_i^\prime$ instead of $X^\prime_{is}$. This way we factor out the nonuniform (and skewed) standard deviation out of the log-counts. Note that now de-noising does make a difference in clustering. Indeed, if we use ${\widetilde X}_{is} = X_{is}/\sigma_i$ (recall that $\sigma_i^2 = \mbox{Cov}(X_{is}, X_{is})$) instead of ${\widetilde X}^\prime_{is} = X^\prime_{is}/\sigma_i^\prime$ in (\ref{k-means}) and (\ref{centers}), the quantity $g$ (and also clusterings) will be different.

\subsection{Fixing Cluster Number}\label{sub.fix.K}

{}Now that we know what to cluster (to wit, ${\widetilde X}^\prime_{is}$) and how to get to the ``unique" clustering, we need to figure out how to fix the (target) number of clusters $K$, which is one of the inputs in our algorithm above.\footnote{\, A variety of methods for fixing the number of clusters have been discussed in other contexts, e.g., \cite{Rousseeuw}, \cite{Pelleg}, \cite{Steinbach}, \cite{Goutte}, \cite{Sugar}, \cite{Hamerly}, \cite{Lleiti}, \cite{DeAmorim}.} In \cite{BioFM} it was argued that in the context of cancer signatures their number can be fixed by building a statistical factor model \cite{StatRM}, i.e., the number of signatures is simply the number of statistical factors.\footnote{\, In the financial context, these are known as statistical risk models \cite{StatRM}. For a discussion and literature on multifactor risk models, see, e.g., \cite{Grinold}, \cite{HetPlus} and references therein. For prior works on fixing the number of statistical risk factors, see, e.g., \cite{Connor} and \cite{Bai}.} So, by the same token, here we identify the (target) number of clusters in our clustering algorithm with the number of statistical factors fixed via the method of \cite{StatRM}.

\subsubsection{Effective Rank}\label{sub.erank}

{}So, following \cite{StatRM} and \cite{BioFM}, we set\footnote{\, Here $\mbox{Round}(\cdot)$ can be replaced by $\mbox{floor}(\cdot) = \lfloor\cdot\rfloor$.}
\begin{equation}\label{eq.eRank}
 K = \mbox{Round}(\mbox{eRank}(\Psi))
\end{equation}
Here $\mbox{eRank}(Z)$ is the effective rank \cite{RV} of a symmetric semi-positive-definite (which suffices for our purposes here) matrix $Z$. It is defined as
\begin{eqnarray}
 &&\mbox{eRank}(Z) = \exp(H)\\
 &&H = -\sum_{a=1}^L p_a~\ln(p_a)\\
 &&p_a = {\lambda^{(a)} \over \sum_{b=1}^L \lambda^{(b)}}
\end{eqnarray}
where $\lambda^{(a)}$ are the $L$ {\em positive} eigenvalues of $Z$, and $H$ has the meaning of the (Shannon a.k.a. spectral) entropy \cite{Campbell60}, \cite{YGH}. Let us emphasize that in (\ref{eq.eRank}) the matrix $\Psi_{ij}$ is computed based on the demeaned log-counts\footnote{\, Note that using normalized demeaned log-counts ${\widetilde X}^\prime_{is}$ gives the same $\Psi_{ij}$.} $X^\prime_{is}$.

{}The meaning of $\mbox{eRank}(\Psi_{ij})$ is that it is a measure of the effective dimensionality of the matrix $\Psi_{ij}$, which is not necessarily the same as the number $L$ of its positive eigenvalues, but often is lower. This is due to the fact that many $d$-vectors $X^\prime_{is}$ can be serially highly correlated (which manifests itself by a large gap in the eigenvalues) thereby further reducing the effective dimensionality of the correlation matrix.

\subsection{How to Compute Weights?}\label{sub.reg}

{}The one remaining thing to accomplish is to figure out how to compute the weights $W_{iA}$. Happily, in the context of clustering we have significant simplifications compared with NMF and computing the weights becomes remarkably simple once we fix the clustering, i.e., the matrix $\Omega_{iA} = \delta_{G(i),A}$ (or, equivalently, the map $G:\{i\}\mapsto\{A\}$, $i=1,\dots,N$, $A=1,\dots,K$, where for the notational convenience we use $K$ to denote the number of clusters in the ``ultimate" clustering -- see above). Just as in NMF, we wish to approximate the matrix $G_{is}$ via a product of the weights matrix $W_{iA}$ and the exposure matrix $H_{As}$, both of which must be nonnegative. More precisely, since we must remove the ``overall" mode, i.e., de-noise the matrix $G_{is}$, following \cite{BioFM}, instead of $G_{is}$ we will approximate the re-exponentiated demeaned log-counts matrix $X^\prime_{is}$:
\begin{equation}
 G^\prime_{is} = \exp(X^\prime_{is})
\end{equation}
We can include an overall normalization by taking $G^\prime_{is} = \exp(\mbox{Mean}(X_{is}) + X^\prime_{is})$, or $G^\prime_{is} = \exp(\mbox{Median}(X_{is}) + X^\prime_{is})$, or $G^\prime_{is} = \exp(\mbox{Median}({\overline X}_s) + X^\prime_{is})$ (recall that ${\overline X}_s$ is the vector of column means of $X_{is}$ -- see Eq. (\ref{Xprime})), etc., to make it look more like the original matrix $G_{is}$; however, this does not affect the extracted signatures.\footnote{\, This is because each column of $W$, being weights, is normalized to add up to 1.} Also, technically speaking, after re-exponentiating we should ``subtract" the extra 1 we added in the definition (\ref{log.def}) (assuming we include one of the aforesaid overall normalizations). However, the inherent noise in the data makes this a moot point.

{}So, we wish to approximate $G^\prime_{is}$ via a product $W~H$. However, with clustering we have $W_{iA} = w_i~\delta_{G(i),A}$, i.e., we have a block (cluster) structure where for a given value of $A$ all $W_{iA}$ are zero except for $i\in J(A) = \{j|G(j) = A\}$, i.e., for the mutation categories labeled by $i$ that belong to the cluster labeled by $A$. Therefore, our matrix factorization of $G_{is}$ into a product $W~H$ now simplifies into a set of $K$ {\em independent} factorizations as follows:
\begin{equation}
 G^\prime_{is} \approx w_i~H_{As},~~~i\in J(A),~~~A=1\dots,K
\end{equation}
So, there is no need to run NMF anymore! Indeed, if we can somehow fix $H_{As}$ for a given cluster, then within this cluster we can determine the corresponding weights $w_i$ ($i\in J(A)$) via a {\em serial} linear regression:
\begin{equation}\label{G.reg}
 G^\prime_{is} = \varepsilon_{is} + w_i~H_{As},~~~i\in J(A),~~~A=1\dots,K
\end{equation}
where $\varepsilon_{is}$ are the regression residuals. I.e., for each $A\in\{1,\dots,K\}$, we regress the $d\times n_A$ matrix\footnote{\, The superscript $T$ denotes matrix transposition.} $[(G^\prime)^T]_{si}$ ($i\in J(A)$, $n_A = |J(A)|$) over the $d$-vector $H_{As}$ ($s=1,\dots,d$), and the regression coefficients are nothing but the $n_A$-vector $w_i$ ($i\in J(A)$), while the residuals are the $d\times n_A$ matrix $[(\varepsilon)^T]_{si}$. Note that this regression is run {\em without} the intercept. Now, this all makes sense as (for each $i \in J(A)$) the regression minimizes the quadratic error term $\sum_{s=1}^d \varepsilon^2_{is}$. Furthermore, if $H_{As}$ are nonnegative, then the weights $w_i$ are {\em automatically nonnegative} as they are given by:
\begin{equation}\label{w.reg}
 w_i = {{\sum_{s=1}^d G^\prime_{is}~H_{G(i),s}}\over {{\sum_{s=1}^d H^2_{G(i),s}}}}
\end{equation}
Now, we wish these weights to be normalized:
\begin{equation}\label{w.norm}
 \sum_{i\in J(A)} w_i = 1
\end{equation}
This can always be achieved by rescaling $H_{As}$. Alternatively, we can pick $H_{As}$ without worrying about the normalization, compute $w_i$ via (\ref{w.reg}), rescale them so that they satisfy (\ref{w.norm}), and simultaneously accordingly rescale $H_{As}$. Mission accomplished!

\subsubsection{Fixing Exposures}

{}Well, almost... We still need to figure out how to fix the exposures $H_{As}$. The simplest way to do this is to note that we can use the matrix $\Omega_{iA} = \delta_{G(i), A}$ to swap the index $i$ in $G^\prime_{is}$ by the index $A$, i.e., we can take
\begin{equation}\label{H}
 H_{As} = \eta_A \sum_{i = 1}^N \Omega_{iA}~G^\prime_{is} = {\widetilde \eta}_A~{1\over n_A}\sum_{i\in J(A)} G^\prime_{is}
\end{equation}
That is, up to the normalization constants ${\widetilde\eta}_A$ (which are fixed via (\ref{w.norm})) we simply take cross-sectional means of $G^\prime_{is}$ in each cluster. (Recall that $n_A = J(A)$.) The so-defined $H_{As}$ are automatically positive as all $G^\prime_{is}$ are positive. Therefore, $w_i$ defined via (\ref{w.reg}) are also all positive. This is a good news -- vanishing $w_i$ would amount to an incomplete weights matrix $W_{iA}$ (i.e., some mutations would belong to no cluster.)

{}So, why does (\ref{H}) make sense? Looking at (\ref{G.reg}), we can observe that, if the residuals $\varepsilon_{is}$ cross-sectionally, within each cluster labeled by $A$, are random, then we expect that $\sum_{i\in J(A)}\varepsilon_{is}\approx 0$. If we had an exact equality here, then we would have (\ref{H}) with $\eta_A  = 1$ (i.e., ${\widetilde \eta}_A = n_A$) assuming the normalization (\ref{w.norm}). In practice, the residuals $\varepsilon_{is}$ are not exactly ``random". First, the number $n_A$ of mutation categories in each cluster is not large. Second, as mentioned above, there is variability in serial standard deviations across mutation types. This leads us to consider variations.

\subsubsection{A Variation}

{}Above we argued that it makes sense to cluster normalized demeaned log-counts ${\widetilde X}^\prime_{is} = X^\prime_{is}/\sigma^\prime_i$ due to the cross-sectional variability (and skewness) in the serial standard deviations $\sigma_i^\prime$. We may worry about similar effects in $G^\prime_{is}$ when computing $H_{As}$ and $w_i$ as we did above. This can be mitigated by using normalized quantities ${\widetilde G}^\prime_{is} = G^\prime_{is} / \omega_i$, where $\omega_i^2 = \mbox{Cov}(G^\prime_{is}, G^\prime_{is})$ are serial variances. That is, we can define\footnote{\, I.e., here we assume that $\varepsilon_{is}/\omega_i$ are approximately random in (\ref{G.reg}).}
\begin{eqnarray}\label{H1}
 &&H_{As} = {\widetilde \eta_A}~{1\over\nu_A} \sum_{i\in J(A)} {\widetilde G}^\prime_{is} = {\widetilde \eta_A}~{1\over\nu_A}\sum_{i\in J(A)} {1 \over \omega_i}~G^\prime_{is}\\
 \label{w.reg.1}
 &&w_i = \omega_i~{{\sum_{s=1}^d {\widetilde G}^\prime_{is}~H_{G(i),s}}\over {{\sum_{s=1}^d H^2_{G(i),s}}}} = {{\sum_{s=1}^d G^\prime_{is}~H_{G(i),s}}\over {{\sum_{s=1}^d H^2_{G(i),s}}}}
\end{eqnarray}
where $\nu_A = \sum_{i\in J(A)}1/\omega_i$.  So, $1/\omega_i$ are the weights in the averages over the clusters.

\subsubsection{Another Variation}

{}Here one may wonder, considering the skewed roughly log-normal distribution of $G_{is}$ and henceforth $G^\prime_{is}$, would it make sense to relate the exposures to within-cluster cross-sectional averages of demeaned log-counts $X^\prime_{is}$ as opposed to those of $G^\prime_{is}$? This is easily achieved. Thus, we can define (this ensures positivity of $H_{As}$):
\begin{equation}
 \ln(H_{As}) = \ln({\widetilde \eta}_A) + {1\over n_A}\sum_{i\in J(A)} X^\prime_{is}
\end{equation}
Exponentiating we get
\begin{equation}\label{H2}
 H_{As} = {\widetilde \eta}_A \left[\prod_{i\in J(A)} G^\prime_{is}\right]^{1/n_A}
\end{equation}
I.e., instead of an arithmetic average as in (\ref{H}), here we have a geometric average.

{}As above, here too we can introduce nontrivial weights. Note that the form of (\ref{w.reg.1}) is the same as (\ref{w.reg}), it is only $H_{As}$ that is affected by the weights. So, we can introduce the weights in the geometric means as follows:
\begin{equation}
 \ln(H_{As}) = \ln({\widetilde \eta}_A) + {1\over \mu_A}\sum_{i\in J(A)} {\widetilde X}^\prime_{is} = \ln({\widetilde \eta}_A) + {1\over \mu_A}\sum_{i\in J(A)} {1\over \sigma^\prime_i}~X^\prime_{is}
\end{equation}
where $\mu_A = \sum_{i\in J(A)}1/\sigma^\prime_i$. Recall that $(\sigma^\prime_i)^2 = \mbox{Cov}(X^\prime_{is}, X^\prime_{is})$. Thus, we have:
\begin{equation}\label{H3}
 H_{As} = {\widetilde \eta}_A \prod_{i\in J(A)} (G^\prime_{is})^{1/\mu_A\sigma_i^\prime}
\end{equation}
So, the weights are the exponents $1/\mu_A\sigma_i^\prime$. Other variations are also possible.

\subsection{Implementation}

{}We are now ready to discuss an actual implementation of the above algorithm, much of the R code for which is already provided in \cite{BioFM} and \cite{StatIndClass}. The R source code is given in Appendix \ref{app.code} hereof.

\section{Empirical Results}\label{sec.3}
\subsection{Data Summary}\label{data.summary}

{}In our empirical analysis below we use the same genome data (from published samples only) as in \cite{BioFM}. This data is summarized in Table \ref{table.genome.summary} (borrowed from \cite{BioFM}), which gives total counts, number of samples and the data sources, which are as follows: A1 = \cite{Alexandrov}, A2 = \cite{Love}, B1 = \cite{Tirode}, C1 = \cite{Zhang}, D1 = \cite{Nik-Zainal}, E1 = \cite{Puente2011}, E2 = \cite{Puente2015}, F1 = \cite{Cheng}, G1 = \cite{Wang}, H1 = \cite{Sung}, H2 = \cite{Fujimoto}, I1 = \cite{Imielinski}, J1 = \cite{Jones}, K1 = \cite{Patch}, L1 = \cite{Waddell}, M1 = \cite{Gundem}, N1 = \cite{Scelo}. Sample IDs with the corresponding publication sources are given in Appendix A of \cite{BioFM}. In our analysis below we aggregate samples by the 14 cancer types. The resulting data is in Tables \ref{table.aggr.data.1} and \ref{table.aggr.data.2}.

\subsubsection{Structure of Data}

{}The underlying data consists of a matrix -- call it $G_{is}$ -- whose elements are occurrence counts of mutation types labeled by $i = 1,\dots, N = 96$ in samples labeled by $s = 1,\dots, d$. More precisely, we can work with one matrix $G_{is}$ which combines data from different cancer types; or, alternatively, we may choose to work with individual matrices $[G(\alpha)]_{is}$, where: $\alpha = 1,\dots,n$ labels $n$ different cancer types; as before, $i=1,\dots,N=96$; and $s = 1,\dots, d(\alpha)$. Here $d(\alpha)$ is the number of samples for the cancer type labeled by $\alpha$. The combined matrix $G_{is}$ is obtained simply by appending (i.e., bootstrapping) the matrices $[G(\alpha)]_{is}$ together column-wise. In the case of the data we use here (see above), this ``big matrix" turns out to have 1389 columns.

{}Generally, individual matrices $[G(\alpha)]_{is}$ and, thereby, the ``big matrix", contain a lot of noise. For some cancer types we can have a relatively small number of samples. We can also have ``sparsely populated" data, i.e., with many zeros for some mutation categories. As mentioned above, different samples are not necessarily uniformly normalized. Etc. The bottom line is that the data is noisy. Furthermore, intuitively it is clear that the larger the matrix we work with, statistically the more ``signatures" (or clusters) we should expect to get with any reasonable algorithm. However, as mentioned above, a large number of signatures would be essentially useless and defy the whole purpose of extracting them in the first place -- we have 96 mutation categories, so it is clear that the number of signatures cannot be more than 96! If we end up with, say, 50+ signatures, what new or useful does this tell us about the underlying cancers? The answer is likely nothing other than that most cancers have not much in common with each other, which would be a disappointing result from the perspective of therapeutic applications. To mitigate the aforementioned issues, at least to a certain extent, following \cite{BioFM}, we can aggregate samples by cancer types. This way we get an $N \times n$ matrix, which we also refer to as $G_{is}$, where the index $s=1,\dots,d$ now takes $d=n$ values corresponding to the cancer types. In the data we use $n=14$, the aggregated matrix $G_{is}$ is much less noisy than the ``big matrix", and we are ready to apply the above machinery to it.

\subsection{Genome Data Results}

{}The $96 \times 14$ matrix $G_{is}$ given in Tables \ref{table.aggr.data.1} and \ref{table.aggr.data.2} is what we pass into the function {\tt\small bio.cl.sigs()} in Appendix \ref{app.code} as the input matrix {\tt\small x}. We use: {\tt\small iter.max = 100} (this is the maximum number of iterations used in the built-in R function {\tt\small kmeans()} -- we note that there was not a single instance in our 150 million runs of {\tt\small kmeans()} where more iterations were required);\footnote{\, The R function {\tt\small kmeans()} produces a warning if it does not converge within {\tt\small iter.max}.} {\tt\small num.try = 1000} (this is the number of individual k-means samplings we aggregate every time); and {\tt\small num.runs = 150000} (which is the number of aggregated clusterings we use to determine the ``ultimate" -- that is, the most frequently occurring -- clustering). So, we ran k-means 150 million times. More precisely, we ran 15 batches with {\tt\small num.runs = 10000} as a sanity check, to make sure that the final result based on 150000 aggregated clusterings was consistent with the results based on smaller batches, i.e., that it was in-sample stable.\footnote{\, We ran these 15 batches consecutively, and each batch produced the same top-10 (by occurrence counts) clusterings as in Table \ref{table.occurrence.cts}; however, the actual occurrence counts are different across the batches with slight variability in the corresponding rankings. The results are pleasantly stable.} Based on Table \ref{table.occurrence.cts}, we identify Clustering-A as the ``ultimate" clustering (cf. Clustering-B/C/D).

{}We give the weights for Clustering-A, Clustering-B, Clustering-C and Clustering-D using unnormalized and normalized regressions with exposures computed based on arithmetic averages (see Subsection \ref{sub.reg}) in Tables \ref{table.weights.A.1}, \ref{table.weights.A.2}, \ref{table.weights.B.1}, \ref{table.weights.B.2}, \ref{table.weights.C.1}, \ref{table.weights.C.2}, \ref{table.weights.D.1}, \ref{table.weights.D.2}, and Figures \ref{Figure1A} through \ref{FigureNorm6D}. We give the weights for Clustering-A using unnormalized and normalized regressions with exposures computed based on geometric averages (see Subsection \ref{sub.reg}) in Tables \ref{table.weights.Z.1}, \ref{table.weights.Z.2}, and Figures \ref{FigureGeom1Z} through \ref{FigureGeomNorm7Z}. The actual mutation categories in each cluster for a given clustering can be read off the aforesaid Tables with the weights (the mutation categories with nonzero weights belong to a given cluster), or from the horizontal axis labels in the aforesaid Figures. It is evident that Clustering-A, Clustering-B, Clustering-C and Clustering-D are essentially variations of each other (Clustering-D has only 6 clusters, while the other 3 have 7 clusters).

\subsection{Reconstruction and Correlations}\label{sub.cor}

{}So, based on genome data, we have constructed clusterings and weights. Do they work? I.e., do they reconstruct the input data well? It is evident from the get-go that the answer to this question may not be binary in the sense that for some cancer types we might have a nice clustering structure, while for others we may not. The aim of the following exercise is to sort this all out. Here come the correlations...

\subsubsection{Within-cluster Correlations}

{}We have our de-noised\footnote{\, De-noising per se does not affect cross-sectional correlations. Adding extra 1 in (\ref{log.def}) (recall that we obtain $G^\prime_{is}$ by cross-sectionally demeaning $X_{is}$ and then re-exponentiating) has a negligible effect. So, in the correlations below we can use the original data matrix $G_{is}$ instead of $G^\prime_{is}$.} matrix $G^\prime_{is}$. We are approximating this matrix via the following factorized matrix:
\begin{equation}\label{fac.st}
 G^*_{is} = \sum_{A=1}^K W_{iA}~H_{As} = w_i~H_{G(i),s}
\end{equation}
We can now compute an $n\times K$ matrix $\Theta_{sA}$ of {\em within-cluster} cross-sectional correlations between $G^\prime_{is}$ and $G^*_{is}$ defined via ($\mbox{xCor}(\cdot,\cdot)$ stands for ``cross-sectional correlation" to distinguish it from ``serial correlation" $\mbox{Cor}(\cdot,\cdot)$ we use above)\footnote{\, Due to the factorized structure (\ref{fac.st}), these correlations do not directly depend on $H_{As}$.}
\begin{equation}
 \Theta_{sA} = \left.\mbox{xCor}(G^\prime_{is}, G^*_{is})\right|_{i\in J(A)} = \left.\mbox{xCor}(G^\prime_{is}, w_i)\right|_{i\in J(A)}
\end{equation}
We give this matrix for Clustering-A with weights using normalized regressions with exposures computed based on arithmetic means (see Subsection \ref{sub.reg}) in Table \ref{table.fit.theta}. Let us mention that, with exposures based on arithmetic means, weights using normalized regressions work a bit better than using unnormalized regressions. Using exposures based on geometric means changes the weights a bit, which in turn slightly affects the within-cluster correlations, but does not alter the qualitative picture.

\subsubsection{Overall Correlations}

{}Another useful metric, which we use as a sanity check, is this. For each value of $s$ (i.e., for each cancer type), we can run a linear cross-sectional regression (without the intercept) of $G^\prime_{is}$ over the matrix $W_{iA}$. So, we have $n=14$ of these regressions. Each regression produces multiple $R^2$ and adjusted $R^2$, which we give in Table \ref{table.fit.theta}. Furthermore, we can compute the {\em fitted} values ${\widehat G}^*_{is}$ based on these regressions, which are given by
\begin{equation}
 {\widehat G}^*_{is} = \sum_{A=1}^K W_{iA}~F_{As} = w_i~F_{G(i),s}
\end{equation}
where (for each value of $s$) $F_{As}$ are the regression coefficients. We can now compute the overall cross-sectional correlations (i.e., the index $i$ runs over all $N=96$ mutation categories)
\begin{equation}
 \Xi_s = \mbox{xCor}(G^\prime_{is}, {\widehat G}^*_{is})
\end{equation}
These correlations are also given in Table \ref{table.fit.theta} and measure the overall fit quality.

\subsubsection{Interpretation}

{}Looking at Table \ref{table.fit.theta} a few things become immediately evident. Clustering works well for 10 out the 14 cancer types we study here. The cancer types for which clustering does not appear to work all that well are Breast Cancer (labeled by X4 in Table \ref{table.fit.theta}), Liver Cancer (X8), Lung Cancer (X9), and Renal Cell Carcinoma (X14). More precisely, for Breast Cancer we do have a high within-cluster correlation for Cl-5 (and also Cl-4), but the overall fit is not spectacular due to low within-cluster correlations in other clusters. Also, above 80\% within-cluster correlations\footnote{\, The 80\% cutoff is somewhat arbitrary, but reasonable.} arise for 5 clusters, to wit, Cl-1, Cl-3, Cl-4, Cl-5 and Cl-6, but not for Cl-2 or Cl-7. Furthermore, remarkably, Cl-1 has high within-cluster correlations for 9 cancer types, and Cl-5 for 6 cancer types. These appear to be the leading clusters. Together they have high within-cluster correlations in 11 cancer types. So what does all this mean?

{}Additional insight is provided by looking at the within-cluster correlations between the 7 cancer signature extracted in \cite{BioFM} and the clusters we find here. Let ${\cal W}_{i\alpha}$ be the weights for the 7 cancer signatures from Tables 13 and 14 of \cite{BioFM}. We can compute the following within-cluster correlations $(\alpha = 1,\dots,7$ labels the cancer signatures of \cite{BioFM}, which we refer to as Sig1 through Sig7):
\begin{equation}
 \Delta_{\alpha A} = \left.\mbox{xCor}({\cal W}_{i\alpha}, W_{iA})\right|_{i\in J(A)}
\end{equation}
These correlations are given in Table \ref{table.cor.delta}. High within-cluster correlations arise for Cl-1 (with Sig1 and Sig7), Cl-5 (with Sig2) and Cl-6 (with Sig4). And this makes perfect sense. Indeed, looking at Figures 14 through 20 of \cite{BioFM}, Sig1, Sig2, Sig4 and Sig7 are precisely the cancer signatures that have ``peaks" (or ``spikes" -- ``tall mountain landscapes"), whereas Sig3, Sig5 and Sig6 do not have such ``peaks" (``flat" or ``rolling hills landscapes"). No wonder such signatures do not have high within-cluster correlations -- they simply do not have cluster-like structures. Looking at Figure 21 in \cite{BioFM}, it becomes evident why clustering does not work well for Liver Cancer (X8) -- it has a whopping 96\% contribution from Sig5! Similarly, Renal Cell Carcinoma (X14) has a 70\% contribution from Sig6. Lung Cancer (X9) is dominated by Sig3, hence no cluster-like structure. Finally, Breast Cancer (X4) is dominated by Sig2, which has a high within-cluster correlation with Cl-5, which is why Breast Cancer has a high within-cluster correlation with Cl-5 (but poor overall correlation in Table \ref{table.fit.theta}). So, it all makes sense. The question is, what does all this tell us about cancer signatures?

{}Quite a bit! It tells us that cancers such as Liver Cancer, Lung Cancer and Renal Cell Carcinoma have little in common with other cancers (and each other)! At least at the level of mutation categories that dominate the genome structure of such cancers. On the other hand, 9 cancers, to wit, Bone Cancer (X2), Brain Lower Grade Glioma (X3), Chronic Lymphocytic Leukemia (X5), Esophageal Cancer (X6), Gastric Cancer (X7), Medulloblastoma (X10), Ovarian Cancer (X11), Pancreatic Cancer (X12) and Prostate Cancer (X13) apparently all have the Cl-1 cluster structure embedded in them substantially. Similarly, 6 cancers, to wit, B Cell Lymphoma (X1), Breast Cancer (X4), Esophageal Cancer(X6), Ovarian Cancer (X11), Pancreatic Cancer (X12) and Prostate Cancer (X13) apparently all have the Cl-5 cluster structure embedded in them substantially. Furthermore, note the overlap between these two lists, to wit, Esophageal Cancer(X6), Ovarian Cancer (X11), Pancreatic Cancer (X12) and Prostate Cancer (X13). We obtained this result purely statistically, with no biologic input, using our clustering algorithm and other statistical methods such as linear regression to obtain the actual weights. It is too early to know whether this insight will aid any therapeutic applications, but that is the hope -- similarities in the underlying genomic structures of different cancer types raise hope that therapeutics for one cancer type could perhaps be applicable to other cancer types. On the other hand, our findings above relating to Liver Cancer, Lung Cancer and Renal Cell Carcinoma (and possibly also Breast Cancer, albeit the latter does appear to have a not-so-insignificant overlap with Cl-5, which differentiates it from the aforesaid 3 cancer types) suggest that these cancer types apparently stand out.

\section{Concluding Remarks}\label{sec.4}

{}Clustering ideas and techniques have been applied in cancer research in various incarnations and contexts aplenty -- for a partial list of works at least to some extent related to our discussion here, see, e.g, \cite{Chen1}, \cite{Chen2}, \cite{Kashuba}, \cite{Nik-Zainal}, \cite{Roberts2012}, \cite {Alexandrov.NMF}, \cite{Alexandrov}, \cite{Burns1}, \cite{Burns2}, \cite{Lawrence}, \cite{Long} \cite{Roberts2013}, \cite{Taylor}, \cite{Xuan}, \cite{AlexStrat}, \cite{Bacolla}, \cite{Bolli}, \cite{Caval}, \cite{Davis}, \cite{Helleday}, \cite{Nik-Zainal2014}, \cite{Poon}, \cite{Qian}, \cite{Roberts1}, \cite{Roberts2}, \cite{Roberts3}, \cite{Sima}, \cite{Chan}, \cite{Pettersen} and references therein. As mentioned above, even in NMF clustering is used at some (perhaps not-so-evident) layer. What is new in our approach -- and hence new results -- is that: i) following \cite{BioFM}, we apply clustering to aggregated by cancer types and de-noised data; ii) we use a tried-and-tested in quantitative finance bag of tricks from \cite{StatIndClass}, which improves clustering; and iii) last but not least, we apply our *K-means algorithm to cancer genome data. As mentioned above, *K-means, unlike vanilla k-means or its other commonly used variations, is essentially deterministic, and it achieves determinism {\em statistically}, not by ``guessing" initial centers or as in agglomerative hierarchical clustering, which basically ``guesses" the initial (e.g., 2-cluster) clustering. Instead, via aggregating a large number of k-means clusterings and statistical examination of the occurrence counts of such aggregations, *K-means takes a mess of myriad vanilla k-means clusterings and systematically reduces randomness and indeterminism without {\em ad hoc} initial ``guesswork".

{}As mentioned above, consistently with the results of \cite{BioFM} obtained via improved NMF techniques, Liver Cancer, Lung Cancer and Renal Cell Carcinoma do not appear to have clustering (sub)structures. This could be both good and bad news. It is a good news because we learned something interesting about these cancer types -- and in two complementary ways. However, it could also be a bad news from the therapeutic standpoint. Since these cancer types appear to have little in common with others, it is likely that they would require specialized therapeutics. On the flipside, we should note that it would make sense to exclude these 3 cancer types when running clustering analysis. However, it would also make sense to include other cancer types by utilizing the International Cancer Genome Consortium data, which we leave for future studies. (For comparative reasons, here we used the same data as in \cite{BioFM}, which was limited to data samples published as of the date thereof.) This paper is not intended to be an exhaustive empirical study but a proof of concept and an opening of a new avenue for extracting and studying cancer signatures beyond the tools that NMF provides.

\begin{figure}[ht]
\centering
\includegraphics[scale=0.7]{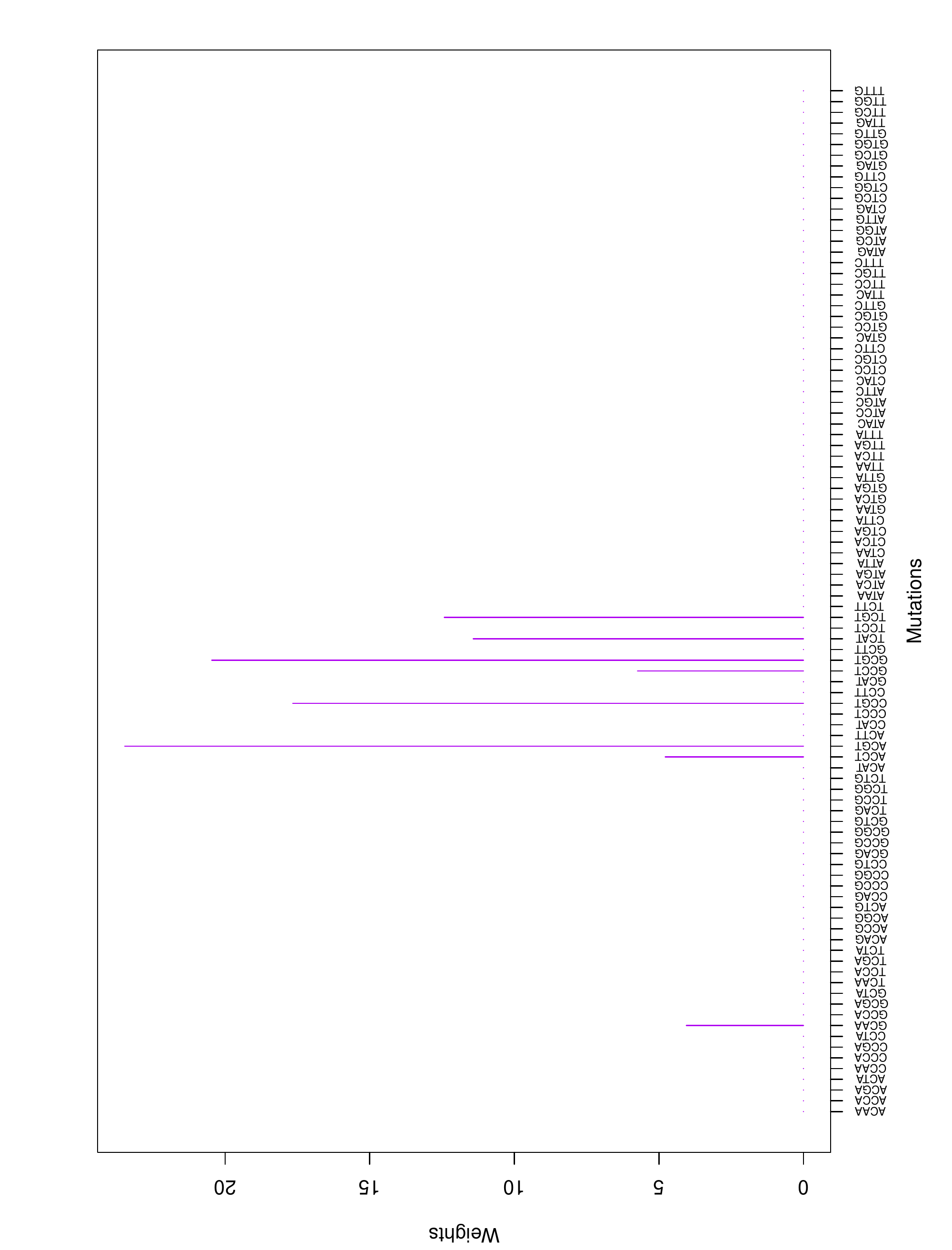}
\caption{Cluster Cl-1 in Clustering-A with weights based on unnormalized regressions with arithmetic means
(see Subsection \ref{sub.reg}). See Tables \ref{table.occurrence.cts}, \ref{table.weights.A.1}, \ref{table.weights.A.2}. Here and in all Figures below, for comparison and visualization convenience, we show all 96 channels on the horizontal axis even though the weights are nonzero only for the mutation categories belonging to a given cluster. Thus, in this cluster, only 8 weights are nonzero, to wit, for GCAA, ACCT, ACGT, CCGT, GCCT, GCGT, TCAT, TCGT.
}
\label{Figure1A}
\end{figure}

\begin{figure}[ht]
\centering
\includegraphics[scale=0.7]{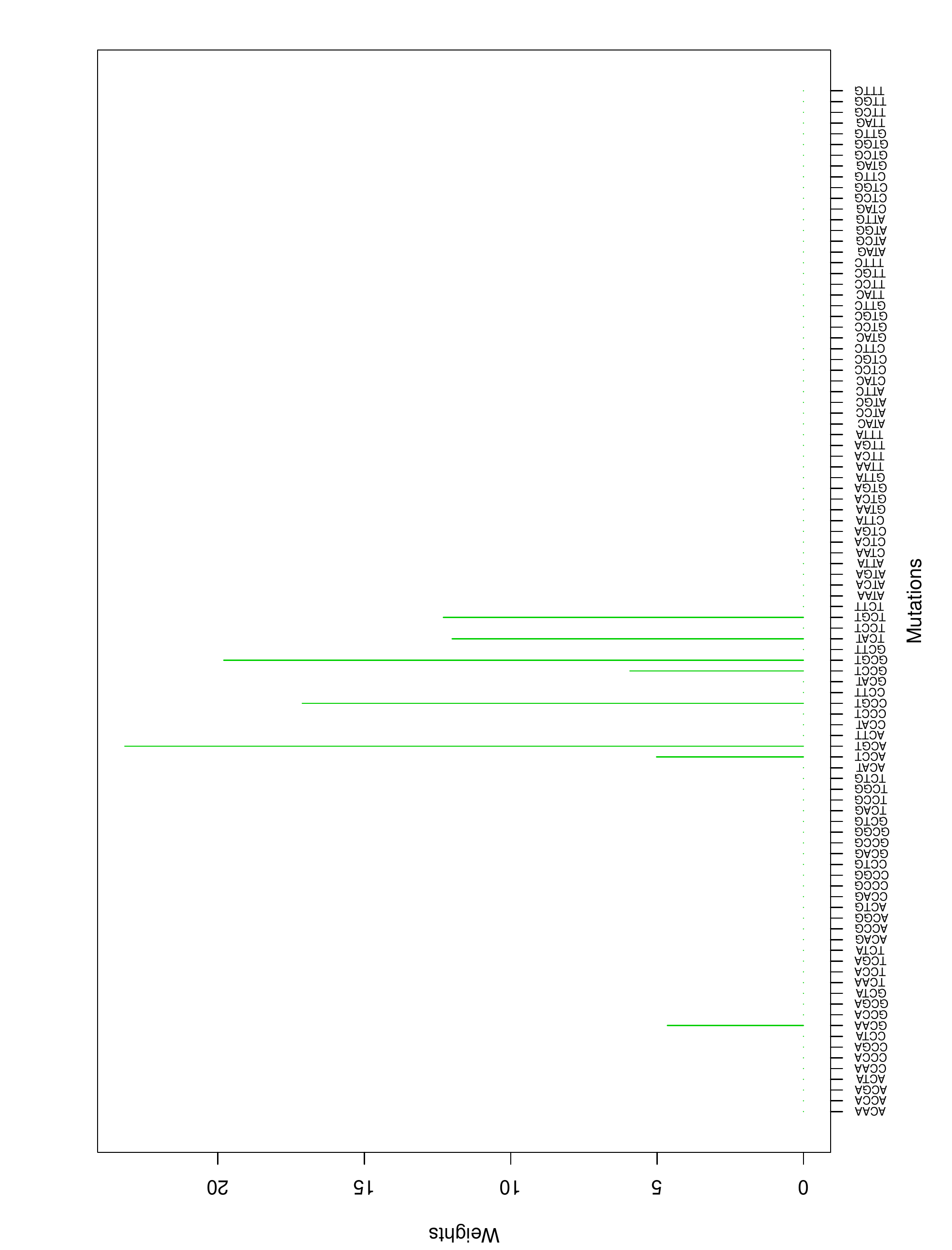}
\caption{Cluster Cl-1 in Clustering-A with weights based on normalized regressions with arithmetic means
(see Subsection \ref{sub.reg}).
See Tables \ref{table.occurrence.cts}, \ref{table.weights.A.1}, \ref{table.weights.A.2}.}
\label{FigureNorm1A}
\end{figure}

\begin{figure}[ht]
\centering
\includegraphics[scale=0.7]{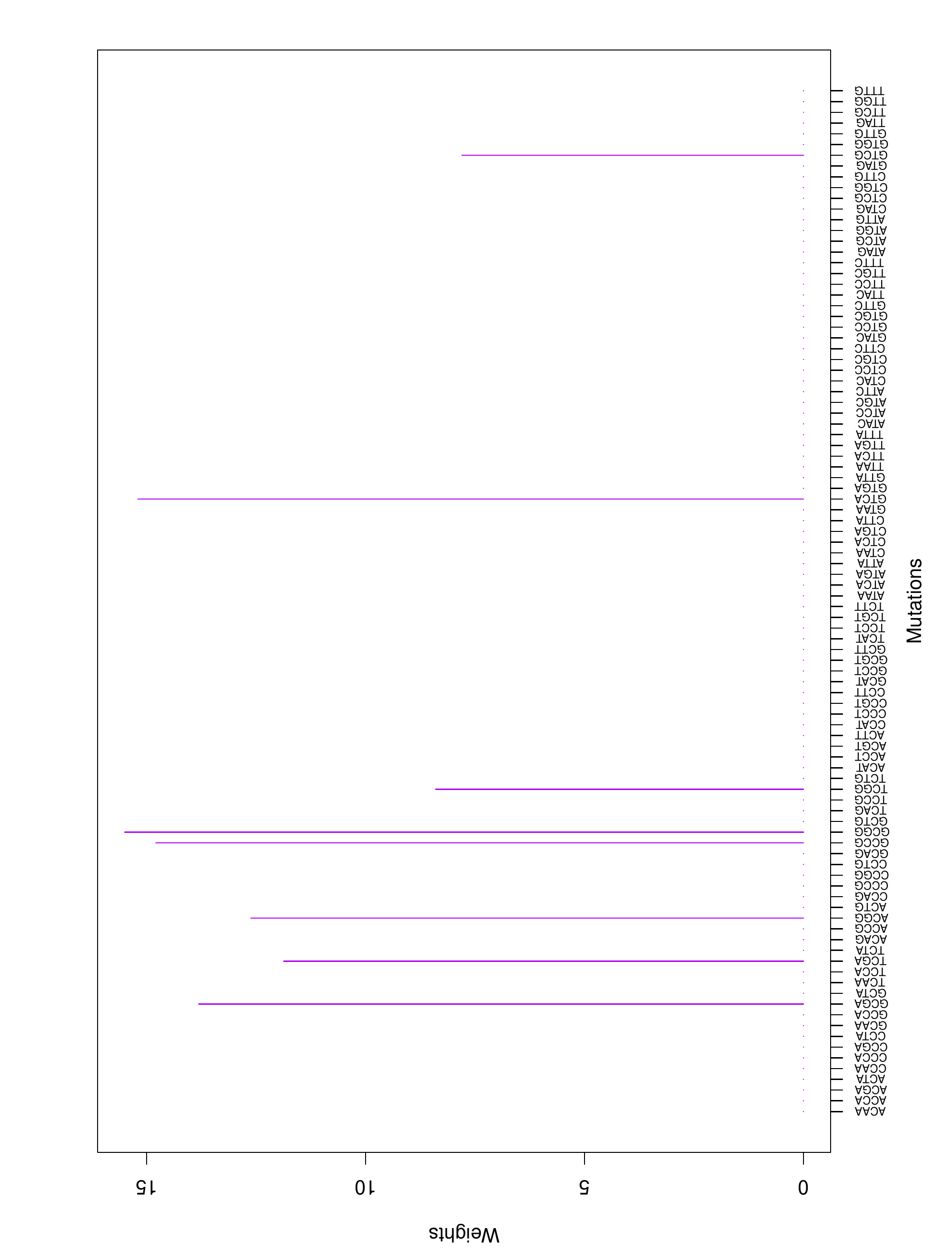}
\caption{Cluster Cl-2 in Clustering-A with weights based on unnormalized regressions with arithmetic means
(see Subsection \ref{sub.reg}).
See Tables \ref{table.occurrence.cts}, \ref{table.weights.A.1}, \ref{table.weights.A.2}.}
\label{Figure2A}
\end{figure}

\begin{figure}[ht]
\centering
\includegraphics[scale=0.7]{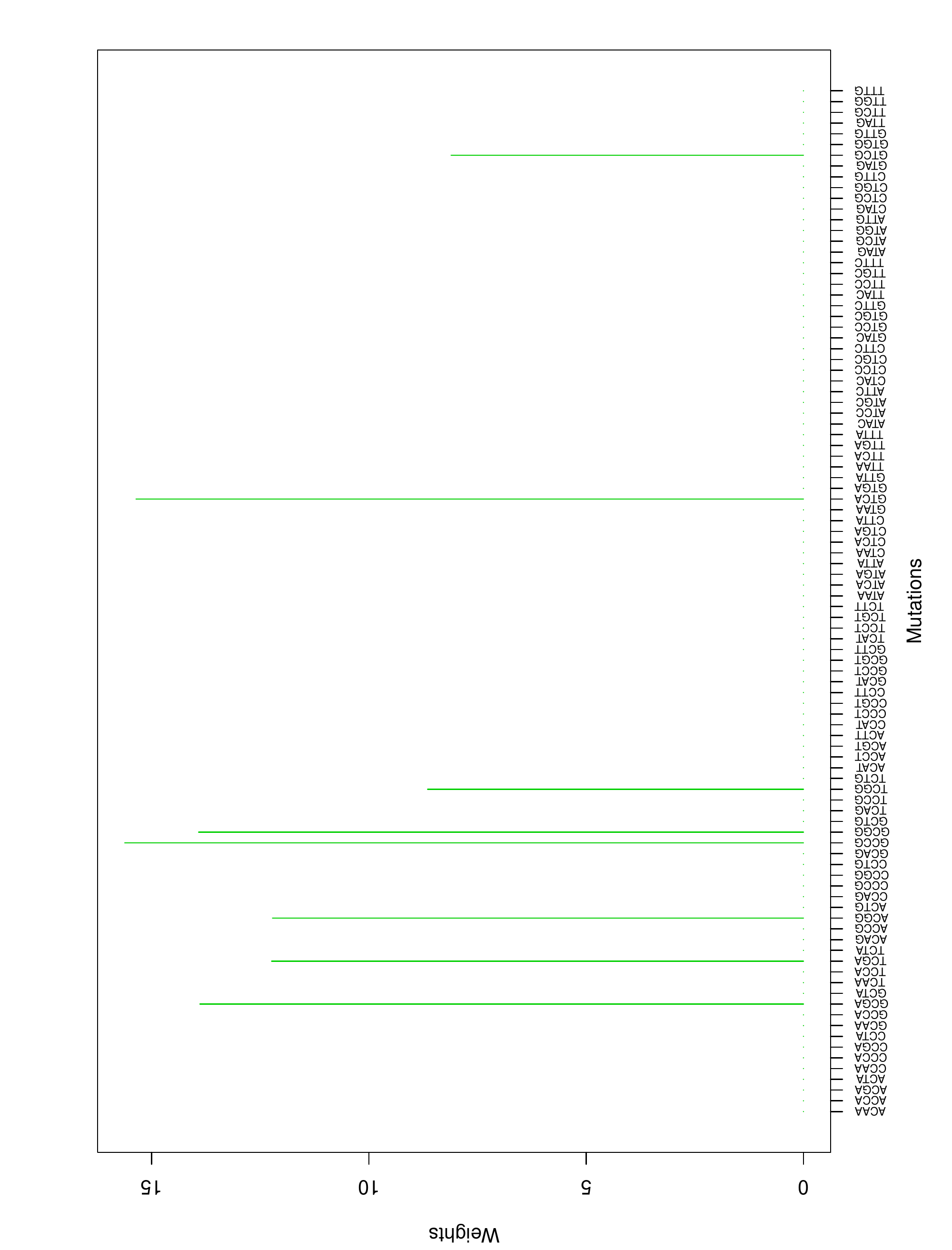}
\caption{Cluster Cl-2 in Clustering-A with weights based on normalized regressions with arithmetic means
(see Subsection \ref{sub.reg}).
See Tables \ref{table.occurrence.cts}, \ref{table.weights.A.1}, \ref{table.weights.A.2}.}
\label{FigureNorm2A}
\end{figure}

\begin{figure}[ht]
\centering
\includegraphics[scale=0.7]{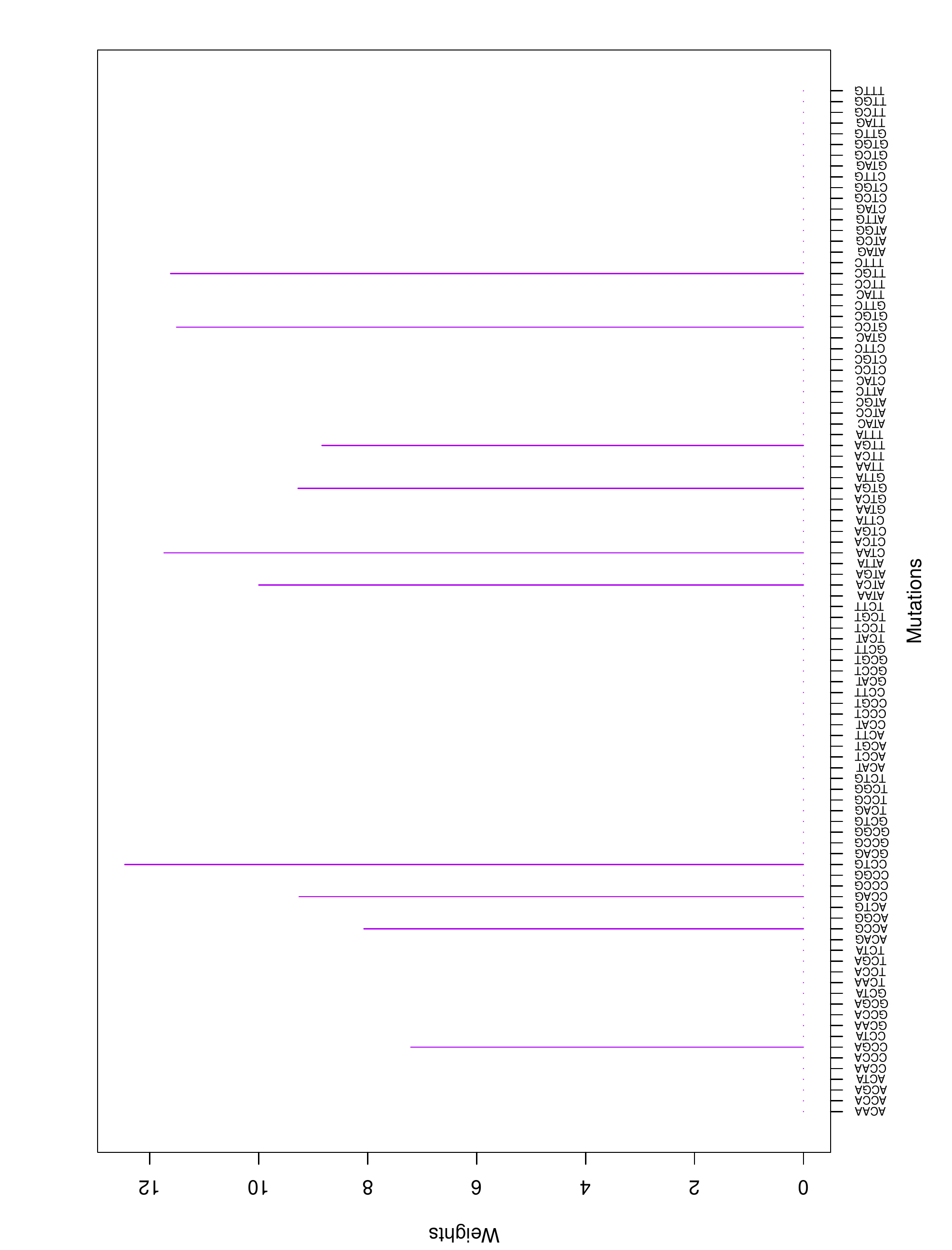}
\caption{Cluster Cl-3 in Clustering-A with weights based on unnormalized regressions with arithmetic means
(see Subsection \ref{sub.reg}).
See Tables \ref{table.occurrence.cts}, \ref{table.weights.A.1}, \ref{table.weights.A.2}.}
\label{Figure3A}
\end{figure}

\begin{figure}[ht]
\centering
\includegraphics[scale=0.7]{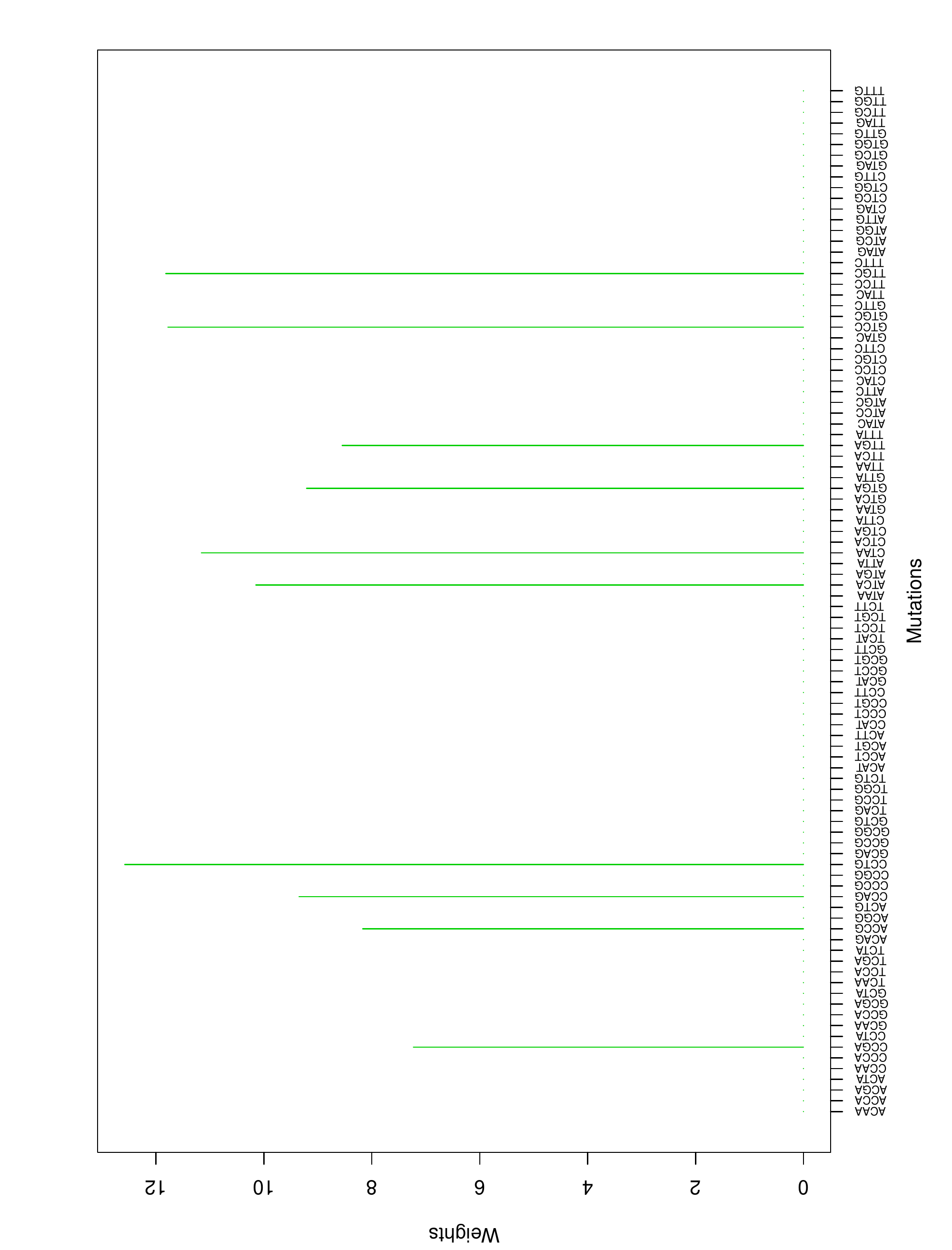}
\caption{Cluster Cl-3 in Clustering-A with weights based on normalized regressions with arithmetic means
(see Subsection \ref{sub.reg}).
See Tables \ref{table.occurrence.cts}, \ref{table.weights.A.1}, \ref{table.weights.A.2}.}
\label{FigureNorm3A}
\end{figure}

\begin{figure}[ht]
\centering
\includegraphics[scale=0.7]{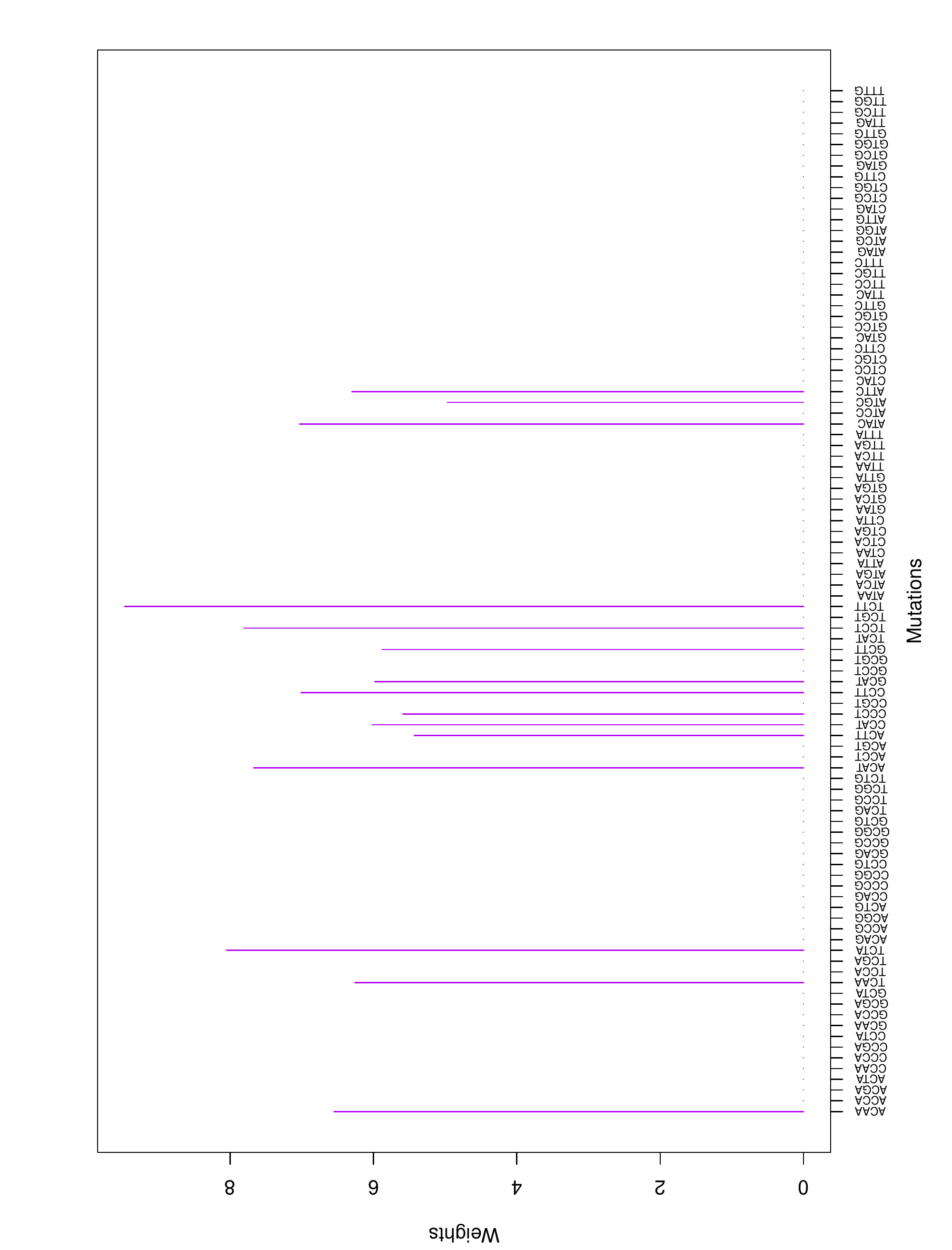}
\caption{Cluster Cl-4 in Clustering-A with weights based on unnormalized regressions with arithmetic means
(see Subsection \ref{sub.reg}).
See Tables \ref{table.occurrence.cts}, \ref{table.weights.A.1}, \ref{table.weights.A.2}.}
\label{Figure4A}
\end{figure}

\begin{figure}[ht]
\centering
\includegraphics[scale=0.7]{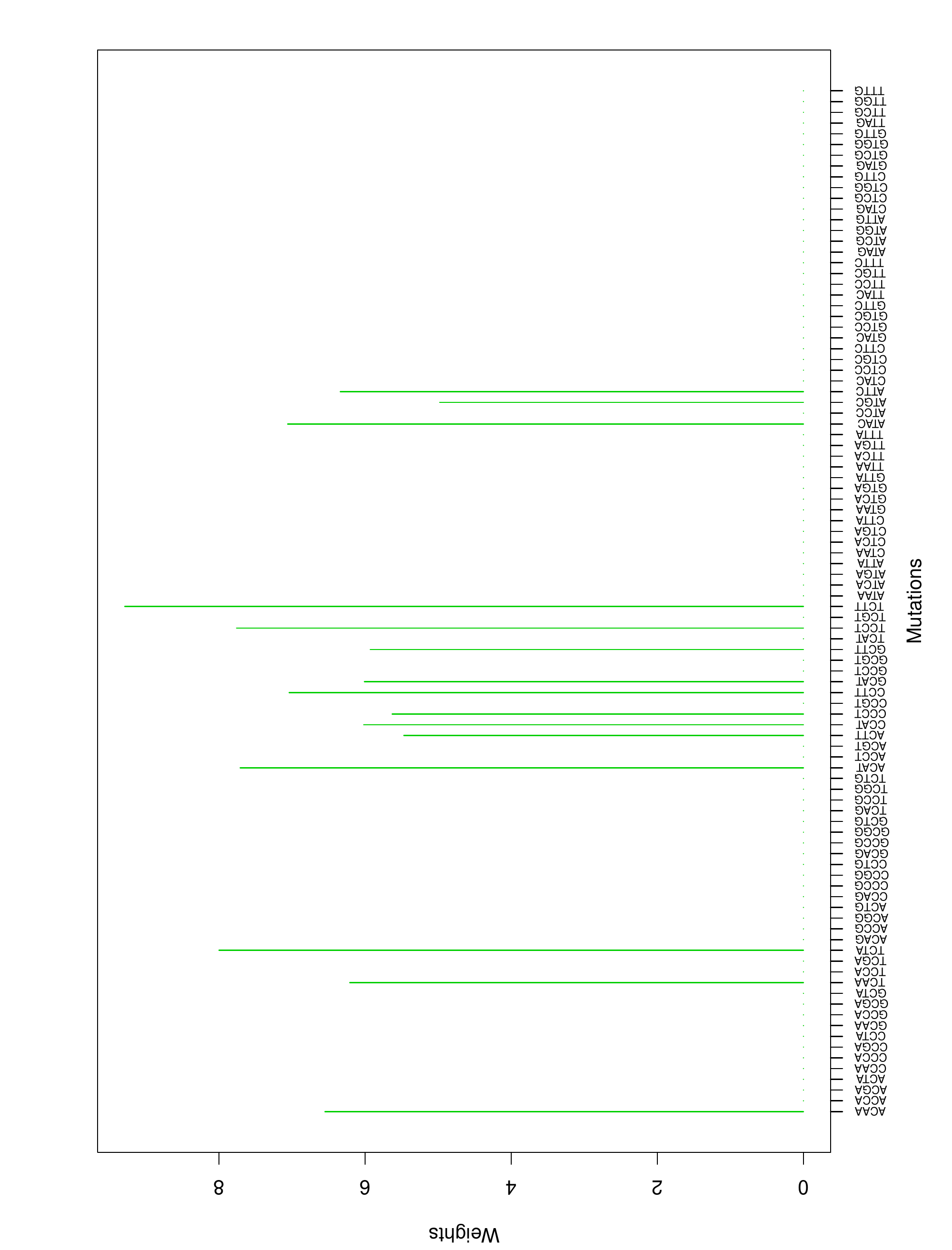}
\caption{Cluster Cl-4 in Clustering-A with weights based on normalized regressions with arithmetic means
(see Subsection \ref{sub.reg}).
See Tables \ref{table.occurrence.cts}, \ref{table.weights.A.1}, \ref{table.weights.A.2}.}
\label{FigureNorm4A}
\end{figure}

\begin{figure}[ht]
\centering
\includegraphics[scale=0.7]{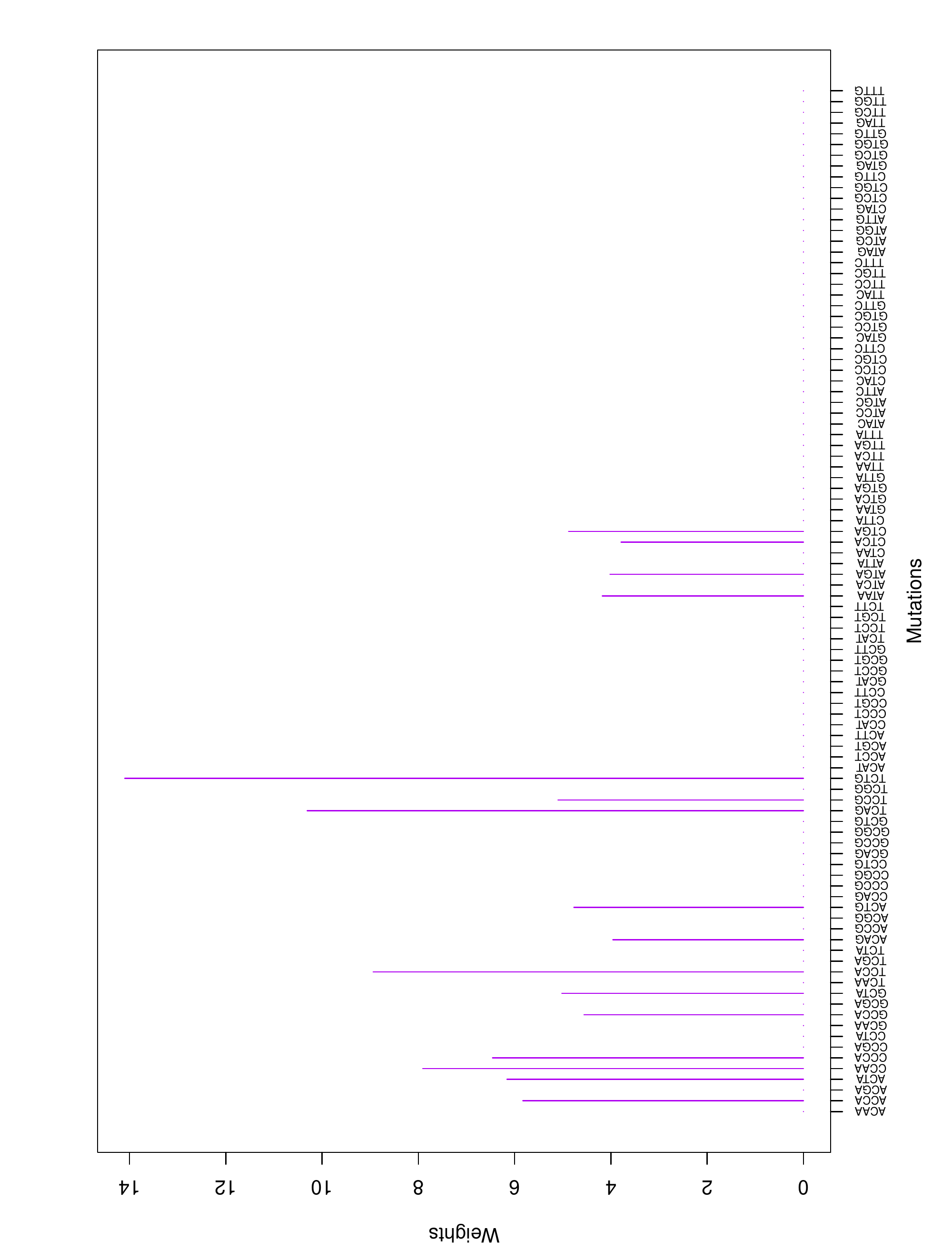}
\caption{Cluster Cl-5 in Clustering-A with weights based on unnormalized regressions with arithmetic means
(see Subsection \ref{sub.reg}).
See Tables \ref{table.occurrence.cts}, \ref{table.weights.A.1}, \ref{table.weights.A.2}.}
\label{Figure5A}
\end{figure}

\begin{figure}[ht]
\centering
\includegraphics[scale=0.7]{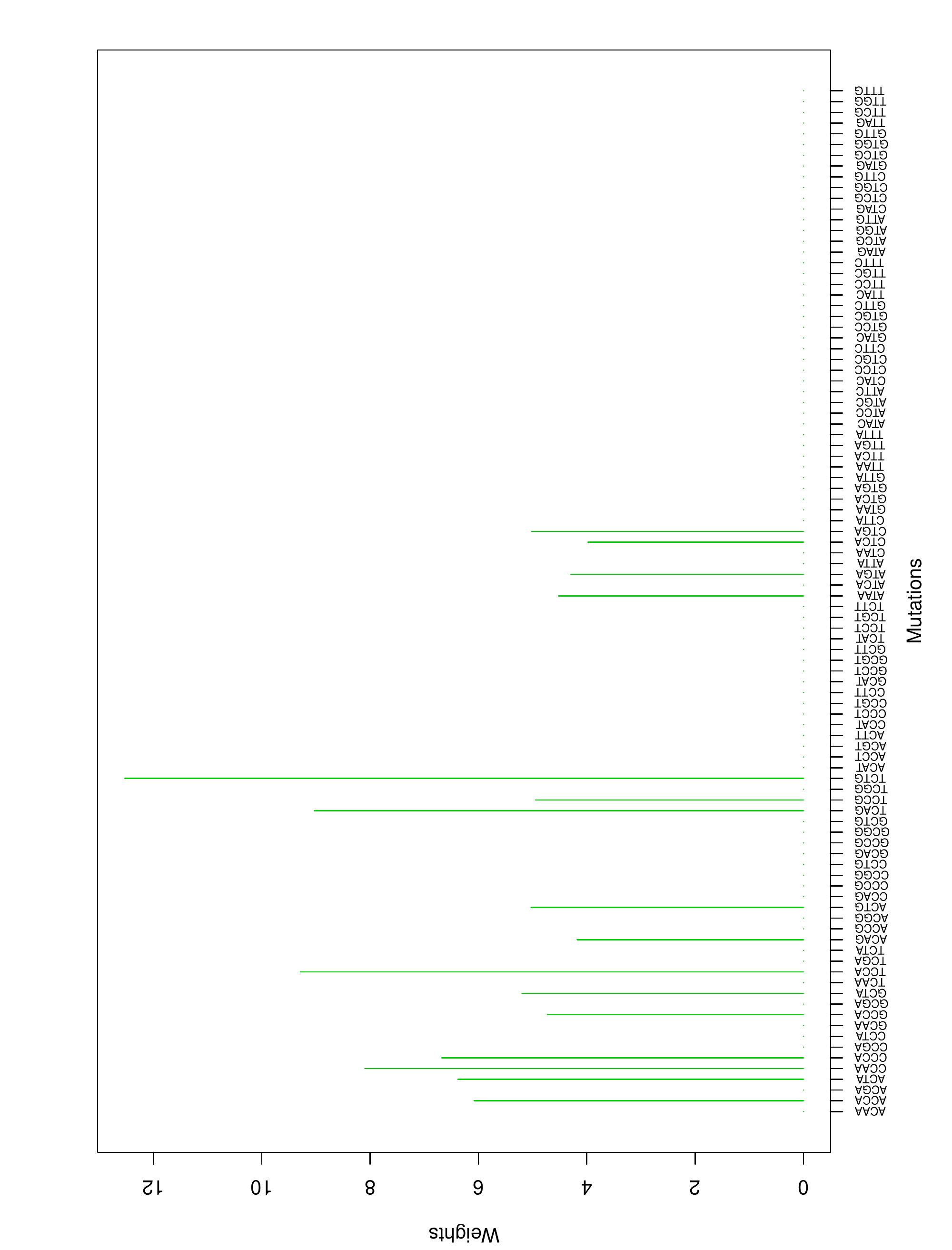}
\caption{Cluster Cl-5 in Clustering-A with weights based on normalized regressions with arithmetic means
(see Subsection \ref{sub.reg}).
See Tables \ref{table.occurrence.cts}, \ref{table.weights.A.1}, \ref{table.weights.A.2}.}
\label{FigureNorm5A}
\end{figure}

\begin{figure}[ht]
\centering
\includegraphics[scale=0.7]{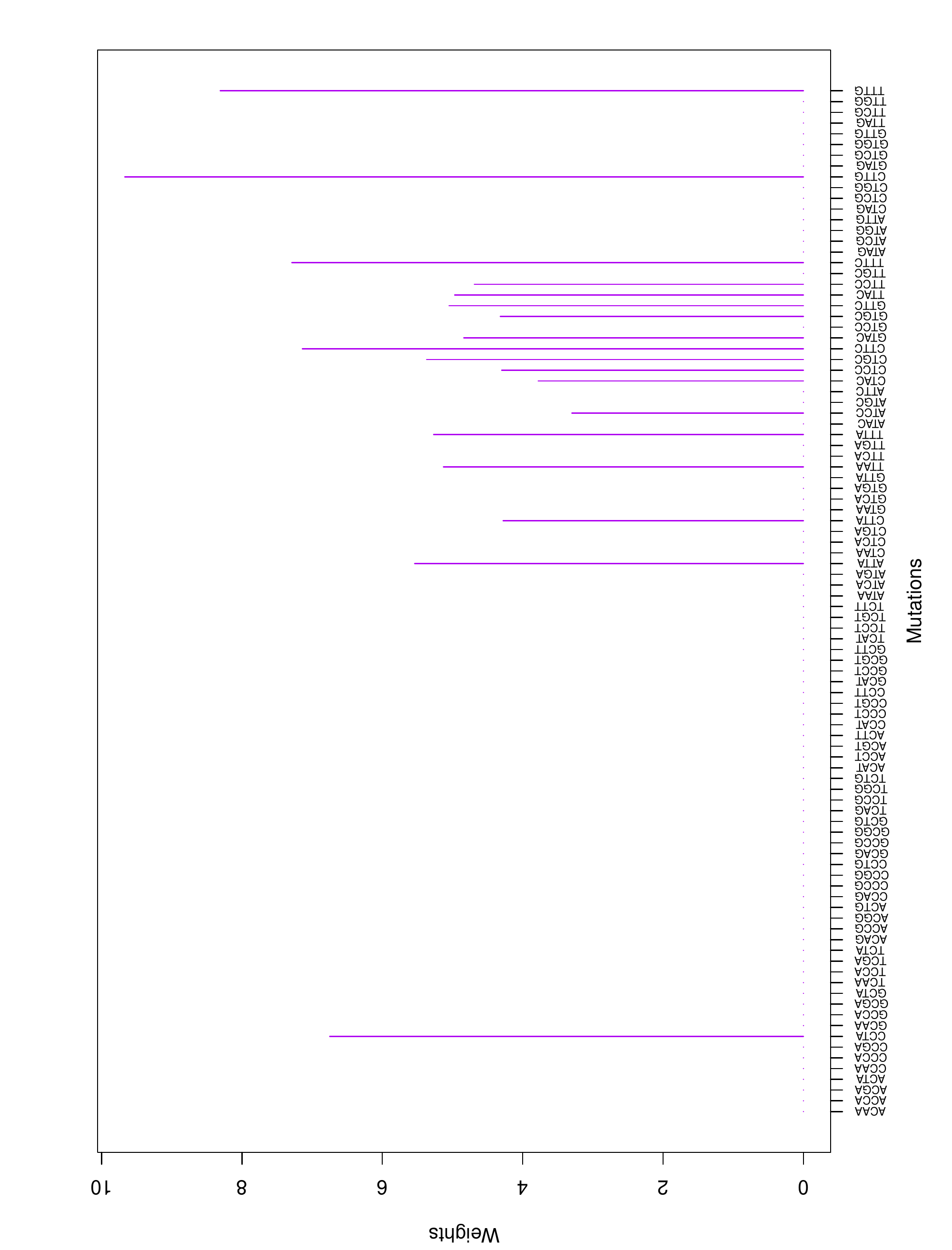}
\caption{Cluster Cl-6 in Clustering-A with weights based on unnormalized regressions with arithmetic means
(see Subsection \ref{sub.reg}).
See Tables \ref{table.occurrence.cts}, \ref{table.weights.A.1}, \ref{table.weights.A.2}.}
\label{Figure6A}
\end{figure}

\begin{figure}[ht]
\centering
\includegraphics[scale=0.7]{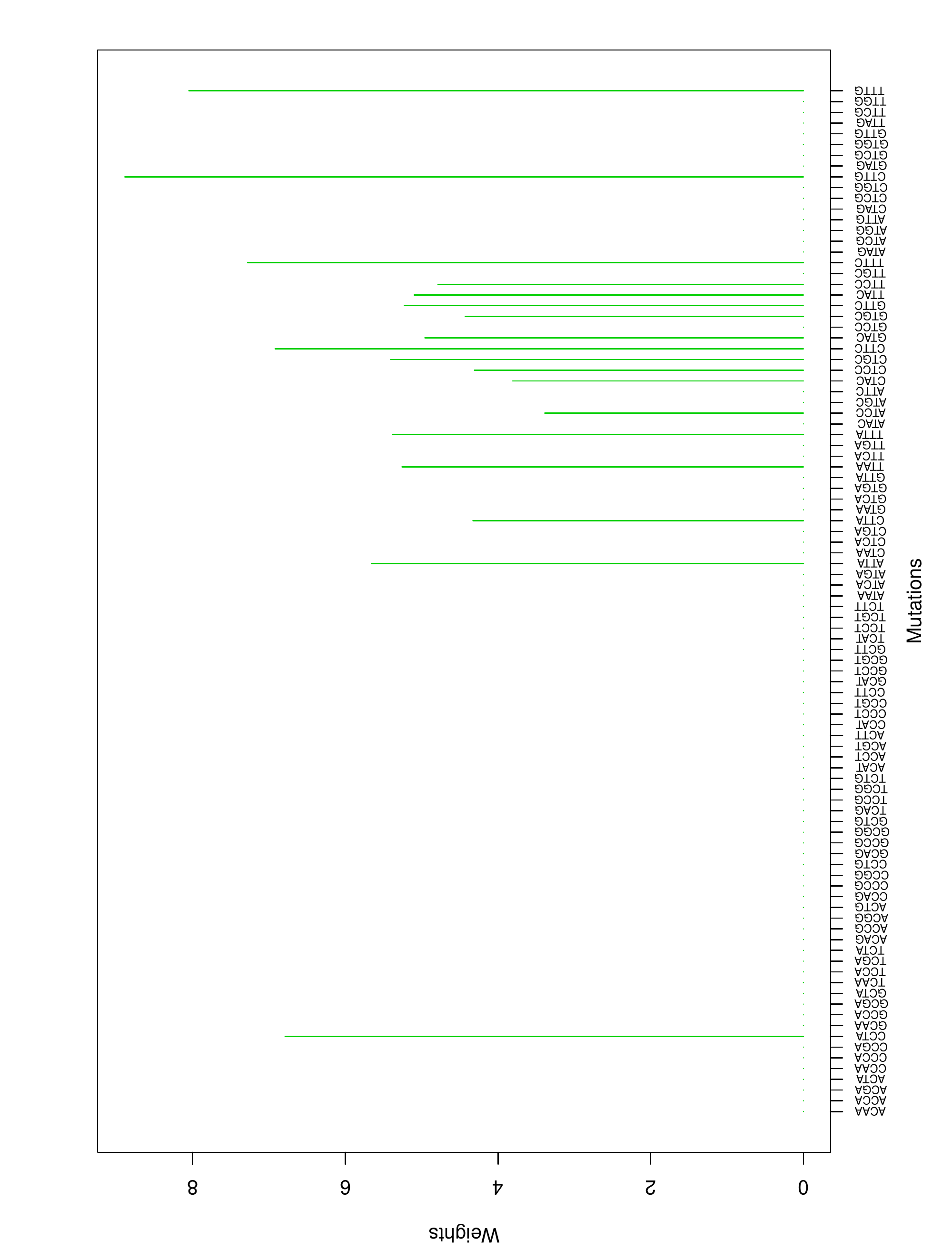}
\caption{Cluster Cl-6 in Clustering-A with weights based on normalized regressions with arithmetic means
(see Subsection \ref{sub.reg}).
See Tables \ref{table.occurrence.cts}, \ref{table.weights.A.1}, \ref{table.weights.A.2}.}
\label{FigureNorm6A}
\end{figure}

\begin{figure}[ht]
\centering
\includegraphics[scale=0.7]{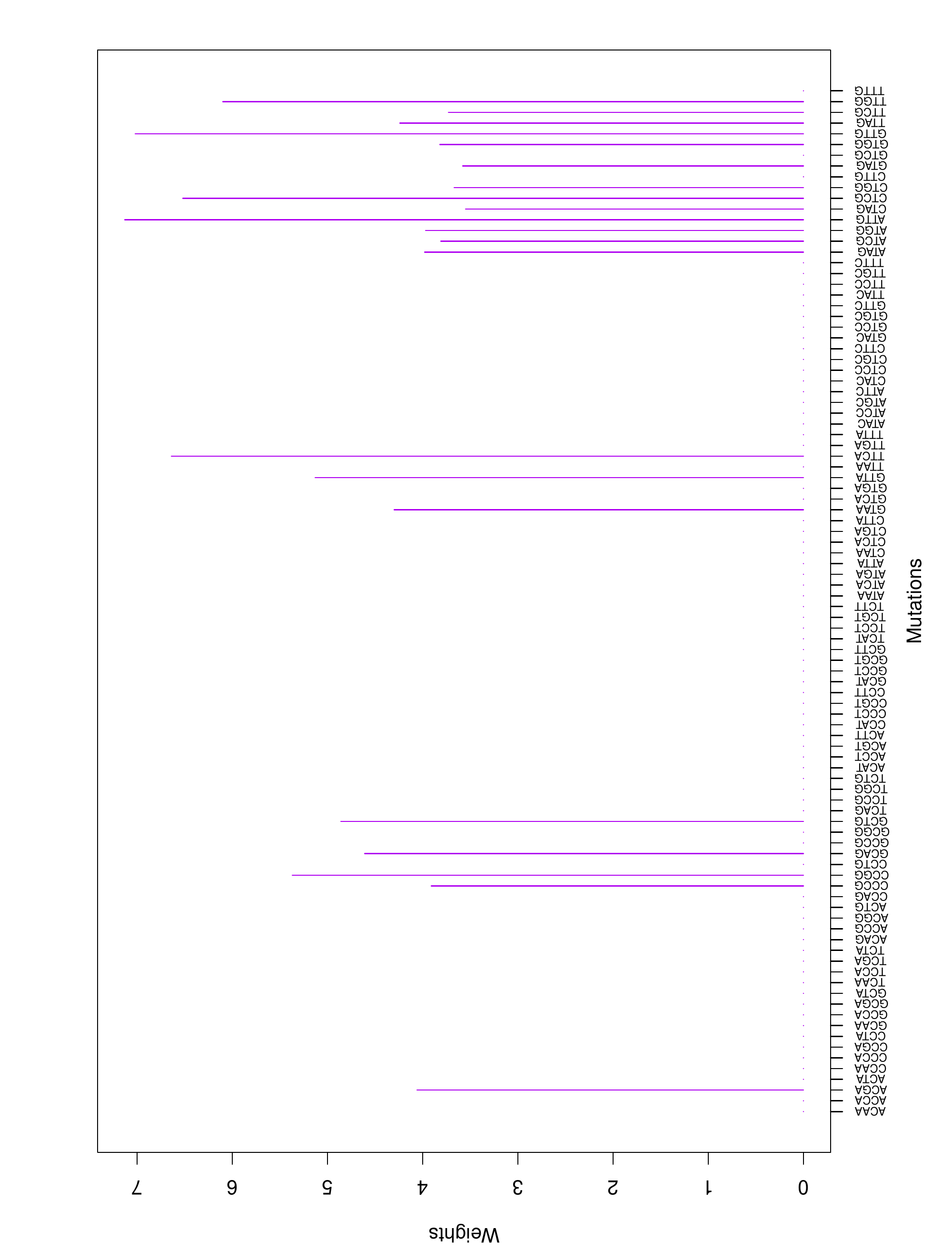}
\caption{Cluster Cl-7 in Clustering-A with weights based on unnormalized regressions with arithmetic means
(see Subsection \ref{sub.reg}).
See Tables \ref{table.occurrence.cts}, \ref{table.weights.A.1}, \ref{table.weights.A.2}.}
\label{Figure7A}
\end{figure}

\begin{figure}[ht]
\centering
\includegraphics[scale=0.7]{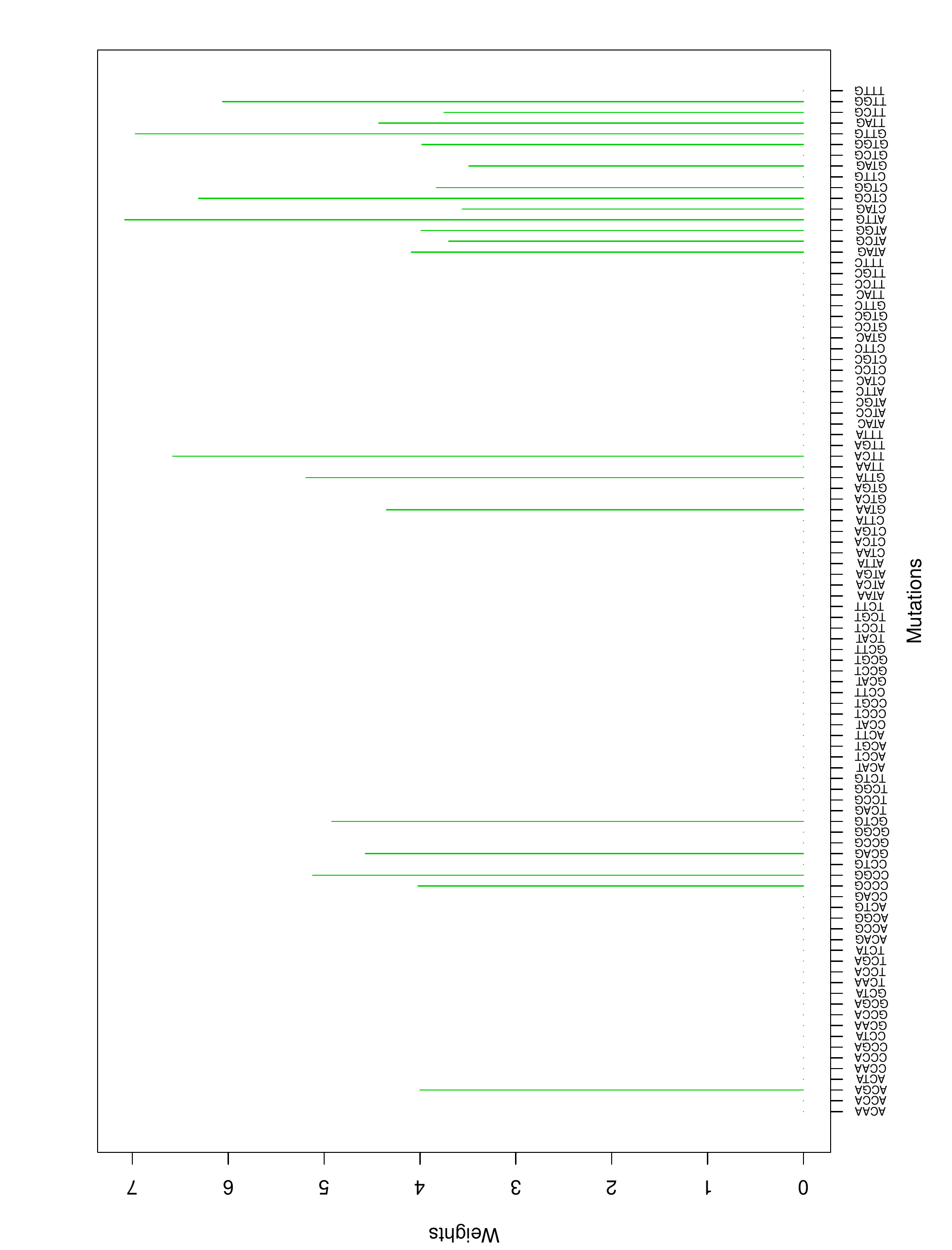}
\caption{Cluster Cl-7 in Clustering-A with weights based on normalized regressions with arithmetic means
(see Subsection \ref{sub.reg}).
See Tables \ref{table.occurrence.cts}, \ref{table.weights.A.1}, \ref{table.weights.A.2}.}
\label{FigureNorm7A}
\end{figure}

\begin{figure}[ht]
\centering
\includegraphics[scale=0.7]{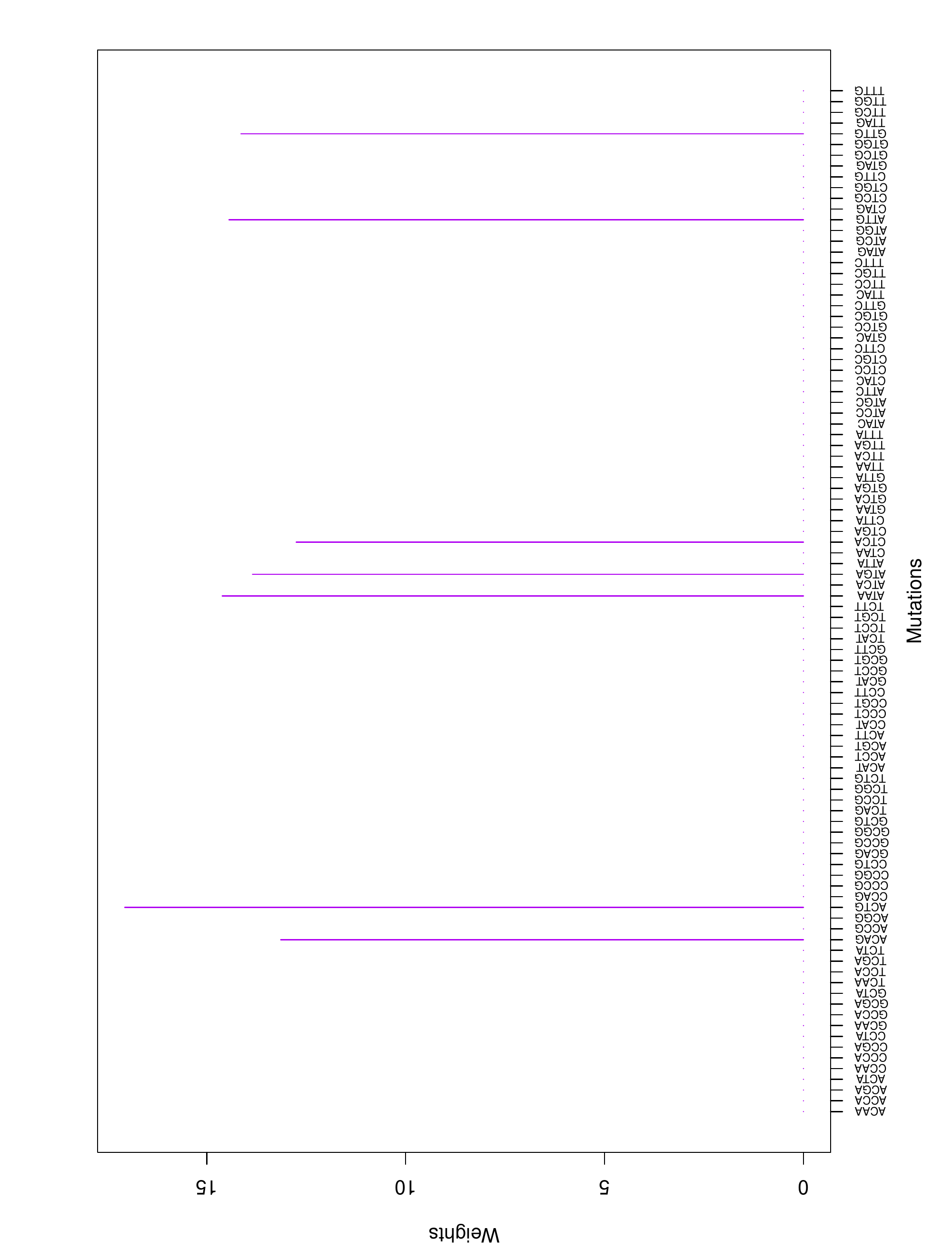}
\caption{Cluster Cl-1 in Clustering-B with weights based on unnormalized regressions with arithmetic means
(see Subsection \ref{sub.reg}).
See Tables \ref{table.occurrence.cts}, \ref{table.weights.B.1}, \ref{table.weights.B.2}.}
\label{Figure1B}
\end{figure}

\begin{figure}[ht]
\centering
\includegraphics[scale=0.7]{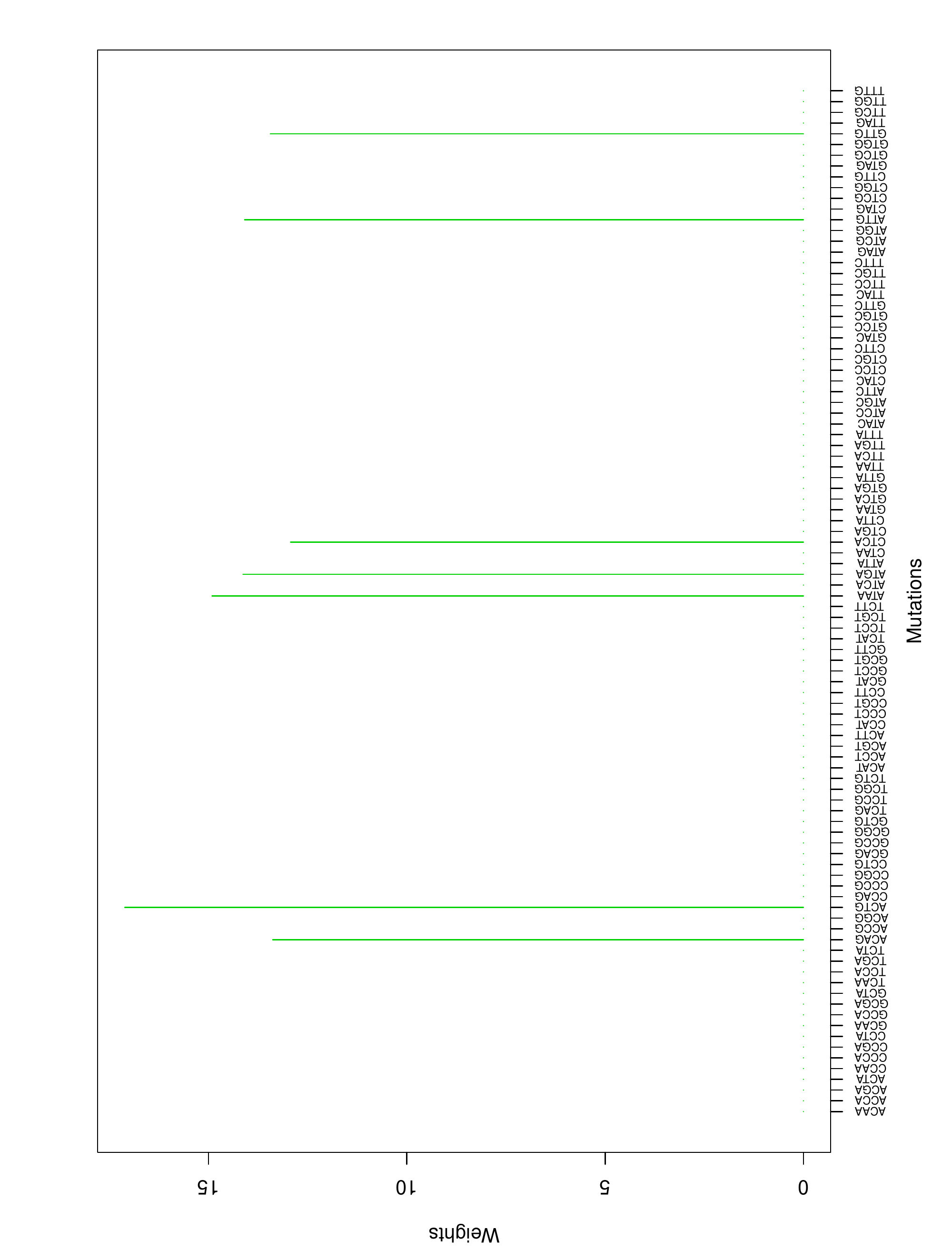}
\caption{Cluster Cl-1 in Clustering-B with weights based on normalized regressions with arithmetic means
(see Subsection \ref{sub.reg}).
See Tables \ref{table.occurrence.cts}, \ref{table.weights.B.1}, \ref{table.weights.B.2}.}
\label{FigureNorm1B}
\end{figure}

\begin{figure}[ht]
\centering
\includegraphics[scale=0.7]{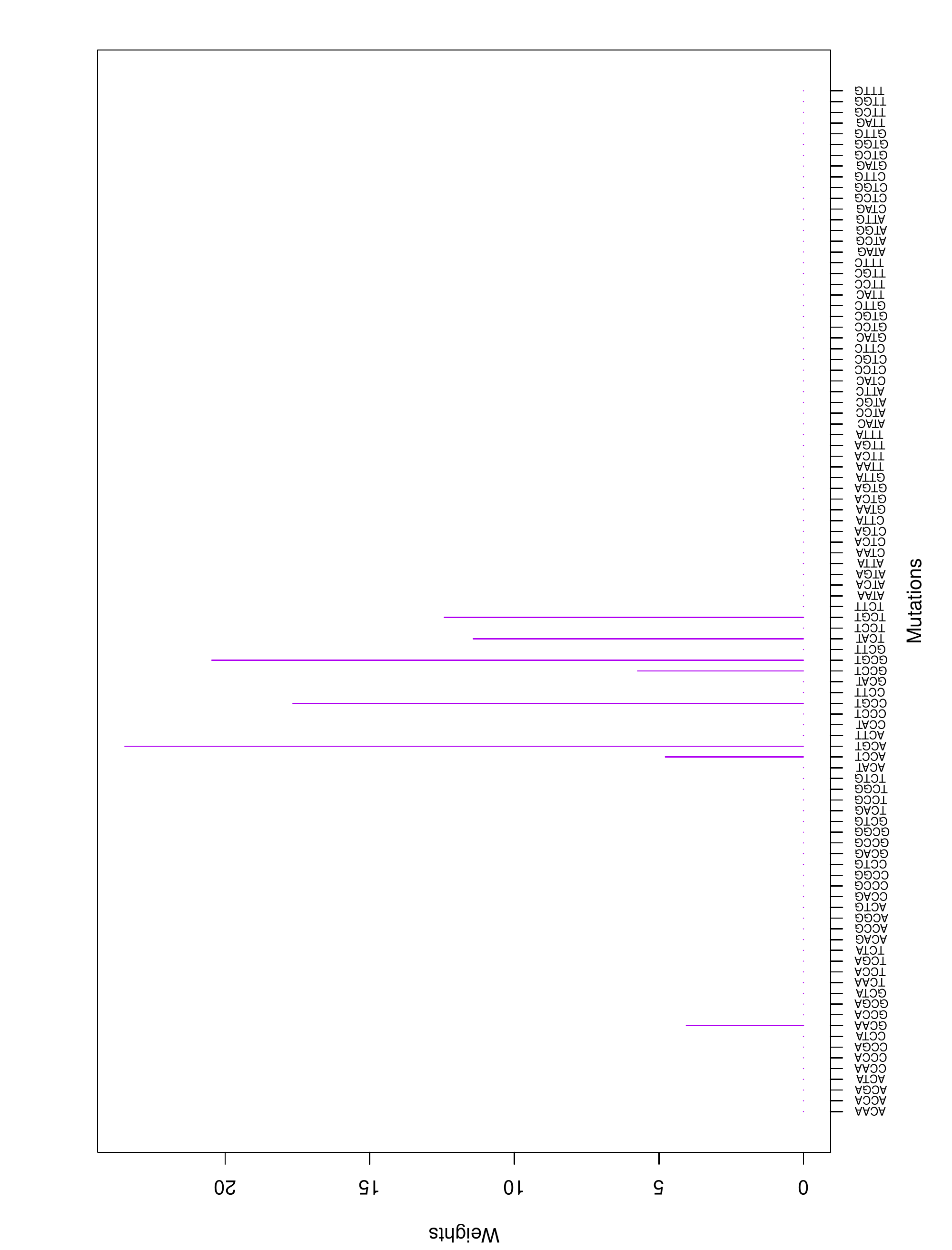}
\caption{Cluster Cl-2 in Clustering-B with weights based on unnormalized regressions with arithmetic means
(see Subsection \ref{sub.reg}).
See Tables \ref{table.occurrence.cts}, \ref{table.weights.B.1}, \ref{table.weights.B.2}.}
\label{Figure2B}
\end{figure}

\begin{figure}[ht]
\centering
\includegraphics[scale=0.7]{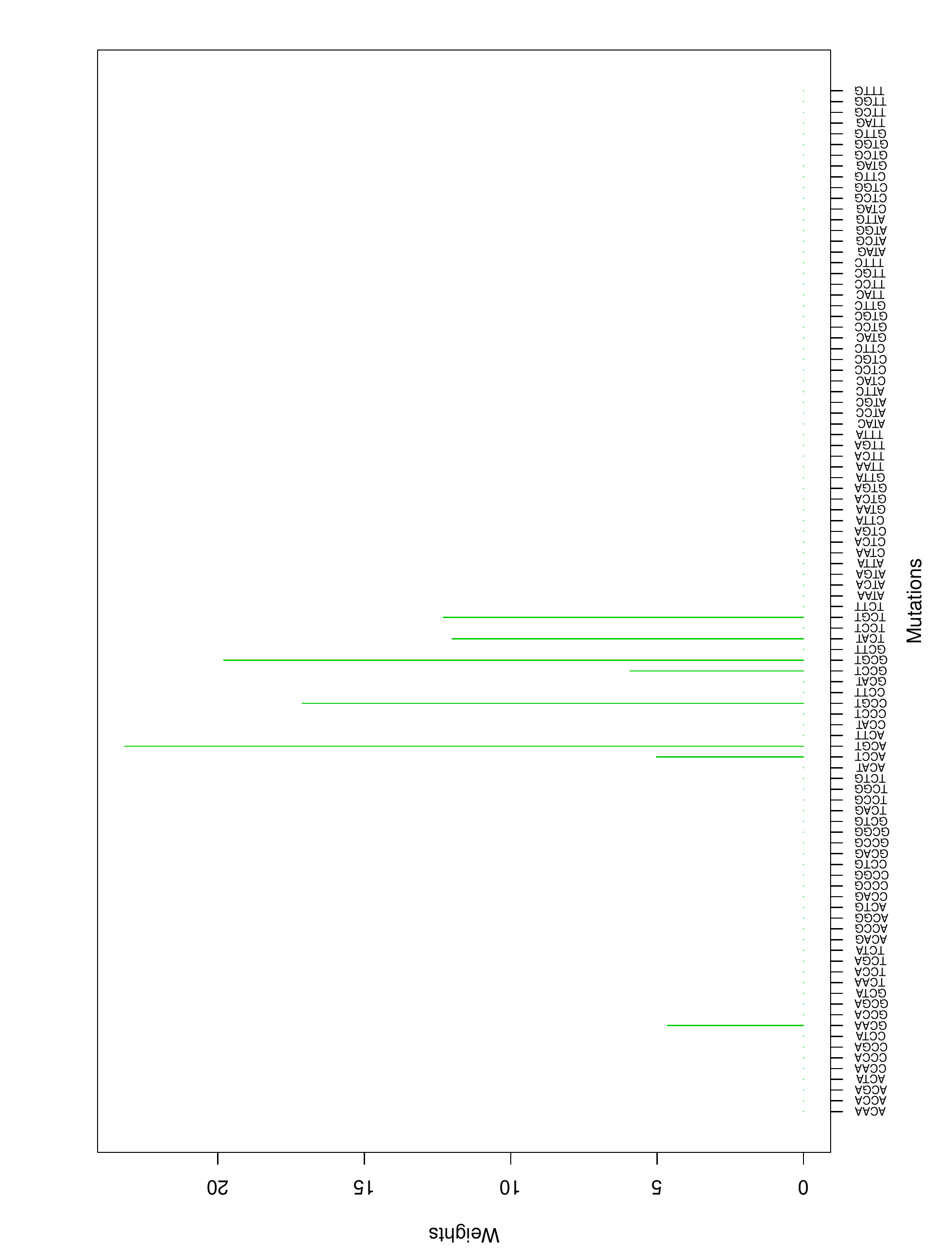}
\caption{Cluster Cl-2 in Clustering-B with weights based on normalized regressions with arithmetic means
(see Subsection \ref{sub.reg}).
See Tables \ref{table.occurrence.cts}, \ref{table.weights.B.1}, \ref{table.weights.B.2}.}
\label{FigureNorm2B}
\end{figure}

\begin{figure}[ht]
\centering
\includegraphics[scale=0.7]{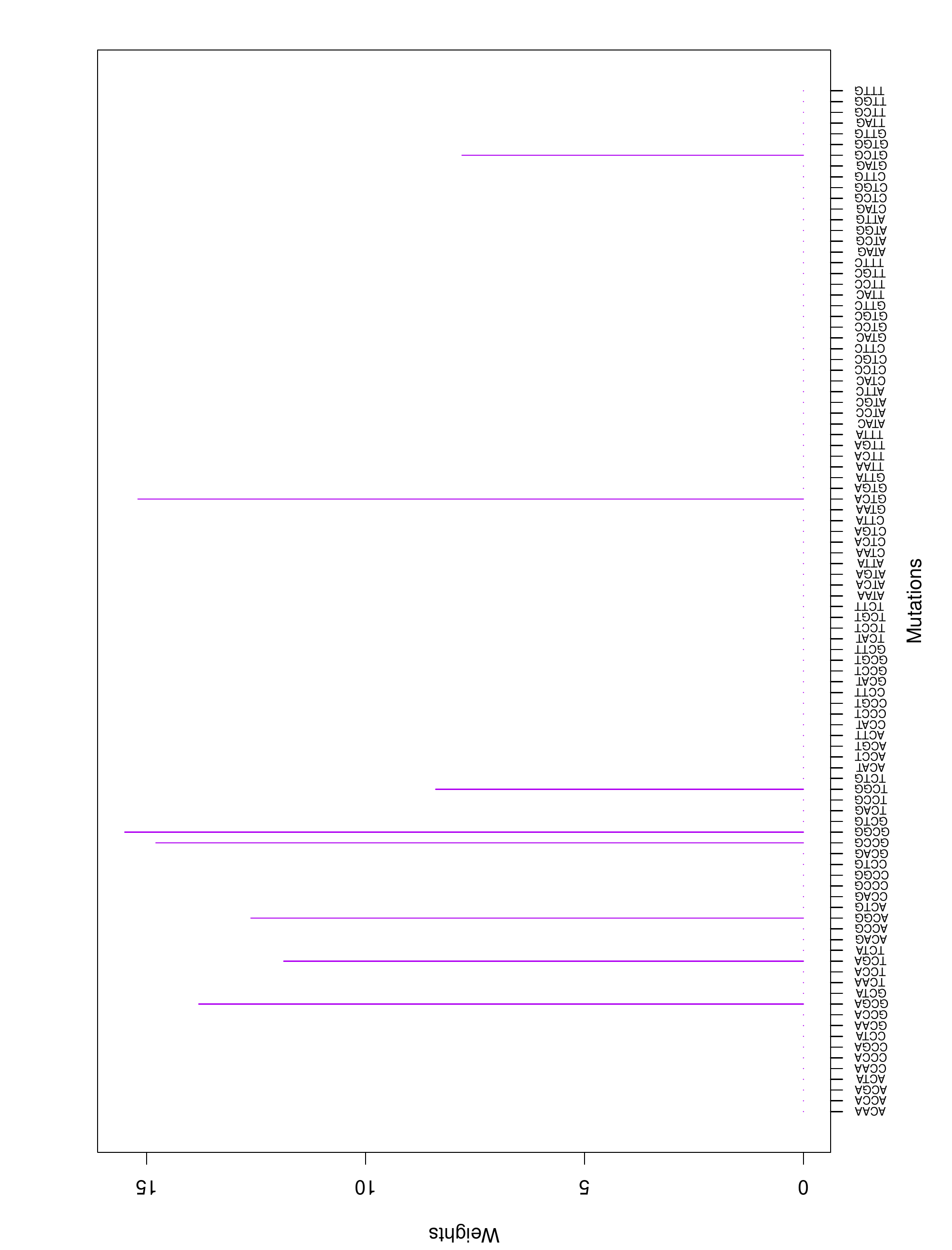}
\caption{Cluster Cl-3 in Clustering-B with weights based on unnormalized regressions with arithmetic means
(see Subsection \ref{sub.reg}).
See Tables \ref{table.occurrence.cts}, \ref{table.weights.B.1}, \ref{table.weights.B.2}.}
\label{Figure3B}
\end{figure}

\begin{figure}[ht]
\centering
\includegraphics[scale=0.7]{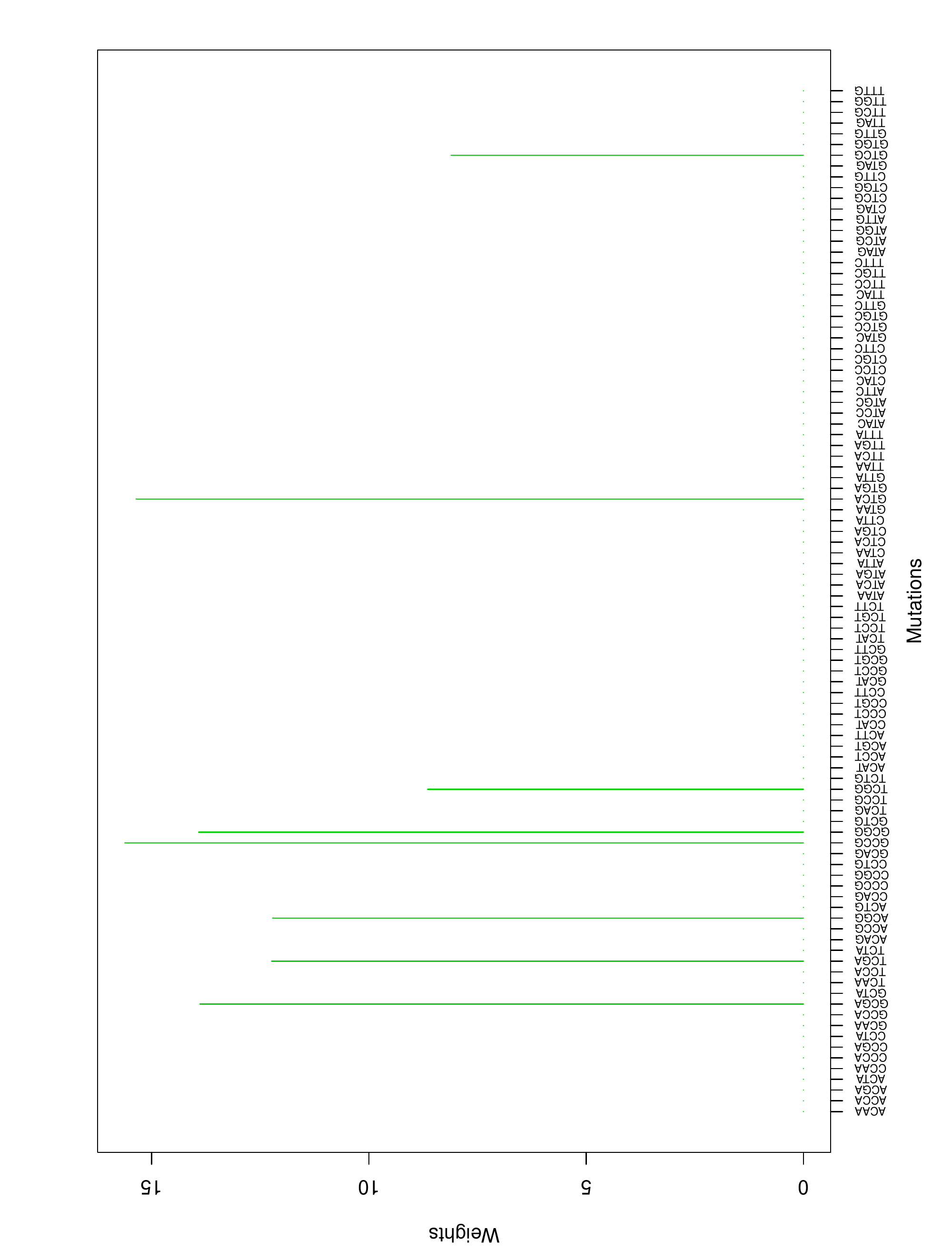}
\caption{Cluster Cl-3 in Clustering-B with weights based on normalized regressions with arithmetic means
(see Subsection \ref{sub.reg}).
See Tables \ref{table.occurrence.cts}, \ref{table.weights.B.1}, \ref{table.weights.B.2}.}
\label{FigureNorm3B}
\end{figure}

\begin{figure}[ht]
\centering
\includegraphics[scale=0.7]{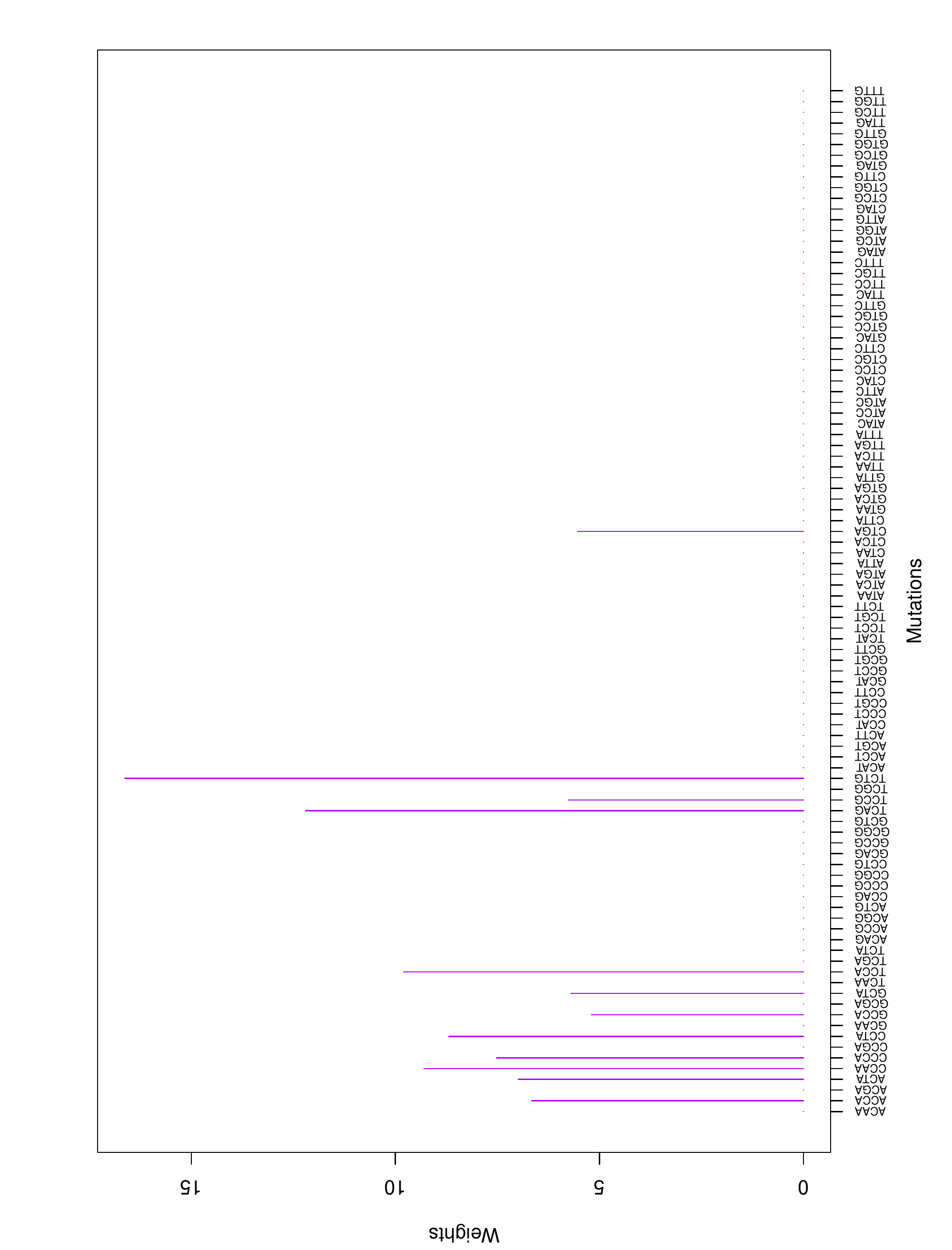}
\caption{Cluster Cl-4 in Clustering-B with weights based on unnormalized regressions with arithmetic means
(see Subsection \ref{sub.reg}).
See Tables \ref{table.occurrence.cts}, \ref{table.weights.B.1}, \ref{table.weights.B.2}.}
\label{Figure4B}
\end{figure}

\begin{figure}[ht]
\centering
\includegraphics[scale=0.7]{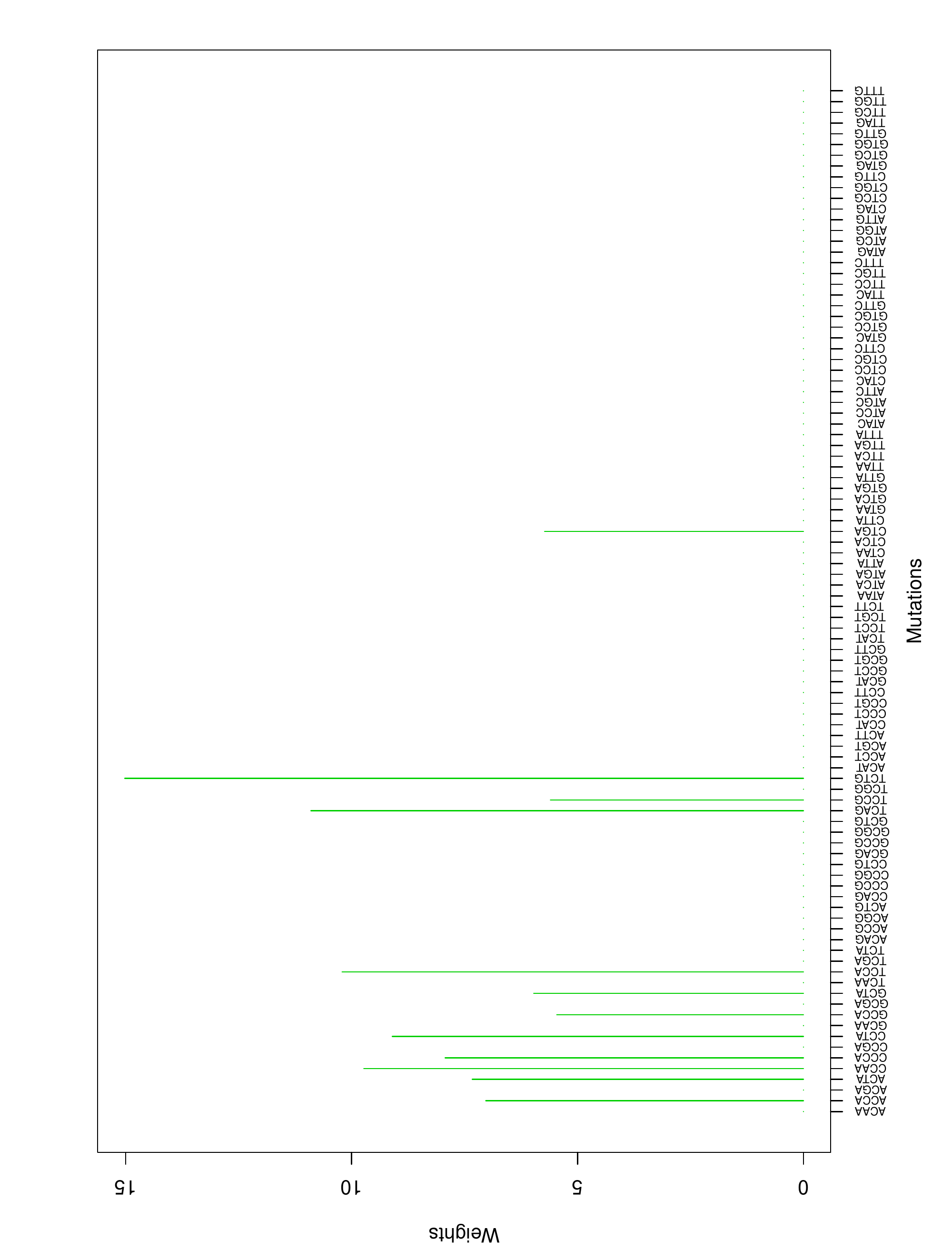}
\caption{Cluster Cl-4 in Clustering-B with weights based on normalized regressions with arithmetic means
(see Subsection \ref{sub.reg}).
See Tables \ref{table.occurrence.cts}, \ref{table.weights.B.1}, \ref{table.weights.B.2}.}
\label{FigureNorm4B}
\end{figure}

\begin{figure}[ht]
\centering
\includegraphics[scale=0.7]{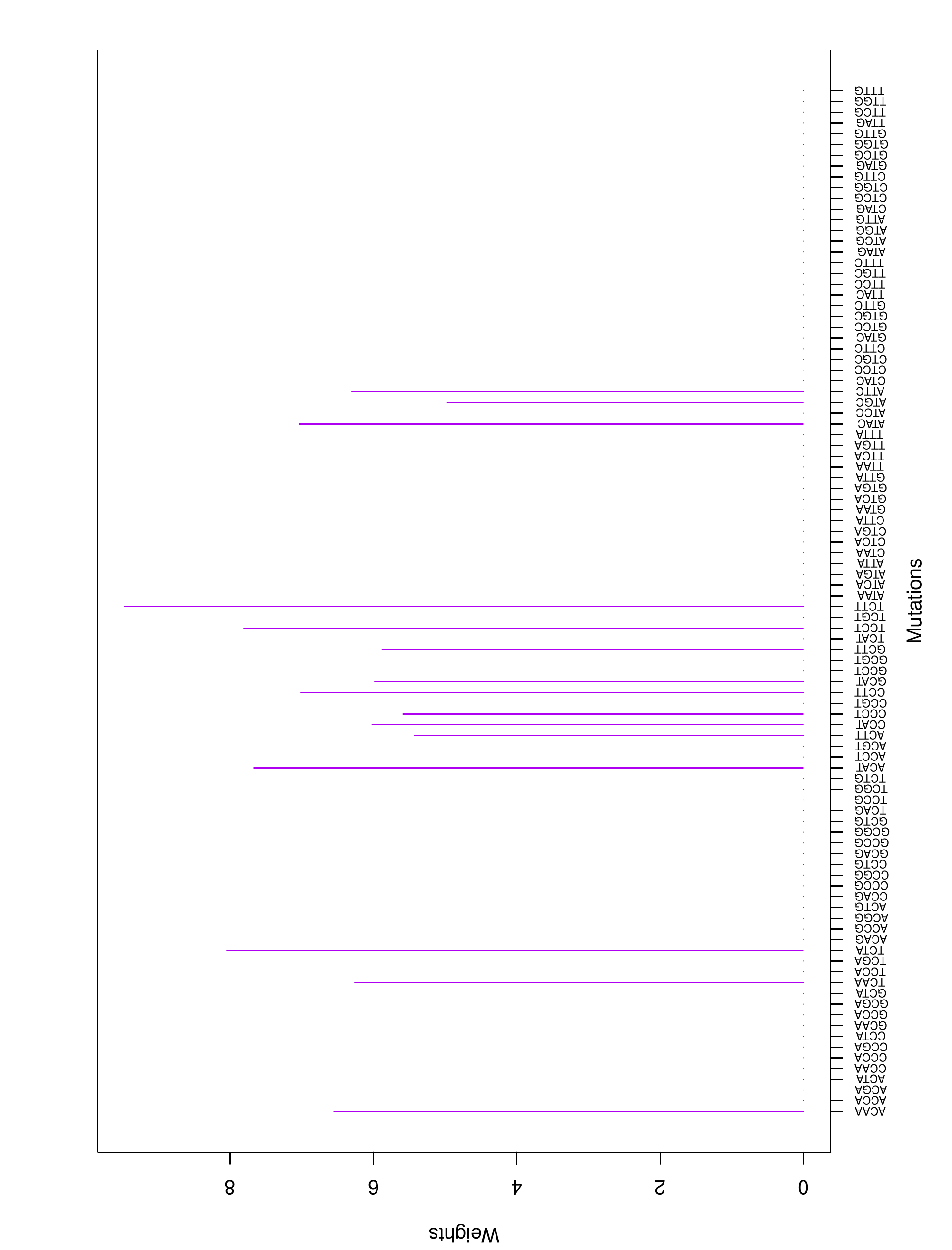}
\caption{Cluster Cl-5 in Clustering-B with weights based on unnormalized regressions with arithmetic means
(see Subsection \ref{sub.reg}).
See Tables \ref{table.occurrence.cts}, \ref{table.weights.B.1}, \ref{table.weights.B.2}.}
\label{Figure5B}
\end{figure}

\begin{figure}[ht]
\centering
\includegraphics[scale=0.7]{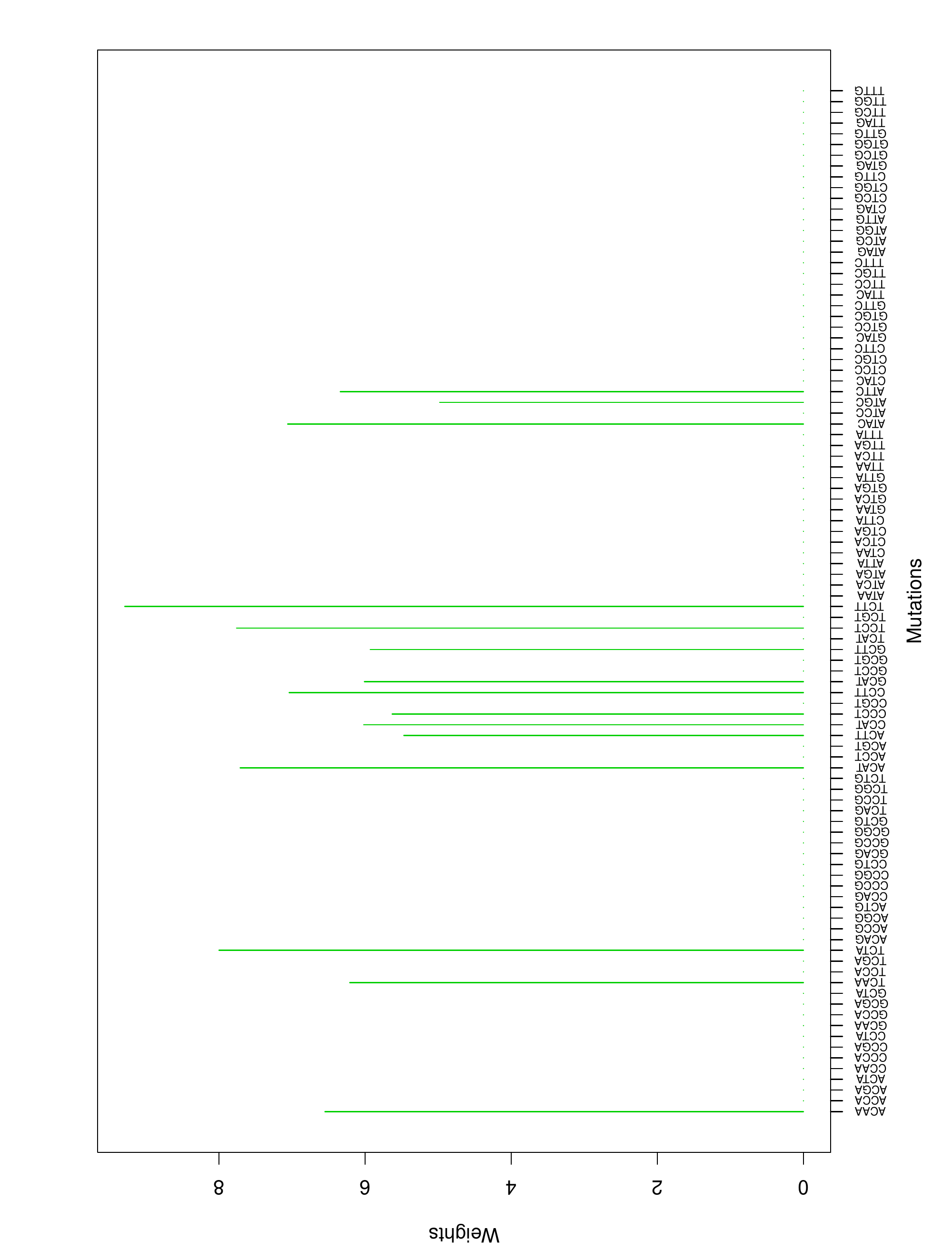}
\caption{Cluster Cl-5 in Clustering-B with weights based on normalized regressions with arithmetic means
(see Subsection \ref{sub.reg}).
See Tables \ref{table.occurrence.cts}, \ref{table.weights.B.1}, \ref{table.weights.B.2}.}
\label{FigureNorm5B}
\end{figure}

\begin{figure}[ht]
\centering
\includegraphics[scale=0.7]{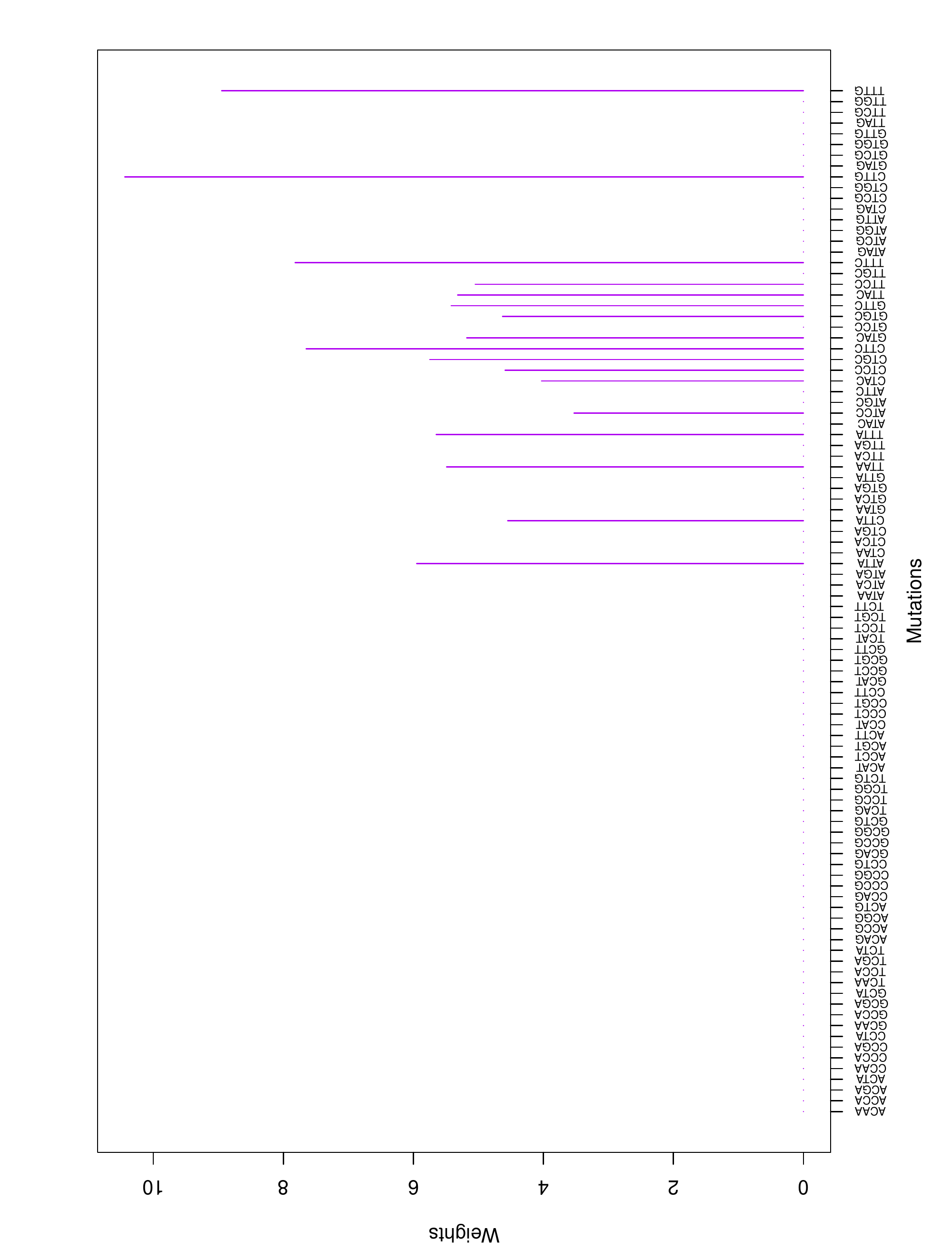}
\caption{Cluster Cl-6 in Clustering-B with weights based on unnormalized regressions with arithmetic means
(see Subsection \ref{sub.reg}).
See Tables \ref{table.occurrence.cts}, \ref{table.weights.B.1}, \ref{table.weights.B.2}.}
\label{Figure6B}
\end{figure}

\begin{figure}[ht]
\centering
\includegraphics[scale=0.7]{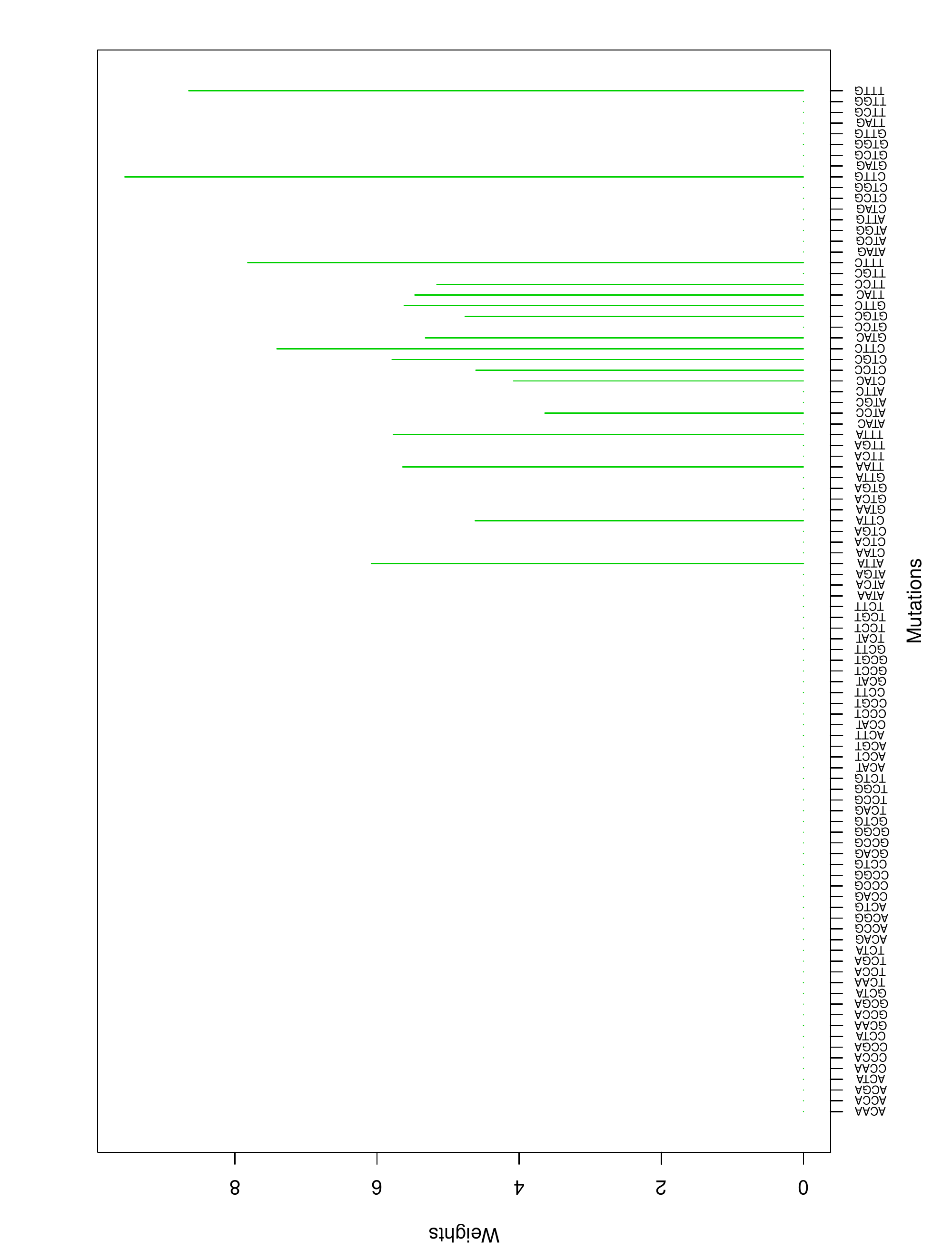}
\caption{Cluster Cl-6 in Clustering-B with weights based on normalized regressions with arithmetic means
(see Subsection \ref{sub.reg}).
See Tables \ref{table.occurrence.cts}, \ref{table.weights.B.1}, \ref{table.weights.B.2}.}
\label{FigureNorm6B}
\end{figure}

\begin{figure}[ht]
\centering
\includegraphics[scale=0.7]{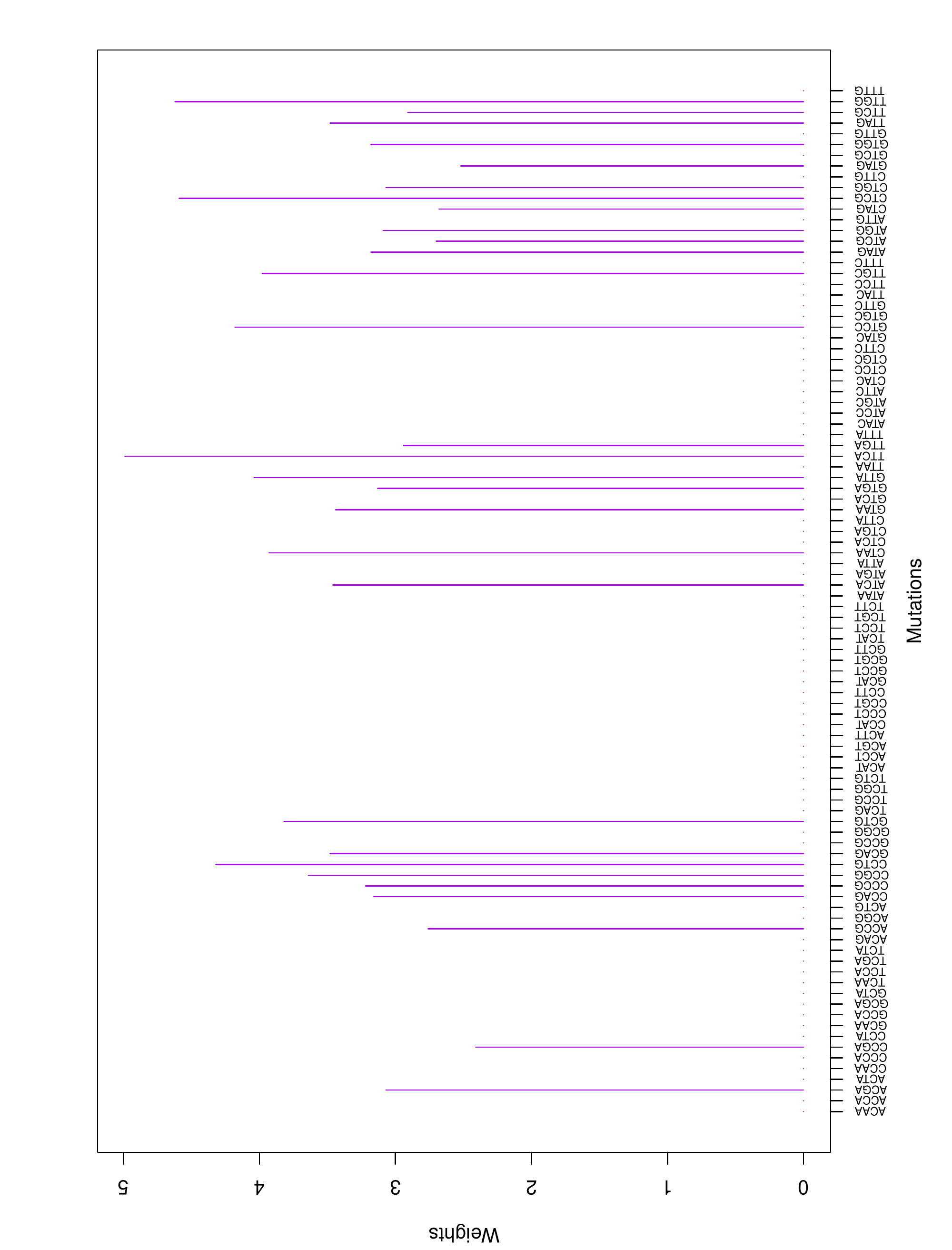}
\caption{Cluster Cl-7 in Clustering-B with weights based on unnormalized regressions with arithmetic means
(see Subsection \ref{sub.reg}).
See Tables \ref{table.occurrence.cts}, \ref{table.weights.B.1}, \ref{table.weights.B.2}.}
\label{Figure7B}
\end{figure}

\begin{figure}[ht]
\centering
\includegraphics[scale=0.7]{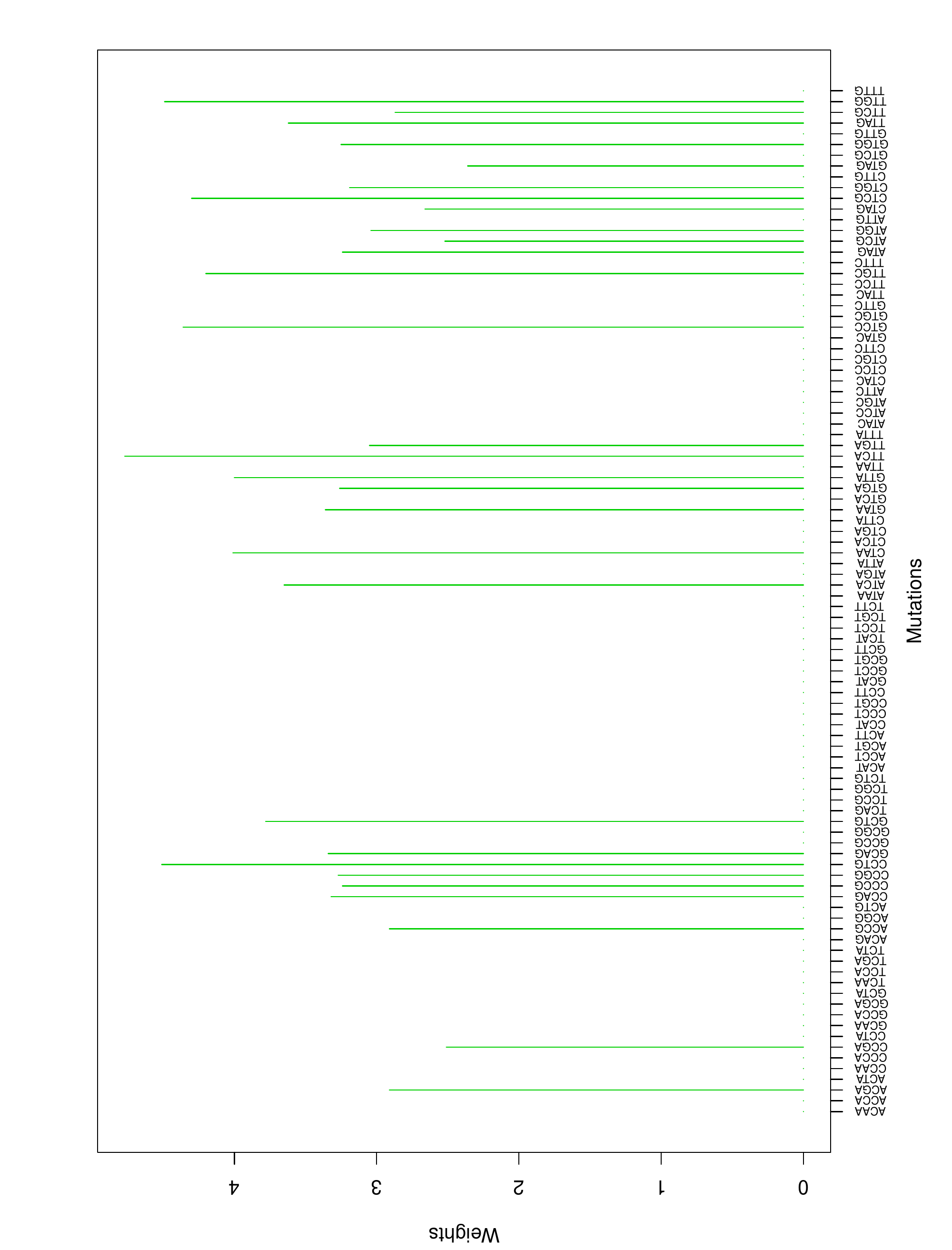}
\caption{Cluster Cl-7 in Clustering-B with weights based on normalized regressions with arithmetic means
(see Subsection \ref{sub.reg}).
See Tables \ref{table.occurrence.cts}, \ref{table.weights.B.1}, \ref{table.weights.B.2}.}
\label{FigureNorm7B}
\end{figure}

\begin{figure}[ht]
\centering
\includegraphics[scale=0.7]{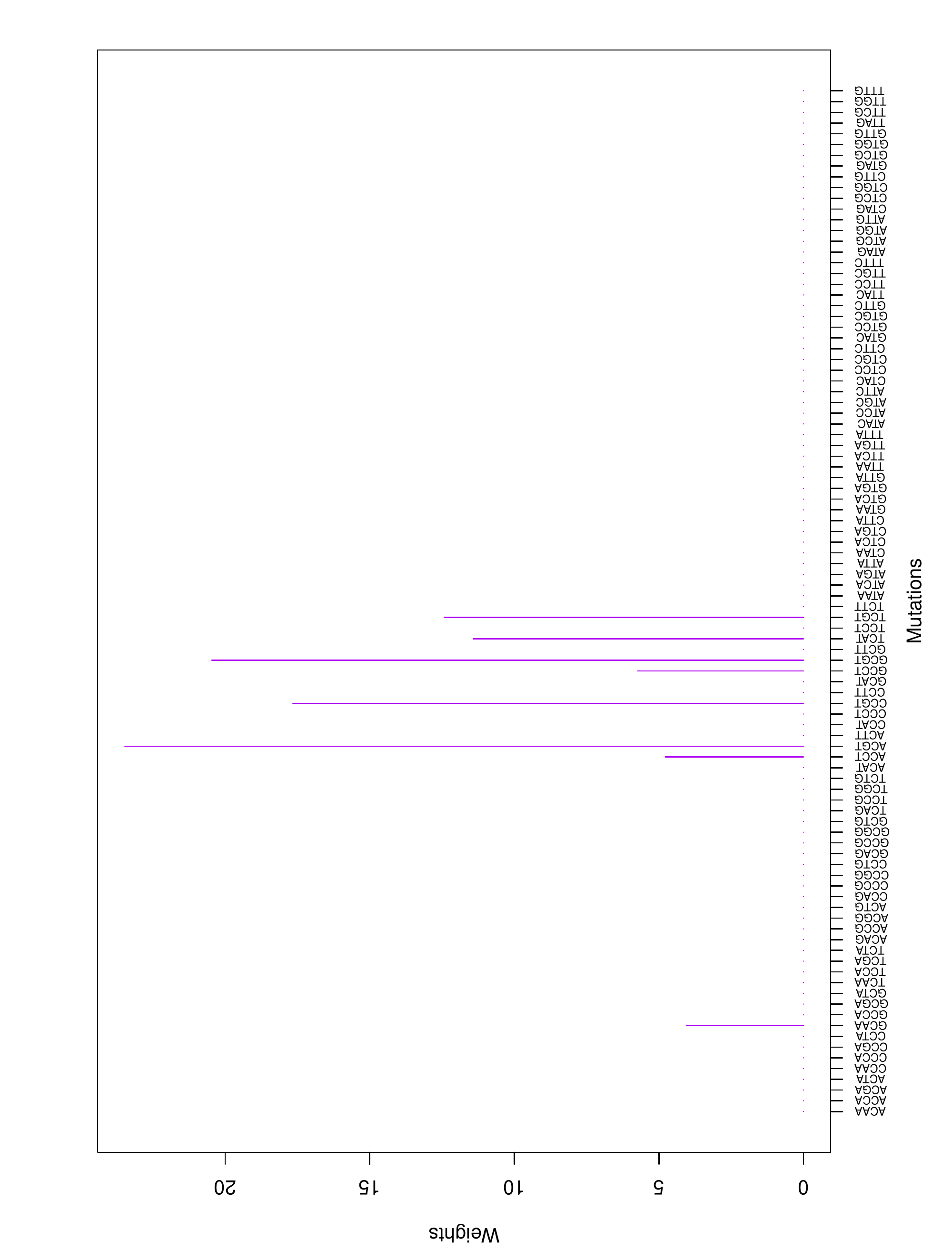}
\caption{Cluster Cl-1 in Clustering-C with weights based on unnormalized regressions with arithmetic means
(see Subsection \ref{sub.reg}).
See Tables \ref{table.occurrence.cts}, \ref{table.weights.C.1}, \ref{table.weights.C.2}.}
\label{Figure1C}
\end{figure}

\begin{figure}[ht]
\centering
\includegraphics[scale=0.7]{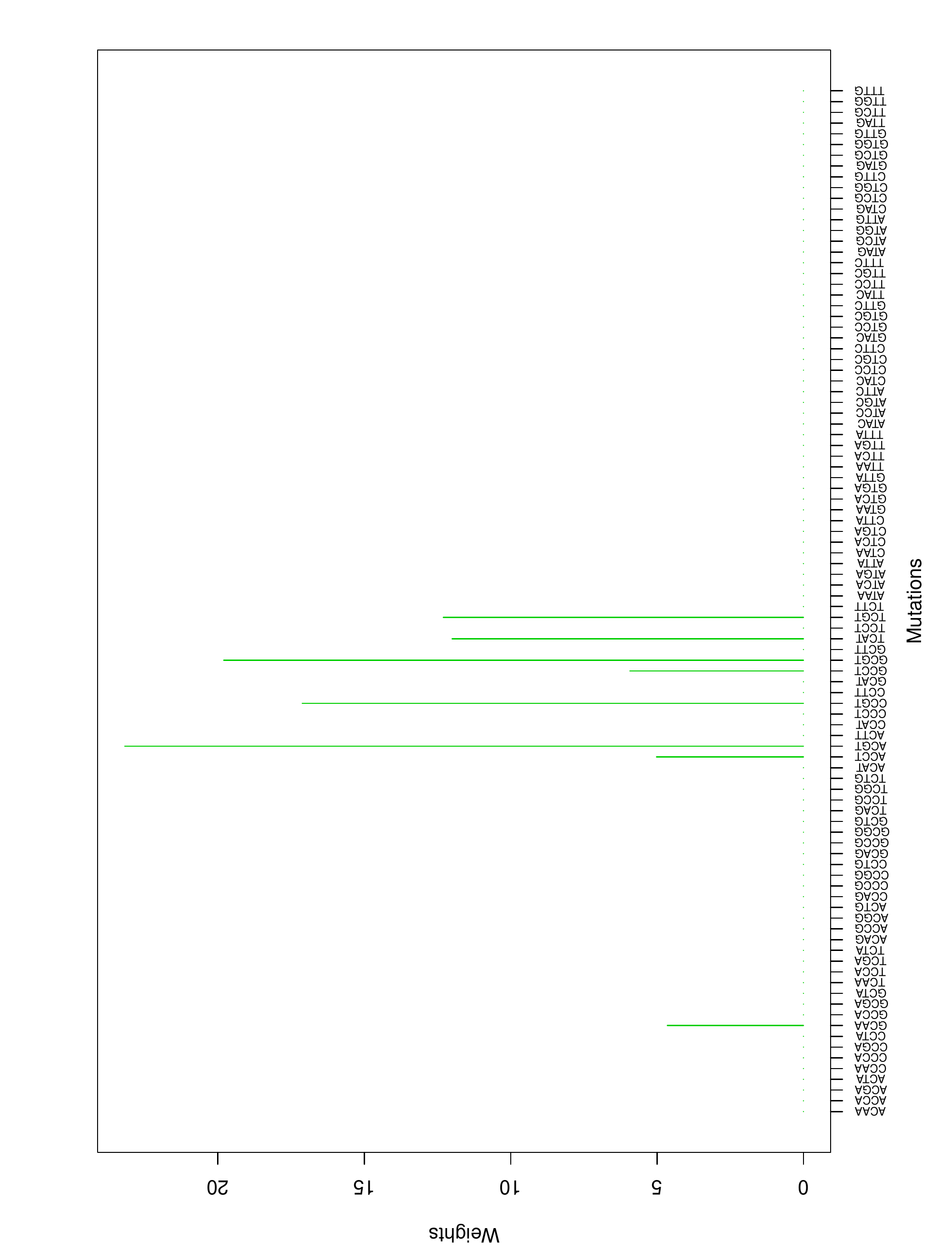}
\caption{Cluster Cl-1 in Clustering-C with weights based on normalized regressions with arithmetic means
(see Subsection \ref{sub.reg}).
See Tables \ref{table.occurrence.cts}, \ref{table.weights.C.1}, \ref{table.weights.C.2}.}
\label{FigureNorm1C}
\end{figure}

\begin{figure}[ht]
\centering
\includegraphics[scale=0.7]{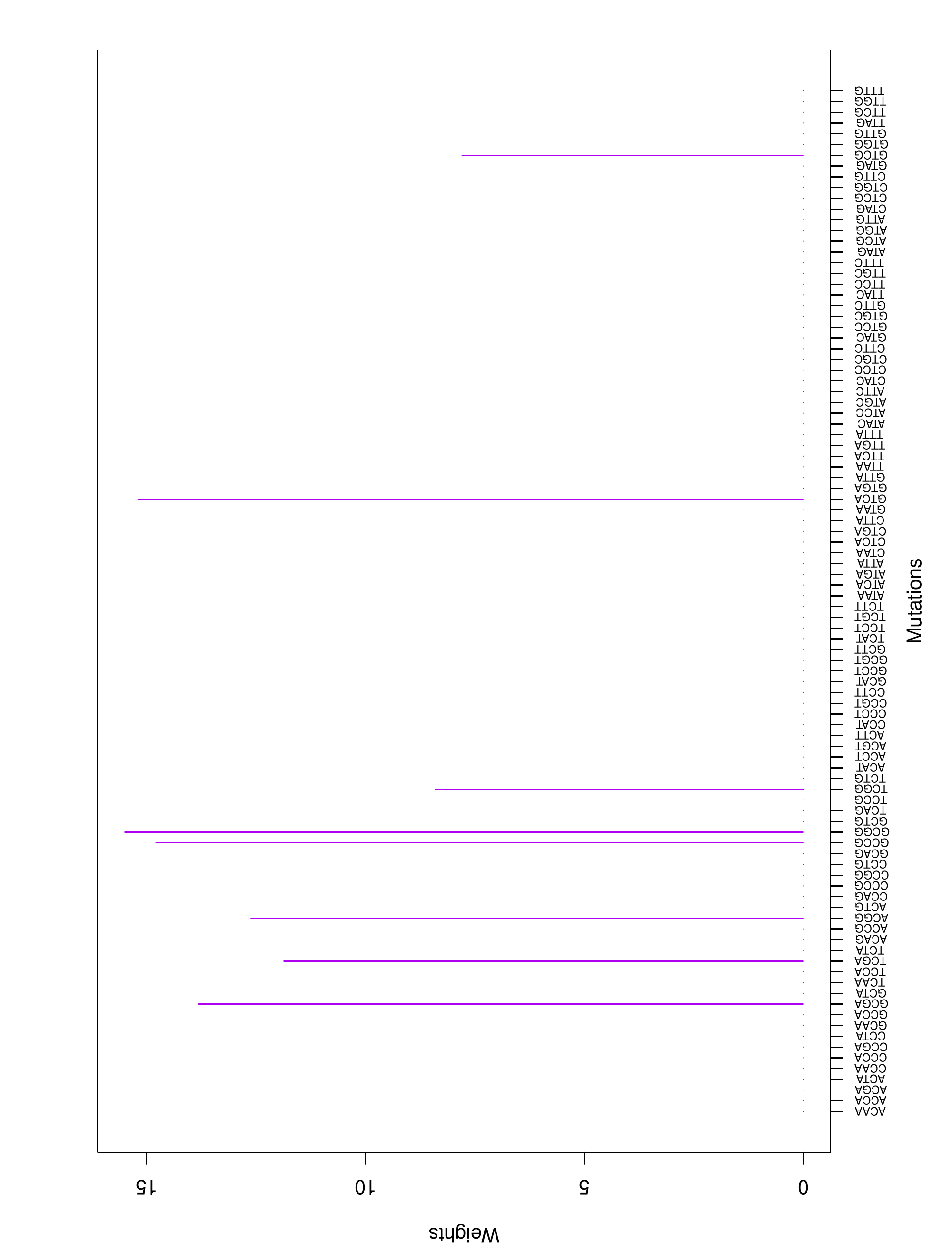}
\caption{Cluster Cl-2 in Clustering-C with weights based on unnormalized regressions with arithmetic means
(see Subsection \ref{sub.reg}).
See Tables \ref{table.occurrence.cts}, \ref{table.weights.C.1}, \ref{table.weights.C.2}.}
\label{Figure2C}
\end{figure}

\begin{figure}[ht]
\centering
\includegraphics[scale=0.7]{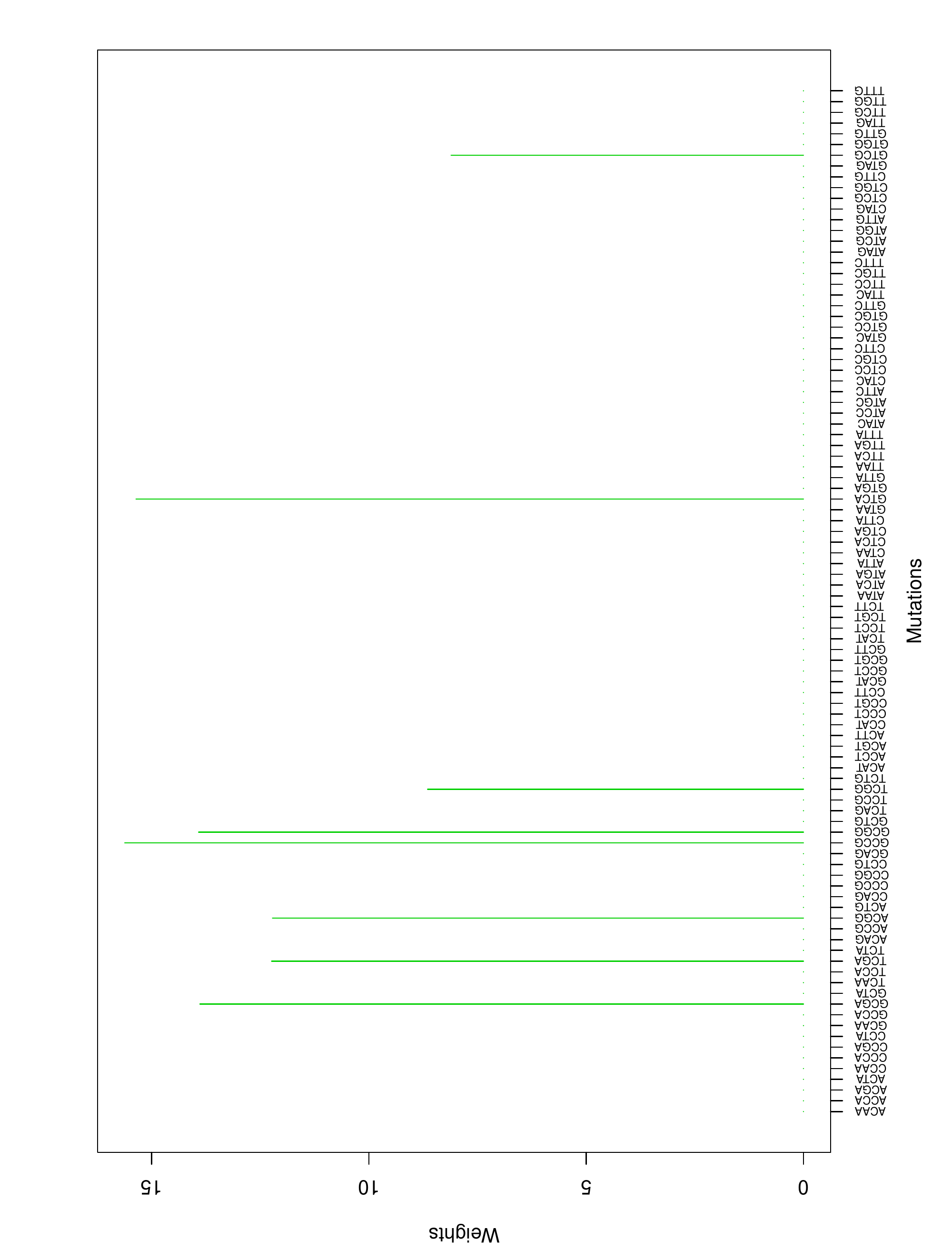}
\caption{Cluster Cl-2 in Clustering-C with weights based on normalized regressions with arithmetic means
(see Subsection \ref{sub.reg}).
See Tables \ref{table.occurrence.cts}, \ref{table.weights.C.1}, \ref{table.weights.C.2}.}
\label{FigureNorm2C}
\end{figure}

\begin{figure}[ht]
\centering
\includegraphics[scale=0.7]{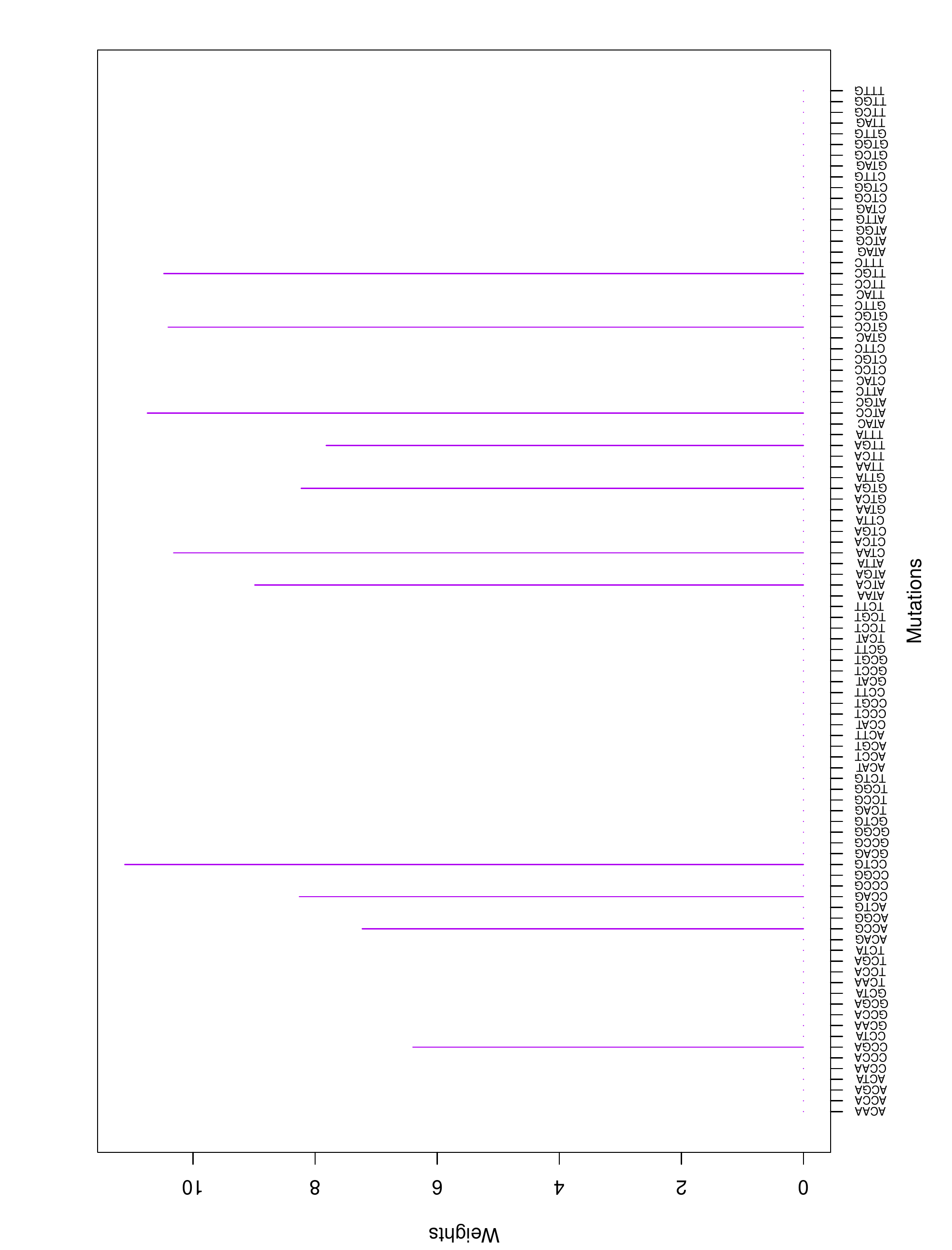}
\caption{Cluster Cl-3 in Clustering-C with weights based on unnormalized regressions with arithmetic means
(see Subsection \ref{sub.reg}).
See Tables \ref{table.occurrence.cts}, \ref{table.weights.C.1}, \ref{table.weights.C.2}.}
\label{Figure3C}
\end{figure}

\begin{figure}[ht]
\centering
\includegraphics[scale=0.7]{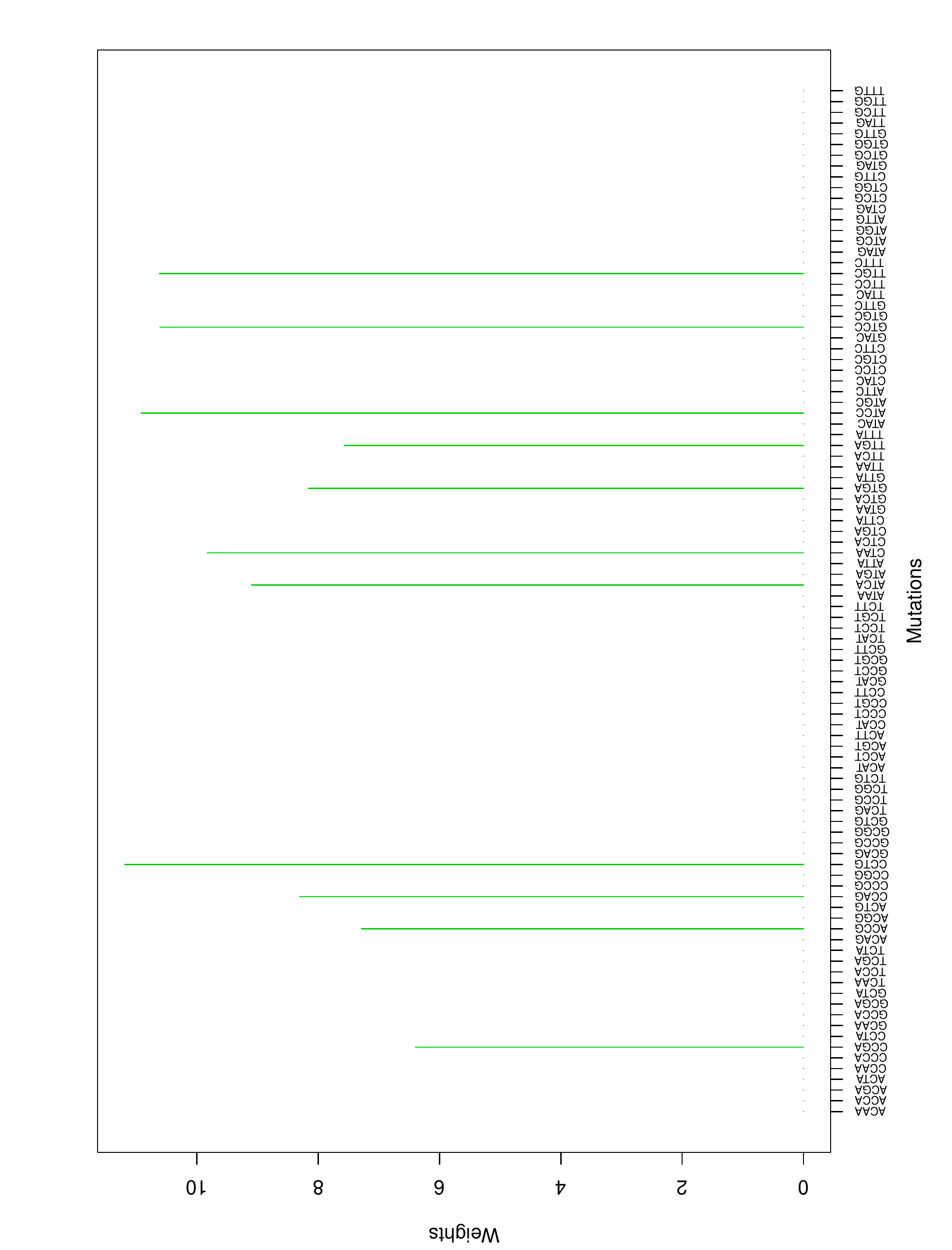}
\caption{Cluster Cl-3 in Clustering-C with weights based on normalized regressions with arithmetic means
(see Subsection \ref{sub.reg}).
See Tables \ref{table.occurrence.cts}, \ref{table.weights.C.1}, \ref{table.weights.C.2}.}
\label{FigureNorm3C}
\end{figure}

\begin{figure}[ht]
\centering
\includegraphics[scale=0.7]{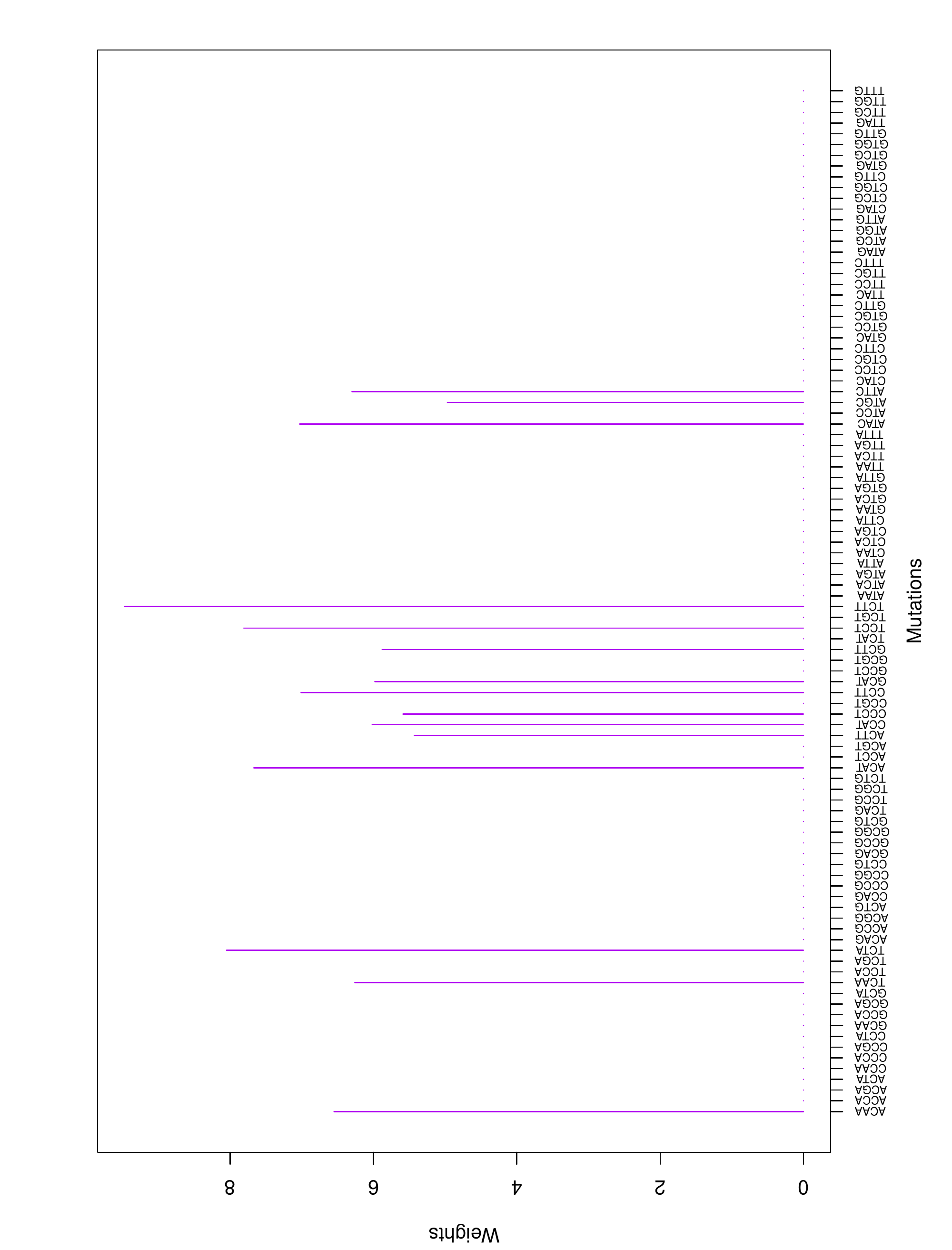}
\caption{Cluster Cl-4 in Clustering-C with weights based on unnormalized regressions with arithmetic means
(see Subsection \ref{sub.reg}).
See Tables \ref{table.occurrence.cts}, \ref{table.weights.C.1}, \ref{table.weights.C.2}.}
\label{Figure4C}
\end{figure}

\begin{figure}[ht]
\centering
\includegraphics[scale=0.7]{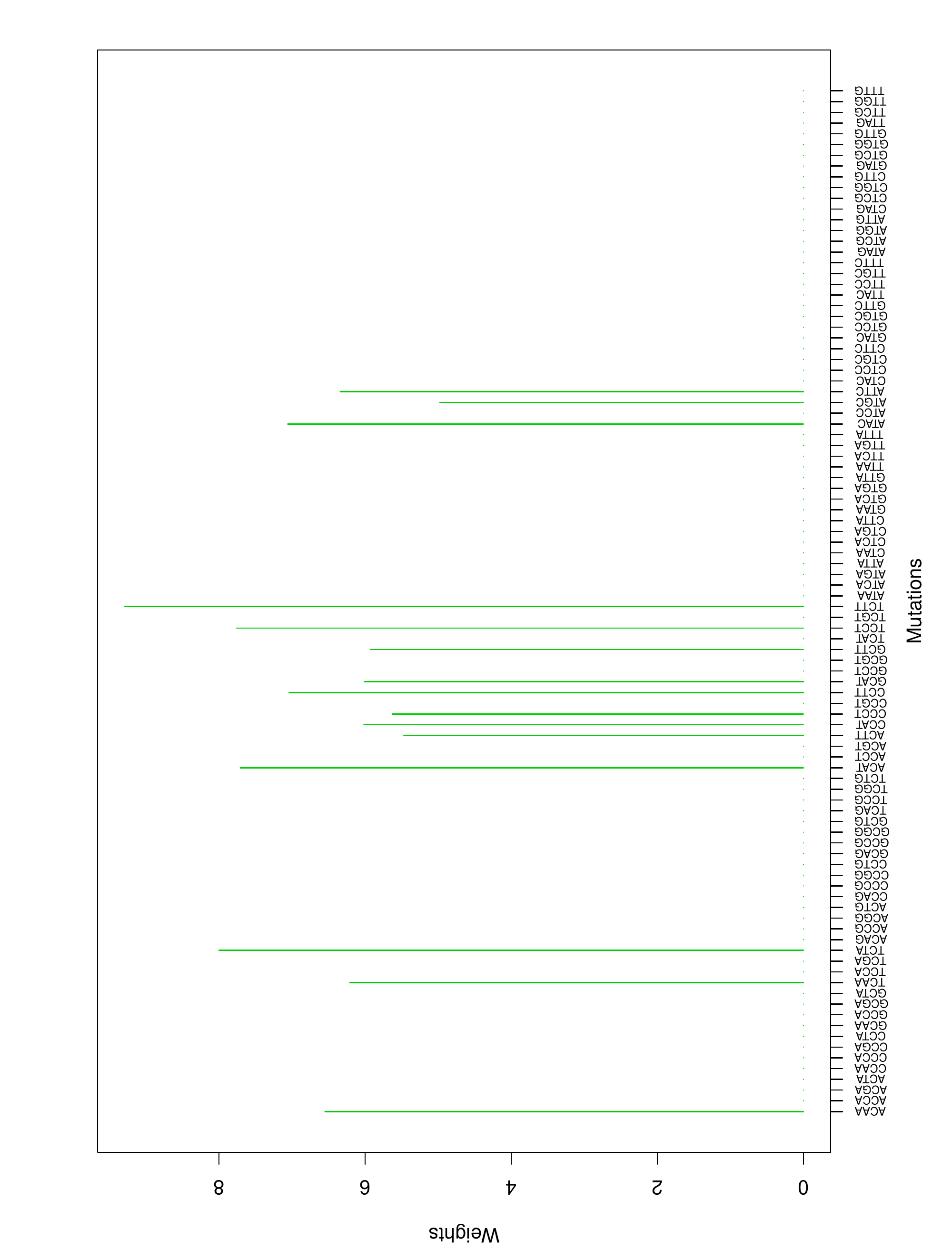}
\caption{Cluster Cl-4 in Clustering-C with weights based on normalized regressions with arithmetic means
(see Subsection \ref{sub.reg}).
See Tables \ref{table.occurrence.cts}, \ref{table.weights.C.1}, \ref{table.weights.C.2}.}
\label{FigureNorm4C}
\end{figure}

\begin{figure}[ht]
\centering
\includegraphics[scale=0.7]{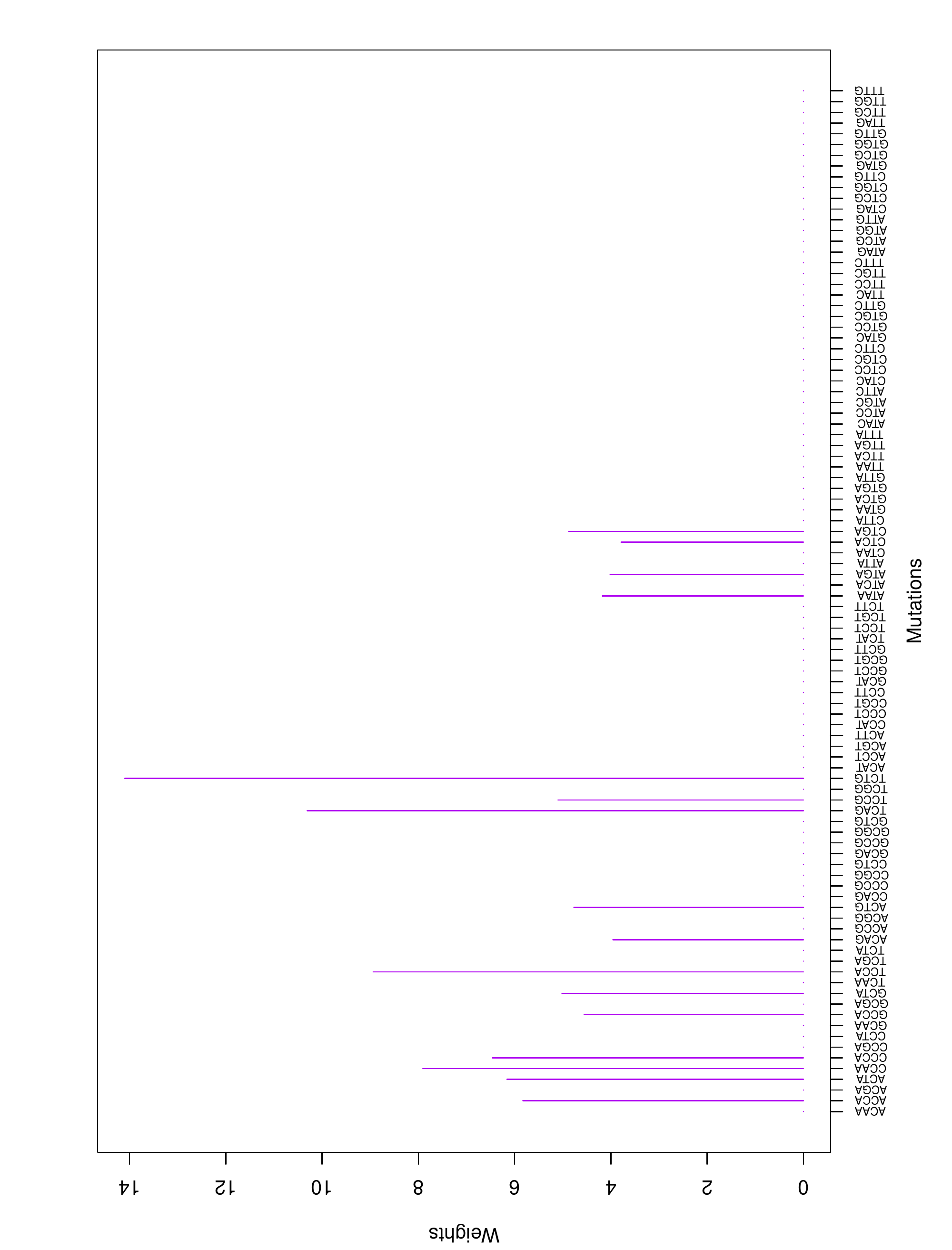}
\caption{Cluster Cl-5 in Clustering-C with weights based on unnormalized regressions with arithmetic means
(see Subsection \ref{sub.reg}).
See Tables \ref{table.occurrence.cts}, \ref{table.weights.C.1}, \ref{table.weights.C.2}.}
\label{Figure5C}
\end{figure}

\begin{figure}[ht]
\centering
\includegraphics[scale=0.7]{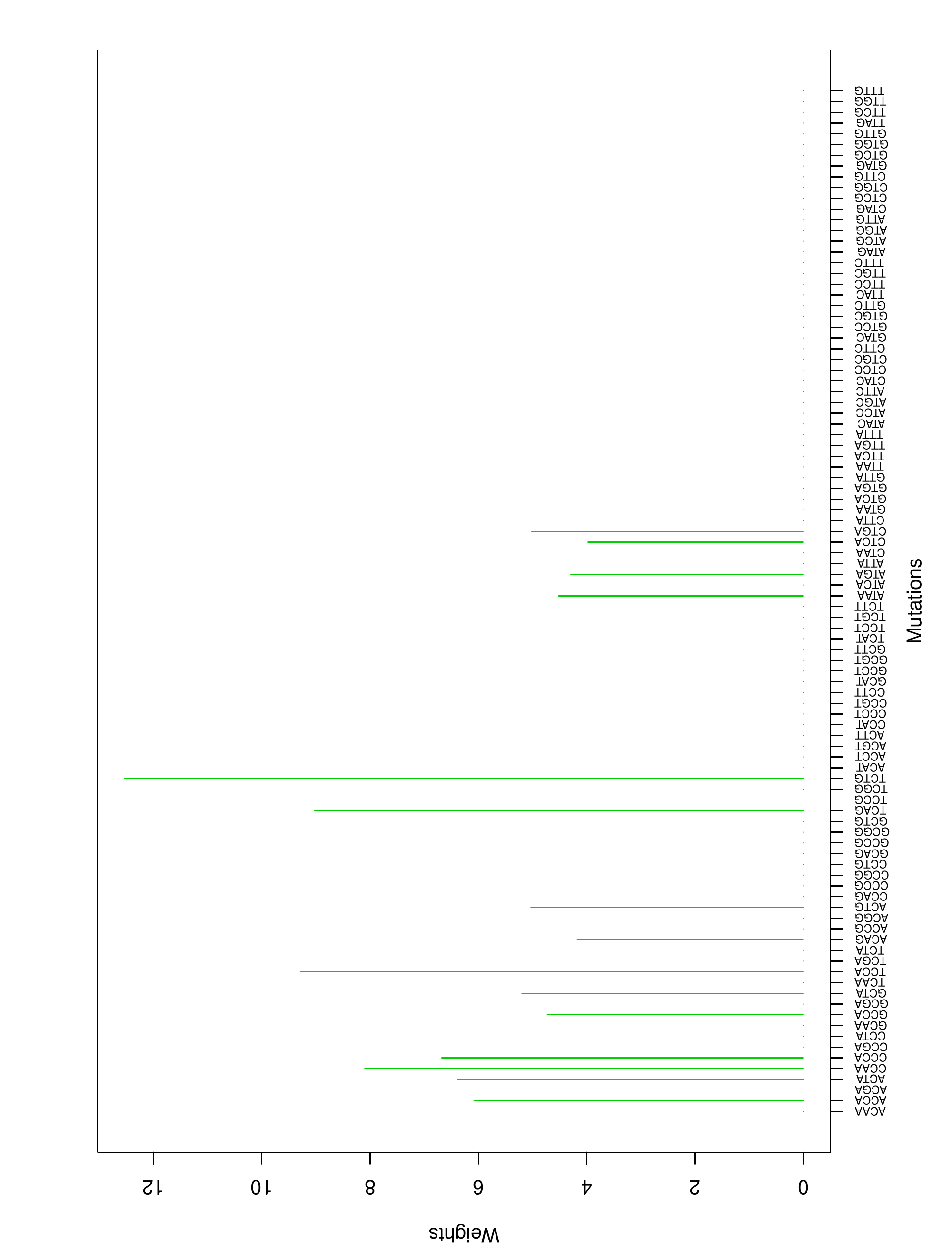}
\caption{Cluster Cl-5 in Clustering-C with weights based on normalized regressions with arithmetic means
(see Subsection \ref{sub.reg}).
See Tables \ref{table.occurrence.cts}, \ref{table.weights.C.1}, \ref{table.weights.C.2}.}
\label{FigureNorm5C}
\end{figure}

\begin{figure}[ht]
\centering
\includegraphics[scale=0.7]{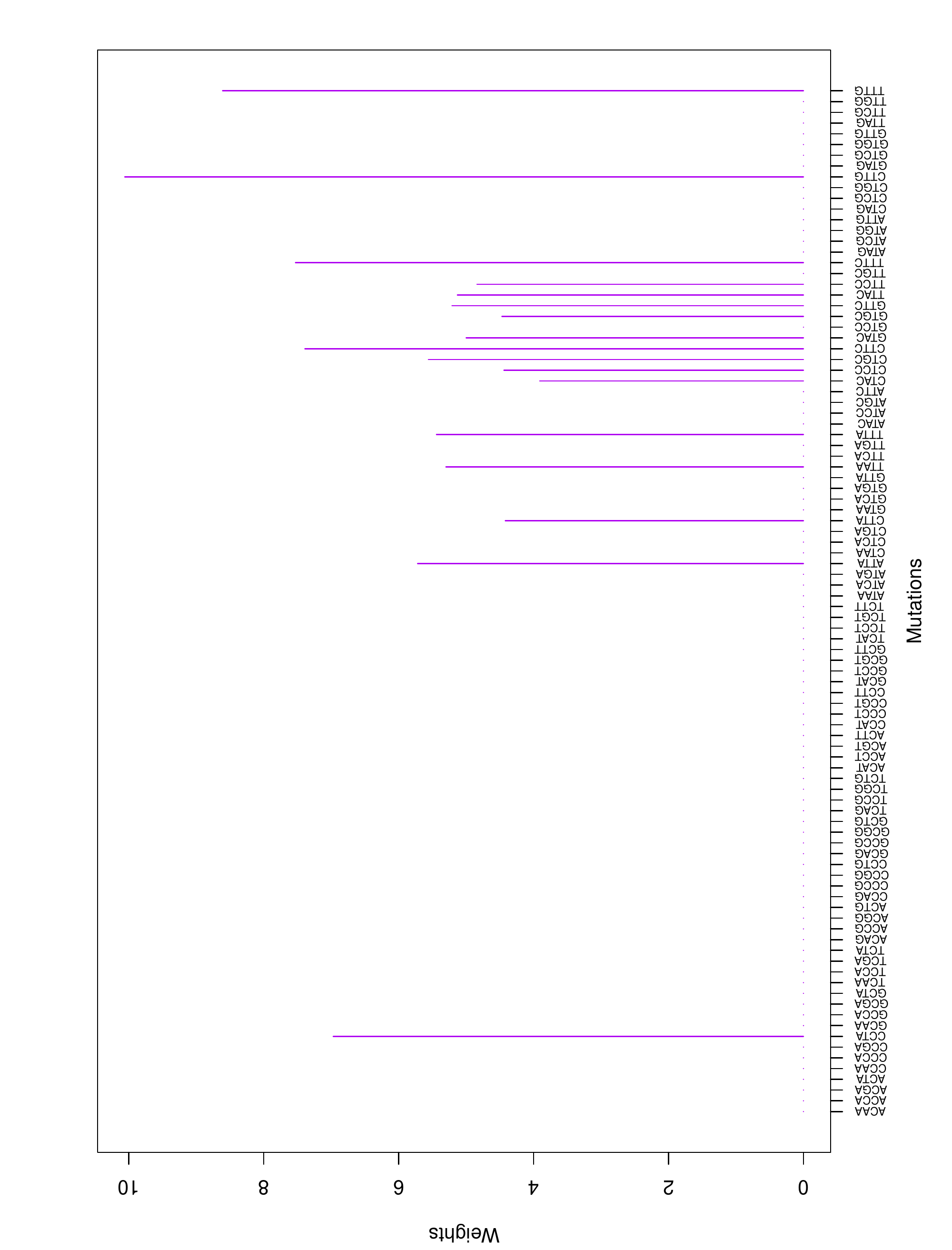}
\caption{Cluster Cl-6 in Clustering-C with weights based on unnormalized regressions with arithmetic means
(see Subsection \ref{sub.reg}).
See Tables \ref{table.occurrence.cts}, \ref{table.weights.C.1}, \ref{table.weights.C.2}.}
\label{Figure6C}
\end{figure}

\begin{figure}[ht]
\centering
\includegraphics[scale=0.7]{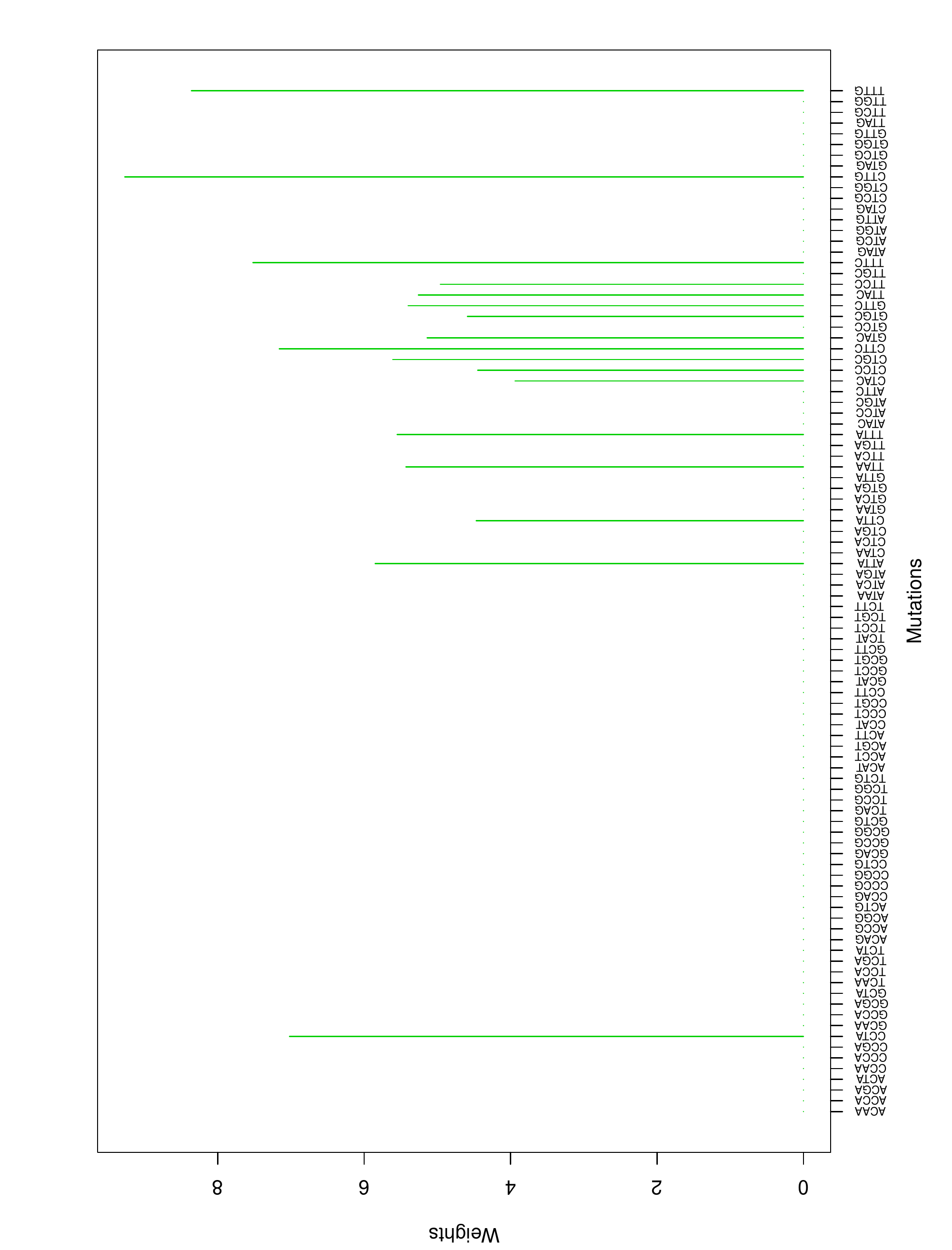}
\caption{Cluster Cl-6 in Clustering-C with weights based on normalized regressions with arithmetic means
(see Subsection \ref{sub.reg}).
See Tables \ref{table.occurrence.cts}, \ref{table.weights.C.1}, \ref{table.weights.C.2}.}
\label{FigureNorm6C}
\end{figure}

\begin{figure}[ht]
\centering
\includegraphics[scale=0.7]{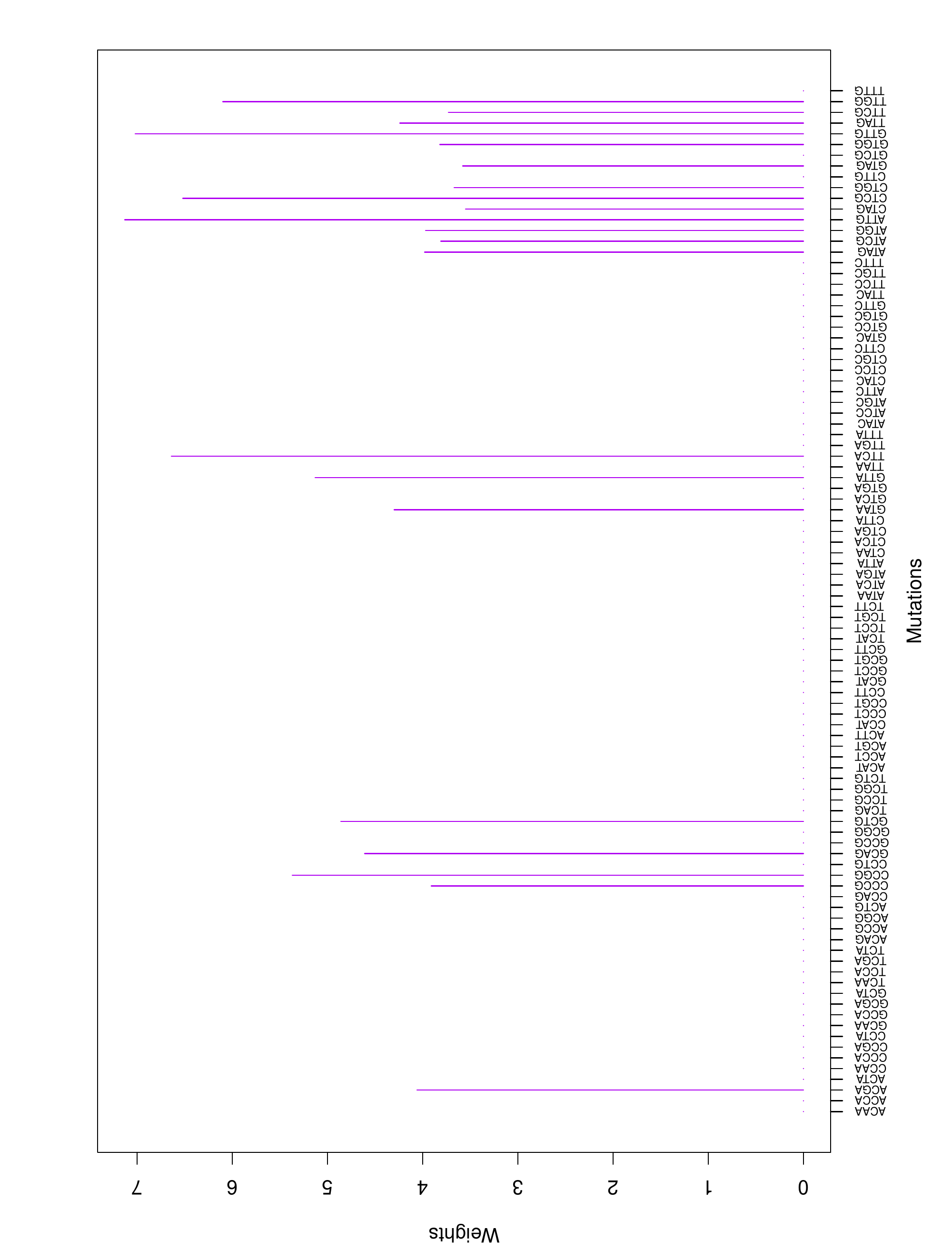}
\caption{Cluster Cl-7 in Clustering-C with weights based on unnormalized regressions with arithmetic means
(see Subsection \ref{sub.reg}).
See Tables \ref{table.occurrence.cts}, \ref{table.weights.C.1}, \ref{table.weights.C.2}.}
\label{Figure7C}
\end{figure}

\begin{figure}[ht]
\centering
\includegraphics[scale=0.7]{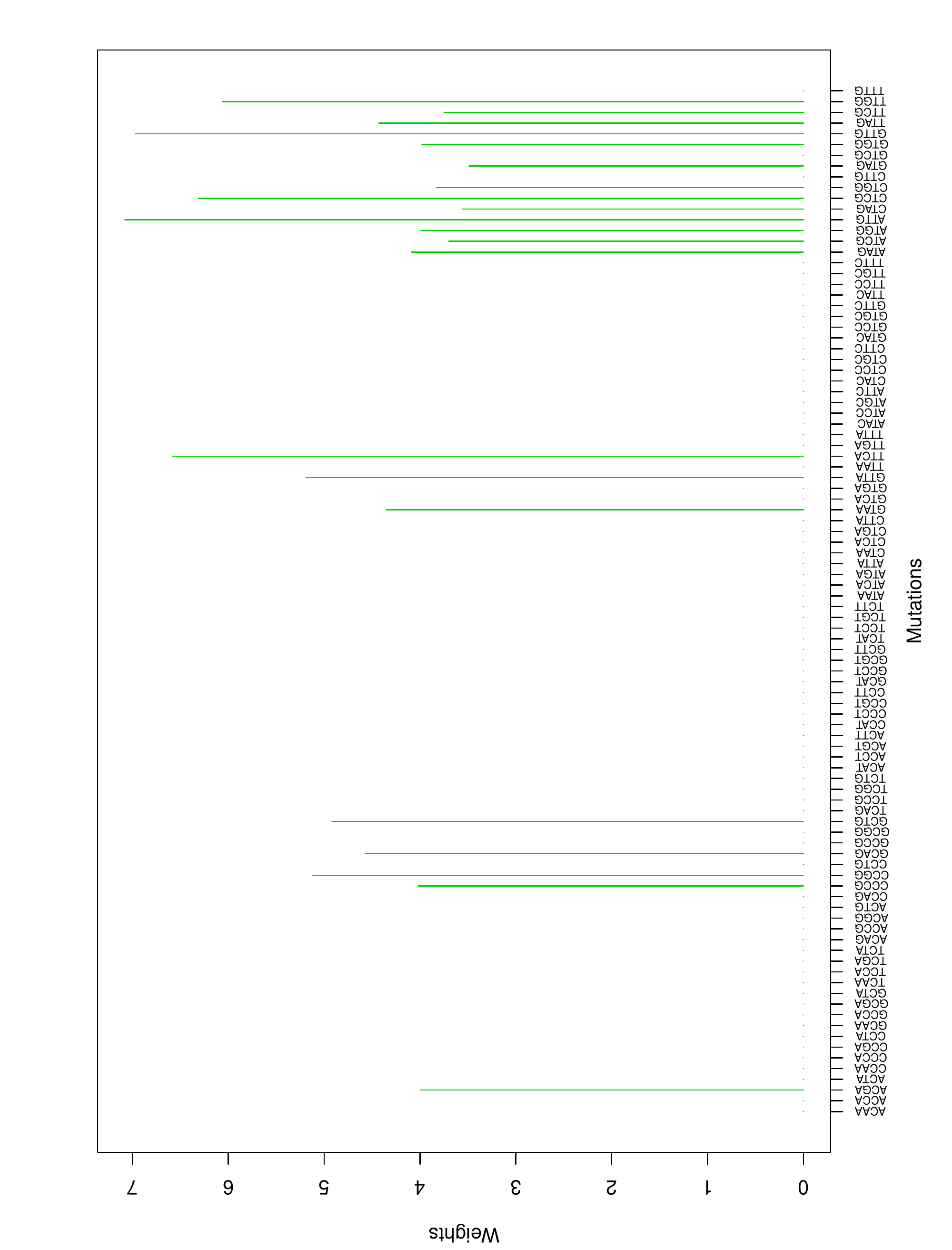}
\caption{Cluster Cl-7 in Clustering-C with weights based on normalized regressions with arithmetic means
(see Subsection \ref{sub.reg}).
See Tables \ref{table.occurrence.cts}, \ref{table.weights.C.1}, \ref{table.weights.C.2}.}
\label{FigureNorm7C}
\end{figure}

\begin{figure}[ht]
\centering
\includegraphics[scale=0.7]{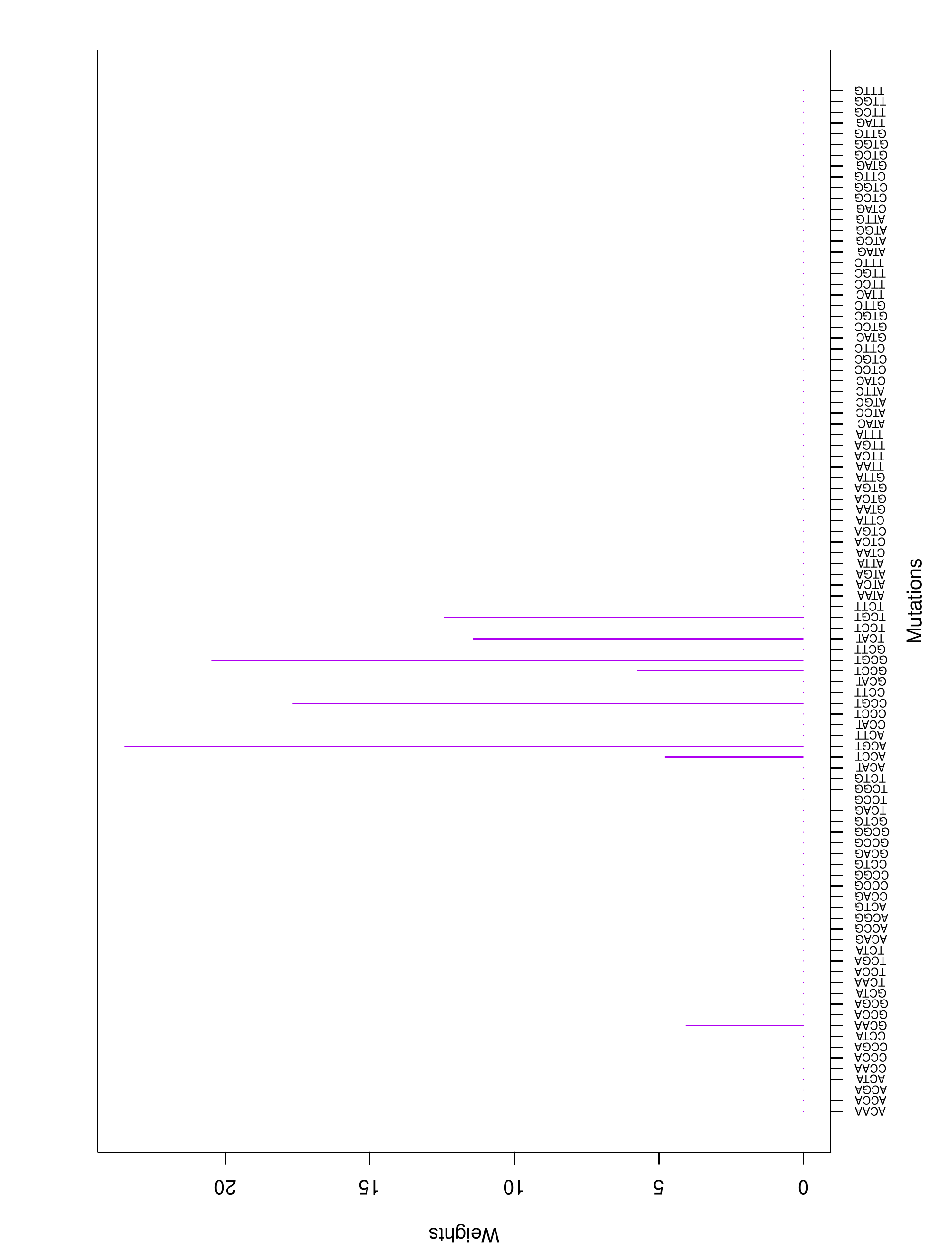}
\caption{Cluster Cl-1 in Clustering-D with weights based on unnormalized regressions with arithmetic means
(see Subsection \ref{sub.reg}).
See Tables \ref{table.occurrence.cts}, \ref{table.weights.D.1}, \ref{table.weights.D.2}.}
\label{Figure1D}
\end{figure}

\begin{figure}[ht]
\centering
\includegraphics[scale=0.7]{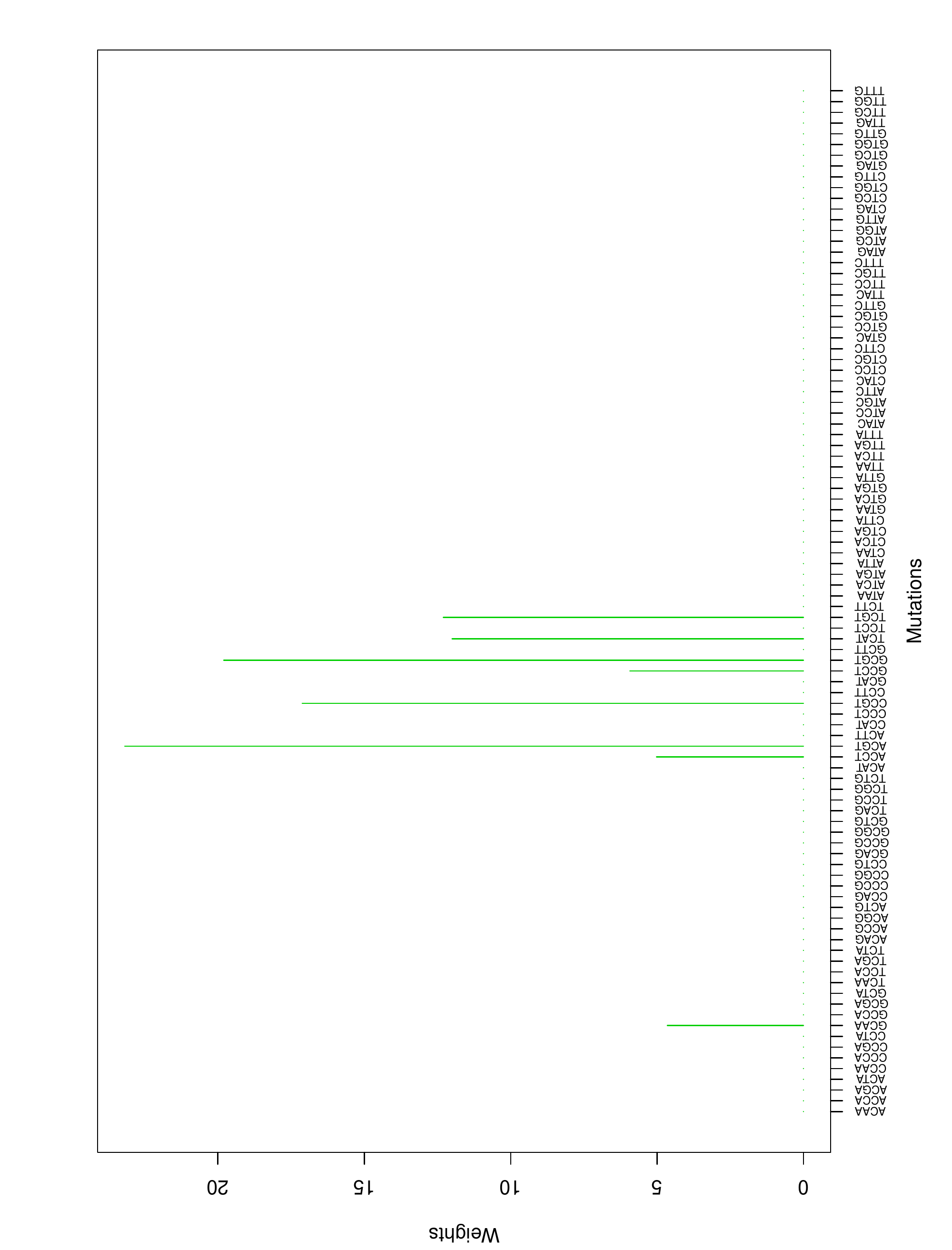}
\caption{Cluster Cl-1 in Clustering-D with weights based on normalized regressions with arithmetic means
(see Subsection \ref{sub.reg}).
See Tables \ref{table.occurrence.cts}, \ref{table.weights.D.1}, \ref{table.weights.D.2}.}
\label{FigureNorm1D}
\end{figure}

\begin{figure}[ht]
\centering
\includegraphics[scale=0.7]{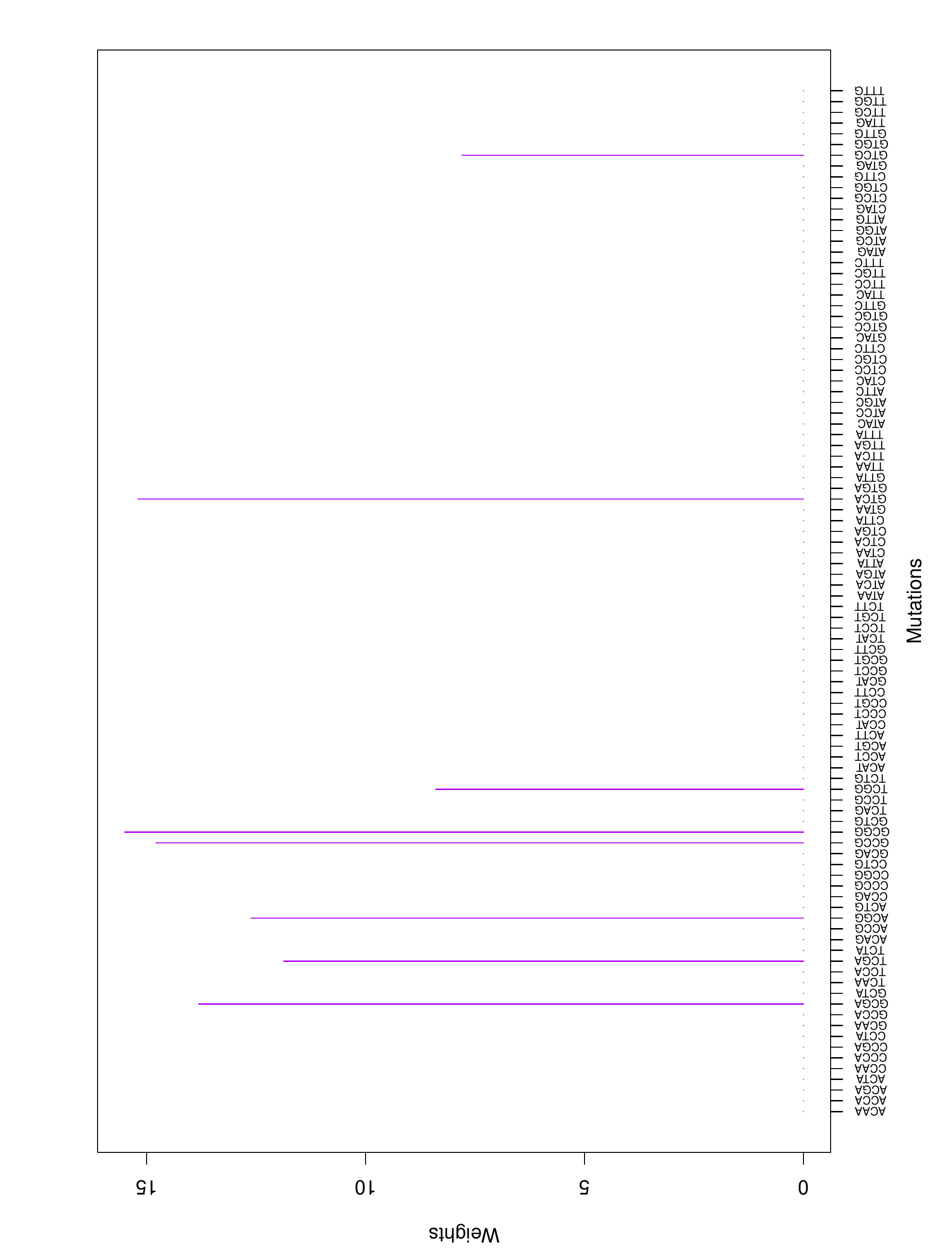}
\caption{Cluster Cl-2 in Clustering-D with weights based on unnormalized regressions with arithmetic means
(see Subsection \ref{sub.reg}).
See Tables \ref{table.occurrence.cts}, \ref{table.weights.D.1}, \ref{table.weights.D.2}.}
\label{Figure2D}
\end{figure}

\begin{figure}[ht]
\centering
\includegraphics[scale=0.7]{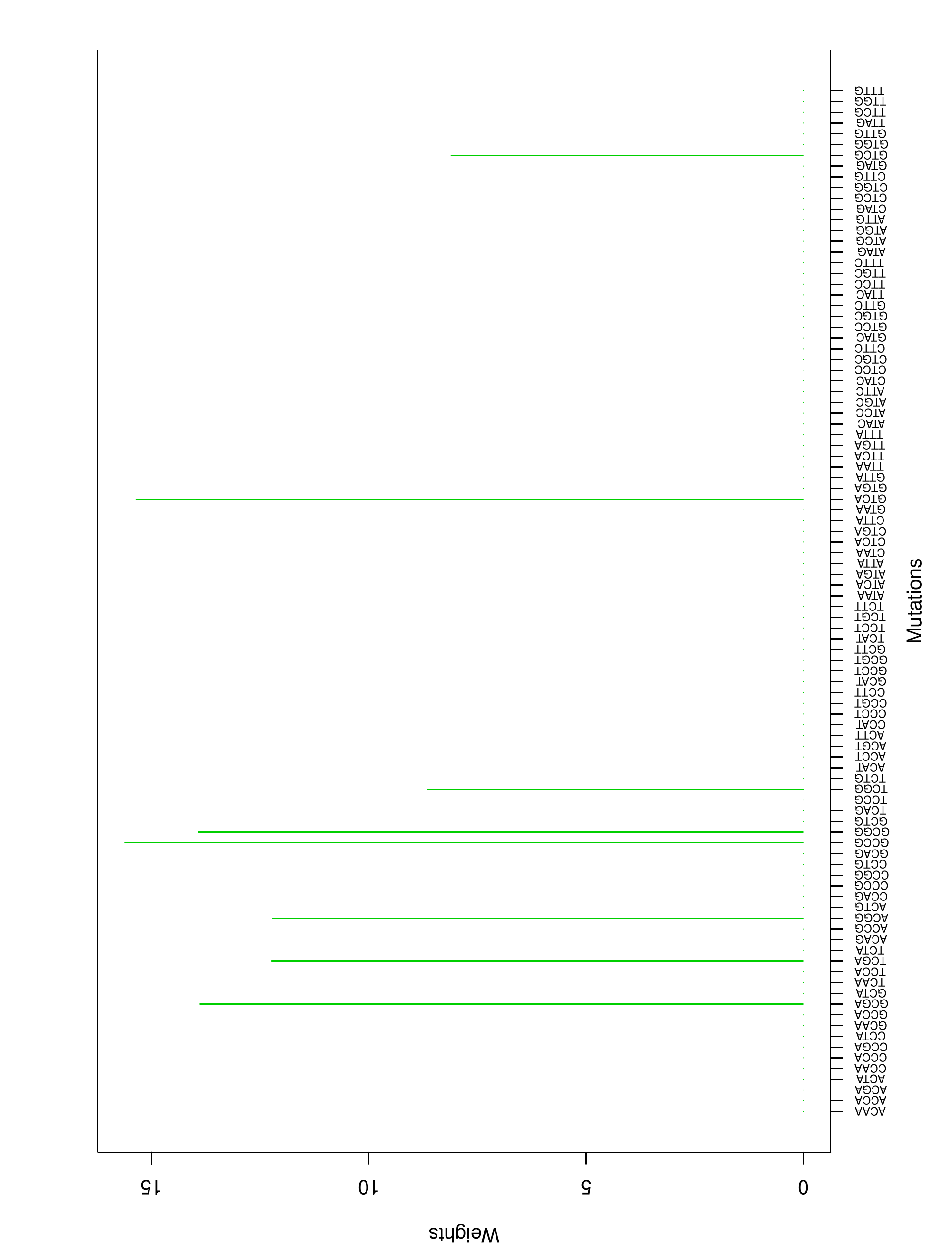}
\caption{Cluster Cl-2 in Clustering-D with weights based on normalized regressions with arithmetic means
(see Subsection \ref{sub.reg}).
See Tables \ref{table.occurrence.cts}, \ref{table.weights.D.1}, \ref{table.weights.D.2}.}
\label{FigureNorm2D}
\end{figure}

\begin{figure}[ht]
\centering
\includegraphics[scale=0.7]{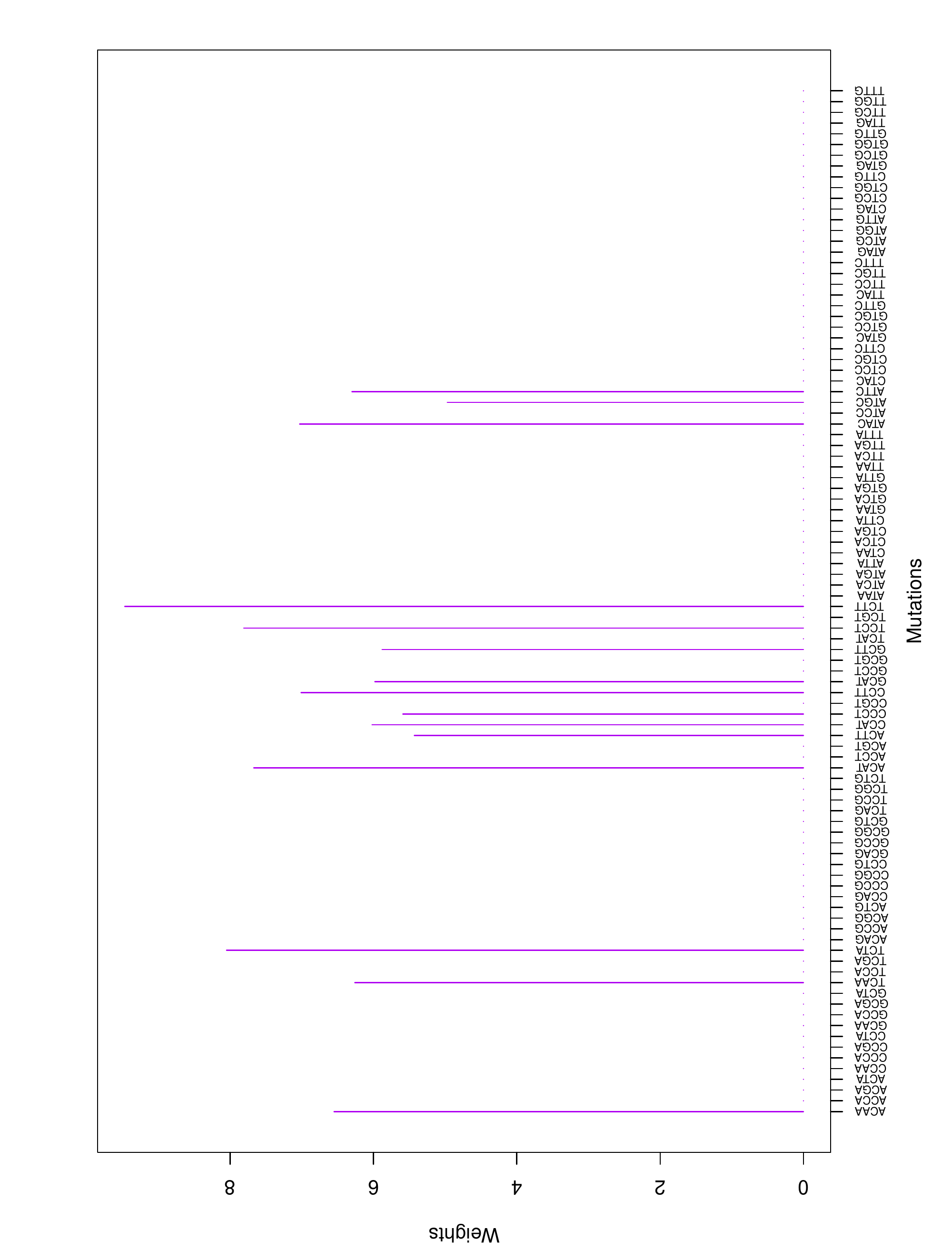}
\caption{Cluster Cl-3 in Clustering-D with weights based on unnormalized regressions with arithmetic means
(see Subsection \ref{sub.reg}).
See Tables \ref{table.occurrence.cts}, \ref{table.weights.D.1}, \ref{table.weights.D.2}.}
\label{Figure3D}
\end{figure}

\begin{figure}[ht]
\centering
\includegraphics[scale=0.7]{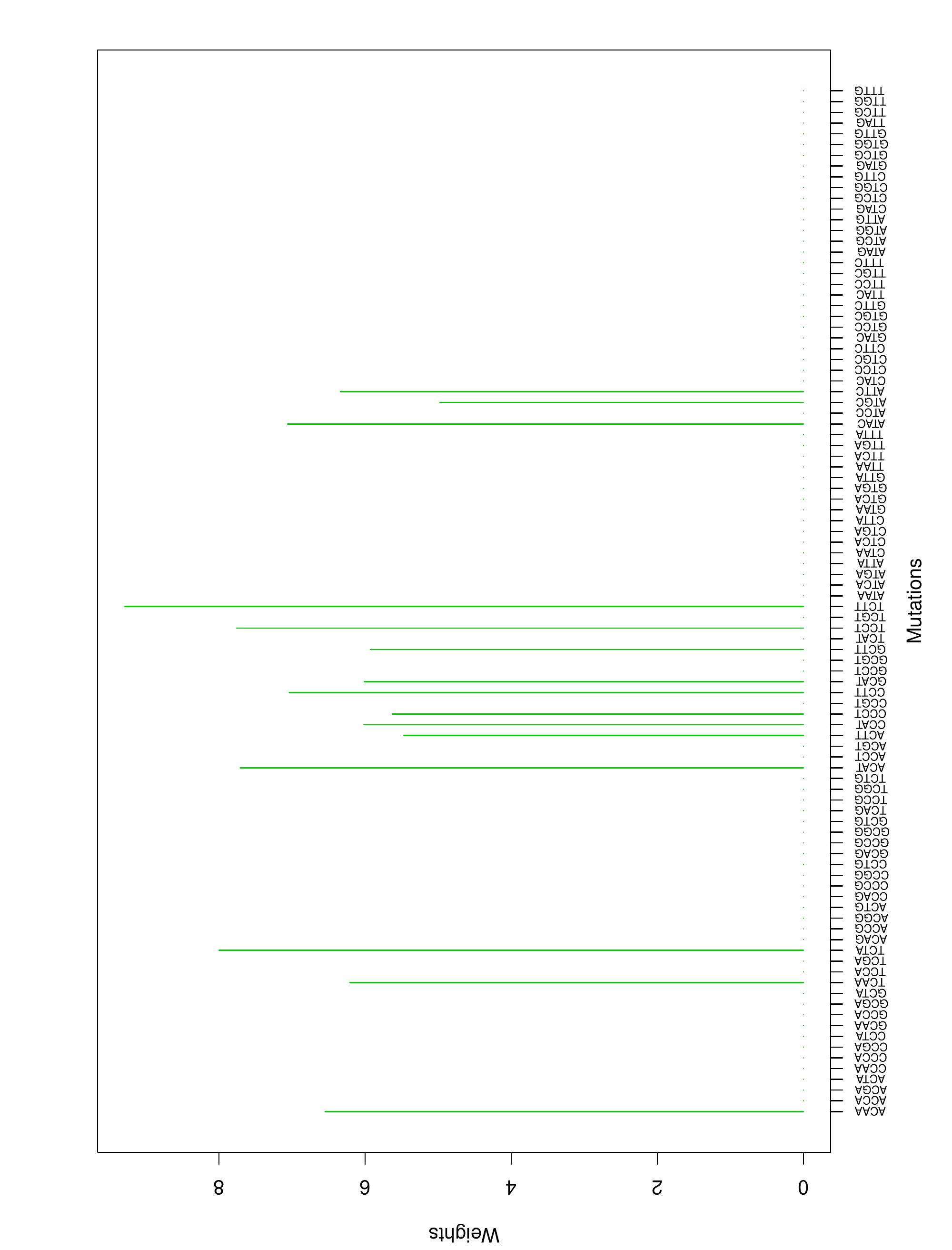}
\caption{Cluster Cl-3 in Clustering-D with weights based on normalized regressions with arithmetic means
(see Subsection \ref{sub.reg}).
See Tables \ref{table.occurrence.cts}, \ref{table.weights.D.1}, \ref{table.weights.D.2}.}
\label{FigureNorm3D}
\end{figure}

\begin{figure}[ht]
\centering
\includegraphics[scale=0.7]{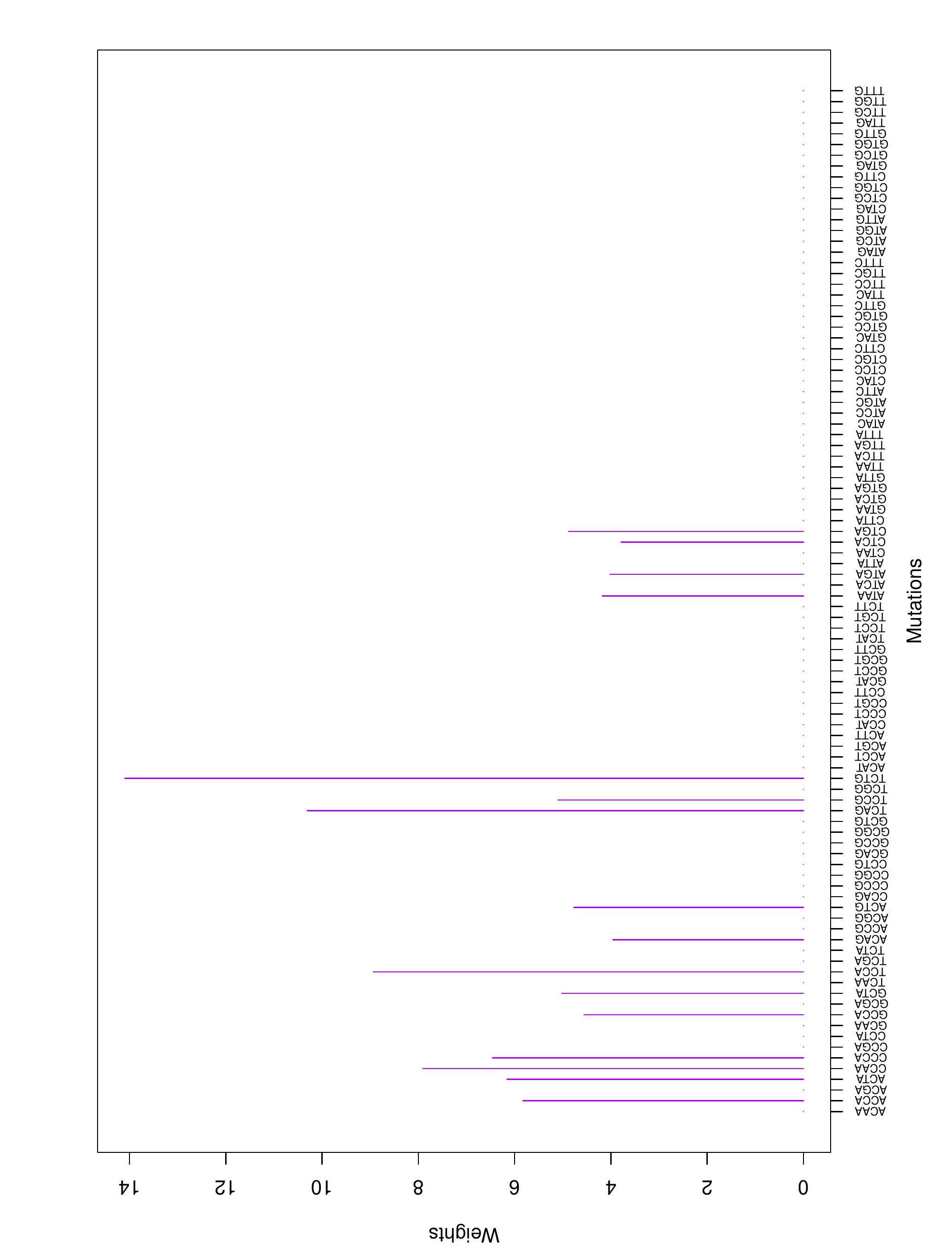}
\caption{Cluster Cl-4 in Clustering-D with weights based on unnormalized regressions with arithmetic means
(see Subsection \ref{sub.reg}).
See Tables \ref{table.occurrence.cts}, \ref{table.weights.D.1}, \ref{table.weights.D.2}.}
\label{Figure4D}
\end{figure}

\begin{figure}[ht]
\centering
\includegraphics[scale=0.7]{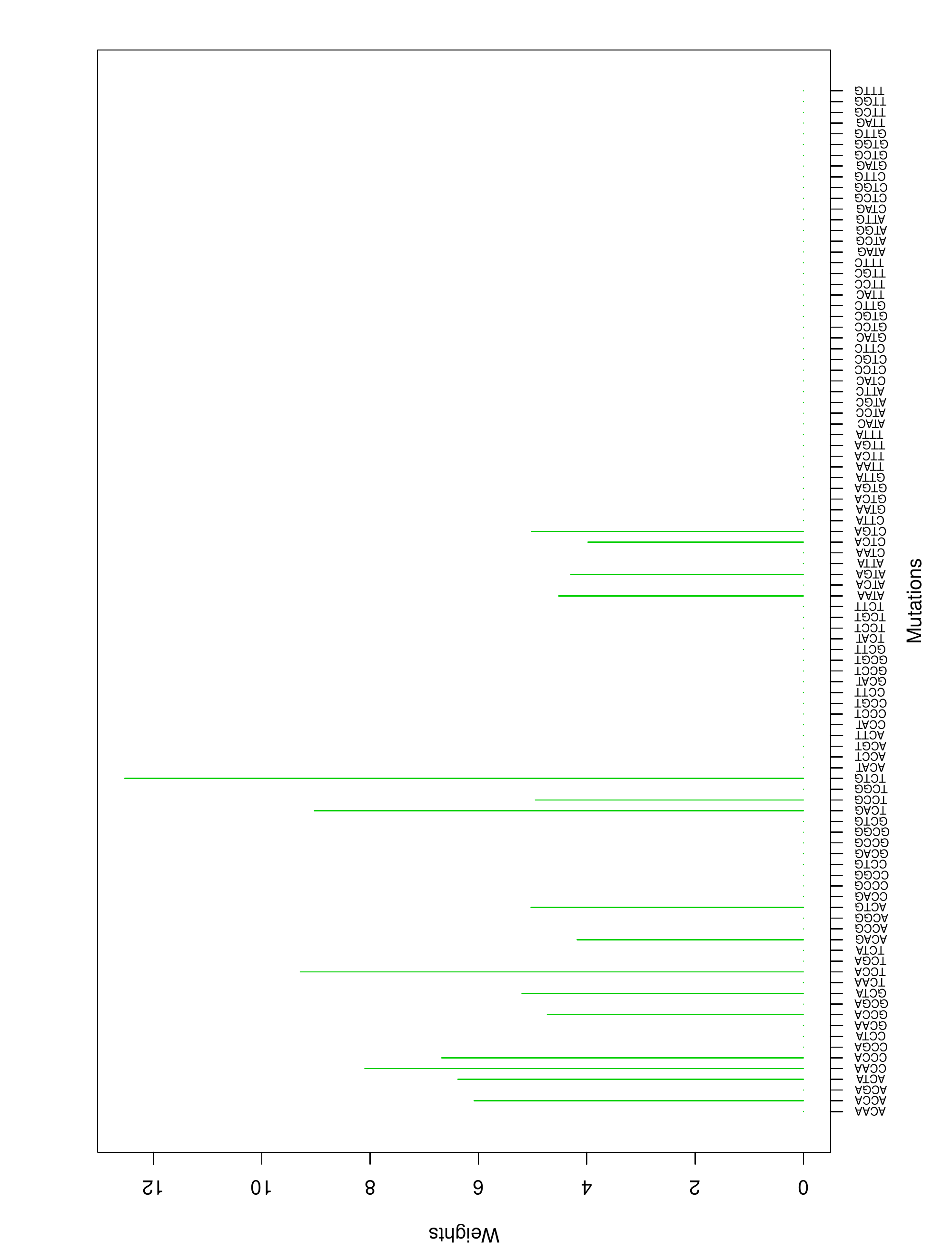}
\caption{Cluster Cl-4 in Clustering-D with weights based on normalized regressions with arithmetic means
(see Subsection \ref{sub.reg}).
See Tables \ref{table.occurrence.cts}, \ref{table.weights.D.1}, \ref{table.weights.D.2}.}
\label{FigureNorm4D}
\end{figure}

\begin{figure}[ht]
\centering
\includegraphics[scale=0.7]{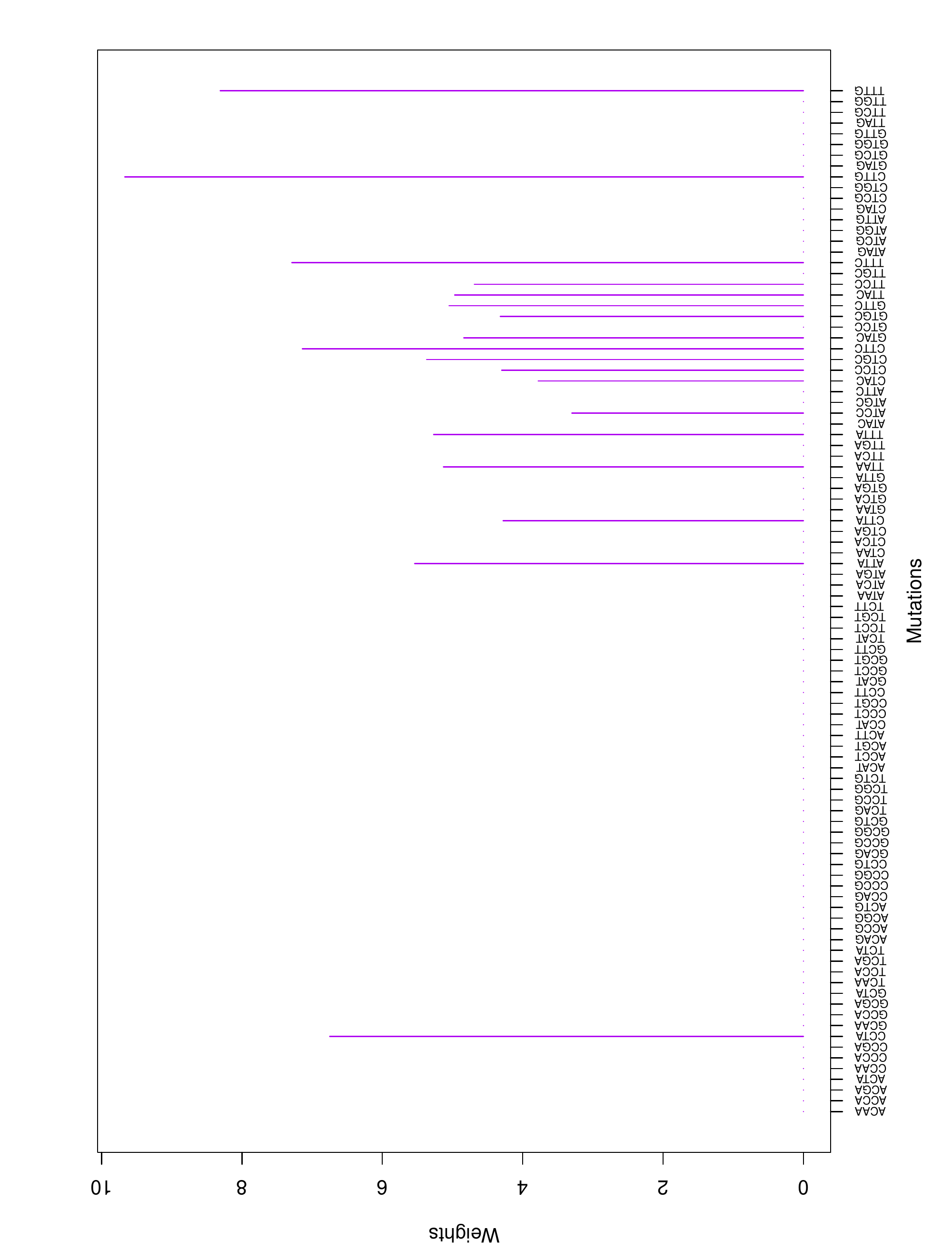}
\caption{Cluster Cl-5 in Clustering-D with weights based on unnormalized regressions with arithmetic means
(see Subsection \ref{sub.reg}).
See Tables \ref{table.occurrence.cts}, \ref{table.weights.D.1}, \ref{table.weights.D.2}.}
\label{Figure5D}
\end{figure}

{}And we do find that 11 out of the 14 cancer types we study here have clustering structures substantially embedded in them and clustering overall works well for at least 10 out of these 11 cancer types.\footnote{\, Breast Cancer possibly being an exception. As mentioned above, it would make sense to exclude Liver Cancer, Lung Cancer and Renal Cell Carcinoma from the analysis, which may affect how well clustering works for Breast Cancer and possibly also the other 10 cancer types.} Now, looking at Figure 14 of \cite{BioFM}, we see that its ``peaks" are located at ACGT, CCGT, GCGT and TCGT. The same ``peaks" are present in our cluster Cl-1 (see Figures \ref{Figure1A} and \ref{FigureNorm1A}). Hence the high within-cluster correlation between Cl-1 and Sig1. On the other hand, Sig1 of \cite{BioFM} is essentially the same as the mutational signature 1 of \cite{Nik-Zainal}, \cite{Alexandrov}, which is due to spontaneous cytosine deamination. So, this is what our cluster Cl-1 describes. Next, looking at Figure 15 of \cite{BioFM}, we see that its ``peaks" are located at TCAG, TCTG, TCAT and TCTT. The first two of these ``peaks" TCAG and TCTG are present in our Cl-5 (see Figures \ref{Figure5A} and \ref{FigureNorm5A}), the third ``peak" TCAT is present in our Cl-1 (see Figures \ref{Figure1A} and \ref{FigureNorm1A}), while the fourth ``peak" TCTT is present in our Cl-4 (see Figures \ref{Figure4A} and \ref{FigureNorm4A}), which is consistent with the high within-cluster correlations between Sig2 and Cl-4 and Cl-5, albeit its within-cluster correlation with Cl-1 is poor. Note that Sig2 of \cite{BioFM} is essentially the same as the mutational signatures 2+13 of \cite{Nik-Zainal}, \cite{Alexandrov}, which are due to APOBEC mediated cytosine deamination. In fact, it was reported as a single signature in \cite{Alexandrov}, however, subsequently, it was split into 2 distinct signatures, which usually appear in the same samples.\footnote{\, For detailed comments, see http://cancer.sanger.ac.uk/cosmic/signatures.} Our clustering results indicate that grouping TCAG and TCTG into one signature makes sense as they belong to the same cluster Cl-5. However, grouping TCAT and TCTT together does not appear to make much sense. Looking at the Figures for Clustering-A, Clustering-B, Clustering-C and Clustering-D, we see that the TCAT ``peak" invariably appears together with the ACGT, CCGT, GCGT and TCGT ``peaks" as in Cl-1 in Clustering-A, Cl-2 in Clustering-B, Cl-1 in Clustering-C, and Cl-1 in Clustering-D, but never with TCTT. So, our clustering approach tells us something new beyond the NMF ``intuition". This may have an important implication for Breast Cancer, which, as mentioned above, is dominated by Sig2. Thus, based on our results in Table \ref{table.fit.theta}, we see that Breast Cancer has high within-cluster correlations with Cl-4 and Cl-5, but not with Cl-1. This may imply that clustering simply does not work well for Breast Cancer, which would appear to put it in the same ``stand-alone" league as Liver Cancer, Lung Cancer and Renal Cell Carcinoma. In any event, clustering invariably suggests that the TCAT ``peak" belongs in Cl-1 with the 4 ``peaks" ACGT, CCGT, GCGT and TCGT related to spontaneous cytosine deamination, rather than those related to APOBEC mediated cytosine deamination.

{}Now, let us check the remaining two signatures of \cite{BioFM} with ``tall mountain landscapes" (see above), to wit, Sig4 and Sig7. Looking at Figure 17 of \cite{BioFM}, we see that its ``peaks" are at CTTC, TTTC, CTTG and TTTG. The same peaks appear in our Cl-6 (see Figures \ref{Figure6A} and \ref{FigureNorm6A}). Hence the high within-cluster correlation between Cl-6 and Sig4. Note that Sig4 is essentially the same as the mutational signature 17 of \cite{Nik-Zainal}, \cite{Alexandrov}, and its underlying mutational process is unknown. Next, looking at Figure 20 of \cite{BioFM}, we see that its ``peaks" for the C $>$ G mutations are essentially the same as in Cl-1. Hence the high within-cluster correlation between Cl-7 and Sig1. So, there are no surprises with Sig1, Sig4 and Sig7. However, based on our clustering results, as we discuss above, with Sig2 we do find -- what we feel is a pleasant -- surprise, that splitting it into two signatures (see above) might be inadequate and the TCAT ``peak" might really belong with the Sig1 ``peaks" (spontaneous v. APOBEC mediated cytosine deamination). This is exciting as it might be an indication of the limitations of NMF (or clustering...).\footnote{\ Or both... Alternatively -- and that would be truly exciting -- perhaps there is a biologic explanation. In any event, it is too early to tell -- yet another possibility is that this is merely an artifact of the dataset we use. More research and analyses on larger datasets (see above) is needed.}

{}In Introduction we promised that we would discuss some potential applications of *K-means in quantitative finance, and so here it is. Let us mention that *K-means is universal, oblivious to the input data and applicable in a variety of fields. In quantitative finance *K-means {\em a priori} can be applied everywhere clustering methods are used with the added bonus of (statistical) determinism.\footnote{\, Albeit with the understanding that it requires additional computational cost.} One evident example is statistical industry classifications discussed in \cite{StatIndClass}, where one uses clustering methods to classify stocks. In fact, *K-means is an extension of the methods discussed in \cite{StatIndClass}. One thing to keep in mind is that in *K-means one sifts through a large number $P$ of aggregations, which can get computationally costly when clustering 2000+ stocks into 100+ clusters.\footnote{\, This can be mitigated by employing top-down clustering \cite{StatIndClass}.} Another potential application is in the context of combining alphas (trading signals) -- see, e.g., \cite{Billion}. Yet another application is when we have a term structure, such as a portfolio of bonds (e.g., U.S. Treasuries or some other bonds) with varying maturities, or futures (e.g., Eurodollar futures) with varying deliveries. These cases resemble the genome data more in the sense that the number $N$ of instruments is relatively small (typically even fewer than the number of mutation categories). Another example with a relatively small number of instruments would be a portfolio of various futures for different FX (foreign exchange) pairs (even with the uniform delivery), e.g., USD/EUR, USD/HKD, EUR/AUD, etc., i.e., FX statistical arbitrage. One approach to optimizing risk in such portfolios is by employing clustering methods and a stable, essentially deterministic algorithm such as *K-means can be useful. Hopefully *K-means will prove a valuable tool in cancer research, quantitative finance as well as various other fields (e.g., image recognition).


\appendix

\section{R Source Code}\label{app.code}

{}In this appendix we give the R (R Package for Statistical Computing, http://www.r-project.org) source code for computing the clusterings and weights using the algorithms of Section \ref{sec.2}. The code is straightforward and self-explanatory.\footnote{\, The source code in Appendix \ref{app.code} hereof is not written to be ``fancy" or optimized for speed or in any other way. Its sole purpose is to illustrate the algorithms described in the main text in a simple-to-understand fashion. See Appendix \ref{app.disc} for some important legalese.} The main function is {\tt\small bio.cl.sigs(x, iter.max = 100, num.try = 1000, num.runs = 10000)}. Here: {\tt\small x} is the $N\times d$ occurrence counts matrix $G_{is}$ (where $N=96$ is the number of mutation categories, and $d$ is the number of samples; or $d=n$, where $n$ is the number of cancer types, when the samples are aggregated by cancer types); {\tt\small iter.max} is the maximum number of iterations that are passed into the R built-in function {\tt\small kmeans()}; {\tt\small num.try} is the number $M$ of aggregated clusterings (see Subsection \ref{sub.aggr}); {\tt\small num.runs} is the number of runs $P$ used to determine the most frequently occurring clustering (the ``ultimate" clustering) obtained via aggregation (see Subsection \ref{sub.ultimate}). The function {\tt\small bio.erank.pc()} is defined in Appendix B of \cite{BioFM}. The function {\tt\small qrm.stat.ind.class()} is defined in Appendix A of \cite{StatIndClass}. This function internally calls another function {\tt\small qrm.calc.norm.ret()}, which we redefine here via the function {\tt\small bio.calc.norm.ret()}.\footnote{\, The definition of {\tt\small qrm.calc.norm.ret()} in \cite{StatIndClass} accounts for some peculiarities and nuances pertinent to quantitative trading, which are not applicable here.} The output is a list, whose elements are as follows: {\tt\small res\$ind} is an $N \times K$ binary matrix $\Omega_{iA} = \delta_{G(i), A}$ ($i=1,\dots,N$, $A=1,\dots,K$, the map $G:\{1,\dots,N\}\mapsto\{1,\dots,K\}$ -- see Section \ref{sec.2}), which defines the $K$ clusters in the ``ultimate" clustering;\footnote{\, The code returns the $K$ clusters ordered such that the number of mutation $n_A$ (i.e., the column sum of $\Omega_{iA}$) in the cluster labeled by $A$ is in the increasing order. It also orders clusters with identical $n_A$. We note, however, that (for presentational convenience reasons) the order of such clusters in the tables and figures below is not necessarily the same as what this code returns.} {\tt\small res\$w} is an $N$-vector of weights obtained via unnormalized regressions using arithmetic means for computing exposures (i.e., via (\ref{w.reg}), (\ref{w.norm}) and (\ref{H})); {\tt\small res\$v} is an $N$-vector of weights obtained via normalized regressions using arithmetic means for computing exposures (i.e., via (\ref{w.reg.1}), (\ref{w.norm}) and (\ref{H1})); {\tt\small res\$w.g} is an $N$-vector of weights obtained via unnormalized regressions using geometric means for computing exposures (i.e., via (\ref{w.reg}), (\ref{w.norm}) and (\ref{H2})); {\tt\small res\$v.g} is an $N$-vector of weights obtained via normalized regressions using geometric means for computing exposures (i.e., via (\ref{w.reg.1}), (\ref{w.norm}) and (\ref{H3})).\\
\\
{\tt{\small
\noindent bio.calc.norm.ret <- function (ret)\\
\{\\
\indent s <- apply(ret, 1, sd)\\
\indent x <- ret / s\\
\indent return(x)\\
\}\\
\\
qrm.calc.norm.ret <- bio.calc.norm.ret\\
\\
bio.cl.sigs <- function(x, iter.max = 100,\\
\indent num.try = 1000, num.runs = 10000)\\
\{\\
\indent cl.ix <- function(x) match(1, x)\\
\\
\indent y <- log(1 + x)\\
\indent y <- t(t(y) - colMeans(y))\\
\indent x.d <- exp(y)\\
\indent k <- ncol(bio.erank.pc(y)\$pc)\\
\\
\indent n <- nrow(x)\\
\indent u <- rnorm(n, 0, 1)\\
\indent q <- matrix(NA, n, num.runs)\\
\indent p <- rep(NA, num.runs)\\
\\
\indent for(i in 1:num.runs)\\
\indent \{\\
\indent \indent z <- qrm.stat.ind.class(y, k, iter.max = iter.max,\\
\indent \indent \indent num.try = num.try, demean.ret = F)\\
\indent \indent p[i] <- sum((residuals(lm(u $\sim$ -1 + z)))\^{}2)\\
\indent \indent q[, i] <- apply(z, 1, cl.ix)\\
\indent \}\\
\\
\indent p1 <- unique(p)\\
\indent ct <- rep(NA, length(p1))\\
\indent for(i in 1:length(p1))\\
\indent \indent ct[i] <- sum(p1[i] == p)\\
\\
\indent p1 <- p1[ct == max(ct)]\\
\indent i <- match(p1, p)[1]\\
\indent ix <- q[, i]\\
\\
\indent k <- max(ix)\\
\indent z <- matrix(NA, n, k)\\
\indent for(j in 1:k)\\
\indent \indent z[, j] <- as.numeric(ix == j)\\
\\
\indent res <- bio.cl.wts(x.d, z)\\
\indent return(res)\\
\}\\
\\
bio.cl.wts <- function (x, ind)\\
\{\\
\indent first.ix <- function(x) match(1, x)[1]\\
\\
\indent calc.wts <- function(x, use.wts = F, use.geom = F)\\
\indent \{\\
\indent \indent if(use.geom)\\
\indent \indent \{\\
\indent \indent \indent if(use.wts)\\
\indent \indent \indent \indent s <- apply(log(x), 1, sd)\\
\indent \indent \indent else\\
\indent \indent \indent \indent s <- rep(1, nrow(x))\\
\indent \indent \indent s <- 1 / s / sum(1 / s)\\
\indent \indent \indent fac <- apply(x\^{}s, 2, prod)\\
\indent \indent \}\\
\indent \indent else\\
\indent \indent \{\\
\indent \indent \indent if(use.wts)\\
\indent \indent \indent \indent s <- apply(x, 1, sd)\\
\indent \indent \indent else\\
\indent \indent \indent \indent s <- rep(1, nrow(x))\\
\indent \indent \indent fac <- colMeans(x / s)\\
\indent \indent \}\\
\indent \indent w <- coefficients(lm(t(x) $\sim$ -1 + fac))\\
\indent \indent w <- 100 * w / sum(w)\\
\indent \indent return(w)\\
\indent \}\\
\\
\indent n <- nrow(x)\\
\indent w <- w.g <- v <- v.g <- rep(NA, n)\\
\\
\indent z <- colSums(ind)\\
\indent z <- as.numeric(paste(z, ".", apply(ind, 2, first.ix), sep = ""))\\
\indent dimnames(ind)[[2]] <- names(z) <- 1:ncol(ind)\\
\indent z <- sort(z)\\
\indent z <- names(z)\\
\indent ind <- ind[, z]\\
\indent dimnames(ind)[[2]] <- NULL\\
\\
\indent for(i in 1:ncol(ind))\\
\indent \{\\
\indent \indent take <- ind[, i] == 1\\
\indent \indent if(sum(take) == 1)\\
\indent \indent \{\\
\indent \indent \indent w[take] <- w.g[take] <- 1\\
\indent \indent \indent v[take] <- v.g[take] <- 1\\
\indent \indent \indent next\\
\indent \indent \}\\
\\
\indent \indent w[take] <- calc.wts(x[take, ], F, F)\\
\indent \indent w.g[take] <- calc.wts(x[take, ], F, T)\\
\indent \indent v[take] <- calc.wts(x[take, ], T, F)\\
\indent \indent v.g[take] <- calc.wts(x[take, ], T, T)\\
\indent \}\\
\\
\indent res <- new.env()\\
\indent res\$ind <- ind\\
\indent res\$w <- w\\
\indent res\$w.g <- w.g\\
\indent res\$v <- v\\
\indent res\$v.g <- v.g\\
\indent return(res)\\
\}
}}

\section{DISCLAIMERS}\label{app.disc}

{}Wherever the context so requires, the masculine gender includes the feminine and/or neuter, and the singular form includes the plural and {\em vice versa}. The author of this paper (``Author") and his affiliates including without limitation Quantigic$^\circledR$ Solutions LLC (``Author's Affiliates" or ``his Affiliates") make no implied or express warranties or any other representations whatsoever, including without limitation implied warranties of merchantability and fitness for a particular purpose, in connection with or with regard to the content of this paper including without limitation any code or algorithms contained herein (``Content").

{}The reader may use the Content solely at his/her/its own risk and the reader shall have no claims whatsoever against the Author or his Affiliates and the Author and his Affiliates shall have no liability whatsoever to the reader or any third party whatsoever for any loss, expense, opportunity cost, damages or any other adverse effects whatsoever relating to or arising from the use of the Content by the reader including without any limitation whatsoever: any direct, indirect, incidental, special, consequential or any other damages incurred by the reader, however caused and under any theory of liability; any loss of profit (whether incurred directly or indirectly), any loss of goodwill or reputation, any loss of data suffered, cost of procurement of substitute goods or services, or any other tangible or intangible loss; any reliance placed by the reader on the completeness, accuracy or existence of the Content or any other effect of using the Content; and any and all other adversities or negative effects the reader might encounter in using the Content irrespective of whether the Author or his Affiliates is or are or should have been aware of such adversities or negative effects.

{}The R code included in Appendix \ref{app.code} hereof is part of the copyrighted R code of Quantigic$^\circledR$ Solutions LLC and is provided herein with the express permission of Quantigic$^\circledR$ Solutions LLC. The copyright owner retains all rights, title and interest in and to its copyrighted source code included in Appendix \ref{app.code} hereof and any and all copyrights therefor.

\newpage\clearpage
\begin{table}[ht]
\noindent
\caption{Genome data summary. All Brain Lower Glioma samples are Pilocytic Astrocytoma samples. See Subsection \ref{data.summary} for the data source definitions.}
\begin{tabular}{l l l l} 
\\
\hline\hline 
Cancer Type & Total Counts & \# of Samples & Source \\[0.5ex] 
\hline 
B Cell Lymphoma & 43626 & 24 & A1-2\\
Bone Cancer & 36374 & 98 & B1\\
Brain Lower Grade Glioma & 3572 & 101 & A1, C1\\
Breast Cancer & 254381 & 119 & A1, D1\\
Chronic Lymphocytic Leukemia & 19489 & 134 & A1, E1-2\\
Esophageal Cancer & 1064 & 17 & F1\\
Gastric Cancer & 1996615 & 100 & G1\\
Liver Cancer & 3017487 & 389 & H1-2\\
Lung Cancer & 449527 & 24 & I1\\
Medulloblastoma & 44689 & 100 & J1\\
Ovarian Cancer & 668918 & 84 & K1\\
Pancreatic Cancer & 5087 & 100 & L1\\
Prostate Cancer & 29142 & 5 & M1\\
Renal Cell Carcinoma & 483329 & 94 & N1\\
All Cancer Types & 7053300 & 1389 & Above\\ [1ex] 
\hline 
\end{tabular}
\label{table.genome.summary} 
\end{table}

\newpage\clearpage
\begin{table}[ht]
\noindent
\caption{Occurrence counts for the first 48 mutation categories for the genome data summarized in Table \ref{table.genome.summary} aggregated by cancer types. X1 = B Cell Lymphoma, X2 = Bone Cancer, X3 = Brain Lower Grade Glioma, X4 = Breast Cancer, X5 = Chronic Lymphocytic Leukemia, X6 = Esophageal Cancer, X7 = Gastric Cancer, X8 = Liver Cancer, X9 = Lung Cancer, X10 = Medulloblastoma, X11 = Ovarian Cancer, X12 = Pancreatic Cancer, X13 = Prostate Cancer, X14 = Renal Cell Carcinoma. The mutations are encoded as follows: XYZW = Y $>$ W: XYZ.}
{\tiny
\begin{tabular}{l l l l l l l l l l l l l l l} 
\\
\hline\hline 
Mutation & X1 & X2 & X3 & X4 & X5 & X6 & X7 & X8 & X9 & X10 & X11 & X12 & X13 & X14\\[0.5ex] 
\hline 
\\
ACAA &  466 &  716 &  59 &  3024 & 286 & 11 &  24884 &  51929 & 15865 &  801 & 13831 &  36 &  577 & 10734 \\
ACCA &  355 &  528 &  41 &  2600 & 249 &  3 &  14361 &  15227 &  9217 &  681 & 11192 &  42 &  422 &  4952 \\
ACGA &   43 &  112 &   6 &   446 &  55 &  4 &   2253 &  51156 &  3567 &  115 &  1804 &   9 &   59 &  1231 \\
ACTA &  309 &  577 &  38 &  2201 & 195 & 12 &  15833 &  33385 & 10038 &  557 & 10508 &  20 &  380 &  5573 \\
CCAA &  420 &  538 &  29 &  2814 & 238 & 11 &  21980 &   9507 & 19569 &  579 & 10943 &  59 &  459 &  7899 \\
CCCA &  245 &  361 &  30 &  2149 & 147 &  6 &  19624 &  26869 & 16889 &  424 &  9720 &  45 &  336 &  6037 \\
CCGA &   77 &   75 &  22 &   470 &  35 &  7 &   4034 &   8695 &  4995 &  105 &  1644 &  29 &   67 &  1069 \\
CCTA &  418 &  502 &  26 &  2380 & 211 & 10 &  43418 &   2967 & 15930 &  544 & 10037 &  80 &  388 &  6825 \\
GCAA &  452 &  624 &  41 &  2144 & 244 & 16 &  26007 &  42508 & 11084 &  532 &  6835 &  33 &  419 &  4795 \\
GCCA &  234 &  346 &  18 &  1677 & 149 &  6 &  10594 &  32252 &  9087 &  337 &  6423 &  33 &  247 &  3563 \\
GCGA &   67 &   66 &   7 &   375 &  29 &  5 &   2649 &  17426 &  3607 &   94 &  1295 &  17 &   67 &   853 \\
GCTA &  351 &  463 &  13 &  1684 & 197 & 11 &  21205 &  33670 &  7502 &  345 &  6543 &  30 &  332 &  3637 \\
TCAA &  584 &  628 &  38 &  5873 & 233 & 11 &  24288 &  33919 & 13358 &  766 &  9816 &  54 &  582 &  8973 \\
TCCA &  313 &  472 &  36 &  3735 & 145 & 19 &  17372 & 139576 & 13140 &  567 &  9697 &  53 &  479 &  7388 \\
TCGA &   66 &   57 &   7 &   543 &  28 &  4 &   3168 &  11569 &  2385 &  128 &  1334 &  17 &   87 &  1097 \\
TCTA &  916 &  849 &  54 &  4938 & 376 & 25 &  43364 &  20943 & 14028 & 1111 & 12202 &  56 &  817 &  8611 \\
ACAG &  232 &  260 &  19 &  2124 & 162 &  1 &  10057 &  42216 &  4369 &  445 &  8653 &  18 &  197 &  4551 \\
ACCG &  143 &  148 &  11 &  1307 &  92 &  4 &   4818 &   6202 &  2088 &  256 &  4506 &  21 &  188 &  2687 \\
ACGG &   48 &   82 &  10 &   641 &  32 &  2 &   1606 &  27957 &   953 &  121 &  2172 &   5 &   46 &   837 \\
ACTG &  290 &  240 &  25 &  2278 & 113 &  8 &   8670 &  90212 &  3091 &  394 &  8460 &  23 &  237 &  4457 \\
CCAG &  153 &  160 &  13 &  1726 &  78 &  7 &   4942 &   8975 &  3233 &  175 &  5559 &  30 &  125 &  2474 \\
CCCG &  131 &  154 &  14 &  1391 &  75 &  2 &   4117 &  27928 &  2902 &  163 &  4857 &  18 &  142 &  2601 \\
CCGG &   31 &   65 &  14 &   597 &  20 &  0 &   1176 &  87975 &  1210 &  122 &  1769 &  11 &   55 &   532 \\
CCTG &  213 &  227 &  19 &  2425 &  92 & 11 &   6840 &  14272 &  3348 &  238 &  7943 &  28 &  204 &  3846 \\
GCAG &  106 &  120 &   8 &  1128 &  52 &  2 &   3844 &  53180 &  2546 &  196 &  3621 &  12 &   95 &  1847 \\
GCCG &  147 &  148 &   7 &  1142 &  80 &  3 &   4174 &  11783 &  2406 &  160 &  3363 &  16 &  112 &  1630 \\
GCGG &   25 &   42 &   2 &   525 &  33 &  1 &    776 &  52658 &  1113 &   52 &   918 &   5 &   27 &   423 \\
GCTG &  225 &  146 &   9 &  1650 &  81 &  5 &   5116 &  43761 &  2187 &  177 &  5418 &  14 &  135 &  2045 \\
TCAG &  391 &  266 &  22 & 16099 &  79 & 46 &  12809 &  12172 &  8882 &  265 & 13033 &  49 &  359 &  3968 \\
TCCG &  279 &  185 &  18 &  4896 &  95 & 14 &   7208 &  41020 &  4566 &  244 &  7977 &  30 &  264 &  4441 \\
TCGG &   30 &   52 &  11 &   794 &  13 &  2 &   1294 &   9554 &   992 &   64 &  1423 &  11 &   33 &   553 \\
TCTG &  660 &  444 &  44 & 21847 & 202 & 53 &  21720 &   3519 & 10841 &  449 & 20292 &  75 &  563 &  7668 \\
ACAT &  950 &  931 &  97 &  3557 & 440 & 14 &  46364 &  28569 &  6530 & 1137 & 13367 &  66 &  542 &  8872 \\
ACCT &  482 &  482 &  49 &  1875 & 265 & 14 &  16724 &  47067 &  3418 &  590 &  6543 & 100 &  285 &  4786 \\
ACGT & 1085 & 2373 & 289 &  4978 & 792 & 70 &  78681 &  17833 &  3830 & 3694 & 17475 & 585 & 1603 &  9034 \\
ACTT &  603 &  628 &  57 &  2570 & 263 & 11 &  21545 &  63195 &  4179 &  827 & 10578 &  54 &  380 &  6567 \\
CCAT &  729 &  542 &  74 &  3344 & 294 & 16 &  19611 &  23743 &  7430 &  801 &  8689 &  79 &  428 &  7106 \\
CCCT &  607 &  545 &  78 &  2443 & 337 & 10 &  18553 &  52051 &  6155 &  810 &  7554 &  70 &  325 &  6038 \\
CCGT &  845 & 1095 & 180 &  3489 & 483 & 70 &  53496 &   7282 &  4362 & 1983 & 11157 & 498 &  888 &  4824 \\
CCTT &  784 &  774 & 124 &  3468 & 380 & 13 &  22137 &  29223 &  7932 &  854 & 11714 &  69 &  430 &  8606 \\
GCAT &  615 &  531 &  65 &  2815 & 285 & 16 &  31548 &  50399 &  4166 &  711 &  8492 & 107 &  420 &  6069 \\
GCCT &  583 &  484 &  64 &  2256 & 295 & 14 &  34372 &  10858 &  3715 &  825 &  7784 & 120 &  402 &  5474 \\
GCGT &  930 & 1382 & 152 &  3870 & 546 & 60 &  75814 &  35864 &  3544 & 2428 & 13447 & 629 & 1294 &  5232 \\
GCTT &  660 &  585 &  78 &  2254 & 316 & 15 &  26554 &  49322 &  3836 &  741 &  8329 &  93 &  378 &  5765 \\
TCAT & 1531 &  761 &  86 & 25251 & 268 & 69 &  30161 &  11611 & 13414 &  780 & 13880 & 131 &  681 &  9723 \\
TCCT & 1172 &  685 &  73 &  7268 & 319 & 25 &  23799 &  31779 &  7136 &  859 & 10287 & 112 &  544 &  8833 \\
TCGT &  628 &  903 & 127 &  4510 & 281 & 58 &  33759 &  53275 &  2998 & 1553 &  8055 & 300 &  708 &  3772 \\
TCTT & 1466 &  813 &  57 & 13615 & 313 & 43 &  29570 &  21922 & 11432 &  859 & 12692 &  79 &  662 &  8933 \\
[1ex] 
\hline 
\end{tabular}
}
\label{table.aggr.data.1} 
\end{table}

\newpage\clearpage
\begin{table}[ht]
\noindent
\caption{Table \ref{table.aggr.data.1} continued: occurrence counts (aggregated by cancer types) for the next 48 mutation categories.}
{\tiny
\begin{tabular}{l l l l l l l l l l l l l l l} 
\\
\hline\hline 
Mutation & X1 & X2 & X3 & X4 & X5 & X6 & X7 & X8 & X9 & X10 & X11 & X12 & X13 & X14\\[0.5ex] 
\hline 
\\
ATAA &  385 &  382 &  22 &  1375 & 248 &  2 &   9613 &  35715 &  3957 &  315 &  5776 &   7 &  313 &  9726 \\
ATCA &  241 &  274 &  13 &  1388 & 191 &  6 &   9419 &   7570 &  1917 &  235 &  4439 &  18 &  214 &  2973 \\
ATGA &  232 &  344 &  21 &  1459 & 211 &  4 &   7666 &  48117 &  3962 &  228 &  5983 &  18 &  197 &  7703 \\
ATTA &  580 &  516 &  29 &  2262 & 301 &  5 &  30067 &  32078 &  2928 &  590 &  9131 &  16 &  523 &  5833 \\
CTAA &  185 &  217 &   9 &  1009 &  98 &  3 &   5551 &   9978 &  4331 &  151 &  4473 &  10 &  175 & 12963 \\
CTCA &  211 &  256 &  17 &  1535 & 118 &  6 &  14584 &  41504 &  3430 &  224 &  6343 &  21 &  184 &  6465 \\
CTGA &  219 &  227 &  26 &  1406 & 146 & 12 &   8654 &   8508 &  7905 &  210 &  6473 &  28 &  207 & 14356 \\
CTTA &  379 &  408 &  27 &  1716 & 162 &  6 &  24273 &   2400 &  4255 &  303 &  8578 &  24 &  221 &  8077 \\
GTAA &  180 &  162 &  10 &   797 &  82 &  1 &   4092 &  36890 &  3147 &  104 &  3254 &   5 &  138 &  5769 \\
GTCA &  174 &  147 &   8 &   791 &  80 &  1 &   5919 &  25736 &  1399 &  113 &  3026 &  10 &   79 &  1954 \\
GTGA &  156 &  145 &   9 &   948 & 102 & 10 &   6717 &   9623 &  4021 &  122 &  4233 &   8 &  135 &  3924 \\
GTTA &  181 &  272 &  14 &  1168 & 129 &  3 &  10065 &  43279 &  2112 &  200 &  5741 &  10 &  156 &  2579 \\
TTAA &  576 &  454 &  29 &  2192 & 256 &  2 &  20211 &  13356 &  4086 &  510 &  6936 &  16 &  499 & 11629 \\
TTCA &  158 &  202 &  17 &  1095 & 110 &  3 &   7296 &  75515 &  1766 &  222 &  4859 &  12 &  142 &  4966 \\
TTGA &  179 &  156 &   4 &   774 & 144 &  3 &   5147 &   5013 &  2912 &  140 &  3493 &  10 &  129 &  7897 \\
TTTA &  434 &  467 &  37 &  2298 & 289 &  3 &  27521 &  12174 &  3117 &  481 & 10786 &   8 &  417 & 10311 \\
ATAC &  910 &  792 &  84 &  3342 & 519 & 10 &  34145 &  35102 &  7884 &  865 & 14628 &  36 &  552 & 11060 \\
ATCC &  382 &  255 &  32 &  1439 & 192 &  7 &  14365 &   4444 &  1952 &  354 &  6071 &  22 &  206 &  2821 \\
ATGC &  572 &  526 &  51 &  2354 & 329 & 14 &  28630 &  16789 &  5325 &  534 & 10017 &  48 &  331 &  6561 \\
ATTC &  792 &  734 &  94 &  3340 & 457 & 11 &  22018 &  48320 &  4344 &  769 & 13211 &  38 &  503 &  6714 \\
CTAC &  456 &  283 &  30 &  1378 & 187 &  1 &  21942 &   9143 &  4198 &  301 &  5972 &  21 &  202 &  4855 \\
CTCC &  427 &  295 &  31 &  1670 & 197 &  5 &  31474 &  13443 &  2361 &  331 &  7202 &  48 &  163 &  3092 \\
CTGC &  531 &  328 &  29 &  1659 & 222 &  6 &  38742 &  59589 &  3845 &  360 &  6553 &  60 &  206 &  4024 \\
CTTC &  795 &  328 &  48 &  1828 & 316 & 12 &  73708 &  10381 &  3566 &  383 &  8088 &  48 &  239 &  4982 \\
GTAC &  452 &  378 &  30 &  1533 & 215 &  4 &  25980 &  46065 &  3605 &  462 &  6804 &  48 &  345 &  4411 \\
GTCC &  404 &  332 &  25 &   981 & 176 & 10 &  15268 &  17372 &  1619 &  299 &  4123 &  20 &  195 &  1823 \\
GTGC &  370 &  284 &  33 &  1233 & 189 &  8 &  24550 &  54718 &  2912 &  332 &  5851 &  48 &  211 &  2962 \\
GTTC &  511 &  447 &  46 &  1680 & 269 &  7 &  22803 &  46663 &  2547 &  532 &  7513 &  36 &  367 &  3575 \\
TTAC &  541 &  428 &  48 &  1819 & 297 &  2 &  22540 &  16834 &  3732 &  497 &  7741 &  25 &  372 &  6232 \\
TTCC &  606 &  383 &  44 &  1610 & 211 &  2 &  25505 &  42903 &  1979 &  473 &  7086 &  24 &  280 &  3847 \\
TTGC &  306 &  242 &  22 &  1079 & 185 &  9 &  17328 &   8016 &  2277 &  323 &  4419 &  24 &  210 &  2892 \\
TTTC &  818 &  437 &  60 &  2187 & 415 &  4 &  48109 &   2717 &  3140 &  617 & 10866 &  28 &  427 & 13766 \\
ATAG &  462 &  133 &   9 &   862 & 229 &  2 &   5956 &  20568 &  1055 &  164 &  3114 &   5 &  158 &  2747 \\
ATCG &  134 &   69 &   3 &   451 &  51 &  0 &   5214 &  50363 &   476 &   92 &  1335 &  10 &   93 &  1949 \\
ATGG &  196 &  136 &  11 &   941 &  76 &  3 &   5505 &  35331 &  1134 &  165 &  3935 &   5 &  116 &  2726 \\
ATTG &  580 &  155 &  17 &  1030 & 166 &  2 &  31252 &  58405 &  1010 &  177 &  3455 &  11 &  221 &  4941 \\
CTAG &  416 &  110 &   4 &   601 & 134 &  3 &   5035 &  29104 &   779 &   85 &  1932 &   3 &   87 &  1381 \\
CTCG &  219 &   91 &  16 &   759 &  71 &  4 &  14376 &  86003 &   695 &   96 &  2754 &  15 &   96 &  1628 \\
CTGG &  342 &  137 &   7 &  1165 & 140 &  4 &  12511 &  13036 &  1684 &  142 &  4488 &  19 &  142 &  2080 \\
CTTG & 1780 &  202 &  21 &  1560 & 344 &  5 & 142206 &  27191 &  1801 &  187 &  4424 &  38 &  171 &  2944 \\
GTAG &  224 &   82 &   3 &   617 &  78 &  0 &   2552 &  45734 &   746 &   87 &  1792 &   3 &   45 &   896 \\
GTCG &  132 &   53 &   4 &   667 &  43 &  0 &   3804 &  10333 &   542 &   53 &  1290 &  10 &   47 &   628 \\
GTGG &  183 &  102 &  12 &  2989 & 116 &  2 &   5256 &  21041 &  2874 &  171 &  5467 &  16 &   58 &  1313 \\
GTTG & 1090 &  113 &  14 &  1194 & 142 &  3 &  37454 &  46981 &  1106 &  174 &  3439 &  13 &  122 &  1423 \\
TTAG &  654 &  199 &  19 &  1021 & 260 &  1 &   7674 &  11096 &  1235 &  187 &  3327 &  14 &  158 &  2799 \\
TTCG &  167 &   96 &  13 &   725 &  79 &  2 &   8315 &  32124 &   740 &  130 &  2507 &  12 &  125 &  3312 \\
TTGG &  265 &  132 &  16 &  1124 & 121 &  7 &  11452 &  62446 &  1693 &  161 &  4480 &   8 &  142 &  2974 \\
TTTG & 1349 &  296 &  43 &  2144 & 403 &  4 &  77262 &  28801 &  2361 &  379 &  7679 &  39 &  353 & 11415 \\
[1ex] 
\hline 
\end{tabular}
}
\label{table.aggr.data.2} 
\end{table}

\newpage\clearpage
\begin{table}[ht]
\noindent
\caption{Top 10 clusterings by occurrence counts (second column) in 150,000 runs (performed as 15 consecutive batches of 10,000 runs in each batch). Each run is based on 1,000 samplings (i.e., {\tt\small num.try = 1000} in the R function {\tt\small qrm.stat.ind.class()}; also, the target number of clusters is {\tt\small k = 7}; see Appendix \ref{app.code} for details). The columns ``Cl-1" through ``Cl-7" give the numbers of mutations in each cluster (the total number of mutations in each clustering is 96). The entries ``--" correspond to clusterings with fewer than 7 clusters. Note that Clustering-C and Clustering-H have the same numbers of mutations in their 7 clusters; however, these two clusterings are different, i.e., equally-sized clusters contain different mutations. While there was slight variability in the placement (by occurrence counts) of the top 10 clusterings within the aforesaid 15 batches of 10,000 runs, Clustering-A through Clustering-J invariably were the top 10 in each batch. The weights are given in Tables \ref{table.weights.A.1} through \ref{table.weights.D.2} (for Clustering-A through Clustering-D) and \ref{table.weights.Z.1} and \ref{table.weights.Z.2} (for Clustering-A).}
\begin{tabular}{l l l l l l l l l} 
\\
\hline\hline 
Name & Count & Cl-1 & Cl-2 & Cl-3 & Cl-4 & Cl-5 & Cl-6 & Cl-7\\[0.5ex] 
\hline 
Clustering-A & 12085 & 8 & 8 & 10 & 15 & 16 & 18 & 21 \\
Clustering-B & 10962 & 7 & 8 & 8 & 12 & 15 & 17 & 29 \\
Clustering-C & 10788 & 8 & 8 & 11 & 15 & 16 & 17 & 21 \\
Clustering-D & 10328 & 8 & 8 & 15 & 16 & 18 & 31 & -- \\
Clustering-E & 6499 & 7 & 8 & 8 & 12 & 15 & 18 & 28 \\
Clustering-F & 5451 & 8 & 8 & 15 & 17 & 17 & 31 & -- \\
Clustering-G & 5421 & 8 & 8 & 10 & 15 & 17 & 17 & 21 \\
Clustering-H & 5302 & 8 & 8 & 11 & 15 & 16 & 17 & 21 \\
Clustering-I & 4602 & 8 & 8 & 10 & 15 & 21 & 34 & -- \\
Clustering-J & 3698 & 8 & 8 & 15 & 31 & 34 & -- & -- \\
 [1ex] 
\hline 
\end{tabular}
\label{table.occurrence.cts} 
\end{table}

\newpage\clearpage
\begin{table}[ht]
\noindent
\caption{Weights (in the units of 1\%, rounded to 2 digits) for the first 48 mutation categories (this Table \ref{table.weights.A.1} is continued in Table \ref{table.weights.A.2} with the next 48 mutation categories) for the 7 clusters in Clustering-A (see Table \ref{table.occurrence.cts}) based on unnormalized (columns 2-8) and normalized (columns 9-15) regressions (see Subsection \ref{sub.reg} for details). Each cluster is defined as containing the mutations with nonzero weights. For instance, cluster Cl-2 contains 8 mutations GCGA, TCGA, ACGG, GCCG, GCGG, TCGG, GTCA, GTCG. In each cluster the weights are normalized to add up to 100\% (up to 2 digits due to the aforesaid rounding). In Tables \ref{table.weights.A.1} through \ref{table.weights.D.2} ``weights based on unnormalized regressions" are given by (\ref{w.reg}), (\ref{w.norm}) and (\ref{H}), while ``weights based on normalized regressions" are given by (\ref{w.reg.1}), (\ref{w.norm}) and (\ref{H1}), i.e., the exposures are calculated based on arithmetic averages (see Subsection \ref{sub.reg} for details).}
{\tiny
\begin{tabular}{l l l l l l l l l l l l l l l} 
\\
\hline\hline 
Mutation & Cl-1 & Cl-2 & Cl-3 & Cl-4 & Cl-5 & Cl-6 & Cl-7 & Cl-1 & Cl-2 & Cl-3 & Cl-4 & Cl-5 & Cl-6 & Cl-7\\[0.5ex] 
\hline 
\\
ACAA &  0.00 &  0.00 &  0.00 & 6.55 &  0.00 & 0.00 & 0.00 &  0.00 &  0.00 &  0.00 & 6.55 &  0.00 & 0.00 & 0.00 \\
ACCA &  0.00 &  0.00 &  0.00 & 0.00 &  5.83 & 0.00 & 0.00 &  0.00 &  0.00 &  0.00 & 0.00 &  6.08 & 0.00 & 0.00 \\
ACGA &  0.00 &  0.00 &  0.00 & 0.00 &  0.00 & 0.00 & 4.06 &  0.00 &  0.00 &  0.00 & 0.00 &  0.00 & 0.00 & 4.00 \\
ACTA &  0.00 &  0.00 &  0.00 & 0.00 &  6.16 & 0.00 & 0.00 &  0.00 &  0.00 &  0.00 & 0.00 &  6.38 & 0.00 & 0.00 \\
CCAA &  0.00 &  0.00 &  0.00 & 0.00 &  7.91 & 0.00 & 0.00 &  0.00 &  0.00 &  0.00 & 0.00 &  8.10 & 0.00 & 0.00 \\
CCCA &  0.00 &  0.00 &  0.00 & 0.00 &  6.46 & 0.00 & 0.00 &  0.00 &  0.00 &  0.00 & 0.00 &  6.68 & 0.00 & 0.00 \\
CCGA &  0.00 &  0.00 &  7.21 & 0.00 &  0.00 & 0.00 & 0.00 &  0.00 &  0.00 &  7.23 & 0.00 &  0.00 & 0.00 & 0.00 \\
CCTA &  0.00 &  0.00 &  0.00 & 0.00 &  0.00 & 6.75 & 0.00 &  0.00 &  0.00 &  0.00 & 0.00 &  0.00 & 6.79 & 0.00 \\
GCAA &  4.05 &  0.00 &  0.00 & 0.00 &  0.00 & 0.00 & 0.00 &  4.65 &  0.00 &  0.00 & 0.00 &  0.00 & 0.00 & 0.00 \\
GCCA &  0.00 &  0.00 &  0.00 & 0.00 &  4.56 & 0.00 & 0.00 &  0.00 &  0.00 &  0.00 & 0.00 &  4.73 & 0.00 & 0.00 \\
GCGA &  0.00 & 13.81 &  0.00 & 0.00 &  0.00 & 0.00 & 0.00 &  0.00 & 13.89 &  0.00 & 0.00 &  0.00 & 0.00 & 0.00 \\
GCTA &  0.00 &  0.00 &  0.00 & 0.00 &  5.02 & 0.00 & 0.00 &  0.00 &  0.00 &  0.00 & 0.00 &  5.20 & 0.00 & 0.00 \\
TCAA &  0.00 &  0.00 &  0.00 & 6.26 &  0.00 & 0.00 & 0.00 &  0.00 &  0.00 &  0.00 & 6.21 &  0.00 & 0.00 & 0.00 \\
TCCA &  0.00 &  0.00 &  0.00 & 0.00 &  8.94 & 0.00 & 0.00 &  0.00 &  0.00 &  0.00 & 0.00 &  9.29 & 0.00 & 0.00 \\
TCGA &  0.00 & 11.87 &  0.00 & 0.00 &  0.00 & 0.00 & 0.00 &  0.00 & 12.24 &  0.00 & 0.00 &  0.00 & 0.00 & 0.00 \\
TCTA &  0.00 &  0.00 &  0.00 & 8.05 &  0.00 & 0.00 & 0.00 &  0.00 &  0.00 &  0.00 & 8.00 &  0.00 & 0.00 & 0.00 \\
ACAG &  0.00 &  0.00 &  0.00 & 0.00 &  3.96 & 0.00 & 0.00 &  0.00 &  0.00 &  0.00 & 0.00 &  4.18 & 0.00 & 0.00 \\
ACCG &  0.00 &  0.00 &  8.07 & 0.00 &  0.00 & 0.00 & 0.00 &  0.00 &  0.00 &  8.17 & 0.00 &  0.00 & 0.00 & 0.00 \\
ACGG &  0.00 & 12.62 &  0.00 & 0.00 &  0.00 & 0.00 & 0.00 &  0.00 & 12.22 &  0.00 & 0.00 &  0.00 & 0.00 & 0.00 \\
ACTG &  0.00 &  0.00 &  0.00 & 0.00 &  4.77 & 0.00 & 0.00 &  0.00 &  0.00 &  0.00 & 0.00 &  5.03 & 0.00 & 0.00 \\
CCAG &  0.00 &  0.00 &  9.26 & 0.00 &  0.00 & 0.00 & 0.00 &  0.00 &  0.00 &  9.35 & 0.00 &  0.00 & 0.00 & 0.00 \\
CCCG &  0.00 &  0.00 &  0.00 & 0.00 &  0.00 & 0.00 & 3.91 &  0.00 &  0.00 &  0.00 & 0.00 &  0.00 & 0.00 & 4.02 \\
CCGG &  0.00 &  0.00 &  0.00 & 0.00 &  0.00 & 0.00 & 5.37 &  0.00 &  0.00 &  0.00 & 0.00 &  0.00 & 0.00 & 5.12 \\
CCTG &  0.00 &  0.00 & 12.46 & 0.00 &  0.00 & 0.00 & 0.00 &  0.00 &  0.00 & 12.58 & 0.00 &  0.00 & 0.00 & 0.00 \\
GCAG &  0.00 &  0.00 &  0.00 & 0.00 &  0.00 & 0.00 & 4.61 &  0.00 &  0.00 &  0.00 & 0.00 &  0.00 & 0.00 & 4.57 \\
GCCG &  0.00 & 14.79 &  0.00 & 0.00 &  0.00 & 0.00 & 0.00 &  0.00 & 15.62 &  0.00 & 0.00 &  0.00 & 0.00 & 0.00 \\
GCGG &  0.00 & 15.50 &  0.00 & 0.00 &  0.00 & 0.00 & 0.00 &  0.00 & 13.92 &  0.00 & 0.00 &  0.00 & 0.00 & 0.00 \\
GCTG &  0.00 &  0.00 &  0.00 & 0.00 &  0.00 & 0.00 & 4.86 &  0.00 &  0.00 &  0.00 & 0.00 &  0.00 & 0.00 & 4.92 \\
TCAG &  0.00 &  0.00 &  0.00 & 0.00 & 10.31 & 0.00 & 0.00 &  0.00 &  0.00 &  0.00 & 0.00 &  9.03 & 0.00 & 0.00 \\
TCCG &  0.00 &  0.00 &  0.00 & 0.00 &  5.10 & 0.00 & 0.00 &  0.00 &  0.00 &  0.00 & 0.00 &  4.95 & 0.00 & 0.00 \\
TCGG &  0.00 &  8.40 &  0.00 & 0.00 &  0.00 & 0.00 & 0.00 &  0.00 &  8.65 &  0.00 & 0.00 &  0.00 & 0.00 & 0.00 \\
TCTG &  0.00 &  0.00 &  0.00 & 0.00 & 14.10 & 0.00 & 0.00 &  0.00 &  0.00 &  0.00 & 0.00 & 12.53 & 0.00 & 0.00 \\
ACAT &  0.00 &  0.00 &  0.00 & 7.67 &  0.00 & 0.00 & 0.00 &  0.00 &  0.00 &  0.00 & 7.71 &  0.00 & 0.00 & 0.00 \\
ACCT &  4.78 &  0.00 &  0.00 & 0.00 &  0.00 & 0.00 & 0.00 &  5.02 &  0.00 &  0.00 & 0.00 &  0.00 & 0.00 & 0.00 \\
ACGT & 23.47 &  0.00 &  0.00 & 0.00 &  0.00 & 0.00 & 0.00 & 23.18 &  0.00 &  0.00 & 0.00 &  0.00 & 0.00 & 0.00 \\
ACTT &  0.00 &  0.00 &  0.00 & 5.43 &  0.00 & 0.00 & 0.00 &  0.00 &  0.00 &  0.00 & 5.47 &  0.00 & 0.00 & 0.00 \\
CCAT &  0.00 &  0.00 &  0.00 & 6.02 &  0.00 & 0.00 & 0.00 &  0.00 &  0.00 &  0.00 & 6.02 &  0.00 & 0.00 & 0.00 \\
CCCT &  0.00 &  0.00 &  0.00 & 5.59 &  0.00 & 0.00 & 0.00 &  0.00 &  0.00 &  0.00 & 5.63 &  0.00 & 0.00 & 0.00 \\
CCGT & 17.66 &  0.00 &  0.00 & 0.00 &  0.00 & 0.00 & 0.00 & 17.12 &  0.00 &  0.00 & 0.00 &  0.00 & 0.00 & 0.00 \\
CCTT &  0.00 &  0.00 &  0.00 & 7.01 &  0.00 & 0.00 & 0.00 &  0.00 &  0.00 &  0.00 & 7.04 &  0.00 & 0.00 & 0.00 \\
GCAT &  0.00 &  0.00 &  0.00 & 5.98 &  0.00 & 0.00 & 0.00 &  0.00 &  0.00 &  0.00 & 6.01 &  0.00 & 0.00 & 0.00 \\
GCCT &  5.74 &  0.00 &  0.00 & 0.00 &  0.00 & 0.00 & 0.00 &  5.93 &  0.00 &  0.00 & 0.00 &  0.00 & 0.00 & 0.00 \\
GCGT & 20.46 &  0.00 &  0.00 & 0.00 &  0.00 & 0.00 & 0.00 & 19.80 &  0.00 &  0.00 & 0.00 &  0.00 & 0.00 & 0.00 \\
GCTT &  0.00 &  0.00 &  0.00 & 5.88 &  0.00 & 0.00 & 0.00 &  0.00 &  0.00 &  0.00 & 5.93 &  0.00 & 0.00 & 0.00 \\
TCAT & 11.42 &  0.00 &  0.00 & 0.00 &  0.00 & 0.00 & 0.00 & 12.00 &  0.00 &  0.00 & 0.00 &  0.00 & 0.00 & 0.00 \\
TCCT &  0.00 &  0.00 &  0.00 & 7.81 &  0.00 & 0.00 & 0.00 &  0.00 &  0.00 &  0.00 & 7.76 &  0.00 & 0.00 & 0.00 \\
TCGT & 12.42 &  0.00 &  0.00 & 0.00 &  0.00 & 0.00 & 0.00 & 12.30 &  0.00 &  0.00 & 0.00 &  0.00 & 0.00 & 0.00 \\
TCTT &  0.00 &  0.00 &  0.00 & 9.47 &  0.00 & 0.00 & 0.00 &  0.00 &  0.00 &  0.00 & 9.29 &  0.00 & 0.00 & 0.00 \\
[1ex] 
\hline 
\end{tabular}
}
\label{table.weights.A.1} 
\end{table}

\newpage\clearpage
\begin{table}[ht]
\noindent
\caption{Table \ref{table.weights.A.1} continued: weights for the next 48 mutation categories.}
{\tiny
\begin{tabular}{l l l l l l l l l l l l l l l} 
\\
\hline\hline 
Mutation & Cl-1 & Cl-2 & Cl-3 & Cl-4 & Cl-5 & Cl-6 & Cl-7 & Cl-1 & Cl-2 & Cl-3 & Cl-4 & Cl-5 & Cl-6 & Cl-7\\[0.5ex] 
\hline 
\\
ATAA &  0.00 &  0.00 &  0.00 & 0.00 &  4.18 & 0.00 & 0.00 &  0.00 &  0.00 &  0.00 & 0.00 &  4.52 & 0.00 & 0.00 \\
ATCA &  0.00 &  0.00 & 10.00 & 0.00 &  0.00 & 0.00 & 0.00 &  0.00 &  0.00 & 10.15 & 0.00 &  0.00 & 0.00 & 0.00 \\
ATGA &  0.00 &  0.00 &  0.00 & 0.00 &  4.02 & 0.00 & 0.00 &  0.00 &  0.00 &  0.00 & 0.00 &  4.30 & 0.00 & 0.00 \\
ATTA &  0.00 &  0.00 &  0.00 & 0.00 &  0.00 & 5.54 & 0.00 &  0.00 &  0.00 &  0.00 & 0.00 &  0.00 & 5.66 & 0.00 \\
CTAA &  0.00 &  0.00 & 11.74 & 0.00 &  0.00 & 0.00 & 0.00 &  0.00 &  0.00 & 11.16 & 0.00 &  0.00 & 0.00 & 0.00 \\
CTCA &  0.00 &  0.00 &  0.00 & 0.00 &  3.79 & 0.00 & 0.00 &  0.00 &  0.00 &  0.00 & 0.00 &  3.98 & 0.00 & 0.00 \\
CTGA &  0.00 &  0.00 &  0.00 & 0.00 &  4.88 & 0.00 & 0.00 &  0.00 &  0.00 &  0.00 & 0.00 &  5.02 & 0.00 & 0.00 \\
CTTA &  0.00 &  0.00 &  0.00 & 0.00 &  0.00 & 4.28 & 0.00 &  0.00 &  0.00 &  0.00 & 0.00 &  0.00 & 4.33 & 0.00 \\
GTAA &  0.00 &  0.00 &  0.00 & 0.00 &  0.00 & 0.00 & 4.30 &  0.00 &  0.00 &  0.00 & 0.00 &  0.00 & 0.00 & 4.35 \\
GTCA &  0.00 & 15.20 &  0.00 & 0.00 &  0.00 & 0.00 & 0.00 &  0.00 & 15.36 &  0.00 & 0.00 &  0.00 & 0.00 & 0.00 \\
GTGA &  0.00 &  0.00 &  9.28 & 0.00 &  0.00 & 0.00 & 0.00 &  0.00 &  0.00 &  9.21 & 0.00 &  0.00 & 0.00 & 0.00 \\
GTTA &  0.00 &  0.00 &  0.00 & 0.00 &  0.00 & 0.00 & 5.13 &  0.00 &  0.00 &  0.00 & 0.00 &  0.00 & 0.00 & 5.19 \\
TTAA &  0.00 &  0.00 &  0.00 & 0.00 &  0.00 & 5.13 & 0.00 &  0.00 &  0.00 &  0.00 & 0.00 &  0.00 & 5.26 & 0.00 \\
TTCA &  0.00 &  0.00 &  0.00 & 0.00 &  0.00 & 0.00 & 6.64 &  0.00 &  0.00 &  0.00 & 0.00 &  0.00 & 0.00 & 6.58 \\
TTGA &  0.00 &  0.00 &  8.84 & 0.00 &  0.00 & 0.00 & 0.00 &  0.00 &  0.00 &  8.55 & 0.00 &  0.00 & 0.00 & 0.00 \\
TTTA &  0.00 &  0.00 &  0.00 & 0.00 &  0.00 & 5.27 & 0.00 &  0.00 &  0.00 &  0.00 & 0.00 &  0.00 & 5.38 & 0.00 \\
ATAC &  0.00 &  0.00 &  0.00 & 7.03 &  0.00 & 0.00 & 0.00 &  0.00 &  0.00 &  0.00 & 7.06 &  0.00 & 0.00 & 0.00 \\
ATCC &  0.00 &  0.00 &  0.00 & 0.00 &  0.00 & 3.30 & 0.00 &  0.00 &  0.00 &  0.00 & 0.00 &  0.00 & 3.39 & 0.00 \\
ATGC &  0.00 &  0.00 &  0.00 & 4.97 &  0.00 & 0.00 & 0.00 &  0.00 &  0.00 &  0.00 & 4.98 &  0.00 & 0.00 & 0.00 \\
ATTC &  0.00 &  0.00 &  0.00 & 6.30 &  0.00 & 0.00 & 0.00 &  0.00 &  0.00 &  0.00 & 6.34 &  0.00 & 0.00 & 0.00 \\
CTAC &  0.00 &  0.00 &  0.00 & 0.00 &  0.00 & 3.78 & 0.00 &  0.00 &  0.00 &  0.00 & 0.00 &  0.00 & 3.81 & 0.00 \\
CTCC &  0.00 &  0.00 &  0.00 & 0.00 &  0.00 & 4.30 & 0.00 &  0.00 &  0.00 &  0.00 & 0.00 &  0.00 & 4.31 & 0.00 \\
CTGC &  0.00 &  0.00 &  0.00 & 0.00 &  0.00 & 5.37 & 0.00 &  0.00 &  0.00 &  0.00 & 0.00 &  0.00 & 5.41 & 0.00 \\
CTTC &  0.00 &  0.00 &  0.00 & 0.00 &  0.00 & 7.14 & 0.00 &  0.00 &  0.00 &  0.00 & 0.00 &  0.00 & 6.92 & 0.00 \\
GTAC &  0.00 &  0.00 &  0.00 & 0.00 &  0.00 & 4.84 & 0.00 &  0.00 &  0.00 &  0.00 & 0.00 &  0.00 & 4.96 & 0.00 \\
GTCC &  0.00 &  0.00 & 11.51 & 0.00 &  0.00 & 0.00 & 0.00 &  0.00 &  0.00 & 11.78 & 0.00 &  0.00 & 0.00 & 0.00 \\
GTGC &  0.00 &  0.00 &  0.00 & 0.00 &  0.00 & 4.32 & 0.00 &  0.00 &  0.00 &  0.00 & 0.00 &  0.00 & 4.43 & 0.00 \\
GTTC &  0.00 &  0.00 &  0.00 & 0.00 &  0.00 & 5.05 & 0.00 &  0.00 &  0.00 &  0.00 & 0.00 &  0.00 & 5.23 & 0.00 \\
TTAC &  0.00 &  0.00 &  0.00 & 0.00 &  0.00 & 4.97 & 0.00 &  0.00 &  0.00 &  0.00 & 0.00 &  0.00 & 5.10 & 0.00 \\
TTCC &  0.00 &  0.00 &  0.00 & 0.00 &  0.00 & 4.69 & 0.00 &  0.00 &  0.00 &  0.00 & 0.00 &  0.00 & 4.79 & 0.00 \\
TTGC &  0.00 &  0.00 & 11.62 & 0.00 &  0.00 & 0.00 & 0.00 &  0.00 &  0.00 & 11.82 & 0.00 &  0.00 & 0.00 & 0.00 \\
TTTC &  0.00 &  0.00 &  0.00 & 0.00 &  0.00 & 7.29 & 0.00 &  0.00 &  0.00 &  0.00 & 0.00 &  0.00 & 7.28 & 0.00 \\
ATAG &  0.00 &  0.00 &  0.00 & 0.00 &  0.00 & 0.00 & 3.98 &  0.00 &  0.00 &  0.00 & 0.00 &  0.00 & 0.00 & 4.09 \\
ATCG &  0.00 &  0.00 &  0.00 & 0.00 &  0.00 & 0.00 & 3.81 &  0.00 &  0.00 &  0.00 & 0.00 &  0.00 & 0.00 & 3.70 \\
ATGG &  0.00 &  0.00 &  0.00 & 0.00 &  0.00 & 0.00 & 3.97 &  0.00 &  0.00 &  0.00 & 0.00 &  0.00 & 0.00 & 3.99 \\
ATTG &  0.00 &  0.00 &  0.00 & 0.00 &  0.00 & 0.00 & 7.13 &  0.00 &  0.00 &  0.00 & 0.00 &  0.00 & 0.00 & 7.08 \\
CTAG &  0.00 &  0.00 &  0.00 & 0.00 &  0.00 & 0.00 & 3.55 &  0.00 &  0.00 &  0.00 & 0.00 &  0.00 & 0.00 & 3.56 \\
CTCG &  0.00 &  0.00 &  0.00 & 0.00 &  0.00 & 0.00 & 6.52 &  0.00 &  0.00 &  0.00 & 0.00 &  0.00 & 0.00 & 6.31 \\
CTGG &  0.00 &  0.00 &  0.00 & 0.00 &  0.00 & 0.00 & 3.67 &  0.00 &  0.00 &  0.00 & 0.00 &  0.00 & 0.00 & 3.83 \\
CTTG &  0.00 &  0.00 &  0.00 & 0.00 &  0.00 & 9.67 & 0.00 &  0.00 &  0.00 &  0.00 & 0.00 &  0.00 & 8.89 & 0.00 \\
GTAG &  0.00 &  0.00 &  0.00 & 0.00 &  0.00 & 0.00 & 3.58 &  0.00 &  0.00 &  0.00 & 0.00 &  0.00 & 0.00 & 3.49 \\
GTCG &  0.00 &  7.80 &  0.00 & 0.00 &  0.00 & 0.00 & 0.00 &  0.00 &  8.11 &  0.00 & 0.00 &  0.00 & 0.00 & 0.00 \\
GTGG &  0.00 &  0.00 &  0.00 & 0.00 &  0.00 & 0.00 & 3.82 &  0.00 &  0.00 &  0.00 & 0.00 &  0.00 & 0.00 & 3.98 \\
GTTG &  0.00 &  0.00 &  0.00 & 0.00 &  0.00 & 0.00 & 7.02 &  0.00 &  0.00 &  0.00 & 0.00 &  0.00 & 0.00 & 6.97 \\
TTAG &  0.00 &  0.00 &  0.00 & 0.00 &  0.00 & 0.00 & 4.24 &  0.00 &  0.00 &  0.00 & 0.00 &  0.00 & 0.00 & 4.43 \\
TTCG &  0.00 &  0.00 &  0.00 & 0.00 &  0.00 & 0.00 & 3.73 &  0.00 &  0.00 &  0.00 & 0.00 &  0.00 & 0.00 & 3.75 \\
TTGG &  0.00 &  0.00 &  0.00 & 0.00 &  0.00 & 0.00 & 6.10 &  0.00 &  0.00 &  0.00 & 0.00 &  0.00 & 0.00 & 6.06 \\
TTTG &  0.00 &  0.00 &  0.00 & 0.00 &  0.00 & 8.31 & 0.00 &  0.00 &  0.00 &  0.00 & 0.00 &  0.00 & 8.05 & 0.00 \\
[1ex] 
\hline 
\end{tabular}
}
\label{table.weights.A.2} 
\end{table}

\newpage\clearpage
\begin{table}[ht]
\noindent
\caption{Weights for the first 48 mutation categories for the 7 clusters in Clustering-B (see Table \ref{table.occurrence.cts}). The conventions are the same as in Table \ref{table.weights.A.1}.}
{\tiny
\begin{tabular}{l l l l l l l l l l l l l l l} 
\\
\hline\hline 
Mutation & Cl-1 & Cl-2 & Cl-3 & Cl-4 & Cl-5 & Cl-6 & Cl-7 & Cl-1 & Cl-2 & Cl-3 & Cl-4 & Cl-5 & Cl-6 & Cl-7\\[0.5ex] 
\hline 
\\
ACAA &  0.00 &  0.00 &  0.00 &  0.00 & 6.55 &  0.00 & 0.00 &  0.00 &  0.00 &  0.00 &  0.00 & 6.55 & 0.00 & 0.00 \\
ACCA &  0.00 &  0.00 &  0.00 &  6.66 & 0.00 &  0.00 & 0.00 &  0.00 &  0.00 &  0.00 &  7.03 & 0.00 & 0.00 & 0.00 \\
ACGA &  0.00 &  0.00 &  0.00 &  0.00 & 0.00 &  0.00 & 3.07 &  0.00 &  0.00 &  0.00 &  0.00 & 0.00 & 0.00 & 2.91 \\
ACTA &  0.00 &  0.00 &  0.00 &  6.99 & 0.00 &  0.00 & 0.00 &  0.00 &  0.00 &  0.00 &  7.33 & 0.00 & 0.00 & 0.00 \\
CCAA &  0.00 &  0.00 &  0.00 &  9.30 & 0.00 &  0.00 & 0.00 &  0.00 &  0.00 &  0.00 &  9.73 & 0.00 & 0.00 & 0.00 \\
CCCA &  0.00 &  0.00 &  0.00 &  7.52 & 0.00 &  0.00 & 0.00 &  0.00 &  0.00 &  0.00 &  7.93 & 0.00 & 0.00 & 0.00 \\
CCGA &  0.00 &  0.00 &  0.00 &  0.00 & 0.00 &  0.00 & 2.41 &  0.00 &  0.00 &  0.00 &  0.00 & 0.00 & 0.00 & 2.51 \\
CCTA &  0.00 &  0.00 &  0.00 &  8.69 & 0.00 &  0.00 & 0.00 &  0.00 &  0.00 &  0.00 &  9.10 & 0.00 & 0.00 & 0.00 \\
GCAA &  0.00 &  4.05 &  0.00 &  0.00 & 0.00 &  0.00 & 0.00 &  0.00 &  4.65 &  0.00 &  0.00 & 0.00 & 0.00 & 0.00 \\
GCCA &  0.00 &  0.00 &  0.00 &  5.20 & 0.00 &  0.00 & 0.00 &  0.00 &  0.00 &  0.00 &  5.46 & 0.00 & 0.00 & 0.00 \\
GCGA &  0.00 &  0.00 & 13.81 &  0.00 & 0.00 &  0.00 & 0.00 &  0.00 &  0.00 & 13.89 &  0.00 & 0.00 & 0.00 & 0.00 \\
GCTA &  0.00 &  0.00 &  0.00 &  5.70 & 0.00 &  0.00 & 0.00 &  0.00 &  0.00 &  0.00 &  5.97 & 0.00 & 0.00 & 0.00 \\
TCAA &  0.00 &  0.00 &  0.00 &  0.00 & 6.26 &  0.00 & 0.00 &  0.00 &  0.00 &  0.00 &  0.00 & 6.21 & 0.00 & 0.00 \\
TCCA &  0.00 &  0.00 &  0.00 &  9.80 & 0.00 &  0.00 & 0.00 &  0.00 &  0.00 &  0.00 & 10.21 & 0.00 & 0.00 & 0.00 \\
TCGA &  0.00 &  0.00 & 11.87 &  0.00 & 0.00 &  0.00 & 0.00 &  0.00 &  0.00 & 12.24 &  0.00 & 0.00 & 0.00 & 0.00 \\
TCTA &  0.00 &  0.00 &  0.00 &  0.00 & 8.05 &  0.00 & 0.00 &  0.00 &  0.00 &  0.00 &  0.00 & 8.00 & 0.00 & 0.00 \\
ACAG & 13.14 &  0.00 &  0.00 &  0.00 & 0.00 &  0.00 & 0.00 & 13.38 &  0.00 &  0.00 &  0.00 & 0.00 & 0.00 & 0.00 \\
ACCG &  0.00 &  0.00 &  0.00 &  0.00 & 0.00 &  0.00 & 2.76 &  0.00 &  0.00 &  0.00 &  0.00 & 0.00 & 0.00 & 2.91 \\
ACGG &  0.00 &  0.00 & 12.62 &  0.00 & 0.00 &  0.00 & 0.00 &  0.00 &  0.00 & 12.22 &  0.00 & 0.00 & 0.00 & 0.00 \\
ACTG & 17.06 &  0.00 &  0.00 &  0.00 & 0.00 &  0.00 & 0.00 & 17.11 &  0.00 &  0.00 &  0.00 & 0.00 & 0.00 & 0.00 \\
CCAG &  0.00 &  0.00 &  0.00 &  0.00 & 0.00 &  0.00 & 3.16 &  0.00 &  0.00 &  0.00 &  0.00 & 0.00 & 0.00 & 3.32 \\
CCCG &  0.00 &  0.00 &  0.00 &  0.00 & 0.00 &  0.00 & 3.22 &  0.00 &  0.00 &  0.00 &  0.00 & 0.00 & 0.00 & 3.24 \\
CCGG &  0.00 &  0.00 &  0.00 &  0.00 & 0.00 &  0.00 & 3.64 &  0.00 &  0.00 &  0.00 &  0.00 & 0.00 & 0.00 & 3.27 \\
CCTG &  0.00 &  0.00 &  0.00 &  0.00 & 0.00 &  0.00 & 4.32 &  0.00 &  0.00 &  0.00 &  0.00 & 0.00 & 0.00 & 4.51 \\
GCAG &  0.00 &  0.00 &  0.00 &  0.00 & 0.00 &  0.00 & 3.48 &  0.00 &  0.00 &  0.00 &  0.00 & 0.00 & 0.00 & 3.34 \\
GCCG &  0.00 &  0.00 & 14.79 &  0.00 & 0.00 &  0.00 & 0.00 &  0.00 &  0.00 & 15.62 &  0.00 & 0.00 & 0.00 & 0.00 \\
GCGG &  0.00 &  0.00 & 15.50 &  0.00 & 0.00 &  0.00 & 0.00 &  0.00 &  0.00 & 13.92 &  0.00 & 0.00 & 0.00 & 0.00 \\
GCTG &  0.00 &  0.00 &  0.00 &  0.00 & 0.00 &  0.00 & 3.82 &  0.00 &  0.00 &  0.00 &  0.00 & 0.00 & 0.00 & 3.78 \\
TCAG &  0.00 &  0.00 &  0.00 & 12.20 & 0.00 &  0.00 & 0.00 &  0.00 &  0.00 &  0.00 & 10.90 & 0.00 & 0.00 & 0.00 \\
TCCG &  0.00 &  0.00 &  0.00 &  5.76 & 0.00 &  0.00 & 0.00 &  0.00 &  0.00 &  0.00 &  5.60 & 0.00 & 0.00 & 0.00 \\
TCGG &  0.00 &  0.00 &  8.40 &  0.00 & 0.00 &  0.00 & 0.00 &  0.00 &  0.00 &  8.65 &  0.00 & 0.00 & 0.00 & 0.00 \\
TCTG &  0.00 &  0.00 &  0.00 & 16.63 & 0.00 &  0.00 & 0.00 &  0.00 &  0.00 &  0.00 & 15.02 & 0.00 & 0.00 & 0.00 \\
ACAT &  0.00 &  0.00 &  0.00 &  0.00 & 7.67 &  0.00 & 0.00 &  0.00 &  0.00 &  0.00 &  0.00 & 7.71 & 0.00 & 0.00 \\
ACCT &  0.00 &  4.78 &  0.00 &  0.00 & 0.00 &  0.00 & 0.00 &  0.00 &  5.02 &  0.00 &  0.00 & 0.00 & 0.00 & 0.00 \\
ACGT &  0.00 & 23.47 &  0.00 &  0.00 & 0.00 &  0.00 & 0.00 &  0.00 & 23.18 &  0.00 &  0.00 & 0.00 & 0.00 & 0.00 \\
ACTT &  0.00 &  0.00 &  0.00 &  0.00 & 5.43 &  0.00 & 0.00 &  0.00 &  0.00 &  0.00 &  0.00 & 5.47 & 0.00 & 0.00 \\
CCAT &  0.00 &  0.00 &  0.00 &  0.00 & 6.02 &  0.00 & 0.00 &  0.00 &  0.00 &  0.00 &  0.00 & 6.02 & 0.00 & 0.00 \\
CCCT &  0.00 &  0.00 &  0.00 &  0.00 & 5.59 &  0.00 & 0.00 &  0.00 &  0.00 &  0.00 &  0.00 & 5.63 & 0.00 & 0.00 \\
CCGT &  0.00 & 17.66 &  0.00 &  0.00 & 0.00 &  0.00 & 0.00 &  0.00 & 17.12 &  0.00 &  0.00 & 0.00 & 0.00 & 0.00 \\
CCTT &  0.00 &  0.00 &  0.00 &  0.00 & 7.01 &  0.00 & 0.00 &  0.00 &  0.00 &  0.00 &  0.00 & 7.04 & 0.00 & 0.00 \\
GCAT &  0.00 &  0.00 &  0.00 &  0.00 & 5.98 &  0.00 & 0.00 &  0.00 &  0.00 &  0.00 &  0.00 & 6.01 & 0.00 & 0.00 \\
GCCT &  0.00 &  5.74 &  0.00 &  0.00 & 0.00 &  0.00 & 0.00 &  0.00 &  5.93 &  0.00 &  0.00 & 0.00 & 0.00 & 0.00 \\
GCGT &  0.00 & 20.46 &  0.00 &  0.00 & 0.00 &  0.00 & 0.00 &  0.00 & 19.80 &  0.00 &  0.00 & 0.00 & 0.00 & 0.00 \\
GCTT &  0.00 &  0.00 &  0.00 &  0.00 & 5.88 &  0.00 & 0.00 &  0.00 &  0.00 &  0.00 &  0.00 & 5.93 & 0.00 & 0.00 \\
TCAT &  0.00 & 11.42 &  0.00 &  0.00 & 0.00 &  0.00 & 0.00 &  0.00 & 12.00 &  0.00 &  0.00 & 0.00 & 0.00 & 0.00 \\
TCCT &  0.00 &  0.00 &  0.00 &  0.00 & 7.81 &  0.00 & 0.00 &  0.00 &  0.00 &  0.00 &  0.00 & 7.76 & 0.00 & 0.00 \\
TCGT &  0.00 & 12.42 &  0.00 &  0.00 & 0.00 &  0.00 & 0.00 &  0.00 & 12.30 &  0.00 &  0.00 & 0.00 & 0.00 & 0.00 \\
TCTT &  0.00 &  0.00 &  0.00 &  0.00 & 9.47 &  0.00 & 0.00 &  0.00 &  0.00 &  0.00 &  0.00 & 9.29 & 0.00 & 0.00 \\
[1ex] 
\hline 
\end{tabular}
}
\label{table.weights.B.1} 
\end{table}

\newpage\clearpage
\begin{table}[ht]
\noindent
\caption{Table \ref{table.weights.B.1} continued: weights for the next 48 mutation categories.}
{\tiny
\begin{tabular}{l l l l l l l l l l l l l l l} 
\\
\hline\hline 
Mutation & Cl-1 & Cl-2 & Cl-3 & Cl-4 & Cl-5 & Cl-6 & Cl-7 & Cl-1 & Cl-2 & Cl-3 & Cl-4 & Cl-5 & Cl-6 & Cl-7\\[0.5ex] 
\hline 
\\
ATAA & 14.61 &  0.00 &  0.00 &  0.00 & 0.00 &  0.00 & 0.00 & 14.91 &  0.00 &  0.00 &  0.00 & 0.00 & 0.00 & 0.00 \\
ATCA &  0.00 &  0.00 &  0.00 &  0.00 & 0.00 &  0.00 & 3.46 &  0.00 &  0.00 &  0.00 &  0.00 & 0.00 & 0.00 & 3.65 \\
ATGA & 13.85 &  0.00 &  0.00 &  0.00 & 0.00 &  0.00 & 0.00 & 14.13 &  0.00 &  0.00 &  0.00 & 0.00 & 0.00 & 0.00 \\
ATTA &  0.00 &  0.00 &  0.00 &  0.00 & 0.00 &  5.95 & 0.00 &  0.00 &  0.00 &  0.00 &  0.00 & 0.00 & 6.08 & 0.00 \\
CTAA &  0.00 &  0.00 &  0.00 &  0.00 & 0.00 &  0.00 & 3.93 &  0.00 &  0.00 &  0.00 &  0.00 & 0.00 & 0.00 & 4.01 \\
CTCA & 12.75 &  0.00 &  0.00 &  0.00 & 0.00 &  0.00 & 0.00 & 12.93 &  0.00 &  0.00 &  0.00 & 0.00 & 0.00 & 0.00 \\
CTGA &  0.00 &  0.00 &  0.00 &  5.54 & 0.00 &  0.00 & 0.00 &  0.00 &  0.00 &  0.00 &  5.73 & 0.00 & 0.00 & 0.00 \\
CTTA &  0.00 &  0.00 &  0.00 &  0.00 & 0.00 &  4.55 & 0.00 &  0.00 &  0.00 &  0.00 &  0.00 & 0.00 & 4.62 & 0.00 \\
GTAA &  0.00 &  0.00 &  0.00 &  0.00 & 0.00 &  0.00 & 3.44 &  0.00 &  0.00 &  0.00 &  0.00 & 0.00 & 0.00 & 3.36 \\
GTCA &  0.00 &  0.00 & 15.20 &  0.00 & 0.00 &  0.00 & 0.00 &  0.00 &  0.00 & 15.36 &  0.00 & 0.00 & 0.00 & 0.00 \\
GTGA &  0.00 &  0.00 &  0.00 &  0.00 & 0.00 &  0.00 & 3.13 &  0.00 &  0.00 &  0.00 &  0.00 & 0.00 & 0.00 & 3.26 \\
GTTA &  0.00 &  0.00 &  0.00 &  0.00 & 0.00 &  0.00 & 4.04 &  0.00 &  0.00 &  0.00 &  0.00 & 0.00 & 0.00 & 4.00 \\
TTAA &  0.00 &  0.00 &  0.00 &  0.00 & 0.00 &  5.49 & 0.00 &  0.00 &  0.00 &  0.00 &  0.00 & 0.00 & 5.64 & 0.00 \\
TTCA &  0.00 &  0.00 &  0.00 &  0.00 & 0.00 &  0.00 & 4.99 &  0.00 &  0.00 &  0.00 &  0.00 & 0.00 & 0.00 & 4.77 \\
TTGA &  0.00 &  0.00 &  0.00 &  0.00 & 0.00 &  0.00 & 2.94 &  0.00 &  0.00 &  0.00 &  0.00 & 0.00 & 0.00 & 3.05 \\
TTTA &  0.00 &  0.00 &  0.00 &  0.00 & 0.00 &  5.65 & 0.00 &  0.00 &  0.00 &  0.00 &  0.00 & 0.00 & 5.77 & 0.00 \\
ATAC &  0.00 &  0.00 &  0.00 &  0.00 & 7.03 &  0.00 & 0.00 &  0.00 &  0.00 &  0.00 &  0.00 & 7.06 & 0.00 & 0.00 \\
ATCC &  0.00 &  0.00 &  0.00 &  0.00 & 0.00 &  3.53 & 0.00 &  0.00 &  0.00 &  0.00 &  0.00 & 0.00 & 3.64 & 0.00 \\
ATGC &  0.00 &  0.00 &  0.00 &  0.00 & 4.97 &  0.00 & 0.00 &  0.00 &  0.00 &  0.00 &  0.00 & 4.98 & 0.00 & 0.00 \\
ATTC &  0.00 &  0.00 &  0.00 &  0.00 & 6.30 &  0.00 & 0.00 &  0.00 &  0.00 &  0.00 &  0.00 & 6.34 & 0.00 & 0.00 \\
CTAC &  0.00 &  0.00 &  0.00 &  0.00 & 0.00 &  4.03 & 0.00 &  0.00 &  0.00 &  0.00 &  0.00 & 0.00 & 4.08 & 0.00 \\
CTCC &  0.00 &  0.00 &  0.00 &  0.00 & 0.00 &  4.59 & 0.00 &  0.00 &  0.00 &  0.00 &  0.00 & 0.00 & 4.61 & 0.00 \\
CTGC &  0.00 &  0.00 &  0.00 &  0.00 & 0.00 &  5.75 & 0.00 &  0.00 &  0.00 &  0.00 &  0.00 & 0.00 & 5.79 & 0.00 \\
CTTC &  0.00 &  0.00 &  0.00 &  0.00 & 0.00 &  7.65 & 0.00 &  0.00 &  0.00 &  0.00 &  0.00 & 0.00 & 7.41 & 0.00 \\
GTAC &  0.00 &  0.00 &  0.00 &  0.00 & 0.00 &  5.18 & 0.00 &  0.00 &  0.00 &  0.00 &  0.00 & 0.00 & 5.32 & 0.00 \\
GTCC &  0.00 &  0.00 &  0.00 &  0.00 & 0.00 &  0.00 & 4.18 &  0.00 &  0.00 &  0.00 &  0.00 & 0.00 & 0.00 & 4.36 \\
GTGC &  0.00 &  0.00 &  0.00 &  0.00 & 0.00 &  4.63 & 0.00 &  0.00 &  0.00 &  0.00 &  0.00 & 0.00 & 4.76 & 0.00 \\
GTTC &  0.00 &  0.00 &  0.00 &  0.00 & 0.00 &  5.42 & 0.00 &  0.00 &  0.00 &  0.00 &  0.00 & 0.00 & 5.62 & 0.00 \\
TTAC &  0.00 &  0.00 &  0.00 &  0.00 & 0.00 &  5.32 & 0.00 &  0.00 &  0.00 &  0.00 &  0.00 & 0.00 & 5.47 & 0.00 \\
TTCC &  0.00 &  0.00 &  0.00 &  0.00 & 0.00 &  5.05 & 0.00 &  0.00 &  0.00 &  0.00 &  0.00 & 0.00 & 5.16 & 0.00 \\
TTGC &  0.00 &  0.00 &  0.00 &  0.00 & 0.00 &  0.00 & 3.98 &  0.00 &  0.00 &  0.00 &  0.00 & 0.00 & 0.00 & 4.20 \\
TTTC &  0.00 &  0.00 &  0.00 &  0.00 & 0.00 &  7.82 & 0.00 &  0.00 &  0.00 &  0.00 &  0.00 & 0.00 & 7.82 & 0.00 \\
ATAG &  0.00 &  0.00 &  0.00 &  0.00 & 0.00 &  0.00 & 3.18 &  0.00 &  0.00 &  0.00 &  0.00 & 0.00 & 0.00 & 3.24 \\
ATCG &  0.00 &  0.00 &  0.00 &  0.00 & 0.00 &  0.00 & 2.70 &  0.00 &  0.00 &  0.00 &  0.00 & 0.00 & 0.00 & 2.52 \\
ATGG &  0.00 &  0.00 &  0.00 &  0.00 & 0.00 &  0.00 & 3.09 &  0.00 &  0.00 &  0.00 &  0.00 & 0.00 & 0.00 & 3.04 \\
ATTG & 14.44 &  0.00 &  0.00 &  0.00 & 0.00 &  0.00 & 0.00 & 14.09 &  0.00 &  0.00 &  0.00 & 0.00 & 0.00 & 0.00 \\
CTAG &  0.00 &  0.00 &  0.00 &  0.00 & 0.00 &  0.00 & 2.68 &  0.00 &  0.00 &  0.00 &  0.00 & 0.00 & 0.00 & 2.66 \\
CTCG &  0.00 &  0.00 &  0.00 &  0.00 & 0.00 &  0.00 & 4.59 &  0.00 &  0.00 &  0.00 &  0.00 & 0.00 & 0.00 & 4.30 \\
CTGG &  0.00 &  0.00 &  0.00 &  0.00 & 0.00 &  0.00 & 3.07 &  0.00 &  0.00 &  0.00 &  0.00 & 0.00 & 0.00 & 3.19 \\
CTTG &  0.00 &  0.00 &  0.00 &  0.00 & 0.00 & 10.44 & 0.00 &  0.00 &  0.00 &  0.00 &  0.00 & 0.00 & 9.55 & 0.00 \\
GTAG &  0.00 &  0.00 &  0.00 &  0.00 & 0.00 &  0.00 & 2.52 &  0.00 &  0.00 &  0.00 &  0.00 & 0.00 & 0.00 & 2.36 \\
GTCG &  0.00 &  0.00 &  7.80 &  0.00 & 0.00 &  0.00 & 0.00 &  0.00 &  0.00 &  8.11 &  0.00 & 0.00 & 0.00 & 0.00 \\
GTGG &  0.00 &  0.00 &  0.00 &  0.00 & 0.00 &  0.00 & 3.18 &  0.00 &  0.00 &  0.00 &  0.00 & 0.00 & 0.00 & 3.25 \\
GTTG & 14.14 &  0.00 &  0.00 &  0.00 & 0.00 &  0.00 & 0.00 & 13.44 &  0.00 &  0.00 &  0.00 & 0.00 & 0.00 & 0.00 \\
TTAG &  0.00 &  0.00 &  0.00 &  0.00 & 0.00 &  0.00 & 3.48 &  0.00 &  0.00 &  0.00 &  0.00 & 0.00 & 0.00 & 3.62 \\
TTCG &  0.00 &  0.00 &  0.00 &  0.00 & 0.00 &  0.00 & 2.91 &  0.00 &  0.00 &  0.00 &  0.00 & 0.00 & 0.00 & 2.87 \\
TTGG &  0.00 &  0.00 &  0.00 &  0.00 & 0.00 &  0.00 & 4.62 &  0.00 &  0.00 &  0.00 &  0.00 & 0.00 & 0.00 & 4.49 \\
TTTG &  0.00 &  0.00 &  0.00 &  0.00 & 0.00 &  8.95 & 0.00 &  0.00 &  0.00 &  0.00 &  0.00 & 0.00 & 8.65 & 0.00 \\
[1ex] 
\hline 
\end{tabular}
}
\label{table.weights.B.2} 
\end{table}

\newpage\clearpage
\begin{table}[ht]
\noindent
\caption{Weights for the first 48 mutation categories for the 7 clusters in Clustering-C (see Table \ref{table.occurrence.cts}). The conventions are the same as in Table \ref{table.weights.A.1}.}
{\tiny
\begin{tabular}{l l l l l l l l l l l l l l l} 
\\
\hline\hline 
Mutation & Cl-1 & Cl-2 & Cl-3 & Cl-4 & Cl-5 & Cl-6 & Cl-7 & Cl-1 & Cl-2 & Cl-3 & Cl-4 & Cl-5 & Cl-6 & Cl-7\\[0.5ex] 
\hline 
\\
ACAA &  0.00 &  0.00 &  0.00 & 6.55 &  0.00 &  0.00 & 0.00 &  0.00 &  0.00 &  0.00 & 6.55 &  0.00 & 0.00 & 0.00 \\
ACCA &  0.00 &  0.00 &  0.00 & 0.00 &  5.83 &  0.00 & 0.00 &  0.00 &  0.00 &  0.00 & 0.00 &  6.08 & 0.00 & 0.00 \\
ACGA &  0.00 &  0.00 &  0.00 & 0.00 &  0.00 &  0.00 & 4.06 &  0.00 &  0.00 &  0.00 & 0.00 &  0.00 & 0.00 & 4.00 \\
ACTA &  0.00 &  0.00 &  0.00 & 0.00 &  6.16 &  0.00 & 0.00 &  0.00 &  0.00 &  0.00 & 0.00 &  6.38 & 0.00 & 0.00 \\
CCAA &  0.00 &  0.00 &  0.00 & 0.00 &  7.91 &  0.00 & 0.00 &  0.00 &  0.00 &  0.00 & 0.00 &  8.10 & 0.00 & 0.00 \\
CCCA &  0.00 &  0.00 &  0.00 & 0.00 &  6.46 &  0.00 & 0.00 &  0.00 &  0.00 &  0.00 & 0.00 &  6.68 & 0.00 & 0.00 \\
CCGA &  0.00 &  0.00 &  6.40 & 0.00 &  0.00 &  0.00 & 0.00 &  0.00 &  0.00 &  6.40 & 0.00 &  0.00 & 0.00 & 0.00 \\
CCTA &  0.00 &  0.00 &  0.00 & 0.00 &  0.00 &  6.97 & 0.00 &  0.00 &  0.00 &  0.00 & 0.00 &  0.00 & 7.02 & 0.00 \\
GCAA &  4.05 &  0.00 &  0.00 & 0.00 &  0.00 &  0.00 & 0.00 &  4.65 &  0.00 &  0.00 & 0.00 &  0.00 & 0.00 & 0.00 \\
GCCA &  0.00 &  0.00 &  0.00 & 0.00 &  4.56 &  0.00 & 0.00 &  0.00 &  0.00 &  0.00 & 0.00 &  4.73 & 0.00 & 0.00 \\
GCGA &  0.00 & 13.81 &  0.00 & 0.00 &  0.00 &  0.00 & 0.00 &  0.00 & 13.89 &  0.00 & 0.00 &  0.00 & 0.00 & 0.00 \\
GCTA &  0.00 &  0.00 &  0.00 & 0.00 &  5.02 &  0.00 & 0.00 &  0.00 &  0.00 &  0.00 & 0.00 &  5.20 & 0.00 & 0.00 \\
TCAA &  0.00 &  0.00 &  0.00 & 6.26 &  0.00 &  0.00 & 0.00 &  0.00 &  0.00 &  0.00 & 6.21 &  0.00 & 0.00 & 0.00 \\
TCCA &  0.00 &  0.00 &  0.00 & 0.00 &  8.94 &  0.00 & 0.00 &  0.00 &  0.00 &  0.00 & 0.00 &  9.29 & 0.00 & 0.00 \\
TCGA &  0.00 & 11.87 &  0.00 & 0.00 &  0.00 &  0.00 & 0.00 &  0.00 & 12.24 &  0.00 & 0.00 &  0.00 & 0.00 & 0.00 \\
TCTA &  0.00 &  0.00 &  0.00 & 8.05 &  0.00 &  0.00 & 0.00 &  0.00 &  0.00 &  0.00 & 8.00 &  0.00 & 0.00 & 0.00 \\
ACAG &  0.00 &  0.00 &  0.00 & 0.00 &  3.96 &  0.00 & 0.00 &  0.00 &  0.00 &  0.00 & 0.00 &  4.18 & 0.00 & 0.00 \\
ACCG &  0.00 &  0.00 &  7.23 & 0.00 &  0.00 &  0.00 & 0.00 &  0.00 &  0.00 &  7.29 & 0.00 &  0.00 & 0.00 & 0.00 \\
ACGG &  0.00 & 12.62 &  0.00 & 0.00 &  0.00 &  0.00 & 0.00 &  0.00 & 12.22 &  0.00 & 0.00 &  0.00 & 0.00 & 0.00 \\
ACTG &  0.00 &  0.00 &  0.00 & 0.00 &  4.77 &  0.00 & 0.00 &  0.00 &  0.00 &  0.00 & 0.00 &  5.03 & 0.00 & 0.00 \\
CCAG &  0.00 &  0.00 &  8.26 & 0.00 &  0.00 &  0.00 & 0.00 &  0.00 &  0.00 &  8.31 & 0.00 &  0.00 & 0.00 & 0.00 \\
CCCG &  0.00 &  0.00 &  0.00 & 0.00 &  0.00 &  0.00 & 3.91 &  0.00 &  0.00 &  0.00 & 0.00 &  0.00 & 0.00 & 4.02 \\
CCGG &  0.00 &  0.00 &  0.00 & 0.00 &  0.00 &  0.00 & 5.37 &  0.00 &  0.00 &  0.00 & 0.00 &  0.00 & 0.00 & 5.12 \\
CCTG &  0.00 &  0.00 & 11.12 & 0.00 &  0.00 &  0.00 & 0.00 &  0.00 &  0.00 & 11.19 & 0.00 &  0.00 & 0.00 & 0.00 \\
GCAG &  0.00 &  0.00 &  0.00 & 0.00 &  0.00 &  0.00 & 4.61 &  0.00 &  0.00 &  0.00 & 0.00 &  0.00 & 0.00 & 4.57 \\
GCCG &  0.00 & 14.79 &  0.00 & 0.00 &  0.00 &  0.00 & 0.00 &  0.00 & 15.62 &  0.00 & 0.00 &  0.00 & 0.00 & 0.00 \\
GCGG &  0.00 & 15.50 &  0.00 & 0.00 &  0.00 &  0.00 & 0.00 &  0.00 & 13.92 &  0.00 & 0.00 &  0.00 & 0.00 & 0.00 \\
GCTG &  0.00 &  0.00 &  0.00 & 0.00 &  0.00 &  0.00 & 4.86 &  0.00 &  0.00 &  0.00 & 0.00 &  0.00 & 0.00 & 4.92 \\
TCAG &  0.00 &  0.00 &  0.00 & 0.00 & 10.31 &  0.00 & 0.00 &  0.00 &  0.00 &  0.00 & 0.00 &  9.03 & 0.00 & 0.00 \\
TCCG &  0.00 &  0.00 &  0.00 & 0.00 &  5.10 &  0.00 & 0.00 &  0.00 &  0.00 &  0.00 & 0.00 &  4.95 & 0.00 & 0.00 \\
TCGG &  0.00 &  8.40 &  0.00 & 0.00 &  0.00 &  0.00 & 0.00 &  0.00 &  8.65 &  0.00 & 0.00 &  0.00 & 0.00 & 0.00 \\
TCTG &  0.00 &  0.00 &  0.00 & 0.00 & 14.10 &  0.00 & 0.00 &  0.00 &  0.00 &  0.00 & 0.00 & 12.53 & 0.00 & 0.00 \\
ACAT &  0.00 &  0.00 &  0.00 & 7.67 &  0.00 &  0.00 & 0.00 &  0.00 &  0.00 &  0.00 & 7.71 &  0.00 & 0.00 & 0.00 \\
ACCT &  4.78 &  0.00 &  0.00 & 0.00 &  0.00 &  0.00 & 0.00 &  5.02 &  0.00 &  0.00 & 0.00 &  0.00 & 0.00 & 0.00 \\
ACGT & 23.47 &  0.00 &  0.00 & 0.00 &  0.00 &  0.00 & 0.00 & 23.18 &  0.00 &  0.00 & 0.00 &  0.00 & 0.00 & 0.00 \\
ACTT &  0.00 &  0.00 &  0.00 & 5.43 &  0.00 &  0.00 & 0.00 &  0.00 &  0.00 &  0.00 & 5.47 &  0.00 & 0.00 & 0.00 \\
CCAT &  0.00 &  0.00 &  0.00 & 6.02 &  0.00 &  0.00 & 0.00 &  0.00 &  0.00 &  0.00 & 6.02 &  0.00 & 0.00 & 0.00 \\
CCCT &  0.00 &  0.00 &  0.00 & 5.59 &  0.00 &  0.00 & 0.00 &  0.00 &  0.00 &  0.00 & 5.63 &  0.00 & 0.00 & 0.00 \\
CCGT & 17.66 &  0.00 &  0.00 & 0.00 &  0.00 &  0.00 & 0.00 & 17.12 &  0.00 &  0.00 & 0.00 &  0.00 & 0.00 & 0.00 \\
CCTT &  0.00 &  0.00 &  0.00 & 7.01 &  0.00 &  0.00 & 0.00 &  0.00 &  0.00 &  0.00 & 7.04 &  0.00 & 0.00 & 0.00 \\
GCAT &  0.00 &  0.00 &  0.00 & 5.98 &  0.00 &  0.00 & 0.00 &  0.00 &  0.00 &  0.00 & 6.01 &  0.00 & 0.00 & 0.00 \\
GCCT &  5.74 &  0.00 &  0.00 & 0.00 &  0.00 &  0.00 & 0.00 &  5.93 &  0.00 &  0.00 & 0.00 &  0.00 & 0.00 & 0.00 \\
GCGT & 20.46 &  0.00 &  0.00 & 0.00 &  0.00 &  0.00 & 0.00 & 19.80 &  0.00 &  0.00 & 0.00 &  0.00 & 0.00 & 0.00 \\
GCTT &  0.00 &  0.00 &  0.00 & 5.88 &  0.00 &  0.00 & 0.00 &  0.00 &  0.00 &  0.00 & 5.93 &  0.00 & 0.00 & 0.00 \\
TCAT & 11.42 &  0.00 &  0.00 & 0.00 &  0.00 &  0.00 & 0.00 & 12.00 &  0.00 &  0.00 & 0.00 &  0.00 & 0.00 & 0.00 \\
TCCT &  0.00 &  0.00 &  0.00 & 7.81 &  0.00 &  0.00 & 0.00 &  0.00 &  0.00 &  0.00 & 7.76 &  0.00 & 0.00 & 0.00 \\
TCGT & 12.42 &  0.00 &  0.00 & 0.00 &  0.00 &  0.00 & 0.00 & 12.30 &  0.00 &  0.00 & 0.00 &  0.00 & 0.00 & 0.00 \\
TCTT &  0.00 &  0.00 &  0.00 & 9.47 &  0.00 &  0.00 & 0.00 &  0.00 &  0.00 &  0.00 & 9.29 &  0.00 & 0.00 & 0.00 \\
[1ex] 
\hline 
\end{tabular}
}
\label{table.weights.C.1} 
\end{table}

\newpage\clearpage
\begin{table}[ht]
\noindent
\caption{Table \ref{table.weights.C.1} continued: weights for the next 48 mutation categories.}
{\tiny
\begin{tabular}{l l l l l l l l l l l l l l l} 
\\
\hline\hline 
Mutation & Cl-1 & Cl-2 & Cl-3 & Cl-4 & Cl-5 & Cl-6 & Cl-7 & Cl-1 & Cl-2 & Cl-3 & Cl-4 & Cl-5 & Cl-6 & Cl-7\\[0.5ex] 
\hline 
\\
ATAA &  0.00 &  0.00 &  0.00 & 0.00 &  4.18 &  0.00 & 0.00 &  0.00 &  0.00 &  0.00 & 0.00 &  4.52 & 0.00 & 0.00 \\
ATCA &  0.00 &  0.00 &  8.99 & 0.00 &  0.00 &  0.00 & 0.00 &  0.00 &  0.00 &  9.10 & 0.00 &  0.00 & 0.00 & 0.00 \\
ATGA &  0.00 &  0.00 &  0.00 & 0.00 &  4.02 &  0.00 & 0.00 &  0.00 &  0.00 &  0.00 & 0.00 &  4.30 & 0.00 & 0.00 \\
ATTA &  0.00 &  0.00 &  0.00 & 0.00 &  0.00 &  5.72 & 0.00 &  0.00 &  0.00 &  0.00 & 0.00 &  0.00 & 5.85 & 0.00 \\
CTAA &  0.00 &  0.00 & 10.32 & 0.00 &  0.00 &  0.00 & 0.00 &  0.00 &  0.00 &  9.83 & 0.00 &  0.00 & 0.00 & 0.00 \\
CTCA &  0.00 &  0.00 &  0.00 & 0.00 &  3.79 &  0.00 & 0.00 &  0.00 &  0.00 &  0.00 & 0.00 &  3.98 & 0.00 & 0.00 \\
CTGA &  0.00 &  0.00 &  0.00 & 0.00 &  4.88 &  0.00 & 0.00 &  0.00 &  0.00 &  0.00 & 0.00 &  5.02 & 0.00 & 0.00 \\
CTTA &  0.00 &  0.00 &  0.00 & 0.00 &  0.00 &  4.42 & 0.00 &  0.00 &  0.00 &  0.00 & 0.00 &  0.00 & 4.47 & 0.00 \\
GTAA &  0.00 &  0.00 &  0.00 & 0.00 &  0.00 &  0.00 & 4.30 &  0.00 &  0.00 &  0.00 & 0.00 &  0.00 & 0.00 & 4.35 \\
GTCA &  0.00 & 15.20 &  0.00 & 0.00 &  0.00 &  0.00 & 0.00 &  0.00 & 15.36 &  0.00 & 0.00 &  0.00 & 0.00 & 0.00 \\
GTGA &  0.00 &  0.00 &  8.23 & 0.00 &  0.00 &  0.00 & 0.00 &  0.00 &  0.00 &  8.16 & 0.00 &  0.00 & 0.00 & 0.00 \\
GTTA &  0.00 &  0.00 &  0.00 & 0.00 &  0.00 &  0.00 & 5.13 &  0.00 &  0.00 &  0.00 & 0.00 &  0.00 & 0.00 & 5.19 \\
TTAA &  0.00 &  0.00 &  0.00 & 0.00 &  0.00 &  5.30 & 0.00 &  0.00 &  0.00 &  0.00 & 0.00 &  0.00 & 5.43 & 0.00 \\
TTCA &  0.00 &  0.00 &  0.00 & 0.00 &  0.00 &  0.00 & 6.64 &  0.00 &  0.00 &  0.00 & 0.00 &  0.00 & 0.00 & 6.58 \\
TTGA &  0.00 &  0.00 &  7.82 & 0.00 &  0.00 &  0.00 & 0.00 &  0.00 &  0.00 &  7.57 & 0.00 &  0.00 & 0.00 & 0.00 \\
TTTA &  0.00 &  0.00 &  0.00 & 0.00 &  0.00 &  5.44 & 0.00 &  0.00 &  0.00 &  0.00 & 0.00 &  0.00 & 5.55 & 0.00 \\
ATAC &  0.00 &  0.00 &  0.00 & 7.03 &  0.00 &  0.00 & 0.00 &  0.00 &  0.00 &  0.00 & 7.06 &  0.00 & 0.00 & 0.00 \\
ATCC &  0.00 &  0.00 & 10.75 & 0.00 &  0.00 &  0.00 & 0.00 &  0.00 &  0.00 & 10.92 & 0.00 &  0.00 & 0.00 & 0.00 \\
ATGC &  0.00 &  0.00 &  0.00 & 4.97 &  0.00 &  0.00 & 0.00 &  0.00 &  0.00 &  0.00 & 4.98 &  0.00 & 0.00 & 0.00 \\
ATTC &  0.00 &  0.00 &  0.00 & 6.30 &  0.00 &  0.00 & 0.00 &  0.00 &  0.00 &  0.00 & 6.34 &  0.00 & 0.00 & 0.00 \\
CTAC &  0.00 &  0.00 &  0.00 & 0.00 &  0.00 &  3.91 & 0.00 &  0.00 &  0.00 &  0.00 & 0.00 &  0.00 & 3.94 & 0.00 \\
CTCC &  0.00 &  0.00 &  0.00 & 0.00 &  0.00 &  4.44 & 0.00 &  0.00 &  0.00 &  0.00 & 0.00 &  0.00 & 4.45 & 0.00 \\
CTGC &  0.00 &  0.00 &  0.00 & 0.00 &  0.00 &  5.56 & 0.00 &  0.00 &  0.00 &  0.00 & 0.00 &  0.00 & 5.61 & 0.00 \\
CTTC &  0.00 &  0.00 &  0.00 & 0.00 &  0.00 &  7.39 & 0.00 &  0.00 &  0.00 &  0.00 & 0.00 &  0.00 & 7.16 & 0.00 \\
GTAC &  0.00 &  0.00 &  0.00 & 0.00 &  0.00 &  5.00 & 0.00 &  0.00 &  0.00 &  0.00 & 0.00 &  0.00 & 5.14 & 0.00 \\
GTCC &  0.00 &  0.00 & 10.41 & 0.00 &  0.00 &  0.00 & 0.00 &  0.00 &  0.00 & 10.61 & 0.00 &  0.00 & 0.00 & 0.00 \\
GTGC &  0.00 &  0.00 &  0.00 & 0.00 &  0.00 &  4.47 & 0.00 &  0.00 &  0.00 &  0.00 & 0.00 &  0.00 & 4.59 & 0.00 \\
GTTC &  0.00 &  0.00 &  0.00 & 0.00 &  0.00 &  5.21 & 0.00 &  0.00 &  0.00 &  0.00 & 0.00 &  0.00 & 5.40 & 0.00 \\
TTAC &  0.00 &  0.00 &  0.00 & 0.00 &  0.00 &  5.13 & 0.00 &  0.00 &  0.00 &  0.00 & 0.00 &  0.00 & 5.26 & 0.00 \\
TTCC &  0.00 &  0.00 &  0.00 & 0.00 &  0.00 &  4.84 & 0.00 &  0.00 &  0.00 &  0.00 & 0.00 &  0.00 & 4.96 & 0.00 \\
TTGC &  0.00 &  0.00 & 10.48 & 0.00 &  0.00 &  0.00 & 0.00 &  0.00 &  0.00 & 10.62 & 0.00 &  0.00 & 0.00 & 0.00 \\
TTTC &  0.00 &  0.00 &  0.00 & 0.00 &  0.00 &  7.53 & 0.00 &  0.00 &  0.00 &  0.00 & 0.00 &  0.00 & 7.52 & 0.00 \\
ATAG &  0.00 &  0.00 &  0.00 & 0.00 &  0.00 &  0.00 & 3.98 &  0.00 &  0.00 &  0.00 & 0.00 &  0.00 & 0.00 & 4.09 \\
ATCG &  0.00 &  0.00 &  0.00 & 0.00 &  0.00 &  0.00 & 3.81 &  0.00 &  0.00 &  0.00 & 0.00 &  0.00 & 0.00 & 3.70 \\
ATGG &  0.00 &  0.00 &  0.00 & 0.00 &  0.00 &  0.00 & 3.97 &  0.00 &  0.00 &  0.00 & 0.00 &  0.00 & 0.00 & 3.99 \\
ATTG &  0.00 &  0.00 &  0.00 & 0.00 &  0.00 &  0.00 & 7.13 &  0.00 &  0.00 &  0.00 & 0.00 &  0.00 & 0.00 & 7.08 \\
CTAG &  0.00 &  0.00 &  0.00 & 0.00 &  0.00 &  0.00 & 3.55 &  0.00 &  0.00 &  0.00 & 0.00 &  0.00 & 0.00 & 3.56 \\
CTCG &  0.00 &  0.00 &  0.00 & 0.00 &  0.00 &  0.00 & 6.52 &  0.00 &  0.00 &  0.00 & 0.00 &  0.00 & 0.00 & 6.31 \\
CTGG &  0.00 &  0.00 &  0.00 & 0.00 &  0.00 &  0.00 & 3.67 &  0.00 &  0.00 &  0.00 & 0.00 &  0.00 & 0.00 & 3.83 \\
CTTG &  0.00 &  0.00 &  0.00 & 0.00 &  0.00 & 10.06 & 0.00 &  0.00 &  0.00 &  0.00 & 0.00 &  0.00 & 9.27 & 0.00 \\
GTAG &  0.00 &  0.00 &  0.00 & 0.00 &  0.00 &  0.00 & 3.58 &  0.00 &  0.00 &  0.00 & 0.00 &  0.00 & 0.00 & 3.49 \\
GTCG &  0.00 &  7.80 &  0.00 & 0.00 &  0.00 &  0.00 & 0.00 &  0.00 &  8.11 &  0.00 & 0.00 &  0.00 & 0.00 & 0.00 \\
GTGG &  0.00 &  0.00 &  0.00 & 0.00 &  0.00 &  0.00 & 3.82 &  0.00 &  0.00 &  0.00 & 0.00 &  0.00 & 0.00 & 3.98 \\
GTTG &  0.00 &  0.00 &  0.00 & 0.00 &  0.00 &  0.00 & 7.02 &  0.00 &  0.00 &  0.00 & 0.00 &  0.00 & 0.00 & 6.97 \\
TTAG &  0.00 &  0.00 &  0.00 & 0.00 &  0.00 &  0.00 & 4.24 &  0.00 &  0.00 &  0.00 & 0.00 &  0.00 & 0.00 & 4.43 \\
TTCG &  0.00 &  0.00 &  0.00 & 0.00 &  0.00 &  0.00 & 3.73 &  0.00 &  0.00 &  0.00 & 0.00 &  0.00 & 0.00 & 3.75 \\
TTGG &  0.00 &  0.00 &  0.00 & 0.00 &  0.00 &  0.00 & 6.10 &  0.00 &  0.00 &  0.00 & 0.00 &  0.00 & 0.00 & 6.06 \\
TTTG &  0.00 &  0.00 &  0.00 & 0.00 &  0.00 &  8.61 & 0.00 &  0.00 &  0.00 &  0.00 & 0.00 &  0.00 & 8.36 & 0.00 \\
[1ex] 
\hline 
\end{tabular}
}
\label{table.weights.C.2} 
\end{table}

\newpage\clearpage
\begin{table}[ht]
\noindent
\caption{Weights for the first 48 mutation categories for the 6 clusters in Clustering-D (see Table \ref{table.occurrence.cts}). The conventions are the same as in Table \ref{table.weights.A.1}.}
{\tiny
\begin{tabular}{l l l l l l l l l l l l l} 
\\
\hline\hline 
Mutation & Cl-1 & Cl-2 & Cl-3 & Cl-4 & Cl-5 & Cl-6 & Cl-1 & Cl-2 & Cl-3 & Cl-4 & Cl-5 & Cl-6 \\[0.5ex] 
\hline 
\\
ACAA &  0.00 &  0.00 & 6.55 &  0.00 & 0.00 & 0.00 &  0.00 &  0.00 & 6.55 &  0.00 & 0.00 & 0.00 \\
ACCA &  0.00 &  0.00 & 0.00 &  5.83 & 0.00 & 0.00 &  0.00 &  0.00 & 0.00 &  6.08 & 0.00 & 0.00 \\
ACGA &  0.00 &  0.00 & 0.00 &  0.00 & 0.00 & 2.75 &  0.00 &  0.00 & 0.00 &  0.00 & 0.00 & 2.63 \\
ACTA &  0.00 &  0.00 & 0.00 &  6.16 & 0.00 & 0.00 &  0.00 &  0.00 & 0.00 &  6.38 & 0.00 & 0.00 \\
CCAA &  0.00 &  0.00 & 0.00 &  7.91 & 0.00 & 0.00 &  0.00 &  0.00 & 0.00 &  8.10 & 0.00 & 0.00 \\
CCCA &  0.00 &  0.00 & 0.00 &  6.46 & 0.00 & 0.00 &  0.00 &  0.00 & 0.00 &  6.68 & 0.00 & 0.00 \\
CCGA &  0.00 &  0.00 & 0.00 &  0.00 & 0.00 & 2.13 &  0.00 &  0.00 & 0.00 &  0.00 & 0.00 & 2.25 \\
CCTA &  0.00 &  0.00 & 0.00 &  0.00 & 6.75 & 0.00 &  0.00 &  0.00 & 0.00 &  0.00 & 6.79 & 0.00 \\
GCAA &  4.05 &  0.00 & 0.00 &  0.00 & 0.00 & 0.00 &  4.65 &  0.00 & 0.00 &  0.00 & 0.00 & 0.00 \\
GCCA &  0.00 &  0.00 & 0.00 &  4.56 & 0.00 & 0.00 &  0.00 &  0.00 & 0.00 &  4.73 & 0.00 & 0.00 \\
GCGA &  0.00 & 13.81 & 0.00 &  0.00 & 0.00 & 0.00 &  0.00 & 13.89 & 0.00 &  0.00 & 0.00 & 0.00 \\
GCTA &  0.00 &  0.00 & 0.00 &  5.02 & 0.00 & 0.00 &  0.00 &  0.00 & 0.00 &  5.20 & 0.00 & 0.00 \\
TCAA &  0.00 &  0.00 & 6.26 &  0.00 & 0.00 & 0.00 &  0.00 &  0.00 & 6.21 &  0.00 & 0.00 & 0.00 \\
TCCA &  0.00 &  0.00 & 0.00 &  8.94 & 0.00 & 0.00 &  0.00 &  0.00 & 0.00 &  9.29 & 0.00 & 0.00 \\
TCGA &  0.00 & 11.87 & 0.00 &  0.00 & 0.00 & 0.00 &  0.00 & 12.24 & 0.00 &  0.00 & 0.00 & 0.00 \\
TCTA &  0.00 &  0.00 & 8.05 &  0.00 & 0.00 & 0.00 &  0.00 &  0.00 & 8.00 &  0.00 & 0.00 & 0.00 \\
ACAG &  0.00 &  0.00 & 0.00 &  3.96 & 0.00 & 0.00 &  0.00 &  0.00 & 0.00 &  4.18 & 0.00 & 0.00 \\
ACCG &  0.00 &  0.00 & 0.00 &  0.00 & 0.00 & 2.46 &  0.00 &  0.00 & 0.00 &  0.00 & 0.00 & 2.61 \\
ACGG &  0.00 & 12.62 & 0.00 &  0.00 & 0.00 & 0.00 &  0.00 & 12.22 & 0.00 &  0.00 & 0.00 & 0.00 \\
ACTG &  0.00 &  0.00 & 0.00 &  4.77 & 0.00 & 0.00 &  0.00 &  0.00 & 0.00 &  5.03 & 0.00 & 0.00 \\
CCAG &  0.00 &  0.00 & 0.00 &  0.00 & 0.00 & 2.81 &  0.00 &  0.00 & 0.00 &  0.00 & 0.00 & 2.98 \\
CCCG &  0.00 &  0.00 & 0.00 &  0.00 & 0.00 & 2.88 &  0.00 &  0.00 & 0.00 &  0.00 & 0.00 & 2.92 \\
CCGG &  0.00 &  0.00 & 0.00 &  0.00 & 0.00 & 3.29 &  0.00 &  0.00 & 0.00 &  0.00 & 0.00 & 2.97 \\
CCTG &  0.00 &  0.00 & 0.00 &  0.00 & 0.00 & 3.84 &  0.00 &  0.00 & 0.00 &  0.00 & 0.00 & 4.05 \\
GCAG &  0.00 &  0.00 & 0.00 &  0.00 & 0.00 & 3.13 &  0.00 &  0.00 & 0.00 &  0.00 & 0.00 & 3.02 \\
GCCG &  0.00 & 14.79 & 0.00 &  0.00 & 0.00 & 0.00 &  0.00 & 15.62 & 0.00 &  0.00 & 0.00 & 0.00 \\
GCGG &  0.00 & 15.50 & 0.00 &  0.00 & 0.00 & 0.00 &  0.00 & 13.92 & 0.00 &  0.00 & 0.00 & 0.00 \\
GCTG &  0.00 &  0.00 & 0.00 &  0.00 & 0.00 & 3.44 &  0.00 &  0.00 & 0.00 &  0.00 & 0.00 & 3.42 \\
TCAG &  0.00 &  0.00 & 0.00 & 10.31 & 0.00 & 0.00 &  0.00 &  0.00 & 0.00 &  9.03 & 0.00 & 0.00 \\
TCCG &  0.00 &  0.00 & 0.00 &  5.10 & 0.00 & 0.00 &  0.00 &  0.00 & 0.00 &  4.95 & 0.00 & 0.00 \\
TCGG &  0.00 &  8.40 & 0.00 &  0.00 & 0.00 & 0.00 &  0.00 &  8.65 & 0.00 &  0.00 & 0.00 & 0.00 \\
TCTG &  0.00 &  0.00 & 0.00 & 14.10 & 0.00 & 0.00 &  0.00 &  0.00 & 0.00 & 12.53 & 0.00 & 0.00 \\
ACAT &  0.00 &  0.00 & 7.67 &  0.00 & 0.00 & 0.00 &  0.00 &  0.00 & 7.71 &  0.00 & 0.00 & 0.00 \\
ACCT &  4.78 &  0.00 & 0.00 &  0.00 & 0.00 & 0.00 &  5.02 &  0.00 & 0.00 &  0.00 & 0.00 & 0.00 \\
ACGT & 23.47 &  0.00 & 0.00 &  0.00 & 0.00 & 0.00 & 23.18 &  0.00 & 0.00 &  0.00 & 0.00 & 0.00 \\
ACTT &  0.00 &  0.00 & 5.43 &  0.00 & 0.00 & 0.00 &  0.00 &  0.00 & 5.47 &  0.00 & 0.00 & 0.00 \\
CCAT &  0.00 &  0.00 & 6.02 &  0.00 & 0.00 & 0.00 &  0.00 &  0.00 & 6.02 &  0.00 & 0.00 & 0.00 \\
CCCT &  0.00 &  0.00 & 5.59 &  0.00 & 0.00 & 0.00 &  0.00 &  0.00 & 5.63 &  0.00 & 0.00 & 0.00 \\
CCGT & 17.66 &  0.00 & 0.00 &  0.00 & 0.00 & 0.00 & 17.12 &  0.00 & 0.00 &  0.00 & 0.00 & 0.00 \\
CCTT &  0.00 &  0.00 & 7.01 &  0.00 & 0.00 & 0.00 &  0.00 &  0.00 & 7.04 &  0.00 & 0.00 & 0.00 \\
GCAT &  0.00 &  0.00 & 5.98 &  0.00 & 0.00 & 0.00 &  0.00 &  0.00 & 6.01 &  0.00 & 0.00 & 0.00 \\
GCCT &  5.74 &  0.00 & 0.00 &  0.00 & 0.00 & 0.00 &  5.93 &  0.00 & 0.00 &  0.00 & 0.00 & 0.00 \\
GCGT & 20.46 &  0.00 & 0.00 &  0.00 & 0.00 & 0.00 & 19.80 &  0.00 & 0.00 &  0.00 & 0.00 & 0.00 \\
GCTT &  0.00 &  0.00 & 5.88 &  0.00 & 0.00 & 0.00 &  0.00 &  0.00 & 5.93 &  0.00 & 0.00 & 0.00 \\
TCAT & 11.42 &  0.00 & 0.00 &  0.00 & 0.00 & 0.00 & 12.00 &  0.00 & 0.00 &  0.00 & 0.00 & 0.00 \\
TCCT &  0.00 &  0.00 & 7.81 &  0.00 & 0.00 & 0.00 &  0.00 &  0.00 & 7.76 &  0.00 & 0.00 & 0.00 \\
TCGT & 12.42 &  0.00 & 0.00 &  0.00 & 0.00 & 0.00 & 12.30 &  0.00 & 0.00 &  0.00 & 0.00 & 0.00 \\
TCTT &  0.00 &  0.00 & 9.47 &  0.00 & 0.00 & 0.00 &  0.00 &  0.00 & 9.29 &  0.00 & 0.00 & 0.00 \\
[1ex] 
\hline 
\end{tabular}
}
\label{table.weights.D.1} 
\end{table}

\newpage\clearpage
\begin{table}[ht]
\noindent
\caption{Table \ref{table.weights.D.1} continued: weights for the next 48 mutation categories.}
{\tiny
\begin{tabular}{l l l l l l l l l l l l l} 
\\
\hline\hline 
Mutation & Cl-1 & Cl-2 & Cl-3 & Cl-4 & Cl-5 & Cl-6 & Cl-1 & Cl-2 & Cl-3 & Cl-4 & Cl-5 & Cl-6 \\[0.5ex] 
\hline 
\\
ATAA &  0.00 &  0.00 & 0.00 &  4.18 & 0.00 & 0.00 &  0.00 &  0.00 & 0.00 &  4.52 & 0.00 & 0.00 \\
ATCA &  0.00 &  0.00 & 0.00 &  0.00 & 0.00 & 3.11 &  0.00 &  0.00 & 0.00 &  0.00 & 0.00 & 3.29 \\
ATGA &  0.00 &  0.00 & 0.00 &  4.02 & 0.00 & 0.00 &  0.00 &  0.00 & 0.00 &  4.30 & 0.00 & 0.00 \\
ATTA &  0.00 &  0.00 & 0.00 &  0.00 & 5.54 & 0.00 &  0.00 &  0.00 & 0.00 &  0.00 & 5.66 & 0.00 \\
CTAA &  0.00 &  0.00 & 0.00 &  0.00 & 0.00 & 3.49 &  0.00 &  0.00 & 0.00 &  0.00 & 0.00 & 3.60 \\
CTCA &  0.00 &  0.00 & 0.00 &  3.79 & 0.00 & 0.00 &  0.00 &  0.00 & 0.00 &  3.98 & 0.00 & 0.00 \\
CTGA &  0.00 &  0.00 & 0.00 &  4.88 & 0.00 & 0.00 &  0.00 &  0.00 & 0.00 &  5.02 & 0.00 & 0.00 \\
CTTA &  0.00 &  0.00 & 0.00 &  0.00 & 4.28 & 0.00 &  0.00 &  0.00 & 0.00 &  0.00 & 4.33 & 0.00 \\
GTAA &  0.00 &  0.00 & 0.00 &  0.00 & 0.00 & 3.09 &  0.00 &  0.00 & 0.00 &  0.00 & 0.00 & 3.04 \\
GTCA &  0.00 & 15.20 & 0.00 &  0.00 & 0.00 & 0.00 &  0.00 & 15.36 & 0.00 &  0.00 & 0.00 & 0.00 \\
GTGA &  0.00 &  0.00 & 0.00 &  0.00 & 0.00 & 2.79 &  0.00 &  0.00 & 0.00 &  0.00 & 0.00 & 2.93 \\
GTTA &  0.00 &  0.00 & 0.00 &  0.00 & 0.00 & 3.65 &  0.00 &  0.00 & 0.00 &  0.00 & 0.00 & 3.63 \\
TTAA &  0.00 &  0.00 & 0.00 &  0.00 & 5.13 & 0.00 &  0.00 &  0.00 & 0.00 &  0.00 & 5.26 & 0.00 \\
TTCA &  0.00 &  0.00 & 0.00 &  0.00 & 0.00 & 4.50 &  0.00 &  0.00 & 0.00 &  0.00 & 0.00 & 4.33 \\
TTGA &  0.00 &  0.00 & 0.00 &  0.00 & 0.00 & 2.63 &  0.00 &  0.00 & 0.00 &  0.00 & 0.00 & 2.74 \\
TTTA &  0.00 &  0.00 & 0.00 &  0.00 & 5.27 & 0.00 &  0.00 &  0.00 & 0.00 &  0.00 & 5.38 & 0.00 \\
ATAC &  0.00 &  0.00 & 7.03 &  0.00 & 0.00 & 0.00 &  0.00 &  0.00 & 7.06 &  0.00 & 0.00 & 0.00 \\
ATCC &  0.00 &  0.00 & 0.00 &  0.00 & 3.30 & 0.00 &  0.00 &  0.00 & 0.00 &  0.00 & 3.39 & 0.00 \\
ATGC &  0.00 &  0.00 & 4.97 &  0.00 & 0.00 & 0.00 &  0.00 &  0.00 & 4.98 &  0.00 & 0.00 & 0.00 \\
ATTC &  0.00 &  0.00 & 6.30 &  0.00 & 0.00 & 0.00 &  0.00 &  0.00 & 6.34 &  0.00 & 0.00 & 0.00 \\
CTAC &  0.00 &  0.00 & 0.00 &  0.00 & 3.78 & 0.00 &  0.00 &  0.00 & 0.00 &  0.00 & 3.81 & 0.00 \\
CTCC &  0.00 &  0.00 & 0.00 &  0.00 & 4.30 & 0.00 &  0.00 &  0.00 & 0.00 &  0.00 & 4.31 & 0.00 \\
CTGC &  0.00 &  0.00 & 0.00 &  0.00 & 5.37 & 0.00 &  0.00 &  0.00 & 0.00 &  0.00 & 5.41 & 0.00 \\
CTTC &  0.00 &  0.00 & 0.00 &  0.00 & 7.14 & 0.00 &  0.00 &  0.00 & 0.00 &  0.00 & 6.92 & 0.00 \\
GTAC &  0.00 &  0.00 & 0.00 &  0.00 & 4.84 & 0.00 &  0.00 &  0.00 & 0.00 &  0.00 & 4.96 & 0.00 \\
GTCC &  0.00 &  0.00 & 0.00 &  0.00 & 0.00 & 3.81 &  0.00 &  0.00 & 0.00 &  0.00 & 0.00 & 3.95 \\
GTGC &  0.00 &  0.00 & 0.00 &  0.00 & 4.32 & 0.00 &  0.00 &  0.00 & 0.00 &  0.00 & 4.43 & 0.00 \\
GTTC &  0.00 &  0.00 & 0.00 &  0.00 & 5.05 & 0.00 &  0.00 &  0.00 & 0.00 &  0.00 & 5.23 & 0.00 \\
TTAC &  0.00 &  0.00 & 0.00 &  0.00 & 4.97 & 0.00 &  0.00 &  0.00 & 0.00 &  0.00 & 5.10 & 0.00 \\
TTCC &  0.00 &  0.00 & 0.00 &  0.00 & 4.69 & 0.00 &  0.00 &  0.00 & 0.00 &  0.00 & 4.79 & 0.00 \\
TTGC &  0.00 &  0.00 & 0.00 &  0.00 & 0.00 & 3.61 &  0.00 &  0.00 & 0.00 &  0.00 & 0.00 & 3.80 \\
TTTC &  0.00 &  0.00 & 0.00 &  0.00 & 7.29 & 0.00 &  0.00 &  0.00 & 0.00 &  0.00 & 7.28 & 0.00 \\
ATAG &  0.00 &  0.00 & 0.00 &  0.00 & 0.00 & 2.91 &  0.00 &  0.00 & 0.00 &  0.00 & 0.00 & 2.95 \\
ATCG &  0.00 &  0.00 & 0.00 &  0.00 & 0.00 & 2.46 &  0.00 &  0.00 & 0.00 &  0.00 & 0.00 & 2.30 \\
ATGG &  0.00 &  0.00 & 0.00 &  0.00 & 0.00 & 2.80 &  0.00 &  0.00 & 0.00 &  0.00 & 0.00 & 2.76 \\
ATTG &  0.00 &  0.00 & 0.00 &  0.00 & 0.00 & 4.93 &  0.00 &  0.00 & 0.00 &  0.00 & 0.00 & 4.79 \\
CTAG &  0.00 &  0.00 & 0.00 &  0.00 & 0.00 & 2.47 &  0.00 &  0.00 & 0.00 &  0.00 & 0.00 & 2.43 \\
CTCG &  0.00 &  0.00 & 0.00 &  0.00 & 0.00 & 4.21 &  0.00 &  0.00 & 0.00 &  0.00 & 0.00 & 3.93 \\
CTGG &  0.00 &  0.00 & 0.00 &  0.00 & 0.00 & 2.80 &  0.00 &  0.00 & 0.00 &  0.00 & 0.00 & 2.90 \\
CTTG &  0.00 &  0.00 & 0.00 &  0.00 & 9.67 & 0.00 &  0.00 &  0.00 & 0.00 &  0.00 & 8.89 & 0.00 \\
GTAG &  0.00 &  0.00 & 0.00 &  0.00 & 0.00 & 2.30 &  0.00 &  0.00 & 0.00 &  0.00 & 0.00 & 2.16 \\
GTCG &  0.00 &  7.80 & 0.00 &  0.00 & 0.00 & 0.00 &  0.00 &  8.11 & 0.00 &  0.00 & 0.00 & 0.00 \\
GTGG &  0.00 &  0.00 & 0.00 &  0.00 & 0.00 & 2.85 &  0.00 &  0.00 & 0.00 &  0.00 & 0.00 & 2.93 \\
GTTG &  0.00 &  0.00 & 0.00 &  0.00 & 0.00 & 4.83 &  0.00 &  0.00 & 0.00 &  0.00 & 0.00 & 4.73 \\
TTAG &  0.00 &  0.00 & 0.00 &  0.00 & 0.00 & 3.20 &  0.00 &  0.00 & 0.00 &  0.00 & 0.00 & 3.30 \\
TTCG &  0.00 &  0.00 & 0.00 &  0.00 & 0.00 & 2.64 &  0.00 &  0.00 & 0.00 &  0.00 & 0.00 & 2.60 \\
TTGG &  0.00 &  0.00 & 0.00 &  0.00 & 0.00 & 4.20 &  0.00 &  0.00 & 0.00 &  0.00 & 0.00 & 4.08 \\
TTTG &  0.00 &  0.00 & 0.00 &  0.00 & 8.31 & 0.00 &  0.00 &  0.00 & 0.00 &  0.00 & 8.05 & 0.00 \\
[1ex] 
\hline 
\end{tabular}
}
\label{table.weights.D.2} 
\end{table}

\newpage\clearpage
\begin{table}[ht]
\noindent
\caption{Weights (in the units of 1\%, rounded to 2 digits) for the first 48 mutation categories for the 7 clusters in Clustering-A (see Table \ref{table.occurrence.cts}) based on unnormalized (columns 2-8) and normalized (columns 9-15) regressions with the exposures computed via geometric means (see Subsection \ref{sub.reg} for details). Here ``weights based on unnormalized regressions" are given by (\ref{w.reg}), (\ref{w.norm}) and (\ref{H2}), while ``weights based on normalized regressions" are given by (\ref{w.reg.1}), (\ref{w.norm}) and (\ref{H3}). Other conventions are the same as in Table \ref{table.weights.A.1}.}
{\tiny
\begin{tabular}{l l l l l l l l l l l l l l l} 
\\
\hline\hline 
Mutation & Cl-1 & Cl-2 & Cl-3 & Cl-4 & Cl-5 & Cl-6 & Cl-7 & Cl-1 & Cl-2 & Cl-3 & Cl-4 & Cl-5 & Cl-6 & Cl-7\\[0.5ex] 
\hline 
\\
ACAA &  0.00 &  0.00 &  0.00 & 6.54 &  0.00 & 0.00 & 0.00 &  0.00 &  0.00 &  0.00 & 6.54 &  0.00 & 0.00 & 0.00 \\
ACCA &  0.00 &  0.00 &  0.00 & 0.00 &  6.16 & 0.00 & 0.00 &  0.00 &  0.00 &  0.00 & 0.00 &  6.20 & 0.00 & 0.00 \\
ACGA &  0.00 &  0.00 &  0.00 & 0.00 &  0.00 & 0.00 & 4.12 &  0.00 &  0.00 &  0.00 & 0.00 &  0.00 & 0.00 & 4.05 \\
ACTA &  0.00 &  0.00 &  0.00 & 0.00 &  6.38 & 0.00 & 0.00 &  0.00 &  0.00 &  0.00 & 0.00 &  6.44 & 0.00 & 0.00 \\
CCAA &  0.00 &  0.00 &  0.00 & 0.00 &  8.27 & 0.00 & 0.00 &  0.00 &  0.00 &  0.00 & 0.00 &  8.27 & 0.00 & 0.00 \\
CCCA &  0.00 &  0.00 &  0.00 & 0.00 &  6.73 & 0.00 & 0.00 &  0.00 &  0.00 &  0.00 & 0.00 &  6.77 & 0.00 & 0.00 \\
CCGA &  0.00 &  0.00 &  7.32 & 0.00 &  0.00 & 0.00 & 0.00 &  0.00 &  0.00 &  7.24 & 0.00 &  0.00 & 0.00 & 0.00 \\
CCTA &  0.00 &  0.00 &  0.00 & 0.00 &  0.00 & 6.77 & 0.00 &  0.00 &  0.00 &  0.00 & 0.00 &  0.00 & 6.76 & 0.00 \\
GCAA &  4.31 &  0.00 &  0.00 & 0.00 &  0.00 & 0.00 & 0.00 &  4.68 &  0.00 &  0.00 & 0.00 &  0.00 & 0.00 & 0.00 \\
GCCA &  0.00 &  0.00 &  0.00 & 0.00 &  4.70 & 0.00 & 0.00 &  0.00 &  0.00 &  0.00 & 0.00 &  4.75 & 0.00 & 0.00 \\
GCGA &  0.00 & 13.79 &  0.00 & 0.00 &  0.00 & 0.00 & 0.00 &  0.00 & 13.76 &  0.00 & 0.00 &  0.00 & 0.00 & 0.00 \\
GCTA &  0.00 &  0.00 &  0.00 & 0.00 &  5.16 & 0.00 & 0.00 &  0.00 &  0.00 &  0.00 & 0.00 &  5.22 & 0.00 & 0.00 \\
TCAA &  0.00 &  0.00 &  0.00 & 6.22 &  0.00 & 0.00 & 0.00 &  0.00 &  0.00 &  0.00 & 6.20 &  0.00 & 0.00 & 0.00 \\
TCCA &  0.00 &  0.00 &  0.00 & 0.00 &  8.86 & 0.00 & 0.00 &  0.00 &  0.00 &  0.00 & 0.00 &  9.08 & 0.00 & 0.00 \\
TCGA &  0.00 & 11.96 &  0.00 & 0.00 &  0.00 & 0.00 & 0.00 &  0.00 & 12.13 &  0.00 & 0.00 &  0.00 & 0.00 & 0.00 \\
TCTA &  0.00 &  0.00 &  0.00 & 8.04 &  0.00 & 0.00 & 0.00 &  0.00 &  0.00 &  0.00 & 8.01 &  0.00 & 0.00 & 0.00 \\
ACAG &  0.00 &  0.00 &  0.00 & 0.00 &  4.08 & 0.00 & 0.00 &  0.00 &  0.00 &  0.00 & 0.00 &  4.16 & 0.00 & 0.00 \\
ACCG &  0.00 &  0.00 &  8.12 & 0.00 &  0.00 & 0.00 & 0.00 &  0.00 &  0.00 &  8.17 & 0.00 &  0.00 & 0.00 & 0.00 \\
ACGG &  0.00 & 12.58 &  0.00 & 0.00 &  0.00 & 0.00 & 0.00 &  0.00 & 12.32 &  0.00 & 0.00 &  0.00 & 0.00 & 0.00 \\
ACTG &  0.00 &  0.00 &  0.00 & 0.00 &  4.73 & 0.00 & 0.00 &  0.00 &  0.00 &  0.00 & 0.00 &  4.88 & 0.00 & 0.00 \\
CCAG &  0.00 &  0.00 &  9.34 & 0.00 &  0.00 & 0.00 & 0.00 &  0.00 &  0.00 &  9.36 & 0.00 &  0.00 & 0.00 & 0.00 \\
CCCG &  0.00 &  0.00 &  0.00 & 0.00 &  0.00 & 0.00 & 3.97 &  0.00 &  0.00 &  0.00 & 0.00 &  0.00 & 0.00 & 4.04 \\
CCGG &  0.00 &  0.00 &  0.00 & 0.00 &  0.00 & 0.00 & 5.47 &  0.00 &  0.00 &  0.00 & 0.00 &  0.00 & 0.00 & 5.24 \\
CCTG &  0.00 &  0.00 & 12.56 & 0.00 &  0.00 & 0.00 & 0.00 &  0.00 &  0.00 & 12.61 & 0.00 &  0.00 & 0.00 & 0.00 \\
GCAG &  0.00 &  0.00 &  0.00 & 0.00 &  0.00 & 0.00 & 4.68 &  0.00 &  0.00 &  0.00 & 0.00 &  0.00 & 0.00 & 4.63 \\
GCCG &  0.00 & 14.96 &  0.00 & 0.00 &  0.00 & 0.00 & 0.00 &  0.00 & 15.53 &  0.00 & 0.00 &  0.00 & 0.00 & 0.00 \\
GCGG &  0.00 & 15.17 &  0.00 & 0.00 &  0.00 & 0.00 & 0.00 &  0.00 & 14.18 &  0.00 & 0.00 &  0.00 & 0.00 & 0.00 \\
GCTG &  0.00 &  0.00 &  0.00 & 0.00 &  0.00 & 0.00 & 4.92 &  0.00 &  0.00 &  0.00 & 0.00 &  0.00 & 0.00 & 4.94 \\
TCAG &  0.00 &  0.00 &  0.00 & 0.00 &  9.40 & 0.00 & 0.00 &  0.00 &  0.00 &  0.00 & 0.00 &  8.99 & 0.00 & 0.00 \\
TCCG &  0.00 &  0.00 &  0.00 & 0.00 &  4.93 & 0.00 & 0.00 &  0.00 &  0.00 &  0.00 & 0.00 &  4.90 & 0.00 & 0.00 \\
TCGG &  0.00 &  8.53 &  0.00 & 0.00 &  0.00 & 0.00 & 0.00 &  0.00 &  8.60 &  0.00 & 0.00 &  0.00 & 0.00 & 0.00 \\
TCTG &  0.00 &  0.00 &  0.00 & 0.00 & 13.10 & 0.00 & 0.00 &  0.00 &  0.00 &  0.00 & 0.00 & 12.56 & 0.00 & 0.00 \\
ACAT &  0.00 &  0.00 &  0.00 & 7.72 &  0.00 & 0.00 & 0.00 &  0.00 &  0.00 &  0.00 & 7.73 &  0.00 & 0.00 & 0.00 \\
ACCT &  4.86 &  0.00 &  0.00 & 0.00 &  0.00 & 0.00 & 0.00 &  5.01 &  0.00 &  0.00 & 0.00 &  0.00 & 0.00 & 0.00 \\
ACGT & 23.50 &  0.00 &  0.00 & 0.00 &  0.00 & 0.00 & 0.00 & 23.33 &  0.00 &  0.00 & 0.00 &  0.00 & 0.00 & 0.00 \\
ACTT &  0.00 &  0.00 &  0.00 & 5.45 &  0.00 & 0.00 & 0.00 &  0.00 &  0.00 &  0.00 & 5.47 &  0.00 & 0.00 & 0.00 \\
CCAT &  0.00 &  0.00 &  0.00 & 6.02 &  0.00 & 0.00 & 0.00 &  0.00 &  0.00 &  0.00 & 6.02 &  0.00 & 0.00 & 0.00 \\
CCCT &  0.00 &  0.00 &  0.00 & 5.60 &  0.00 & 0.00 & 0.00 &  0.00 &  0.00 &  0.00 & 5.62 &  0.00 & 0.00 & 0.00 \\
CCGT & 17.45 &  0.00 &  0.00 & 0.00 &  0.00 & 0.00 & 0.00 & 17.08 &  0.00 &  0.00 & 0.00 &  0.00 & 0.00 & 0.00 \\
CCTT &  0.00 &  0.00 &  0.00 & 7.03 &  0.00 & 0.00 & 0.00 &  0.00 &  0.00 &  0.00 & 7.05 &  0.00 & 0.00 & 0.00 \\
GCAT &  0.00 &  0.00 &  0.00 & 5.98 &  0.00 & 0.00 & 0.00 &  0.00 &  0.00 &  0.00 & 6.00 &  0.00 & 0.00 & 0.00 \\
GCCT &  5.85 &  0.00 &  0.00 & 0.00 &  0.00 & 0.00 & 0.00 &  5.97 &  0.00 &  0.00 & 0.00 &  0.00 & 0.00 & 0.00 \\
GCGT & 20.08 &  0.00 &  0.00 & 0.00 &  0.00 & 0.00 & 0.00 & 19.63 &  0.00 &  0.00 & 0.00 &  0.00 & 0.00 & 0.00 \\
GCTT &  0.00 &  0.00 &  0.00 & 5.90 &  0.00 & 0.00 & 0.00 &  0.00 &  0.00 &  0.00 & 5.92 &  0.00 & 0.00 & 0.00 \\
TCAT & 11.55 &  0.00 &  0.00 & 0.00 &  0.00 & 0.00 & 0.00 & 12.00 &  0.00 &  0.00 & 0.00 &  0.00 & 0.00 & 0.00 \\
TCCT &  0.00 &  0.00 &  0.00 & 7.77 &  0.00 & 0.00 & 0.00 &  0.00 &  0.00 &  0.00 & 7.75 &  0.00 & 0.00 & 0.00 \\
TCGT & 12.39 &  0.00 &  0.00 & 0.00 &  0.00 & 0.00 & 0.00 & 12.30 &  0.00 &  0.00 & 0.00 &  0.00 & 0.00 & 0.00 \\
TCTT &  0.00 &  0.00 &  0.00 & 9.35 &  0.00 & 0.00 & 0.00 &  0.00 &  0.00 &  0.00 & 9.27 &  0.00 & 0.00 & 0.00 \\
[1ex] 
\hline 
\end{tabular}
}
\label{table.weights.Z.1} 
\end{table}

\newpage\clearpage
\begin{table}[ht]
\noindent
\caption{Table \ref{table.weights.Z.1} continued: weights for the next 48 mutation categories.}
{\tiny
\begin{tabular}{l l l l l l l l l l l l l l l} 
\\
\hline\hline 
Mutation & Cl-1 & Cl-2 & Cl-3 & Cl-4 & Cl-5 & Cl-6 & Cl-7 & Cl-1 & Cl-2 & Cl-3 & Cl-4 & Cl-5 & Cl-6 & Cl-7\\[0.5ex] 
\hline 
\\
ATAA &  0.00 &  0.00 &  0.00 & 0.00 &  4.41 & 0.00 & 0.00 &  0.00 &  0.00 &  0.00 & 0.00 &  4.51 & 0.00 & 0.00 \\
ATCA &  0.00 &  0.00 & 10.06 & 0.00 &  0.00 & 0.00 & 0.00 &  0.00 &  0.00 & 10.15 & 0.00 &  0.00 & 0.00 & 0.00 \\
ATGA &  0.00 &  0.00 &  0.00 & 0.00 &  4.15 & 0.00 & 0.00 &  0.00 &  0.00 &  0.00 & 0.00 &  4.25 & 0.00 & 0.00 \\
ATTA &  0.00 &  0.00 &  0.00 & 0.00 &  0.00 & 5.59 & 0.00 &  0.00 &  0.00 &  0.00 & 0.00 &  0.00 & 5.64 & 0.00 \\
CTAA &  0.00 &  0.00 & 11.34 & 0.00 &  0.00 & 0.00 & 0.00 &  0.00 &  0.00 & 11.10 & 0.00 &  0.00 & 0.00 & 0.00 \\
CTCA &  0.00 &  0.00 &  0.00 & 0.00 &  3.87 & 0.00 & 0.00 &  0.00 &  0.00 &  0.00 & 0.00 &  3.94 & 0.00 & 0.00 \\
CTGA &  0.00 &  0.00 &  0.00 & 0.00 &  5.08 & 0.00 & 0.00 &  0.00 &  0.00 &  0.00 & 0.00 &  5.07 & 0.00 & 0.00 \\
CTTA &  0.00 &  0.00 &  0.00 & 0.00 &  0.00 & 4.33 & 0.00 &  0.00 &  0.00 &  0.00 & 0.00 &  0.00 & 4.31 & 0.00 \\
GTAA &  0.00 &  0.00 &  0.00 & 0.00 &  0.00 & 0.00 & 4.33 &  0.00 &  0.00 &  0.00 & 0.00 &  0.00 & 0.00 & 4.36 \\
GTCA &  0.00 & 15.17 &  0.00 & 0.00 &  0.00 & 0.00 & 0.00 &  0.00 & 15.40 &  0.00 & 0.00 &  0.00 & 0.00 & 0.00 \\
GTGA &  0.00 &  0.00 &  9.30 & 0.00 &  0.00 & 0.00 & 0.00 &  0.00 &  0.00 &  9.24 & 0.00 &  0.00 & 0.00 & 0.00 \\
GTTA &  0.00 &  0.00 &  0.00 & 0.00 &  0.00 & 0.00 & 5.18 &  0.00 &  0.00 &  0.00 & 0.00 &  0.00 & 0.00 & 5.22 \\
TTAA &  0.00 &  0.00 &  0.00 & 0.00 &  0.00 & 5.21 & 0.00 &  0.00 &  0.00 &  0.00 & 0.00 &  0.00 & 5.21 & 0.00 \\
TTCA &  0.00 &  0.00 &  0.00 & 0.00 &  0.00 & 0.00 & 6.73 &  0.00 &  0.00 &  0.00 & 0.00 &  0.00 & 0.00 & 6.66 \\
TTGA &  0.00 &  0.00 &  8.62 & 0.00 &  0.00 & 0.00 & 0.00 &  0.00 &  0.00 &  8.51 & 0.00 &  0.00 & 0.00 & 0.00 \\
TTTA &  0.00 &  0.00 &  0.00 & 0.00 &  0.00 & 5.36 & 0.00 &  0.00 &  0.00 &  0.00 & 0.00 &  0.00 & 5.35 & 0.00 \\
ATAC &  0.00 &  0.00 &  0.00 & 7.07 &  0.00 & 0.00 & 0.00 &  0.00 &  0.00 &  0.00 & 7.08 &  0.00 & 0.00 & 0.00 \\
ATCC &  0.00 &  0.00 &  0.00 & 0.00 &  0.00 & 3.38 & 0.00 &  0.00 &  0.00 &  0.00 & 0.00 &  0.00 & 3.40 & 0.00 \\
ATGC &  0.00 &  0.00 &  0.00 & 4.99 &  0.00 & 0.00 & 0.00 &  0.00 &  0.00 &  0.00 & 4.99 &  0.00 & 0.00 & 0.00 \\
ATTC &  0.00 &  0.00 &  0.00 & 6.34 &  0.00 & 0.00 & 0.00 &  0.00 &  0.00 &  0.00 & 6.36 &  0.00 & 0.00 & 0.00 \\
CTAC &  0.00 &  0.00 &  0.00 & 0.00 &  0.00 & 3.82 & 0.00 &  0.00 &  0.00 &  0.00 & 0.00 &  0.00 & 3.81 & 0.00 \\
CTCC &  0.00 &  0.00 &  0.00 & 0.00 &  0.00 & 4.31 & 0.00 &  0.00 &  0.00 &  0.00 & 0.00 &  0.00 & 4.32 & 0.00 \\
CTGC &  0.00 &  0.00 &  0.00 & 0.00 &  0.00 & 5.27 & 0.00 &  0.00 &  0.00 &  0.00 & 0.00 &  0.00 & 5.35 & 0.00 \\
CTTC &  0.00 &  0.00 &  0.00 & 0.00 &  0.00 & 7.09 & 0.00 &  0.00 &  0.00 &  0.00 & 0.00 &  0.00 & 7.01 & 0.00 \\
GTAC &  0.00 &  0.00 &  0.00 & 0.00 &  0.00 & 4.82 & 0.00 &  0.00 &  0.00 &  0.00 & 0.00 &  0.00 & 4.90 & 0.00 \\
GTCC &  0.00 &  0.00 & 11.65 & 0.00 &  0.00 & 0.00 & 0.00 &  0.00 &  0.00 & 11.80 & 0.00 &  0.00 & 0.00 & 0.00 \\
GTGC &  0.00 &  0.00 &  0.00 & 0.00 &  0.00 & 4.26 & 0.00 &  0.00 &  0.00 &  0.00 & 0.00 &  0.00 & 4.36 & 0.00 \\
GTTC &  0.00 &  0.00 &  0.00 & 0.00 &  0.00 & 5.08 & 0.00 &  0.00 &  0.00 &  0.00 & 0.00 &  0.00 & 5.18 & 0.00 \\
TTAC &  0.00 &  0.00 &  0.00 & 0.00 &  0.00 & 5.06 & 0.00 &  0.00 &  0.00 &  0.00 & 0.00 &  0.00 & 5.09 & 0.00 \\
TTCC &  0.00 &  0.00 &  0.00 & 0.00 &  0.00 & 4.69 & 0.00 &  0.00 &  0.00 &  0.00 & 0.00 &  0.00 & 4.76 & 0.00 \\
TTGC &  0.00 &  0.00 & 11.69 & 0.00 &  0.00 & 0.00 & 0.00 &  0.00 &  0.00 & 11.81 & 0.00 &  0.00 & 0.00 & 0.00 \\
TTTC &  0.00 &  0.00 &  0.00 & 0.00 &  0.00 & 7.37 & 0.00 &  0.00 &  0.00 &  0.00 & 0.00 &  0.00 & 7.31 & 0.00 \\
ATAG &  0.00 &  0.00 &  0.00 & 0.00 &  0.00 & 0.00 & 3.94 &  0.00 &  0.00 &  0.00 & 0.00 &  0.00 & 0.00 & 4.03 \\
ATCG &  0.00 &  0.00 &  0.00 & 0.00 &  0.00 & 0.00 & 3.83 &  0.00 &  0.00 &  0.00 & 0.00 &  0.00 & 0.00 & 3.74 \\
ATGG &  0.00 &  0.00 &  0.00 & 0.00 &  0.00 & 0.00 & 4.00 &  0.00 &  0.00 &  0.00 & 0.00 &  0.00 & 0.00 & 4.01 \\
ATTG &  0.00 &  0.00 &  0.00 & 0.00 &  0.00 & 0.00 & 6.98 &  0.00 &  0.00 &  0.00 & 0.00 &  0.00 & 0.00 & 7.00 \\
CTAG &  0.00 &  0.00 &  0.00 & 0.00 &  0.00 & 0.00 & 3.50 &  0.00 &  0.00 &  0.00 & 0.00 &  0.00 & 0.00 & 3.52 \\
CTCG &  0.00 &  0.00 &  0.00 & 0.00 &  0.00 & 0.00 & 6.53 &  0.00 &  0.00 &  0.00 & 0.00 &  0.00 & 0.00 & 6.37 \\
CTGG &  0.00 &  0.00 &  0.00 & 0.00 &  0.00 & 0.00 & 3.63 &  0.00 &  0.00 &  0.00 & 0.00 &  0.00 & 0.00 & 3.76 \\
CTTG &  0.00 &  0.00 &  0.00 & 0.00 &  0.00 & 9.36 & 0.00 &  0.00 &  0.00 &  0.00 & 0.00 &  0.00 & 9.13 & 0.00 \\
GTAG &  0.00 &  0.00 &  0.00 & 0.00 &  0.00 & 0.00 & 3.59 &  0.00 &  0.00 &  0.00 & 0.00 &  0.00 & 0.00 & 3.51 \\
GTCG &  0.00 &  7.84 &  0.00 & 0.00 &  0.00 & 0.00 & 0.00 &  0.00 &  8.08 &  0.00 & 0.00 &  0.00 & 0.00 & 0.00 \\
GTGG &  0.00 &  0.00 &  0.00 & 0.00 &  0.00 & 0.00 & 3.87 &  0.00 &  0.00 &  0.00 & 0.00 &  0.00 & 0.00 & 3.97 \\
GTTG &  0.00 &  0.00 &  0.00 & 0.00 &  0.00 & 0.00 & 6.71 &  0.00 &  0.00 &  0.00 & 0.00 &  0.00 & 0.00 & 6.77 \\
TTAG &  0.00 &  0.00 &  0.00 & 0.00 &  0.00 & 0.00 & 4.17 &  0.00 &  0.00 &  0.00 & 0.00 &  0.00 & 0.00 & 4.32 \\
TTCG &  0.00 &  0.00 &  0.00 & 0.00 &  0.00 & 0.00 & 3.74 &  0.00 &  0.00 &  0.00 & 0.00 &  0.00 & 0.00 & 3.76 \\
TTGG &  0.00 &  0.00 &  0.00 & 0.00 &  0.00 & 0.00 & 6.11 &  0.00 &  0.00 &  0.00 & 0.00 &  0.00 & 0.00 & 6.09 \\
TTTG &  0.00 &  0.00 &  0.00 & 0.00 &  0.00 & 8.22 & 0.00 &  0.00 &  0.00 &  0.00 & 0.00 &  0.00 & 8.12 & 0.00 \\
[1ex] 
\hline 
\end{tabular}
}
\label{table.weights.Z.2} 
\end{table}

\newpage\clearpage
\begin{table}[ht]
\noindent
\caption{The within-cluster cross-sectional correlations $\Theta_{sA}$ (columns 2-8), the overall correlations $\Xi_s$ (column 11) based on the overall cross-sectional regressions, and multiple $R^2$ and adjusted $R^2$ of these regressions (columns 9 and 10). See Subsection \ref{sub.cor} for details. Cancer types are labeled by X1 through X14 as in Table \ref{table.aggr.data.1}. All quantities are in the units of 1\% rounded to 2 digits. The values above 80\% are given in bold font.}
{\scriptsize
\begin{tabular}{l l l l l l l l l l l} 
\\
\hline\hline 
Cancer Type & Cl-1 & Cl-2 & Cl-3 & Cl-4 & Cl-5 & Cl-6 & Cl-7 & r.sq & adj.r.sq & Overall Cor\\[0.5ex] 
\hline 
X1 & 57.66 & 31.8 & 75.04 & {\bf 88.43} & {\bf 81.27} & {\bf 84.82} & 41.7 & {\bf 89.05} & {\bf 88.19} & {\bf 83.84} \\
X2 & {\bf 90.57} & 66.35 & {\bf 81.97} & 79.64 & 41.42 & -2.87 & 25.43 & {\bf 94.77} & {\bf 94.35} & {\bf 93.82} \\
X3 & {\bf 93.29} & -12.6 & 39.19 & 12.59 & 68.65 & 17.06 & 68.74 & {\bf 93.86} & {\bf 93.38} & {\bf 94.19} \\
X4 & 9.88 & 16.97 & 52.94 & 79.11 & {\bf 81.85} & 46.74 & 7.34 & 58.18 & 54.9 & 61.53 \\
X5 & {\bf 89.52} & 63.31 & 50.79 & 28.58 & 5.12 & {\bf 80.88} & 13.66 & {\bf 93.26} & {\bf 92.73} & {\bf 88.62} \\
X6 & {\bf 86.53} & 34.07 & 48.92 & 76.77 & {\bf 85.01} & 19.59 & 34.54 & {\bf 89.57} & {\bf 88.75} & {\bf 91.28} \\
X7 & {\bf 92.78} & 34.69 & 64.65 & 48.79 & 63.79 & {\bf 86.55} & 72.56 & {\bf 86.72} & {\bf 85.67} & {\bf 86.04} \\
X8 & -31.6 & 39.99 & 65.56 & -46.21 & -6.95 & -3.36 & 61.8 & 69.52 & 67.12 & 41.88 \\
X9 & -28.63 & 53.86 & -34.26 & 46.93 & 59.88 & 13.59 & -12.39 & 77.76 & 76.02 & 70.18 \\
X10 & {\bf 93.97} & 61.59 & 63.06 & 67.15 & 41.13 & 4.11 & 43.87 & {\bf 95.17} & {\bf 94.79} & {\bf 95.47} \\
X11 & {\bf 88.16} & 56.6 & 66.76 & 55.12 & {\bf 90.27} & 16.33 & 26.3 & {\bf 95.02} & {\bf 94.63} & {\bf 89.62} \\
X12 & {\bf 94.75} & 17.48 & 5.1 & 16.5 & {\bf 90} & 27.74 & 21.63 & {\bf 94.04} & {\bf 93.57} & {\bf 96.11} \\
X13 & {\bf 97.05} & 58.21 & 75.77 & 78.67 & {\bf 88.42} & 20.28 & 44.07 & {\bf 96.31} & {\bf 96.02} & {\bf 95.35} \\
X14 & 38.93 & 65.92 & 17.23 & 58.54 & 4.73 & 35.72 & 31.27 & {\bf 82.52} & {\bf 81.14} & 65.4 \\
 [1ex] 
\hline 
\end{tabular}
}
\label{table.fit.theta} 
\end{table}

\newpage\clearpage
\begin{table}[ht]
\noindent
\caption{The within-cluster cross-sectional correlations $\Delta_{\alpha A}$ between the weights for 7 cancer signatures Sig1 through Sig7 of \cite{BioFM} and the weights (using normalized regressions with exposures based on arithmetic averages) for 7 clusters in Clustering A (see Subsection \ref{sub.cor} for details). All quantities are in the units of 1\% rounded to 2 digits. The values above 80\% are given in bold font.}
{
\begin{tabular}{l l l l l l l l} 
\\
\hline\hline 
Signature & Cl-1 & Cl-2 & Cl-3 & Cl-4 & Cl-5 & Cl-6 & Cl-7 \\[0.5ex] 
\hline 
Sig1 & {\bf 92.05} & 10.29 & -6.42 & -8.33 & 51.12 & 29.06 & 20.61 \\
Sig2 & -0.37 & 1.75 & 42.13 & 75.58 & {\bf 80.12} & -27.92 & -3.34 \\
Sig3 & -51.53 & 54.4 & -37.16 & 28.19 & 32.98 & 12.37 & -17.7 \\
Sig4 & 31.56 & 11.97 & 54.43 & 56.83 & -1.17 & {\bf 84.25} & 60.41 \\
Sig5 & -42.53 & 40.31 & 62.96 & -47.62 & -8.34 & -8.39 & 61.61 \\
Sig6 & 47.79 & 40.62 & 17.8 & 27.45 & -27.96 & 16.87 & 16.97 \\
Sig7 & {\bf 80.94} & 19.87 & 55.03 & 33.4 & 13.89 & -29.59 & 13.93 \\
 [1ex] 
\hline 
\end{tabular}
}
\label{table.cor.delta} 
\end{table}

\newpage\clearpage

\newpage\clearpage

\newpage\clearpage

\newpage\clearpage

\newpage\clearpage

\newpage\clearpage

\newpage\clearpage

\newpage\clearpage

\newpage\clearpage

\newpage\clearpage

\newpage\clearpage

\newpage\clearpage

\newpage\clearpage

\newpage\clearpage

\newpage\clearpage

\newpage\clearpage

\newpage\clearpage

\newpage\clearpage

\newpage\clearpage

\newpage\clearpage

\newpage\clearpage

\newpage\clearpage

\newpage\clearpage

\newpage\clearpage

\newpage\clearpage

\newpage\clearpage

\newpage\clearpage

\newpage\clearpage

\newpage\clearpage

\newpage\clearpage

\newpage\clearpage

\newpage\clearpage

\newpage\clearpage

\newpage\clearpage

\newpage\clearpage

\newpage\clearpage

\newpage\clearpage

\newpage\clearpage

\newpage\clearpage

\newpage\clearpage

\newpage\clearpage

\newpage\clearpage

\newpage\clearpage

\newpage\clearpage

\newpage\clearpage

\newpage\clearpage

\newpage\clearpage

\newpage\clearpage

\newpage\clearpage

\newpage\clearpage

\newpage\clearpage

\newpage\clearpage

\newpage\clearpage
\begin{figure}[ht]
\centering
\includegraphics[scale=0.7]{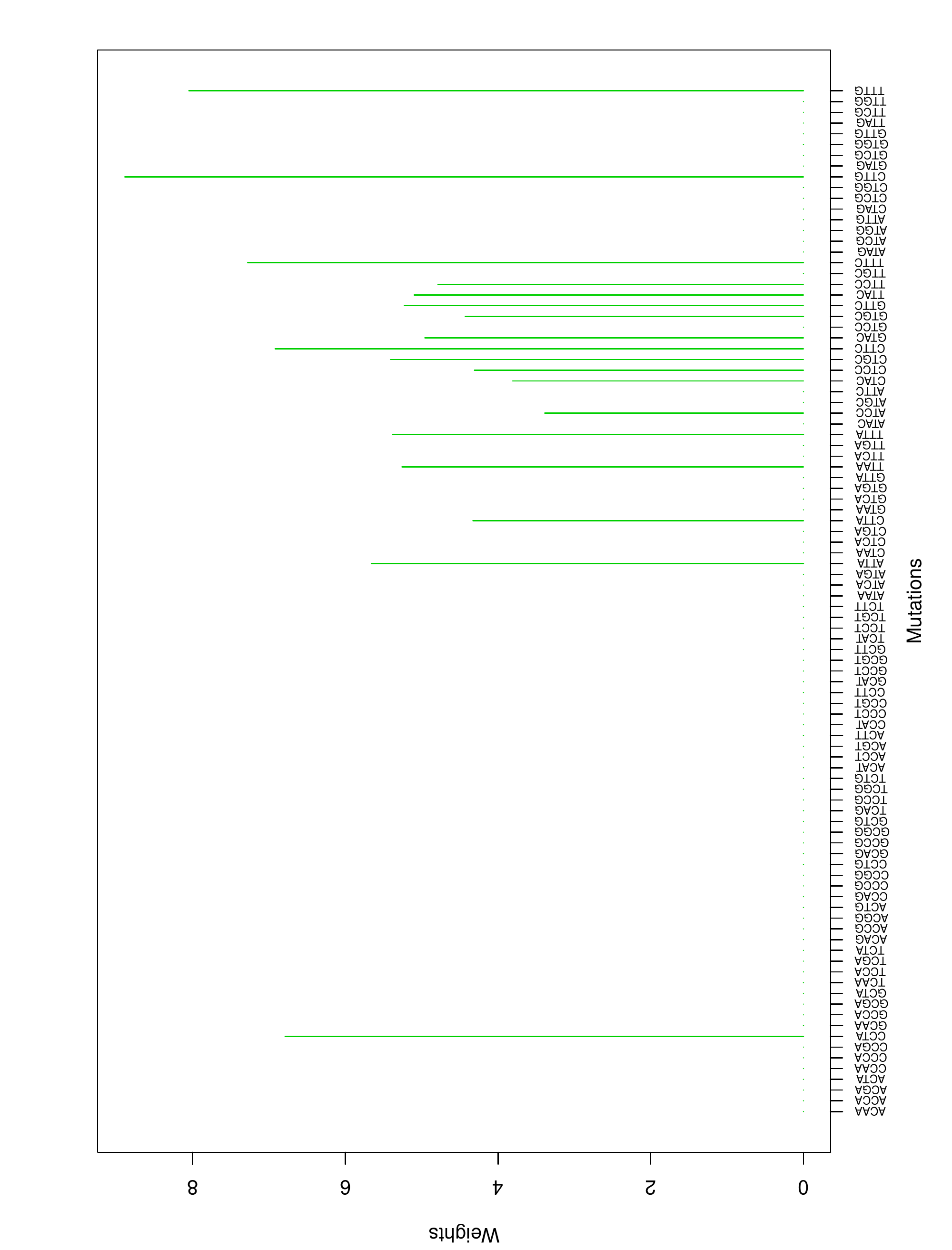}
\caption{Cluster Cl-5 in Clustering-D with weights based on normalized regressions with arithmetic means
(see Subsection \ref{sub.reg}).
See Tables \ref{table.occurrence.cts}, \ref{table.weights.D.1}, \ref{table.weights.D.2}.}
\label{FigureNorm5D}
\end{figure}

\newpage\clearpage
\begin{figure}[ht]
\centering
\includegraphics[scale=0.7]{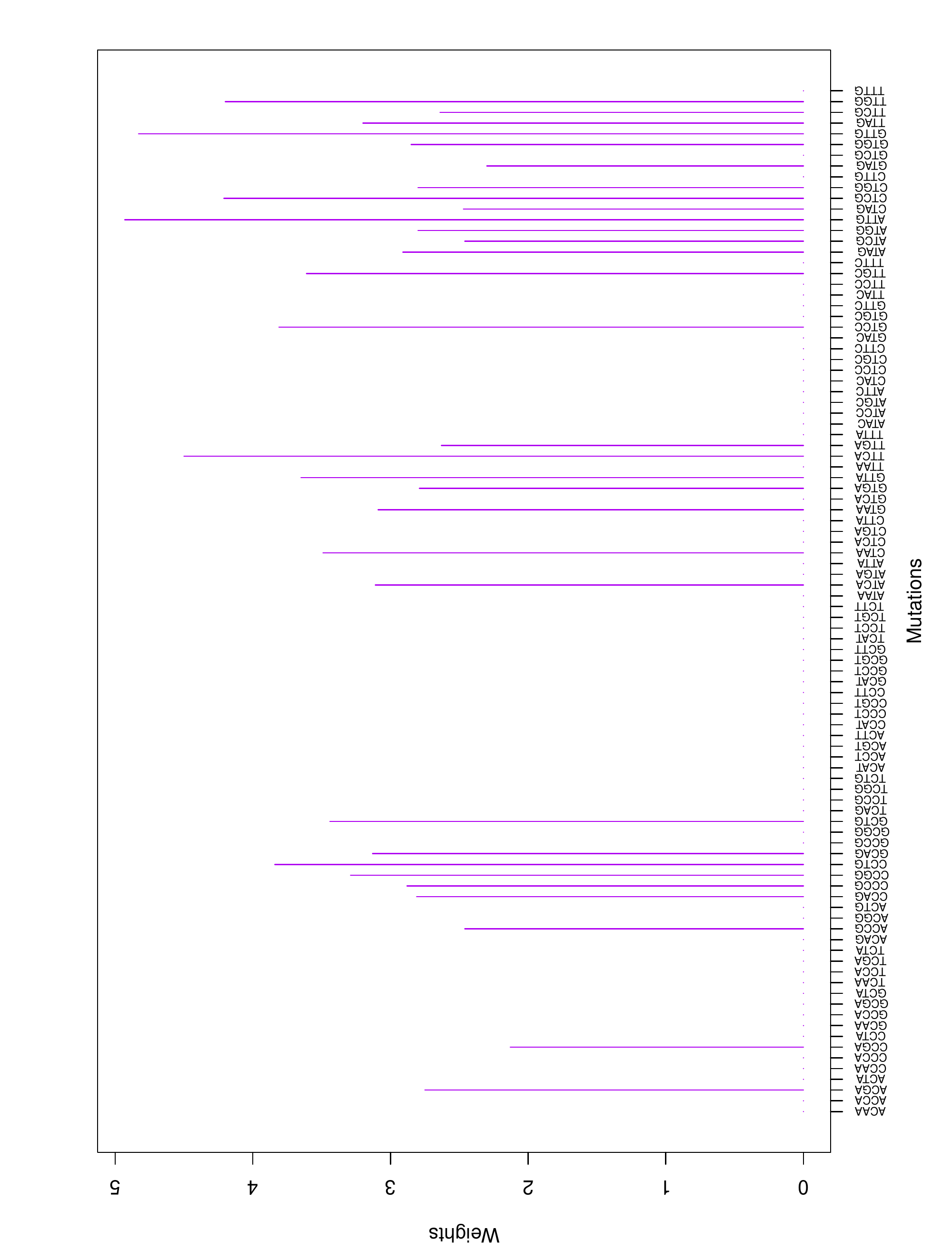}
\caption{Cluster Cl-6 in Clustering-D with weights based on unnormalized regressions with arithmetic means
(see Subsection \ref{sub.reg}).
See Tables \ref{table.occurrence.cts}, \ref{table.weights.D.1}, \ref{table.weights.D.2}.}
\label{Figure6D}
\end{figure}

\newpage\clearpage
\begin{figure}[ht]
\centering
\includegraphics[scale=0.7]{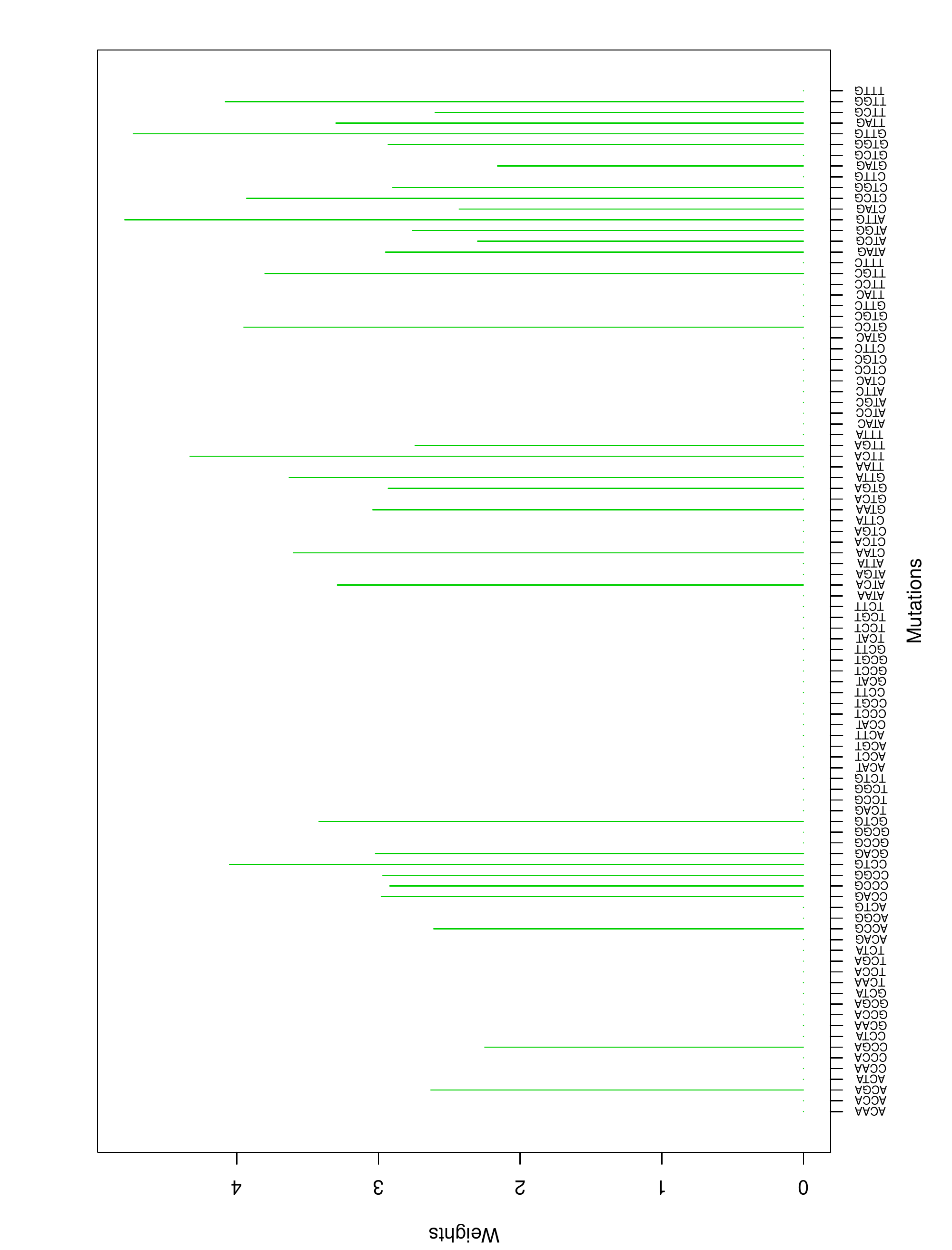}
\caption{Cluster Cl-6 in Clustering-D with weights based on normalized regressions with arithmetic means
(see Subsection \ref{sub.reg}).
See Tables \ref{table.occurrence.cts}, \ref{table.weights.D.1}, \ref{table.weights.D.2}.}
\label{FigureNorm6D}
\end{figure}

\newpage\clearpage
\begin{figure}[ht]
\centering
\includegraphics[scale=0.7]{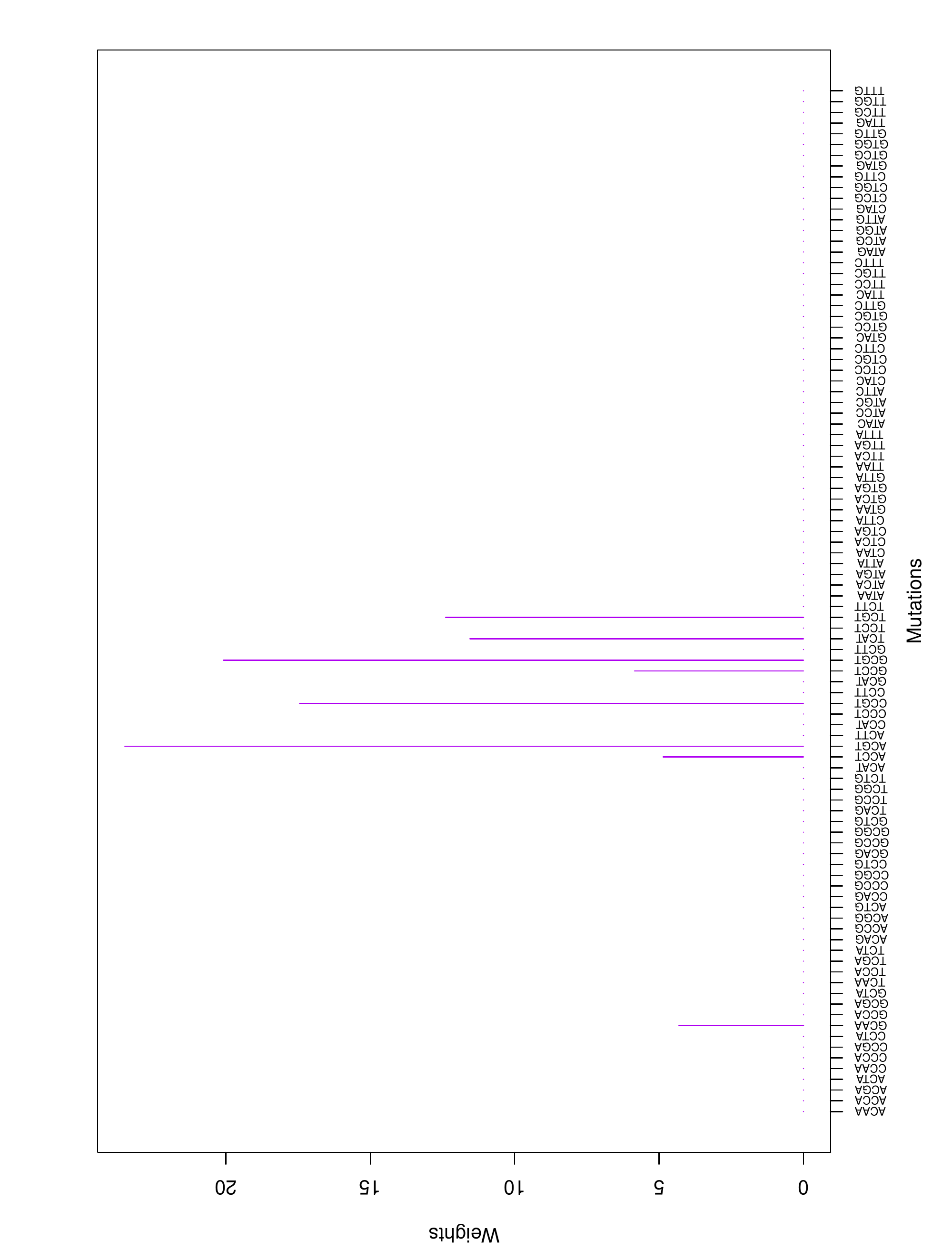}
\caption{Cluster Cl-1 in Clustering-A with weights based on unnormalized regressions with geometric means
(see Subsection \ref{sub.reg}).
See Tables \ref{table.occurrence.cts}, \ref{table.weights.Z.1}, \ref{table.weights.Z.2}.}
\label{FigureGeom1Z}
\end{figure}

\newpage\clearpage
\begin{figure}[ht]
\centering
\includegraphics[scale=0.7]{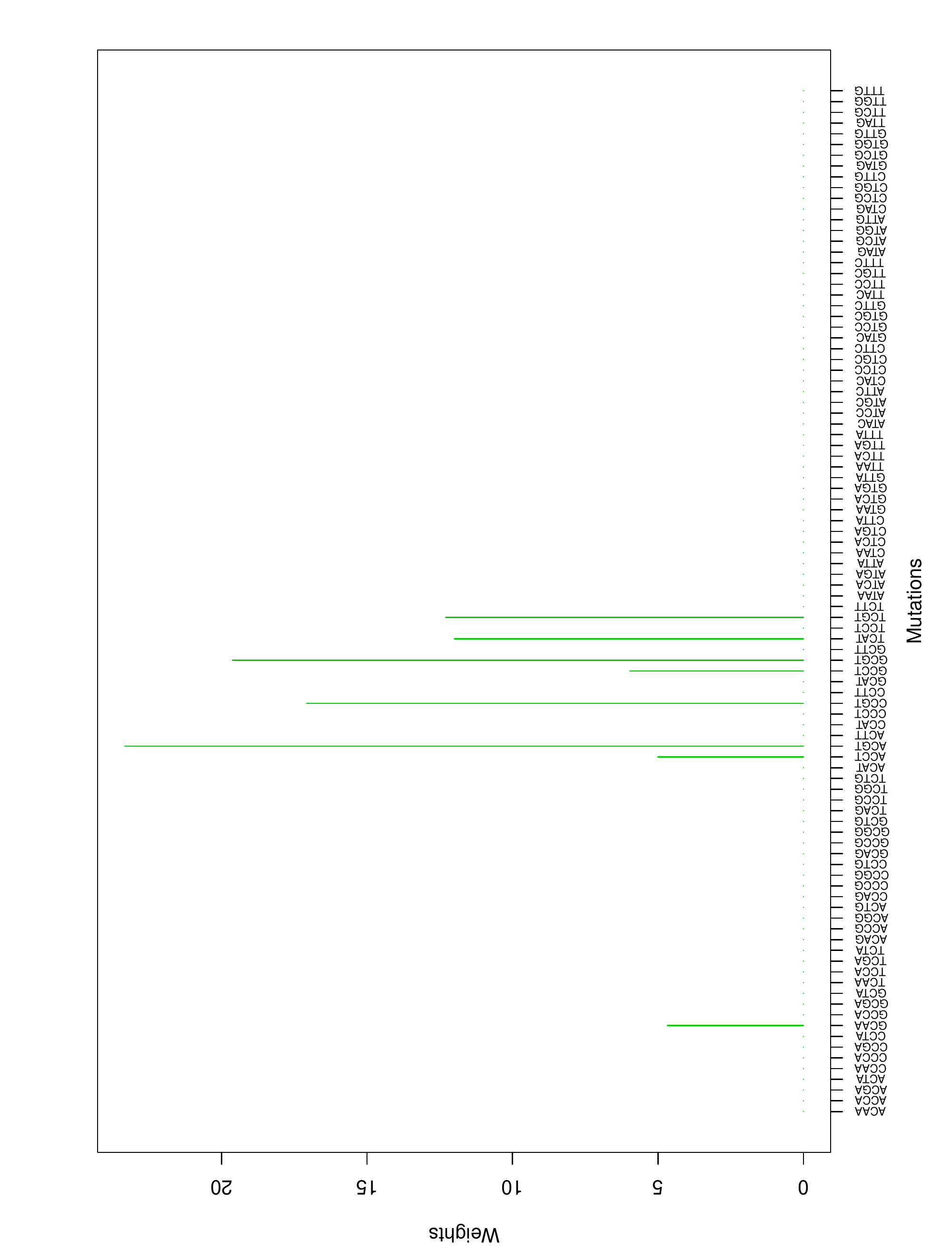}
\caption{Cluster Cl-1 in Clustering-A with weights based on normalized regressions with geometric means
(see Subsection \ref{sub.reg}).
See Tables \ref{table.occurrence.cts}, \ref{table.weights.Z.1}, \ref{table.weights.Z.2}.}
\label{FigureGeomNorm1Z}
\end{figure}

\newpage\clearpage
\begin{figure}[ht]
\centering
\includegraphics[scale=0.7]{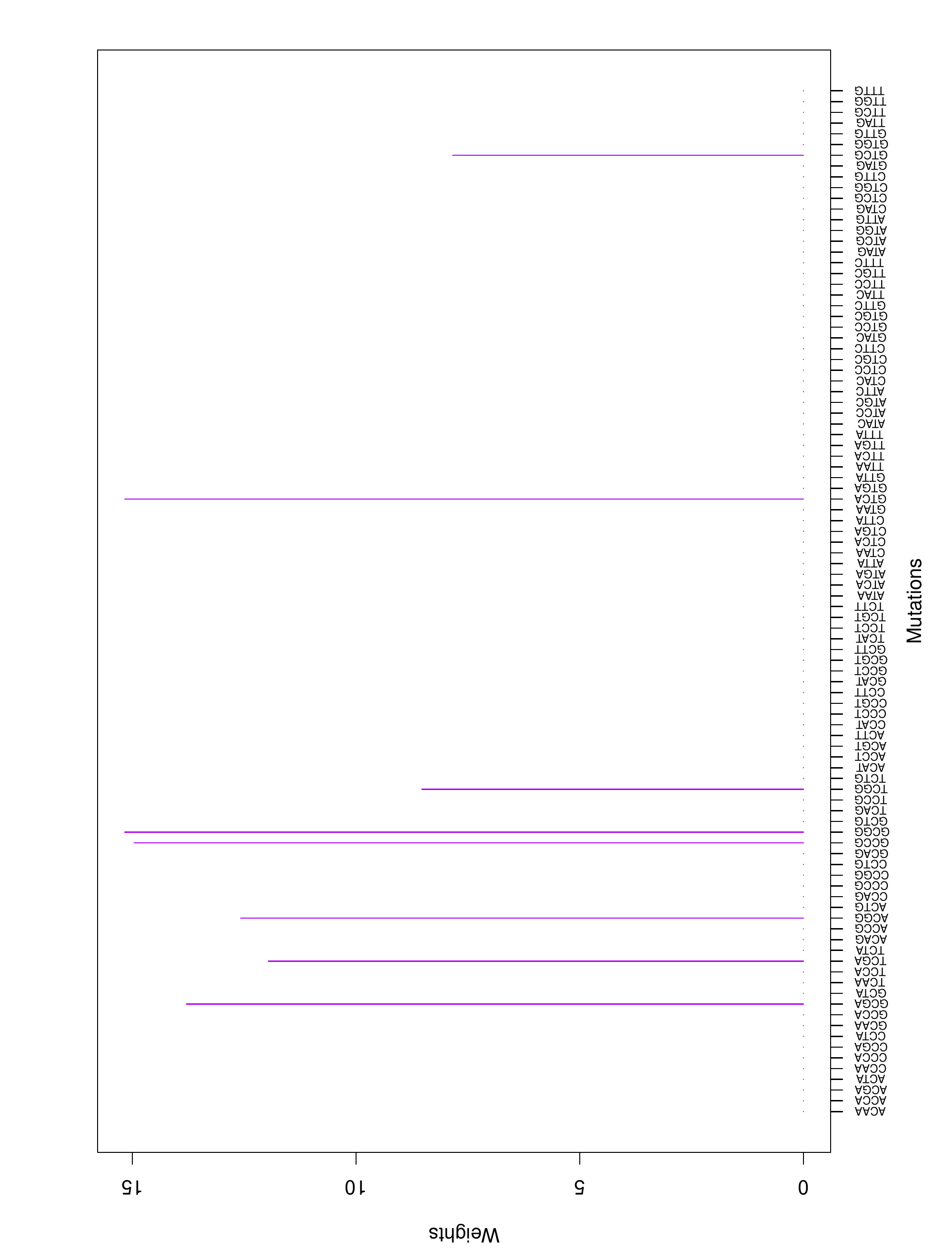}
\caption{Cluster Cl-2 in Clustering-A with weights based on unnormalized regressions with geometric means
(see Subsection \ref{sub.reg}).
See Tables \ref{table.occurrence.cts}, \ref{table.weights.Z.1}, \ref{table.weights.Z.2}.}
\label{FigureGeom2Z}
\end{figure}

\newpage\clearpage
\begin{figure}[ht]
\centering
\includegraphics[scale=0.7]{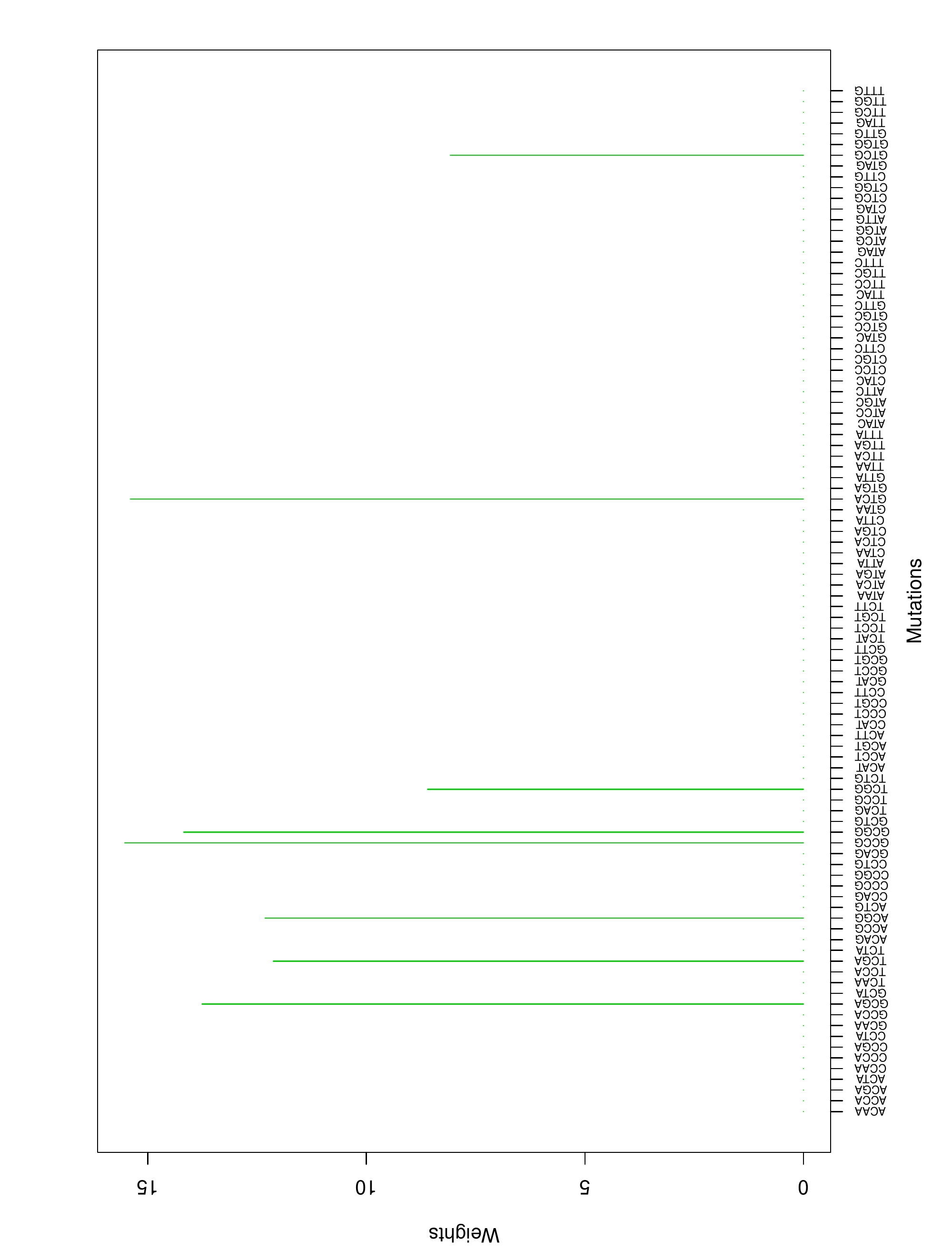}
\caption{Cluster Cl-2 in Clustering-A with weights based on normalized regressions with geometric means
(see Subsection \ref{sub.reg}).
See Tables \ref{table.occurrence.cts}, \ref{table.weights.Z.1}, \ref{table.weights.Z.2}.}
\label{FigureGeomNorm2Z}
\end{figure}

\newpage\clearpage
\begin{figure}[ht]
\centering
\includegraphics[scale=0.7]{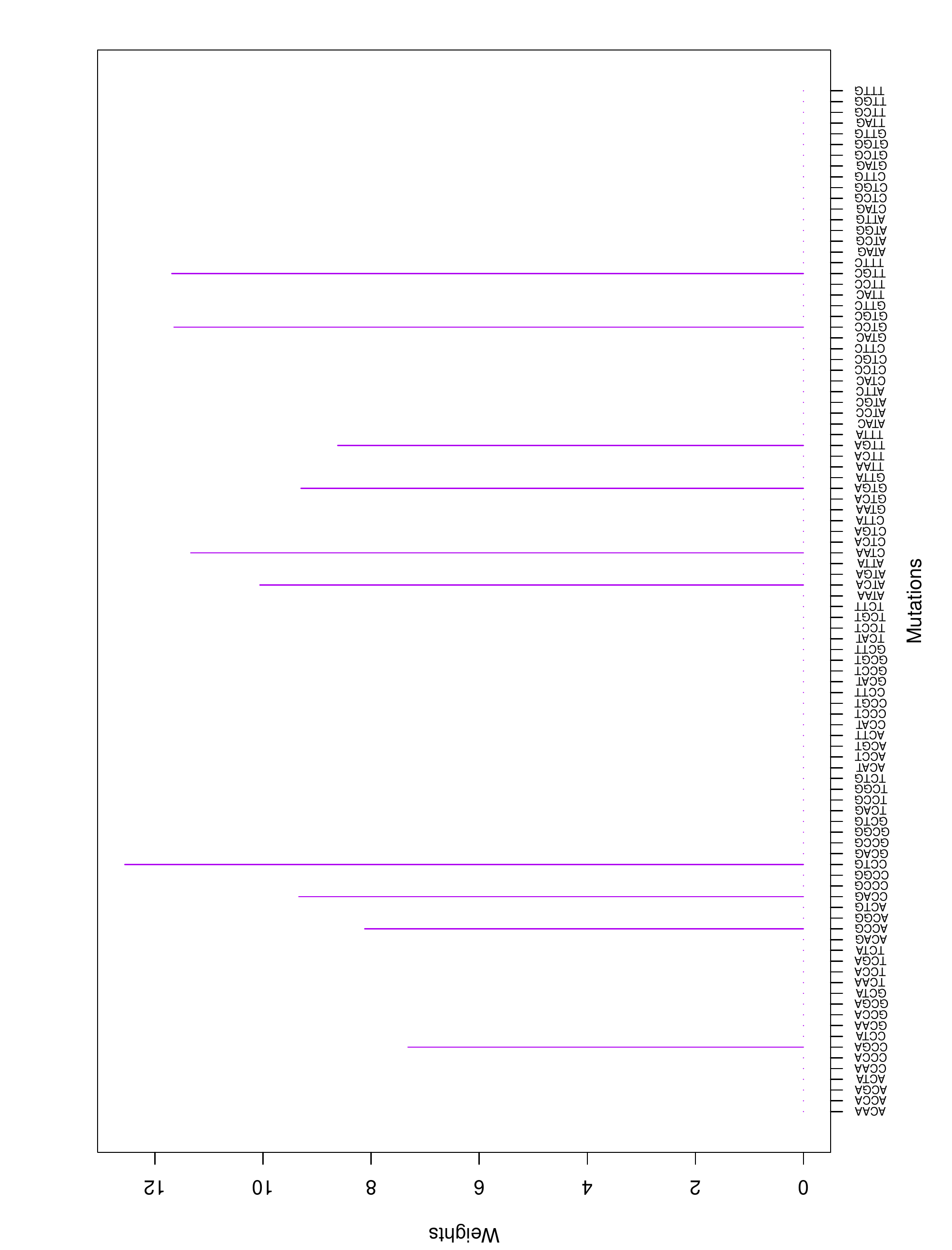}
\caption{Cluster Cl-3 in Clustering-A with weights based on unnormalized regressions with geometric means
(see Subsection \ref{sub.reg}).
See Tables \ref{table.occurrence.cts}, \ref{table.weights.Z.1}, \ref{table.weights.Z.2}.}
\label{FigureGeom3Z}
\end{figure}

\newpage\clearpage
\begin{figure}[ht]
\centering
\includegraphics[scale=0.7]{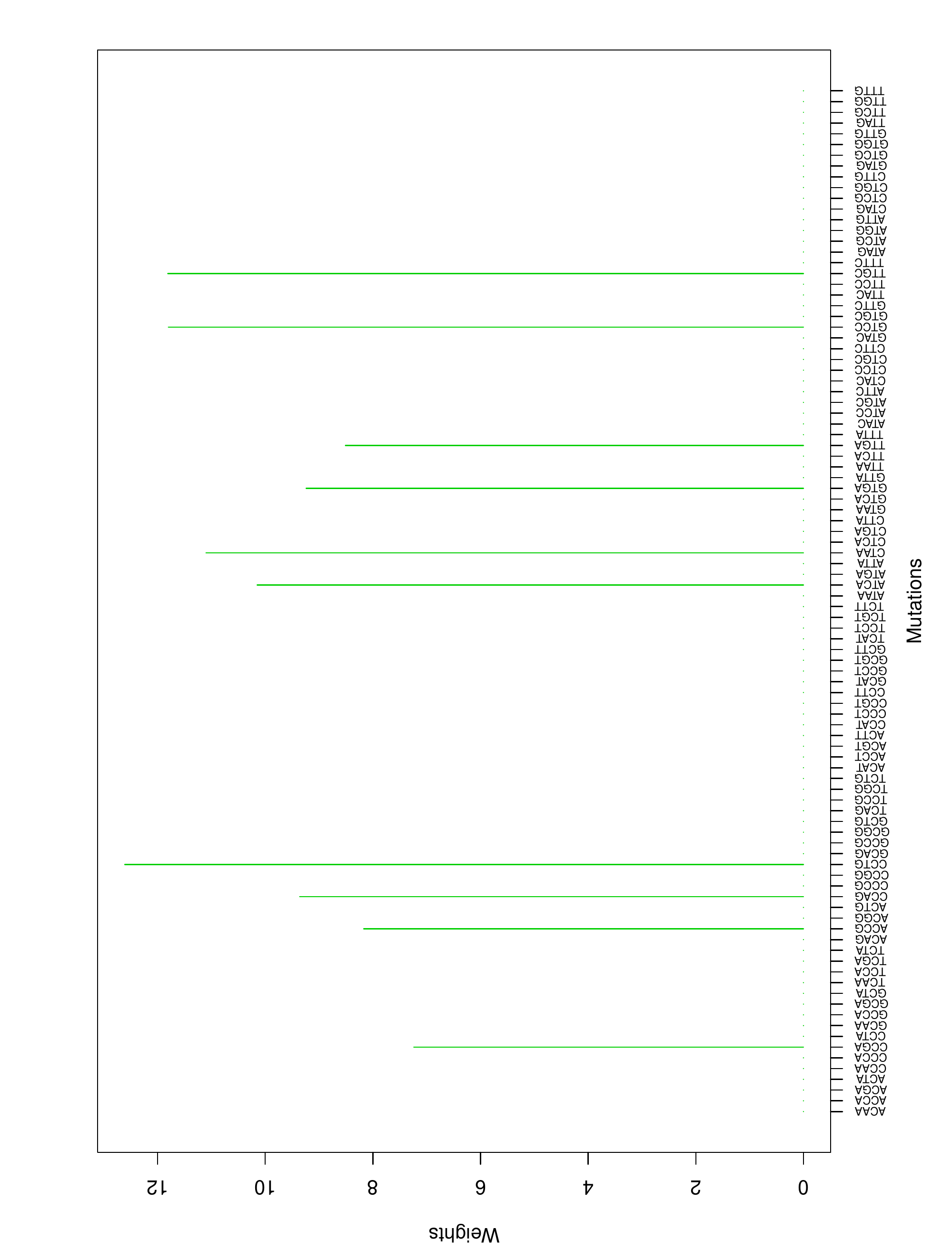}
\caption{Cluster Cl-3 in Clustering-A with weights based on normalized regressions with geometric means
(see Subsection \ref{sub.reg}).
See Tables \ref{table.occurrence.cts}, \ref{table.weights.Z.1}, \ref{table.weights.Z.2}.}
\label{FigureGeomNorm3Z}
\end{figure}

\newpage\clearpage
\begin{figure}[ht]
\centering
\includegraphics[scale=0.7]{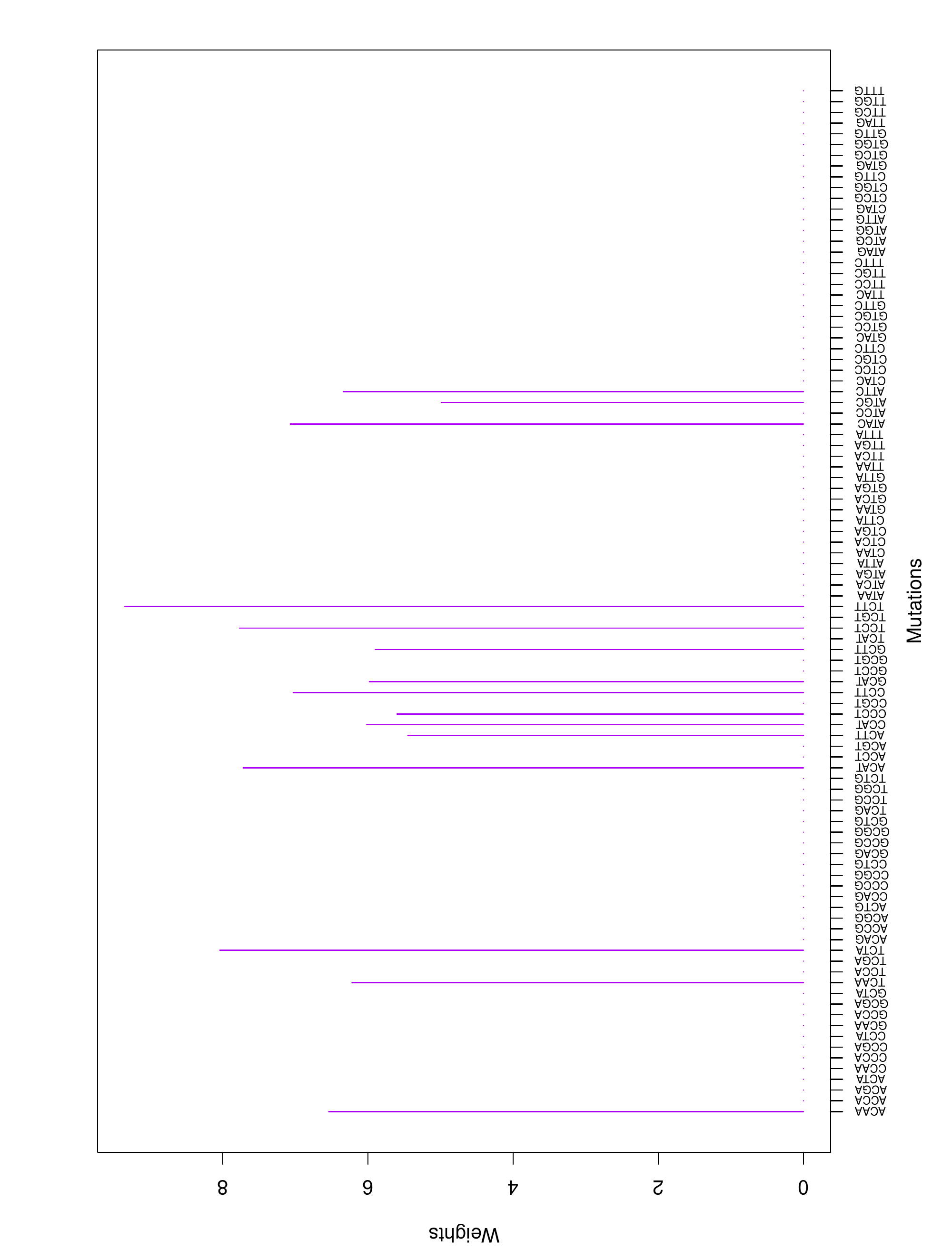}
\caption{Cluster Cl-4 in Clustering-A with weights based on unnormalized regressions with geometric means
(see Subsection \ref{sub.reg}).
See Tables \ref{table.occurrence.cts}, \ref{table.weights.Z.1}, \ref{table.weights.Z.2}.}
\label{FigureGeom4Z}
\end{figure}

\newpage\clearpage
\begin{figure}[ht]
\centering
\includegraphics[scale=0.7]{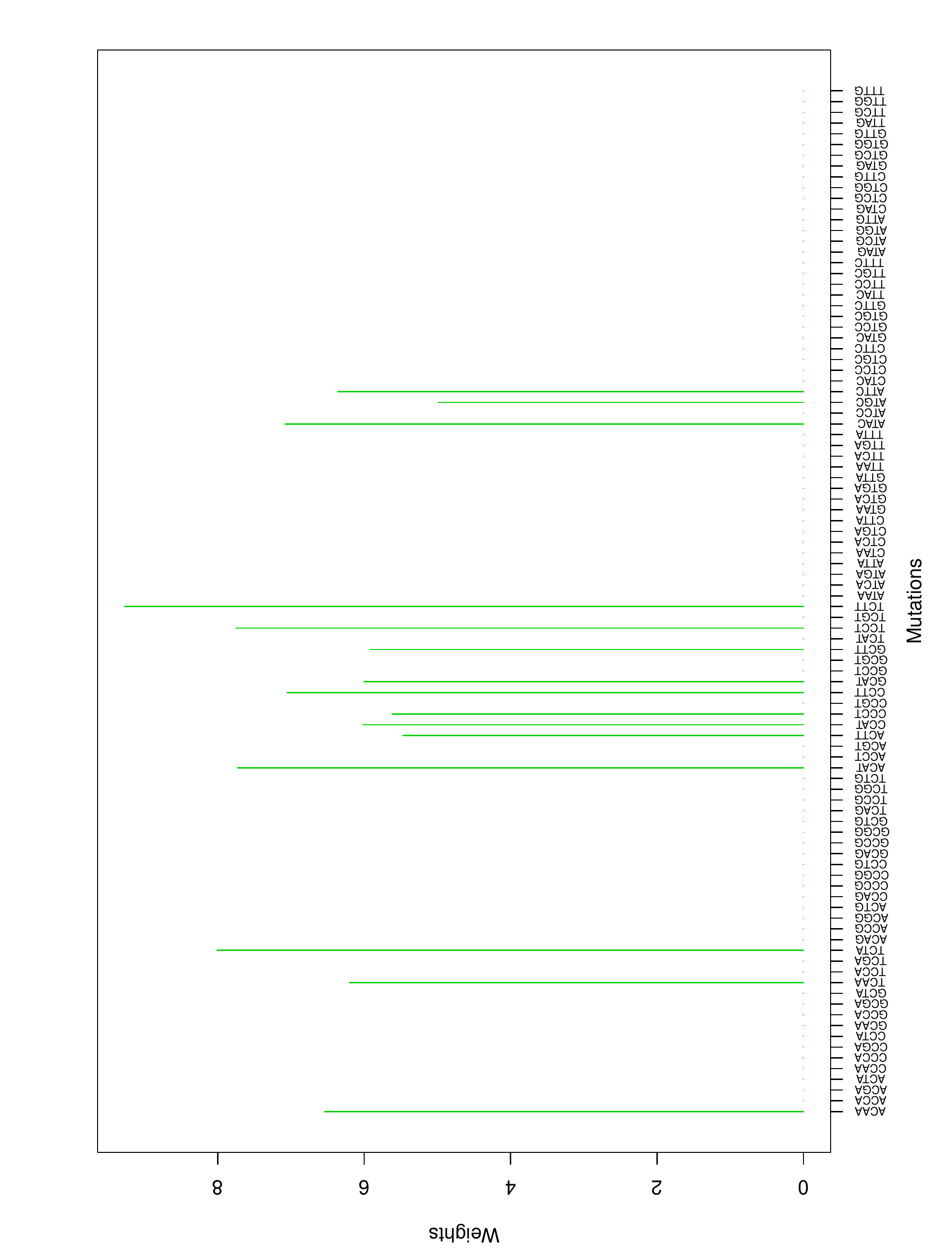}
\caption{Cluster Cl-4 in Clustering-A with weights based on normalized regressions with geometric means
(see Subsection \ref{sub.reg}).
See Tables \ref{table.occurrence.cts}, \ref{table.weights.Z.1}, \ref{table.weights.Z.2}.}
\label{FigureGeomNorm4Z}
\end{figure}

\newpage\clearpage
\begin{figure}[ht]
\centering
\includegraphics[scale=0.7]{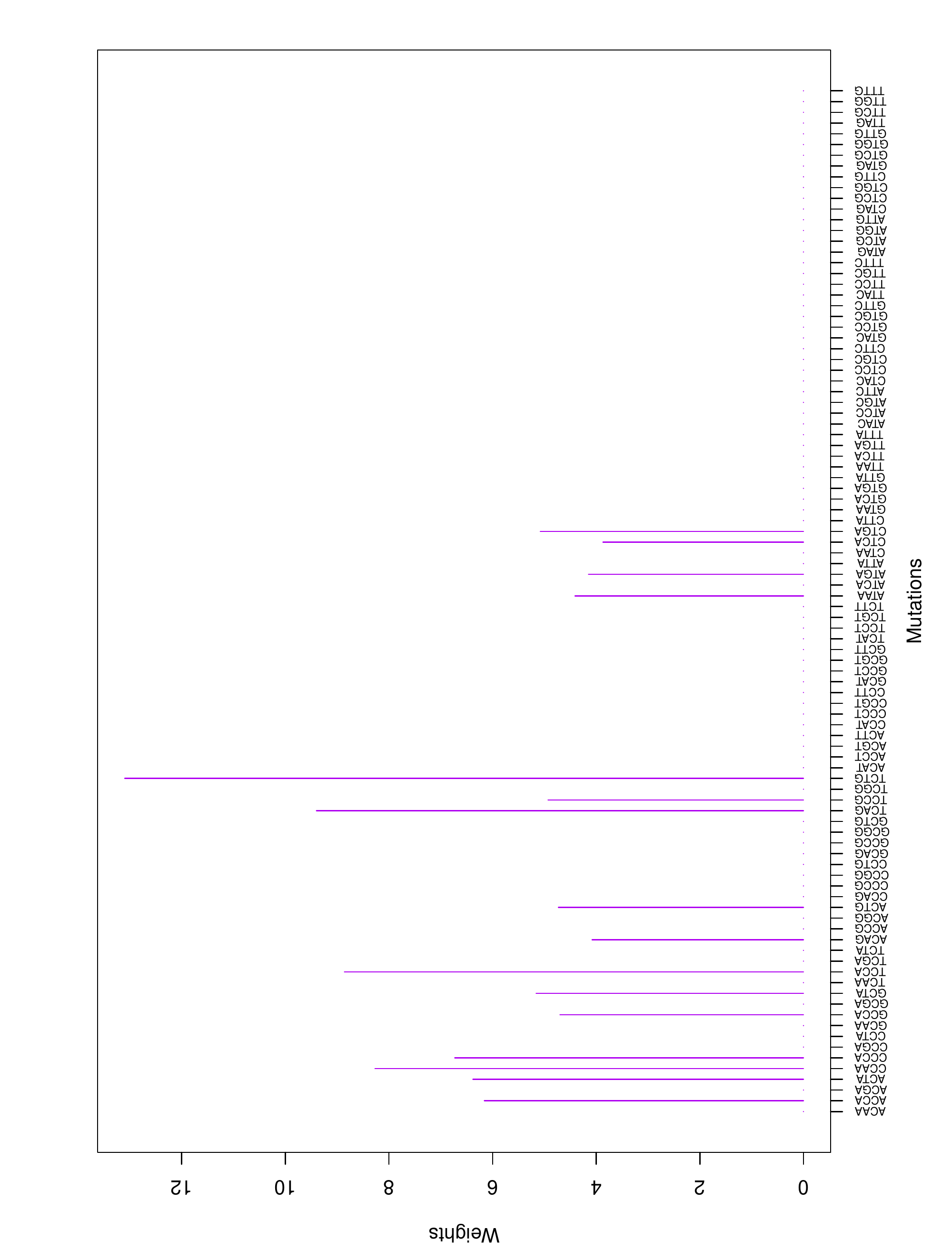}
\caption{Cluster Cl-5 in Clustering-A with weights based on unnormalized regressions with geometric means
(see Subsection \ref{sub.reg}).
See Tables \ref{table.occurrence.cts}, \ref{table.weights.Z.1}, \ref{table.weights.Z.2}.}
\label{FigureGeom5Z}
\end{figure}

\newpage\clearpage
\begin{figure}[ht]
\centering
\includegraphics[scale=0.7]{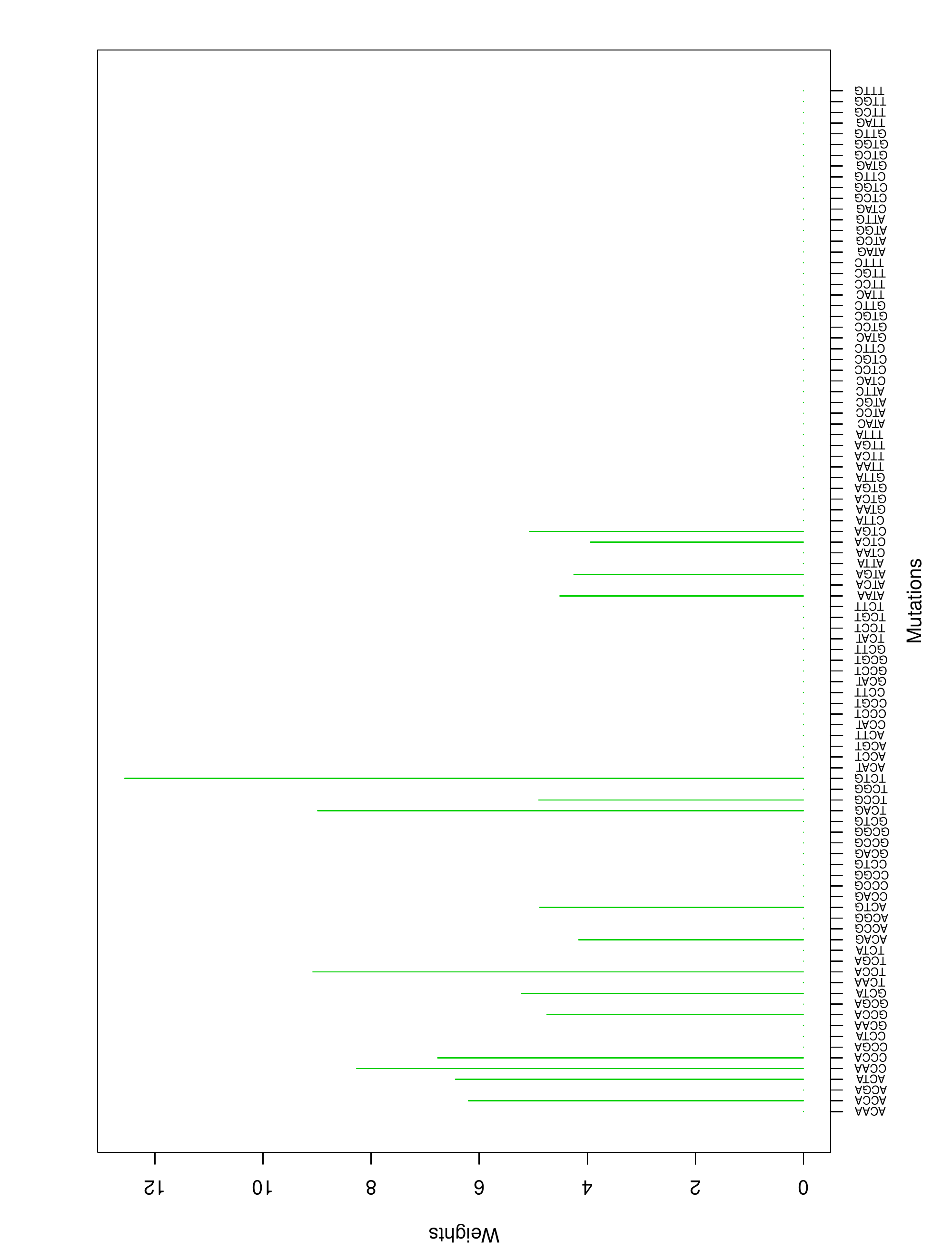}
\caption{Cluster Cl-5 in Clustering-A with weights based on normalized regressions with geometric means
(see Subsection \ref{sub.reg}).
See Tables \ref{table.occurrence.cts}, \ref{table.weights.Z.1}, \ref{table.weights.Z.2}.}
\label{FigureGeomNorm5Z}
\end{figure}

\newpage\clearpage
\begin{figure}[ht]
\centering
\includegraphics[scale=0.7]{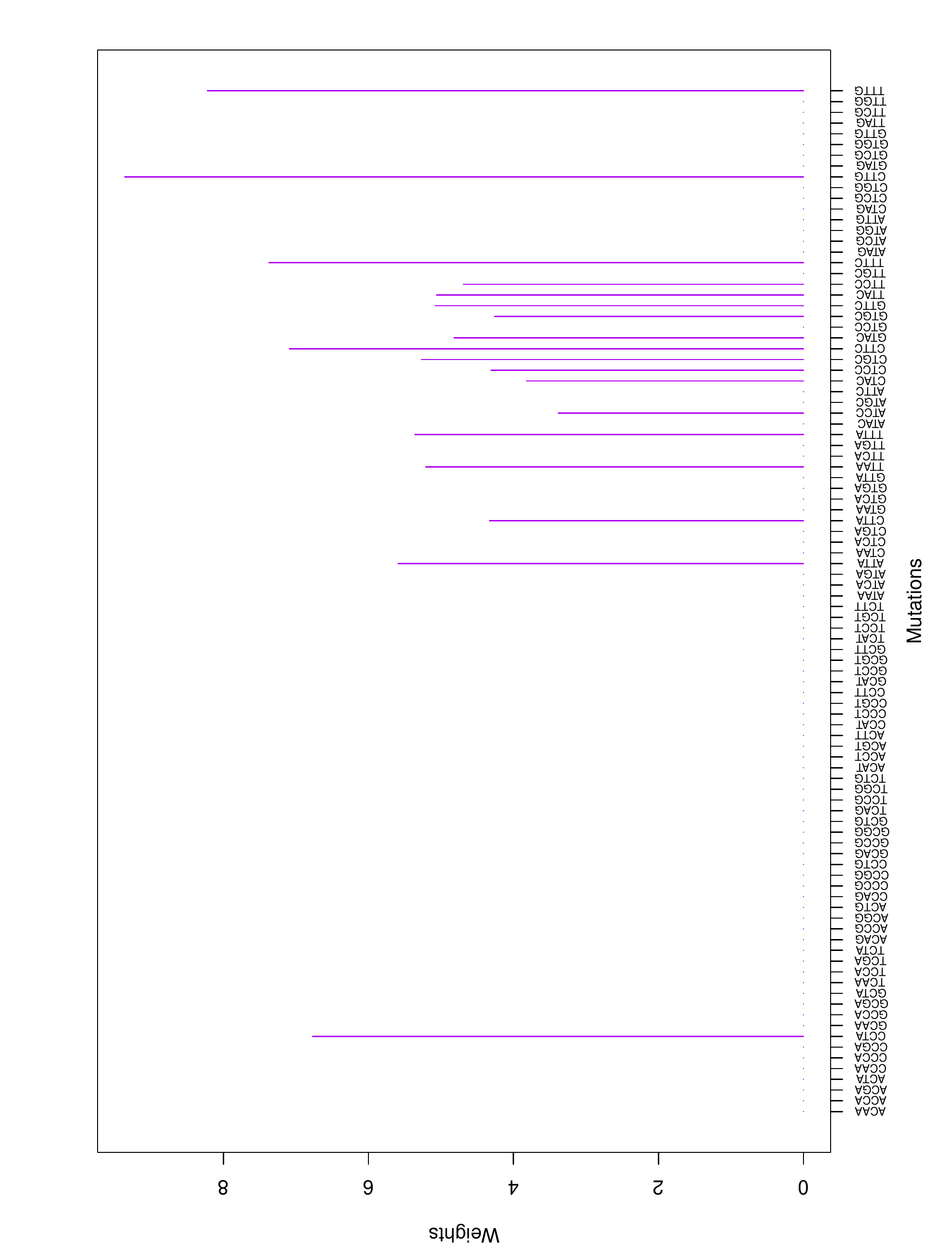}
\caption{Cluster Cl-6 in Clustering-A with weights based on unnormalized regressions with geometric means
(see Subsection \ref{sub.reg}).
See Tables \ref{table.occurrence.cts}, \ref{table.weights.Z.1}, \ref{table.weights.Z.2}.}
\label{FigureGeom6Z}
\end{figure}

\newpage\clearpage
\begin{figure}[ht]
\centering
\includegraphics[scale=0.7]{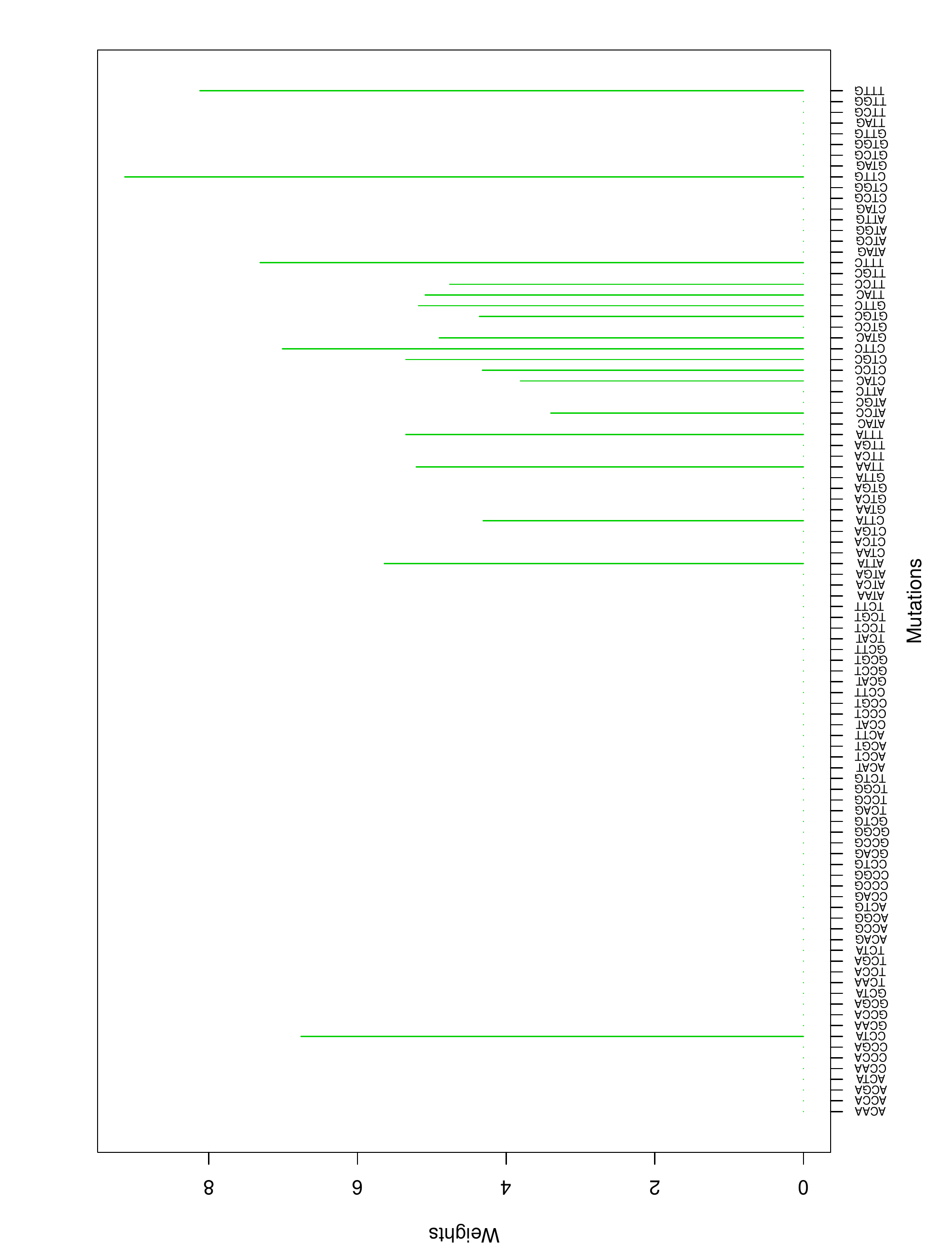}
\caption{Cluster Cl-6 in Clustering-A with weights based on normalized regressions with geometric means
(see Subsection \ref{sub.reg}).
See Tables \ref{table.occurrence.cts}, \ref{table.weights.Z.1}, \ref{table.weights.Z.2}.}
\label{FigureGeomNorm6Z}
\end{figure}

\newpage\clearpage
\begin{figure}[ht]
\centering
\includegraphics[scale=0.7]{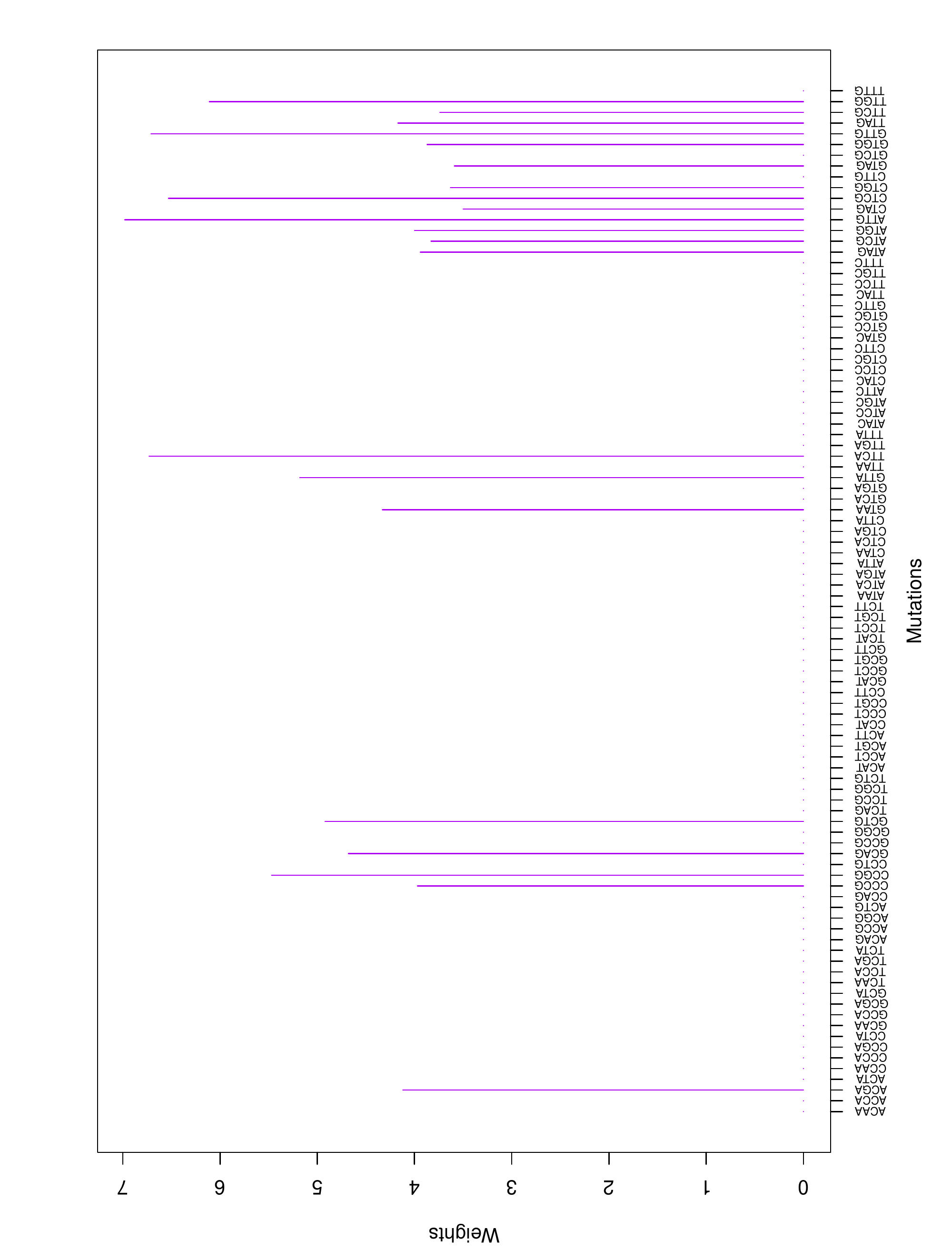}
\caption{Cluster Cl-7 in Clustering-A with weights based on unnormalized regressions with geometric means
(see Subsection \ref{sub.reg}).
See Tables \ref{table.occurrence.cts}, \ref{table.weights.Z.1}, \ref{table.weights.Z.2}.}
\label{FigureGeom7Z}
\end{figure}

\newpage\clearpage
\begin{figure}[ht]
\centering
\includegraphics[scale=0.7]{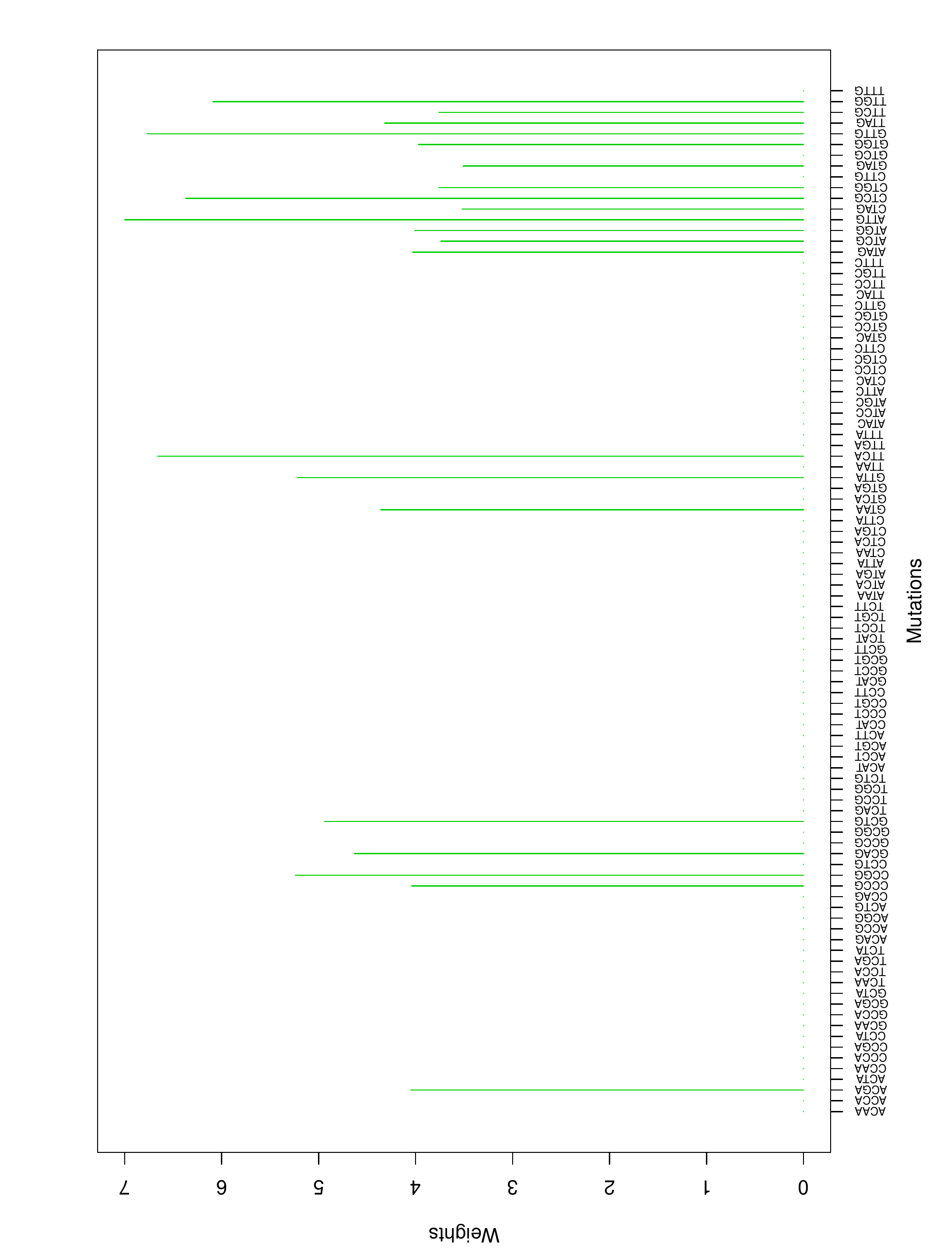}
\caption{Cluster Cl-7 in Clustering-A with weights based on normalized regressions with geometric means
(see Subsection \ref{sub.reg}).
See Tables \ref{table.occurrence.cts}, \ref{table.weights.Z.1}, \ref{table.weights.Z.2}.}
\label{FigureGeomNorm7Z}
\end{figure}

\end{document}